%% file: paper.tex
\documentclass[preprint]{elsarticle}
\usepackage[utf8]{inputenc}
\usepackage{makeidx}
\makeindex
\usepackage[T1]{fontenc}
\usepackage{epstopdf}
\usepackage{url}
\usepackage[font=footnotesize]{caption}
\usepackage{xcolor}
\usepackage{amsmath}
\usepackage{amssymb}
\usepackage{multirow}
\usepackage{eso-pic}
\usepackage{balance}
\usepackage{listings}
\usepackage{xspace}
\usepackage{subcaption}
\usepackage{tabularx,booktabs}
\usepackage{blindtext}
\usepackage{multicol}
\usepackage[normalem]{ulem}
\usepackage{listings}
\usepackage{comment}

\lstdefinelanguage{P4}
{
	morekeywords={
		action,
		actions,
		apply,
		bit,
		bool,
		const,
		control,
		default, else,
		enum, error, extern, exit,
		false, header, header\_union, if,
		in, inout, int, match\_kind,
		package, parser, ou,t return,
		select, state, struct, switch,
		table, transition, true, tuple,
		typedef, varbit, verify, void
	},
	sensitive=false,
	morecomment=[l]{//},
	morecomment=[s]{/*}{*/},
	morestring=[b]"
}
\usepackage{color}
\definecolor{eclipseBlue}{RGB}{42,0.0,255}
\definecolor{eclipseGreen}{RGB}{63,127,95}
\definecolor{eclipsePurple}{RGB}{127,0,85}

\newcolumntype{Y}{>{\centering\arraybackslash}X}

\newcommand\fig[1]{Figure~\ref{fig:#1}}
\newcommand\sect[1]{Section~\ref{sec:#1}}
\newcommand\tbl[1]{Table~\ref{tbl:#1}}
\newcommand\pfour[0]{P4\textsubscript{14}\xspace}
\newcommand\psix[0]{P4\textsubscript{16}\xspace}
\newcommand\tapas[0]{T\textsubscript{4}P\textsubscript{4}S\xspace}

\newcommand\totalReferences[0]{519\xspace}
\newcommand\totalPapers[0]{377\xspace}
\newcommand\totalApplications[0]{245\xspace}

\newcommand{\paperTable}[4]{#3 \cite{#1} & #2 & #4 & \\}
\newcommand{\paperTableSource}[5]{#3 \cite{#1} & #2 & #4 & \cite{#5} \\}
\newcommand{\venueStat}[2]{#1 & #2 \\}

\usepackage{siunitx}
\DeclareSIUnit{\Bit}{Bit}
\DeclareSIUnit{\Byte}{Byte}

\usepackage[printonlyused]{acronym}

\graphicspath{{figures/}}

\journal{JNCA}

\begin{document}

\begin{frontmatter}

\title{A Survey on Data Plane Programming with P4: Fundamentals, Advances, and Applied Research}

\author[tu]{Frederik Hauser}
\ead{frederik.hauser@uni-tuebingen.de}
\author[tu]{Marco~Häberle}
\ead{marco.haeberle@uni-tuebingen.de}
\author[tu]{Daniel Merling}
\ead{daniel.merling@uni-tuebingen.de}
\author[tu]{Steffen Lindner}
\ead{steffen.lindner@uni-tuebingen.de}
\author[bf]{Vladimir~Gurevich}
\ead{vladimir.gurevich@intel.com}
\author[si]{Florian Zeiger}
\ead{florian.zeiger@siemens.com}
\author[si]{Reinhard Frank}
\ead{reinhard.frank@siemens.com}
\author[tu]{Michael Menth}
\ead{menth@uni-tuebingen.de}
\address[tu]{University of Tuebingen, Department of Computer Science, Chair of 
	Communication Networks, Tuebingen, Germany}
\address[bf]{Intel, Barefoot Division (BXD), United States of America}
\address[si]{Siemens AG, Corporate Technology, Munich, Germany}

\begin{abstract}
		\input{chapters/00-abstract}
\end{abstract}


\end{frontmatter}

\input{chapters/01_introduction}
\input{chapters/02_network_programmability}

\input{chapters/03_p4_programming}
\input{chapters/04_p4_deployment}
\input{chapters/05_p4_targets}
\input{chapters/06_p4_control_plane}
\input{chapters/07_p4_further_development}

\input{chapters/08-use-cases}
\input{chapters/09-uc_monitoring}
\input{chapters/10-uc_traffic_management}
\input{chapters/11-uc_forwarding_routing}
\input{chapters/12-uc_advanced_networking}
\input{chapters/13-uc_network-security}
\input{chapters/14-uc_others}

\input{chapters/15-discussion.tex}
\input{chapters/16-conclusion.tex}
\input{chapters/17-acknowledgement}

\bibliographystyle{elsarticle-num}
\bibliography{literature}

\end{document}

%% file: chapters/00-abstract.tex
Programmable data planes allow users to define their own data plane algorithms for network devices including appropriate data plane \acp{API} which may be leveraged by user-defined \ac{SDN} control.
This offers great flexibility for network customization, be it for specialized, commercial appliances, e.g., in 5G or data center networks, or for rapid prototyping in industrial and academic research. 
Programming protocol-independent packet processors (P4) has emerged as the currently most widespread abstraction, programming language, and concept for data plane programming.
It is developed and standardized by an open community, and it is supported by various software and hardware platforms.

In the first part of this paper we give a tutorial of data plane programming models, the P4 programming language, architectures, compilers, targets, and data plane \acsp{API}.
We also consider research efforts to advance P4 technology.
In the second part, we categorize a large body of literature of P4-based applied research into different research domains, summarize the contributions of these papers, and extract prototypes, target platforms, and source code availability.
For each research domain, we analyze how the reviewed works benefit from P4's core features.
Finally, we discuss potential next steps based on our findings.

%% file: chapters/01_introduction.tex
\section{Introduction}
\label{sec:introduction}

Traditional networking devices such as routers and switches process packets using data and control plane algorithms.
Users can configure control plane features and protocols, e.g., via \acsp{CLI}, web interfaces, or management \acsp{API}, but the underlying algorithms can be changed only by the vendor.
This limitation has been broken up by \ac{SDN} and even more by data plane programming.

\ac{SDN} makes network devices programmable by introducing an \acs{API} that allows users to bypass the built-in control plane algorithms and to replace them with self-defined algorithms.
Those algorithms are expressed in software and typically run on an \ac{SDN} controller with an overall view of the network.
Thereby, complex control plane algorithms designed for distributed control can be replaced by simpler algorithms designed for centralized control.
This is beneficial for use cases that are demanding with regard to flexibility, efficiency and security, e.g., massive data centers or 5G networks.

Programmable data planes enable users to implement their own data plane algorithms on forwarding devices.
Users, e.g., programmers, practitioners, or operators, may define new protocol headers and forwarding behavior, which is without programmable data planes only possible for a vendor.
They may also add data plane \acp{API} for \ac{SDN} control.

Data plane programming changes the power of the users as they can build custom network equipment without any compromise in performance, scalability, speed, or power on appropriate platforms.
There are different data plane programming models, each with many implementations and programming languages.
Examples are Click \cite{KoMo00}, VPP \cite{VPP}, NPL \cite{Broadcom}, and SDNet \cite{Xilinx}.

Programming protocol-independent packet processors (P4) is currently the most widespread abstraction, programming language, and concept for data plane programming.
First published as a research paper in 2014 \cite{BoDa14}, it is now developed and standardized in the P4 Language Consortium, it is supported by various software- and hardware-based target platforms, and it is widely applied in academia and industry.

In the following, we clarify the contribution of this survey, point out its novelty, explain its organization, and provide a table with acronyms frequently used in this work.

\subsection{Contributions}
This survey pursues two objectives.
First, it provides a comprehensive introduction and overview of P4.
Second, it surveys publications describing applied research based on P4 technology.
Its main contributions are the following: 

\begin{itemize}
    \item We explain the evolution of data plane programming with P4, relate it to prior developments such as \ac{SDN}, and compare it to other data plane programming models.
    \item We give an overview of data plane programming with P4. It comprises the P4 programming language, architectures, compilers, targets, and data plane \acsp{API}. These sections do not only include foundations but also present related work on advancements, extensions, or experiences.
    \item We summarize research efforts to advance P4 data planes. It comprises optimization of development and deployment, testing and debugging, research on P4 targets, and advances on control plane operation.
    \item We analyze a large body of literature considering P4-based applied research. We categorize \totalApplications research papers into different application domains, summarize their key contributions, and characterize them with respect to prototypes, target platforms, and source code availability. For each research domain, we analyze how the reviewed works benefit from P4's core features.
\end{itemize}

We consider publications on P4 that were published until the end of 2020 and selected paper from 2021.
Beside journal, conference, and workshop papers, we also include contents from standards, websites, and source code repositories.
The paper comprises \totalReferences references out of which \totalPapers are scientific publications: 73 are from 2017 and before, 66 from 2018, 113 from 2019, 116 from 2020, and 9 from 2021.

\subsection{Novelty}
There are numerous surveys on \ac{SDN} published in 2014 \cite{NuMe14,JaMa14}, 2015 \cite{XiWe15,MaGu15,KrRa15}, and 2016 \cite{MaGh16,TrFa16} as well as surveys on \ac{OF} from 2014 \cite{BrMe14,HuHa14,LaKo14}. Only one of them \cite{TrFa16} mentions P4 in a single sentence.
Two surveys of data plane programming from 2015 \cite{KrRa15,MaGu15} were published shortly after the release of P4, one conference paper from 2018 \cite{BiRe18} and a survey from 2019 \cite{KaMa19} present P4 just as one among other data plane programming languages.
Likewise, Michel et al. \cite{MiBi21} gives an overview of data plane programming in general and P4 is one among other examined abstractions and programming languages.
Our survey is dedicated to P4 only.
It covers more details of P4 and a many more papers of P4-based applied research which have mostly emerged only within the last two years.

A recent survey focusing on P4 data plane programming has been published in \cite{KaKu21}.
The authors introduce data plane programming with P4, review 33 research works from four research domains, and discuss research issues.
Another recent technical report \cite{KfCr21} reviews 150 research papers from seven research domains.
While typical research areas of P4 are covered, others (e.g., industrial networking, novel routing and forwarding schemes, and time-sensitive networking) are not part of the literature review.
The different aspects of P4, e.g., the programming language, architectures, compilers, targets, data plane \acp{API}, and their advancements are not treated in the paper.
In contrast to both surveys on P4, we cover a greater level of detail of P4 technology and their advancements, and our literature review is more comprehensive.
\color{black}

\subsection{Paper Organization}
\begin{figure}[ht]
  \centering
  \includegraphics[width=.8\linewidth]{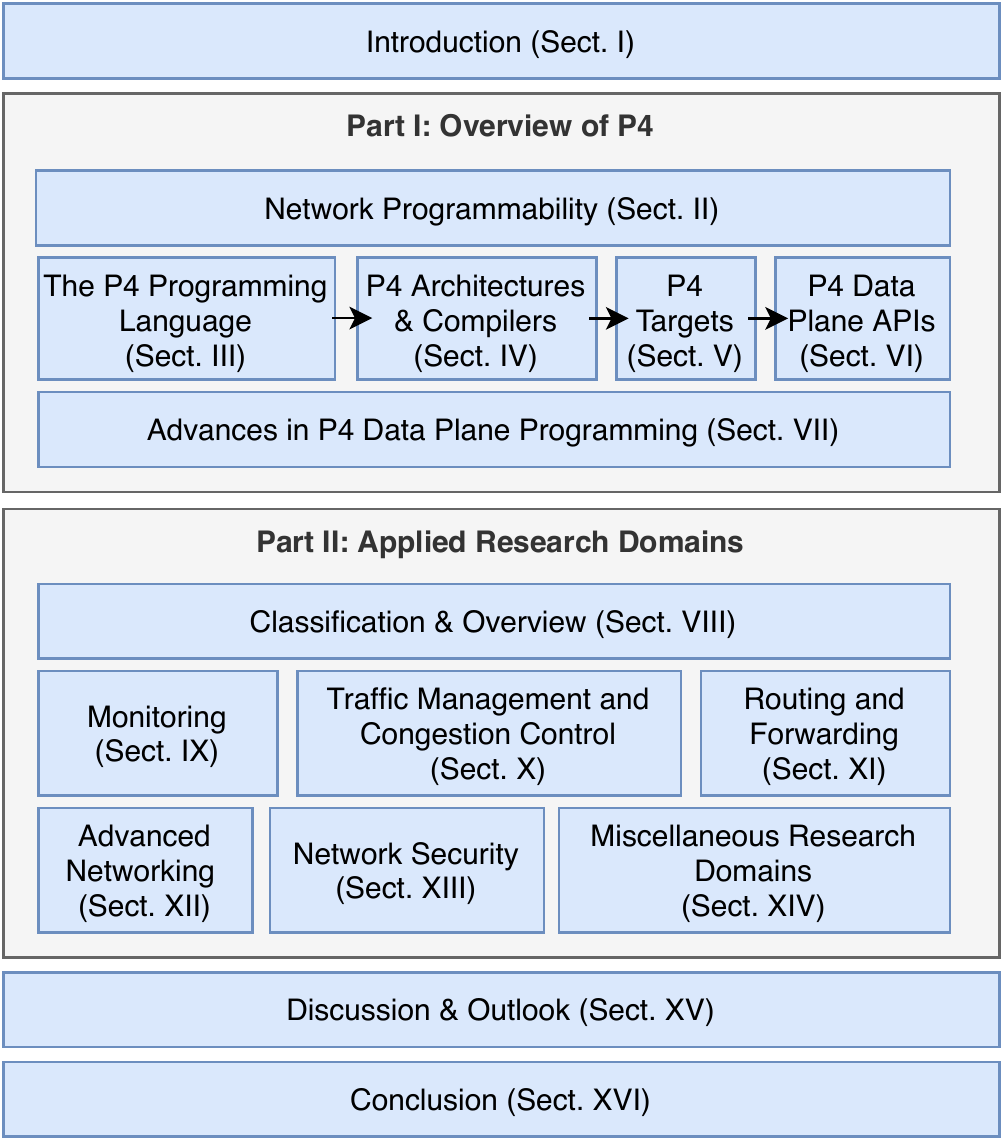}
  \caption{Organization of the paper.}
  \label{fig:paper-organization}
\end{figure}

\fig{paper-organization} depicts the structure of this paper which is divided into two main parts: an \emph{overview of P4} and a \emph{survey of research publications}. 

In the first part,
\sect{network-programmability} gives an introduction to network programmability.
We describe the development from traditional networking and \ac{SDN} to data plane programming and present the two most common data plane programming models.
In \sect{p4-programming}, we give a technology-oriented tutorial of P4 based on its latest version \psix.
We introduce the P4 programming language and describe how user-provided P4 programs are compiled and executed on P4 targets.
\sect{p4-architectures-compilers} presents the concept of P4 architectures as intermediate layer between the P4 programs and the targets.
We introduce the four most common architectures in detail and describe P4 compilers.
In \sect{p4-targets}, we categorize and present platforms that execute P4 programs, so-called P4 targets that are based on software, \acsp{FPGA}, \acsp{ASIC}, or \acsp{NPU}.
\sect{p4-data-plane-apis} gives an introduction to data plane \acsp{API}.
We describe their functions, present a characterization, introduce the four main P4 data plane \acsp{API} that serve as interfaces for \ac{SDN} controllers, and point out controller use case patterns.
In \sect{development-debugging-testing}, we summarize research efforts that aim to improve P4 data plane programming.

The second part of the paper surveys P4-based applied research in communication networks.
In \sect{use-cases}, we classify core features of P4 that make it attractive for the implementation of data plane algorithms.
We use these properties in later sections to effectively reason about P4's value for the implementation of various prototypes.
We present an overview of the research domains and compile statistics about the included publications.
The super-ordinate research domains are monitoring (\sect{use-cases-monitoring}), traffic management and congestion control (\sect{use-cases-traffic-management}), routing and forwarding (\sect{use-cases-routing}), advanced networking (\sect{use-cases-advanced-networking}), network security (\sect{use-cases-network-security}), and miscellaneous (\sect{uses-cases-others}) to cover additional, different topics.
Each category includes a table to give a quick overview of the analyzed papers with regard to prototype implementations, target platforms, and source code availability.
At the end of each section, we analyze how the reviewed works benefit from P4's core features.

In \sect{discussion} we discuss insights from this survey and give an outlook on potential next steps.
\sect{conclusion} concludes this work.

\subsection{List of Acronyms}
The following acronyms are used in this paper.
\input{acronyms.tex}

%% file: acronyms.tex
\begin{acronym}[SMGW-PP]

\acro{ACL}{access control list} 
\acro{ALU}{arithmetic logic unit}
\acro{API}{application programming interface}
\acro{AQM}{active queue management}
\acro{ASIC}{application-specific integrated circuit}
\acro{AWW}{adjusting advertised windows}
\acro{bmv2}{Behavioral Model version 2}
\acro{BGP}{Border Gateway Protocol}
\acro{BPF}{Berkeley Packet Filter}
\acro{CLI}{command line interface}
\acro{DAG}{directed acyclic graph}
\acro{DDoS}{distributed denial of service}
\acro{DPI}{deep packet inspection}
\acro{DPDK}{Data Plane Development Kit}
\acro{DSL}{domain-specific language}
\acro{eBPF}{Extended Berkeley Packet Filter}
\acro{ECN}{Explicit Congestion Notification}
\acro{FPGA}{field programmable gate array}
\acro{FSM}{finite state machine}
\acro{GTP}{GPRS tunneling protocol}
\acro{HDL}{hardware description language}
\acro{HLIR}{high-level intermediate representation}
\acro{IDE}{integrated development environment}
\acro{IDL}{Intent Definition Language}
\acro{IDS}{intrusion detection system}
\acro{INT}{in-band network telemetry}
\acro{LDWG}{Language Design Working Group}
\acro{LPM}{longest prefix matching}
\acro{LUT}{look up table}
\acro{MAT}{match-action-table}
\acro{ML}{machine learning}
\acro{NDN}{named data networking}
\acro{NF}{network function}
\acro{NFP}{network flow processing}
\acro{NFV}{network function virtualization}
\acro{NIC}{network interface card}
\acro{NPU}{network processing unit}
\acro{ODM}{original design manufacturer}
\acro{ODP}{Open Data Plane}
\acro{OEM}{original equipment manufacturer}
\acro{OF}{OpenFlow}
\acro{ONF}{Open Networking Foundation}
\acro{ONL}{Open Network Linux}
\acro{OVS}{Open vSwitch}
\acro{PISA}{Protocol Independent Switching Architecture}
\acro{PSA}{Portable Switch Architecture}
\acro{REG}{register}
\acro{RPC}{remote procedure call}
\acro{RTL}{register-transfer level}
\acro{SDK}{software development kit}
\acro{SDN}{software-defined networking}
\acro{SF}{service function}
\acro{SFC}{service function chain}
\acro{SRAM}{static random-access memory}
\acro{TCAM}{ternary content-addressable memory}
\acro{TSN}{Time-Sensitive Networking}
\acro{TNA}{Tofino Native Architecture}
\acro{uBPF}{user-space BPF}
\acro{VM}{virtual machine}
\acro{VNF}{virtual network function}
\acro{VPP}{Vector Packet Processors}
\acro{WG}{working group}
\acro{XDP}{eXpress Data Path}

\setlength{\parskip}{0ex}
\setlength{\itemsep}{0ex}

\end{acronym}

%% file: chapters/02_network_programmability.tex
\section{Network Programmability}
\label{sec:network-programmability}

In this section, we first define the notion of network programmability and related terms.
Then, we discuss control plane programmability and data plane programming, elaborate on data plane programming models, and point out the benefits of data plane programming.

\subsection{Definition of Terms}
We define \emph{programmability} as the ability of the software or the hardware to execute an externally defined processing algorithm.
This ability separates programmable entities from \emph{flexible} (or \emph{configurable}) ones; the latter only allow changing different parameters of the internally defined algorithm which stays the same.

Thus, the term \emph{network programmability} means the ability to define the processing algorithm executed in a network and specifically in individual processing nodes, such as switches, routers, load balancers, etc.
It is usually assumed that no special processing happens in the links connecting network nodes.
If necessary, such processing can be described as if it takes place on the nodes that are the endpoints of the links or by adding a "bump-in-the-wire" node with one input and one output. 

Traditionally, the algorithms, executed by telecommunication devices, are split into three distinct classes: the data plane, the control plane, and the management plane.
Out of these three classes, the management plane algorithms have the smallest effect on both the overall packet processing and network behavior.
Moreover, they have been programmable for decades, e.g., SNMPv1 was standardized in 1988 and created even earlier than that.
Therefore, management plane algorithms will not be further discussed in this section.

True network programmability implies the ability to specify and change both the control plane and data plane algorithms.
In practice this means the ability of network operators (users) to define both data and control plane algorithms on their own, without the need to involve the original designers of the network equipment.
For the network equipment vendors (who typically design their own control plane anyway), network programmability mostly means the ability to define data plane algorithms without the need to involve the original designers of the chosen packet processing \ac{ASIC}.

Network programmability is a powerful concept that allows both the network equipment vendors and the users to build networks ideally suited to their needs.
In addition, they can do it much faster and often cheaper than ever before and without compromising the performance or quality of the equipment.

For a variety of technical reasons, different layers became programmable at different point in time.
While the management plane became programmable in the 1980s, control plane programmability was not achieved until late 2000s to early 2010s and a programmable switching \acp{ASIC} did not appear till the end of 2015.

Thus, despite the focus on data plane programmability, we will start by discussing control plane programmability and its most well-known embodiment, called \acf{SDN}.
This discussion will also better prepare us to understand the significance of data plane programmability.

\subsection{Control Plane Programmability and SDN}

Traditional networking devices such as routers or switches have complex data and control plane algorithms.
They are built into them and generally cannot be replaced by the users.
Thus, the functionality of a device is defined by its vendor who is the only one who can change it. In industry parlance, vendors are often called \acp{OEM}.

Software-defined networking (\acs{SDN}) was historically the first attempt to make the devices, and \emph{specifically their control plane}, programmable.
On selected systems, device manufacturers allowed users to bypass built-in control plane algorithms so that the users can introduce their own.
These algorithms could then directly supply the necessary forwarding information to the data plane which was still non-replaceable and remained under the control of the device vendor or their chosen silicon provider.

\begin{figure}[ht]
  \centering
  \includegraphics[width=.7\linewidth]{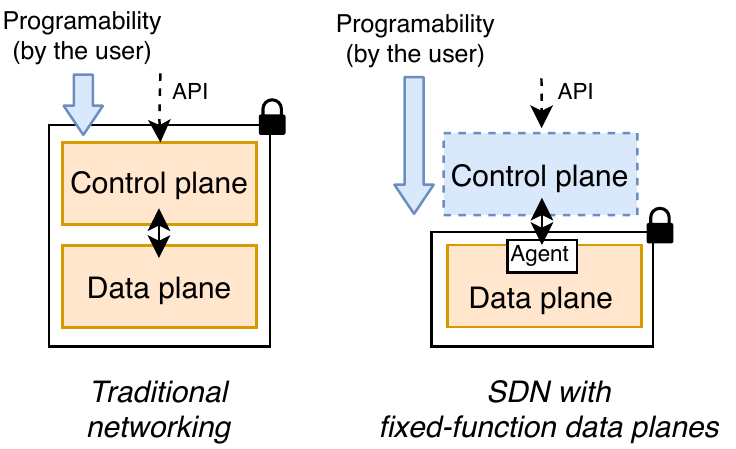}
  \caption{Distinction between traditional networking and \ac{SDN} with fixed-function data planes.}
  \label{fig:data-plane-programming-differences}
\end{figure}

For a variety of technical reasons, it was decided to provide an \acp{API} that could be called remotely and that is how \ac{SDN} was born.
\fig{data-plane-programming-differences} depicts \ac{SDN} in comparison to traditional networking.
Not only the control plane became programmable, but it also became possible to implement network-wide control plane algorithms in a centralized controller.
In several important use cases, such as tightly controlled, massive data centers, these centralized, network-wide algorithms proved to be a lot simpler and more efficient, than the traditional algorithms (e.g. \ac{BGP}) designed for decentralized control of many autonomous networks.

The effort to standardize this approach resulted in the development of \acf{OF} \cite{McAn08}.
The hope was that once \ac{OF} standardized the messaging \ac{API} to control the data plane functionality, \ac{SDN} applications will be able to leverage the functions offered by this \ac{API} to implement network control.
There is a huge body of literature giving an overview of \ac{OF} \cite{BrMe14,HuHa14,LaKo14} and \ac{SDN} \cite{NuMe14,JaMa14,XiWe15,MaGu15,MaGh16,KrRa15,TrFa16}.

However, it soon became apparent that \ac{OF} assumed a specific data plane functionality which was not formally specified.
Moreover, the specific data plane, that served as the basis for \ac{OF}, could not be changed.
It executed the sole, although relatively flexible, algorithm defined by the \ac{OF} specifications.

In part, it was this realization that led to the development of modern data plane programming that we discuss in the following section. 

\subsection{Data Plane Programming}
As mentioned above, data plane programmability means that the data plane with its algorithms can be defined by the users, be they network operators or equipment designers working with a packet processing \ac{ASIC}.
In fact, data plane programmability existed during most of the networking industry history because data plane algorithms were typically executed on general-purpose CPUs.
It is only with the advent of high-speed links, exceeding the CPU processing capabilities, and the subsequent introduction of packet processing (switching) \acp{ASIC} that data plane programmability (or lack thereof) became an issue.

The data plane algorithms are responsible for processing all the packets that pass through a telecommunication system.
Thus, they ultimately define the functionality, performance, and the scalability of such systems.
Any attempt to implement data plane functionality in the control plane typically leads to significant performance degradation.
When data plane programming is provided to users, it qualitatively changes their power.
They can build custom network equipment without any compromise in performance, scalability, speed, or energy consumption.

\begin{figure}[ht]
  \centering
  \includegraphics[width=.75\linewidth]{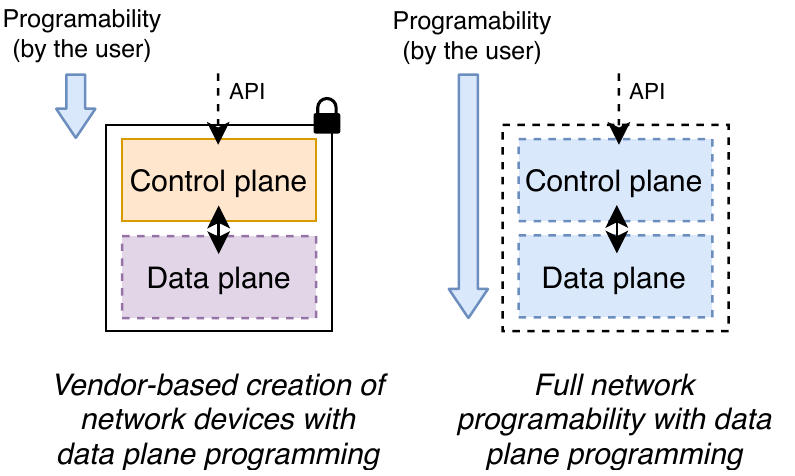}
  \caption{Data plane programmability may be used by vendors for more efficient development or by users to provide own data and control plane algorithms.}
  \label{fig:data-plane-programming}
\end{figure}

For custom networks, new control planes and \ac{SDN} applications can be designed and for them users can design data plane algorithms that fit them ideally.
Data plane programming does not necessarily imply any provision of \acp{API} for users nor does it require support for outside control planes as in \ac{OF}.
Device vendors might still decide to develop a proprietary control plane and use data plane programming only for their own benefit without necessarily making their systems more open (although many do open their systems now).
\fig{data-plane-programming} visualizes both options.

Four surveys from \cite{KrRa15,MaGu15,BiRe18,KaMa19} give an overview on data plane programming, but do not set a particular focus to P4.

\subsection{Data Plane Programming Models}
Data plane algorithms can and often are expressed using standard programming languages.
However, they do not map very well onto specialized hardware such as high-speed \acp{ASIC}.
Therefore, several data plane models have been proposed as abstractions of the hardware. 
Data plane programming languages are tailored to those data plane models and provide ways to express algorithms for them in an abstract way.
The resulting code is then compiled for execution on a specific packet processing node supporting the respective data plane programming model. 

Data flow graph abstractions and the \ac{PISA} are examples for data plane models. We give an overview of the first and elaborate in-depths on the second as \ac{PISA} is the data plane programming model for P4.

\subsubsection{Data Flow Graph Abstractions}
\label{sec:data-flow-graph-abstractions}
\begin{figure}[ht]
  \begin{center}
    \includegraphics[width=0.8\linewidth]{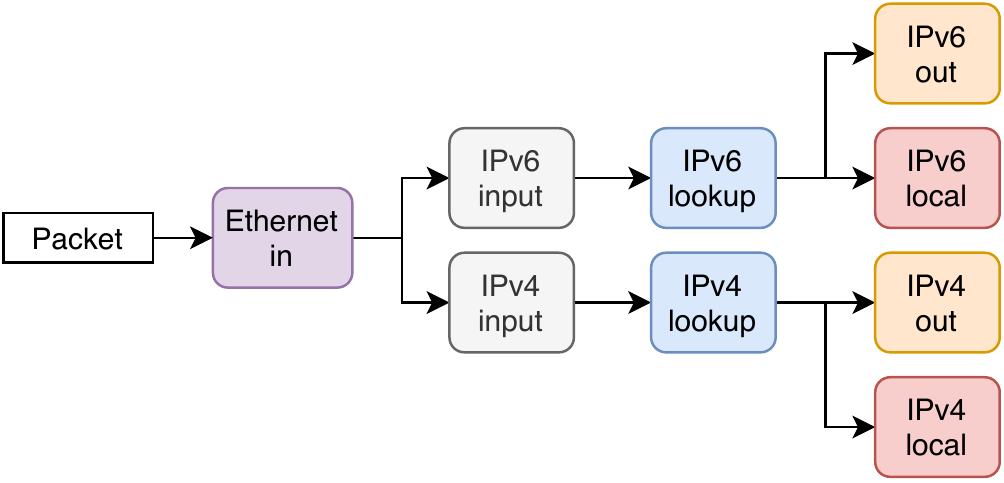}
  \end{center}
  \caption{Data flow graph abstraction: example graph for IPv4 and IPv6 forwarding.}
  \label{fig:data-flow-graph-abstraction}
\end{figure}

In these data plane programming models, packet processing is described by a directed graph.
The nodes of the graph represent simple, reusable primitives that can be applied to packets, e.g., packet header modifications.
The directed edges of the graph represent packet traversals where traversal decisions are performed in nodes on a per-packet basis.
\fig{data-flow-graph-abstraction} shows an exemplary graph for IPv4 and IPv6 packet forwarding.

Examples for programming languages that implement this data plane programming model are Click \cite{KoMo00}, \ac{VPP} \cite{VPP}, and BESS \cite{BESS}.

\subsubsection{Protocol-Independent Switching Architecture (\acs{PISA})}
\begin{figure}[ht]
  \centering
  \includegraphics[width=.85\linewidth]{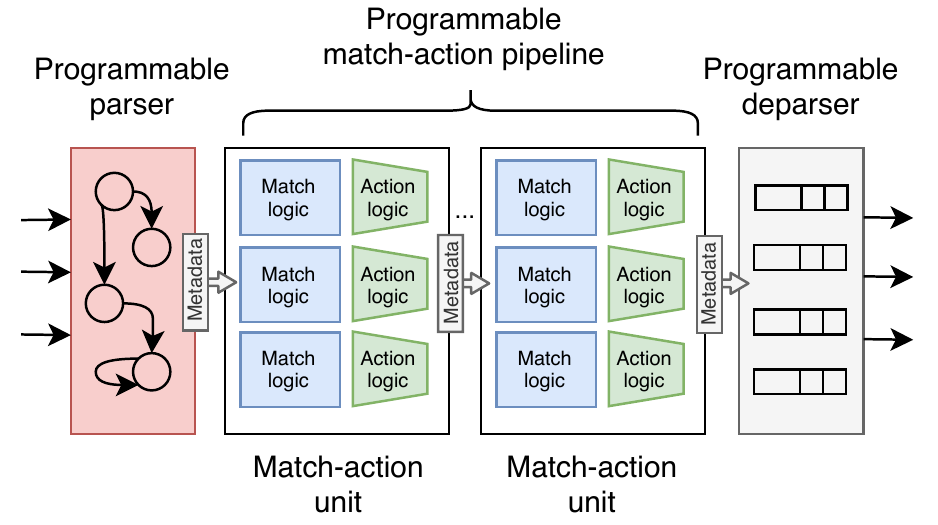}
  \caption{Protocol-Independent Switch Architecture (\acs{PISA}).}
  \label{fig:pisa}
\end{figure}

\fig{pisa} depicts the \ac{PISA}.
It is based on the concept of a programmable match-action pipeline that well matches modern switching hardware.
It is a generalization of reconfigurable match-action tables (RMTs) \cite{BoGi13} and disaggregated reconfigurable match-action tables (dRMTs) \cite{ChFi17}.

\ac{PISA} consists of a programmable parser, a programmable deparser, and a programmable match-action pipeline in between consisting of multiple stages.

\begin{itemize}
    \item The \emph{programmable parser} allows programmers to declare arbitrary headers together with a finite state machine that defines the order of the headers within packets. It converts the serialized packet headers into a well-structured form.
    \item The \emph{programmable match-action pipeline} consists of multiple match-action units. Each unit includes one or more \acp{MAT} to match packets and perform match-specific actions with supplied action data. The bulk of a packet processing algorithm is defined in the form of such \acp{MAT}. Each \ac{MAT} includes matching logic coupled with the memory (\ac{SRAM} or \ac{TCAM}) to store lookup keys and the corresponding action data. The action logic, e.g., arithmetic operations or header modifications, is implemented by \acp{ALU}. Additional action logic can be implemented using stateful objects, e.g., counters, meters, or registers, that are stored in the \ac{SRAM}. A control plane manages the matching logic by writing entries in the \acp{MAT} to influence the runtime behavior.
    \item In the \emph{programmable deparser}, programmers declare how packets are serialized.
\end{itemize}

A packet, processed by a PISA pipeline, consists of packet payload and packet metadata.
PISA only processes packet metadata that travels from the parser all the way to the deparser but not the packet payload that travels separately.

Packet metadata can be divided into packet headers, user-defined and intrinsic metadata.

\begin{itemize}
    \item \emph{Packet headers} is metadata that corresponds to the network protocol headers. They are usually extracted in the parser, emitted in the deparser or both.
    \item \emph{Intrinsic metadata} is metadata that relates to the fixed-function components. P4-programmable components may receive information from the fixed-function components by reading the intrinsic metadata they produce or control their behavior by setting the intrinsic metadata they consume.
    \item \emph{User-defined metadata} (often referred as simply \emph{metadata}) is a temporary storage, similar to local variables in other programming languages. It allows the developers to add information to packets that can be used throughout the processing pipeline.
\end{itemize}

All metadata, be it packet headers, user-defined or intrinsic metadata is \emph{transient}, meaning that it is discarded when the corresponding packet leaves the processing pipeline (e.g., is sent out of an egress port or dropped).

\ac{PISA} provides an abstract model that is applied in various ways to create concrete architectures.
For example, it allows specifying pipelines containing different combinations of programmable components, e.g., a pipeline with no parser or deparser, a pipeline with two parsers and deparsers, and additional match-action pipelines between them.
\ac{PISA} also allows for specialized components that are required for advanced processing, e.g., hash/checksum calculations.
Besides the programmable components of \ac{PISA}, switch architectures typically also include configurable fixed-function components.
Examples are ingress/egress port blocks that receive or send packets, packet replication engines that implements multicasting or cloning/mirroring of packets, and traffic managers, responsible for packet buffering, queuing, and scheduling.

The fixed-function components communicate with the programmable ones by generating and/or consuming intrinsic metadata.
For example, the ingress port block generates ingress metadata that represents the ingress port number that might be used within the match-action units.
To output a packet, the match-action units generates intrinsic metadata that represents an egress port number; this intrinsic metadata is then consumed by the traffic manager and/or egress port block.

\begin{figure}[ht]
  \centering
  \includegraphics[width=.85\linewidth]{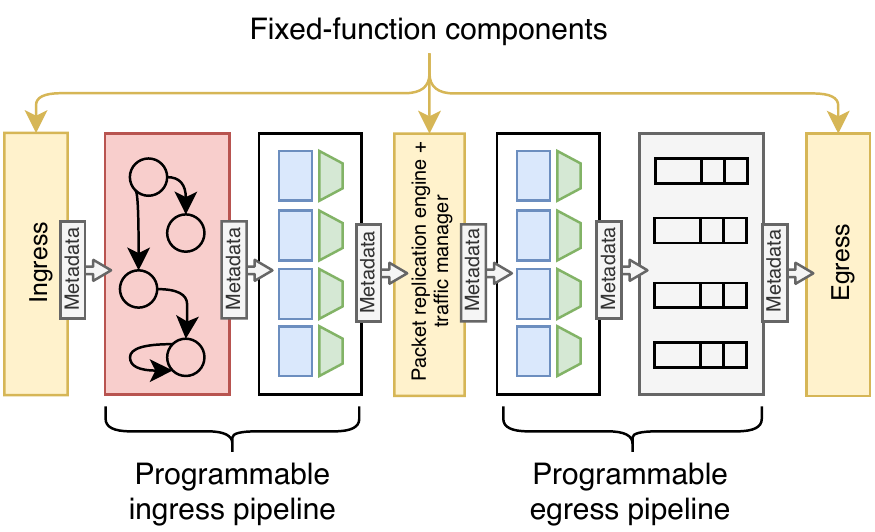}
  \caption{Exemplary switch architecture based on \ac{PISA}.}
  \label{fig:pisa-ff-pf}
\end{figure}

\fig{pisa-ff-pf} depicts a typical switch architecture based on \ac{PISA}.
It comprises a programmable ingress and egress pipeline and three fixed-function components: an ingress block, an egress block, and a packet replication engine together with a traffic manager between ingress and egress pipeline.

P4 (Programming Protocol-Independent Packet Processors) \cite{BoDa14} is the most widely used domain-specific programming language for describing data plane algorithms for \ac{PISA}.
Its initial idea and name were introduced in 2013 \cite{p4-tutorial-slides} and it was published as a research paper in 2014 \cite{BoDa14}.
Since then, P4 has been further developed and standardized by the P4 Language Consortium \cite{p4-language-consortium} that is part of the \ac{ONF} since 2019.
The P4 Language Consortium is managed by a technical steering committee and hosts five \acp{WG}.
\pfour \cite{p4_14} was the first standardized version of the language.
The current specification is \psix \cite{p4_16} which was first introduced in 2016.

Other data plane programming languages for \ac{PISA} are FAST \cite{MoBh14}, OpenState \cite{BiBo14}, Domino \cite{SiCh16}, FlowBlaze \cite{PoBi19}, Protocol-Oblivious Forwarding \cite{So13}, and NetKAT \cite{AnFo14}.
In addition, Broadcom \cite{Broadcom} and Xilinx \cite{Xilinx} offer vendor-specific programmable data planes based on match-action tables.

\subsection{Benefits}
Data plane programmability entails multiple benefits.
In the following, we summarize key benefits.

Data plane programming introduces full flexibility to network packet processing, i.e., algorithms, protocols, features can be added, modified, or removed by the user.
In addition, programmable data planes can be equipped with a user-defined \ac{API} for control plane programmability and \ac{SDN}. 
To keep complexity low, only components needed for a particular use case might be included in the code.
This improves security and efficiency compared to multi-purpose appliances.

In conjunction with suitable hardware platforms, data plane programming allows network equipment designers and even users to experiment with new protocols and design unique applications; both do no longer depend on vendors of specialized packet-processing \acp{ASIC} to implement custom algorithms.
Compared to long development circles of new silicon-based solutions, new algorithms can be programmed and deployed in a matter of days.

Data plane programming is also beneficial for network equipment developers that can easily create differentiated products despite using the same packet processing \ac{ASIC}.
In addition, they can keep their know-how to themselves without the need to share the details with the \ac{ASIC} vendor and potentially disclose it to their competitors that will use the same \ac{ASIC}.

So far, modern data plane programs and programming languages have not yet achieved the degree of portability attained by the general-purpose programming languages.
However, expressing data plane algorithms in a high-level language has the potential to make telecommunication systems significantly more target-independent.
Also, data plane programming does not require but encourages full transparency.
If the source code is shared, all definitions for protocols and behaviors can be viewed, analyzed, and reasoned about, so that data plane programs benefit from community development and review.
As a result, users could choose cost-efficient hardware that is well suited for their purposes and run their algorithms on top of it.
This trend has been fueled by \ac{SDN} and is commonly known as network disaggregation.

%% file: chapters/03_p4_programming.tex
\section{The P4 Programming Language}
\label{sec:p4-programming}

We give an overview of the P4 programming language.
We briefly recap its specification history and describe how P4 programs are deployed.
We introduce the P4 processing pipeline and data types.
We discuss parsers, match-action controls, and deparsers.
Finally, we give an overview of tutorials and guides to P4.

\subsection{Specification History}
The P4 \ac{LDWG} of the P4 Language Consortium has standardized so far two distinct standards of P4: \pfour and \psix.
Table~\ref{tab:p4-specifications} depicts their specification history.

\begin{table}[!htb]
    \caption{Specification history of \pfour and \psix.}
    \begin{minipage}{.5\linewidth}
      \centering
        \begin{tabular}{|l|l|}
            \hline
            \multicolumn{2}{|l|}{\textbf{\pfour}}\\ \hline
            Version 1.0.2 & 03/2015 \\ \hline
            Version 1.1.0 & 01/2016 \\ \hline
            Version 1.0.3 & 11/2016 \\ \hline
            Version 1.0.4 & 05/2017 \\ \hline            
            Version 1.0.5 & 11/2018 \\ \hline
        \end{tabular}
    \end{minipage}%
    \begin{minipage}{.5\linewidth}
      \centering
        \begin{tabular}{|l|l|}
            \hline
            \multicolumn{2}{|l|}{\textbf{\psix}}\\ \hline
            Version 1.0.0 & 05/2017\\ \hline            
            Version 1.1.0 & 11/2018\\ \hline         
            Version 1.2.0 & 11/2018\\ \hline
            Version 1.2.1 & 06/2020\\ \hline
        \end{tabular}
    \end{minipage} 
\label{tab:p4-specifications}
\end{table}

The \pfour programming language dialect allows the programmers to describe data plane algorithms using a combination of familiar, general-purpose imperative constructs and more specialized declarative ones that provide support for the typical data-plane-specific functionality, e.g., counters, meters, checksum calculations, etc. 
As a result, the \pfour language core includes more than 70 keywords.
It further assumed a specific pipeline architecture based on \ac{PISA}. 

\psix has been introduced to address several \pfour limitations that became apparent in the course of its use. Those include the lack of means to describe various targets and architectures, weak typing and generally loose semantics (caused, in part, by the above-mentioned mix of imperative and declarative programming constructs), relatively low-level constructs, and weak support for program modularity.

\begin{figure}[ht]
  \centering
  \includegraphics[width=0.8\linewidth]{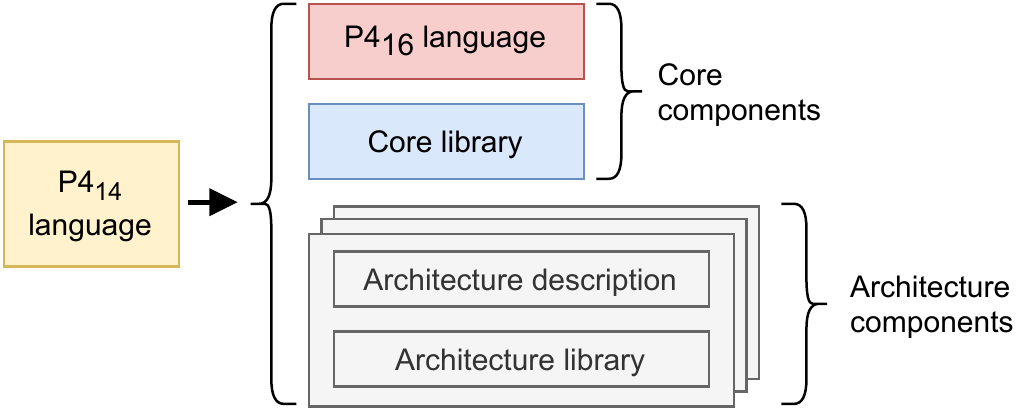}
  \caption{Comparison of the \pfour and \psix language according to \cite{p4_16}.}
  \label{fig:14-to-16}
\end{figure}

Support for multiple different targets and pipeline architecture is the major contribution of the \psix standard and is achieved by separating the core language from the specifics of a given architecture, thus making it architecture-agnostic. The structure, capabilities and interfaces of a specific pipeline are now encapsulated into an architecture description, while the architecture- or target-specific functions are accessible through an architecture library, typically provided by the target vendor.
The core components are further structured into a small set of language constructs and a core library that is useful for most P4 programs. Compared to \pfour, \psix introduced strict typing, expressions, nested data structures, several modularity mechanisms, and also removed declarative constructs, making it possible to better reason about the programs, written in the language. \fig{14-to-16} illustrates the concept which is subdivided into core components and architecture components.

Due to the obvious advantages of \psix, \pfour development has been discontinued, although it is still supported on a number of targets. Therefore, we focus on \psix in the remainder of this paper where P4 implicitly stands for \psix.

\subsection{Development and Deployment Process}
\begin{figure}[ht]
  \centering
  \includegraphics[width=.9\linewidth]{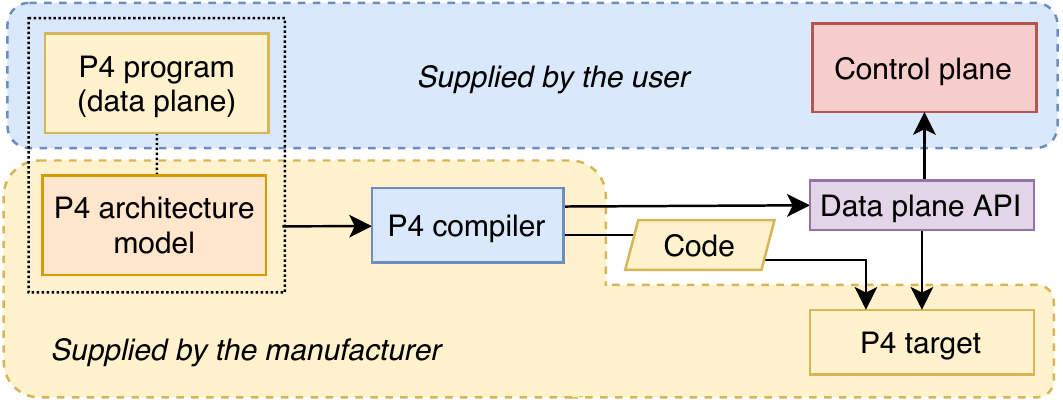}
  \caption{P4 deployment process according to \cite{p4_16}.}
  \label{fig:p4_components}
\end{figure}

\fig{p4_components} illustrates the development and deployment process of P4 programs.

P4-programmable nodes, so-called P4 targets, are available as software or specialized hardware (see \sect{p4-targets}).
They feature packet processing pipelines consisting of both P4-programmable and fixed-function components.
The exact structure of these pipelines is target-specific and is described by a corresponding P4 architecture model (see \sect{p4-architectures-compilers}) which is provided by the manufacturer of the target.

P4 programs are supplied by the user and are implemented for a particular P4 architecture model.
They define algorithms that will be executed by the P4-programmable components and their interaction with the ones implemented in the fixed-function logic.
The composition of the P4 programs and the fixed-function logic constitutes the full data plane algorithm.

P4 compilers (see \sect{p4-architectures-compilers}) are also provided by the manufacturers.
They translate P4 programs into target-specific code which is loaded and executed by the P4 target.

The P4 compiler also generates a data plane \ac{API} that can be used by a user-supplied control plane (see \sect{p4-data-plane-apis}) to manage the runtime behavior of the P4 target.

\subsection{Information Flow}
\begin{figure}[ht]
  \centering
  \includegraphics[width=.85\linewidth]{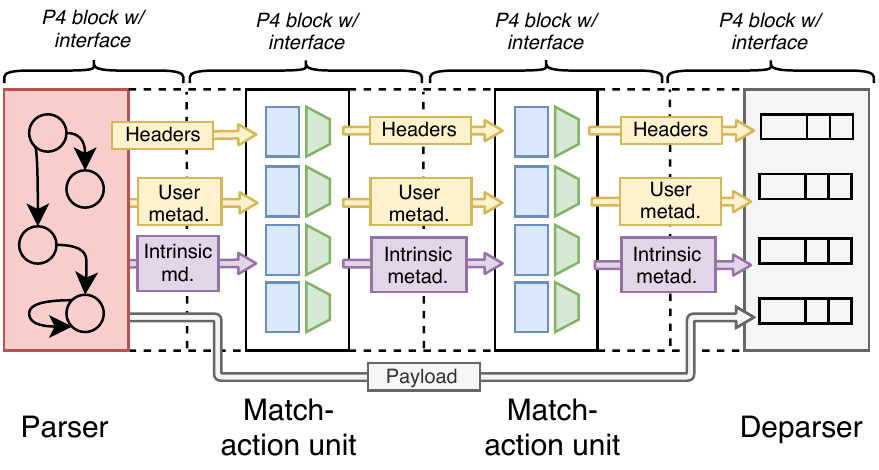}
  \caption{Information flow.}
  \label{fig:information-flow}
\end{figure}

\psix adopts \ac{PISA}'s concept of packet metadata.
\fig{information-flow} illustrates the information flow in the P4 processing pipeline.
It comprises different blocks, where packet metadata (be it headers, user-defined or intrinsic metadata) is used to pass the information between them, therefore representing a uniform interface.

The parser splits up the received packet into individual headers and the remaining payload.
Intrinsic metadata from the ingress block, e.g., the ingress port number or the ingress timestamp, is often provided by the hardware and can be made available for further processing.
Many targets allow the user metadata to be initialized in the parser as well.
Then, the headers and metadata are passed to the match-action pipeline that consists of one or more match-action units.
The remaining payload travels separately and cannot be directly affected by the match-action pipeline processing.

While traversing the individual match-action pipeline units, the headers can be added, modified, or removed and additional metadata can be generated.

The deparser assembles the packet back by emitting the specified headers followed by the original packet payload.
Packet output is configured with intrinsic metadata that includes information such as a drop flag, desired egress port, queue number, etc.

\subsection{Data Types}
\psix is a statically typed language that supports a rich set of data types for data plane programming.

\subsubsection{Basic Data Types}
\psix includes common basic types such as Boolean (\texttt{bool}), signed (\texttt{int}), and unsigned (\texttt{bit}) integers which are also known as bit strings. 
Unlike many common programming languages, the size of these integers is specified at \emph{bit} granularity, with a wide range of supported widths.
For example, types such as \texttt{bit<1>}, \texttt{int<3>}, \texttt{bit<128>} and wider are allowed.

In addition, P4 supports bit strings of variable width, represented by a special varbit type.
For example, IPv4 options can be represented as \texttt{varbit<320>} since the size of IPv4 options ranges from zero to 10 32-bit words.

\psix also supports enumeration types that can be serializable (with the actual representation specified as \texttt{bit<N>} or \texttt{int<N>} during the type definition) or non-serializable, where the type representation is chosen by the compiler and hidden from the user.

\subsubsection{Derived Data Types}
Basic data types can be composed to construct derived data types. The most common derived data types are \texttt{header}, \texttt{header stack}, and \texttt{struct}.

The \texttt{header} data type facilitates the definition of packet protocol headers, e.g., IPv4 or TCP.
A header consists of one more fields of the serializable types described above, typically \texttt{bit<N>}, serializable \texttt{enum}, or \texttt{varbit}.
A header also has an implicit validity field indicating whether the header is part of a packet. The field is accessible through standard methods such as \emph{setvalid()}, \emph{setInvalid()}, and \emph{isValid()}.
Packet parsing starts with all headers being invalid.
If the parser determines that a header is present in the packet, the header fields are extracted and the header's validity field is set valid.
The standard packet \emph{emit()} method used by a deparser equips packets only with valid headers.
Thus, P4 programs can easily add and remove headers by manipulating their validity bits.
A sample header declaration is shown in \fig{p4-sample-header}.

A \texttt{header stack} is used to define repeating headers, e.g., VLAN tags or MPLS labels. It supports special operations allowing headers to be ``pushed'' onto the stack or ``popped'' from it.

\texttt{Struct} in P4 is a composed data type similar to structs in programming languages like C.
Unlike the \texttt{header} data type, they can contain fields of any type including other structs, headers, and others.

\begin{figure}[h]
\begin{lstlisting}[mathescape=true]
typedef bit<48> macAddr_t;

header ethernet_t {
    macAddr_t dstAddr;
    macAddr_t srcAddr;
    bit<16>   etherType;
}
\end{lstlisting}
\caption{Sample declaration of the Ethernet header.}
\label{fig:p4-sample-header}
\end{figure}

\subsection{Parsers}
\label{sec:p4-parser}

Parsers extract header fields from ingress packets into header data and metadata.
P4 does not include predefined packet formats, i.e., all required header formats including parsing mechanisms need to be part of the P4 program.
Parsers are defined as \ac{FSM} with an explicit \emph{Start} state, two ending states (\emph{Accept} and \emph{Reject}), and custom states in between.

\begin{figure}[ht]
  \centering
  \includegraphics[width=.85\linewidth]{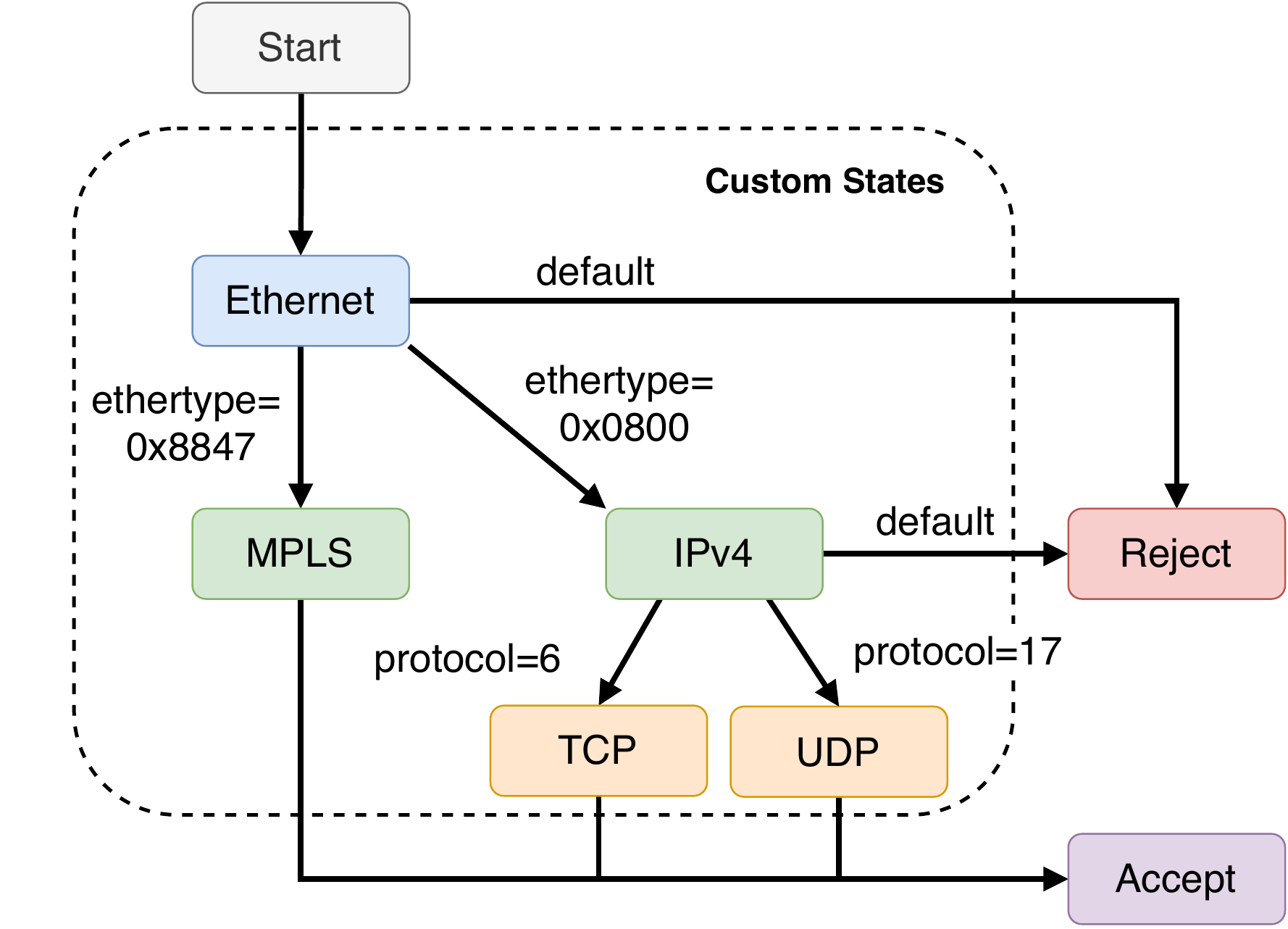}
  \caption{Example for the \ac{FSM} of a P4 parser that parses packets with Ethernet, MPLS, IPv4, TCP, and UDP headers.}
  \label{fig:p4_parsers}
\end{figure}

\fig{p4_parsers} depicts the structure of a typical P4 parser for Ethernet, MPLS, IPv4, TCP, and UDP headers.
\fig{p4-compiler-src} shows the source code fragment of the example parser in a \psix program.
The process starts in the \emph{Start} state and switches to the \emph{Ethernet} state.
In this state and the following states, information from the packet headers is extracted according to the defined header structure.

State transitions may be either conditional or unconditional.
In the given example, the transition from the \emph{Start} state to the \emph{Ethernet} state is unconditional while in the \emph{Ethernet} state the transition to the \emph{MPLS}, \emph{IPv4}, or \emph{Reject} state depends on the value of the \emph{EtherType} field of the extracted Ethernet header.
Based on previously parsed header information, any number of further headers can be extracted from the packet.
If the header order does not comply with the expected order, a packet can be discarded by switching to the \emph{Reject} state.
The parser can also implicitly transition into the \emph{Reject} state in case of a parser exception, e.g., if a packet is too short.

\begin{figure}[ht]
\begin{lstlisting}[mathescape=true]
parser SampleParser(packet_in p, out headers h) {

    state start {
        transition parse_ethernet;
    }

    state parse_ethernet {
        p.extract(h.ethernet);
        transition select(h.ethernet.etherType) {
             0x8847: parse_mpls;
             0x0800: parse_ipv4;
            default: reject;
        };
    }
    
    state parse_ipv4 {
        p.extract(h.ipv4);
        transition select(h.ipv4.protocol) {
                  6: parse_tcp;
                 17: parse_udp;
            default: accept;
        }
    }
    
    state parse_udp {
        p.extract(h.udp);
        transition accept;
    }
    /* Other states follow */
}
\end{lstlisting}
\caption{Sample parser implementation of the FSM in \fig{p4_parsers}.}
\label{fig:p4-compiler-src}
\end{figure}

\subsection{Match-Action Controls}
\label{sec:p4-controls}

Match-action controls express the bulk of the packet processing algorithm and resemble traditional imperative programs. They are executed after successful parsing of a packet. In some architectures they are also called match-action pipeline units. In the following, we give an overview of control blocks, actions, and match-action tables.

\subsubsection{Control Blocks}
Control blocks, or just \texttt{controls}, are similar to functions in general-purpose languages.
They are called by an \emph{apply()} method.
They have parameters and can call also other control blocks.
The body of a control block contains the definition of resources, such as tables, actions, and externs that will be used for processing.
Furthermore, a single \emph{apply()} method is defined that expresses the processing algorithm.

P4 offers statements to express the program flow within a control block.
Unlike common programming languages, P4 does not provide any statements that would allow the programmer to create loops.
This ensures that all the algorithms that can be coded in P4 can be expressed as \acp{DAG} and thus are guaranteed to complete within a predictable time interval. 
Specific control statements include:

\begin{itemize}
    \item a block statement \texttt{\{\}} that expresses sequential execution of instructions.
    \item an \texttt{if()} statement that expresses an execution predicated on a Boolean condition
    \item a \texttt{switch()} statement that expresses a choice from multiple alternatives
    \item an \texttt{exit()} statement that ends the control flow within a control block and passes the control to the end of the top-level control
\end{itemize}

Transformations are performed by several constructs, such as

\begin{itemize}
    \item An assignment statement which evaluates the expression on its right-hand-side and assigns the result to a header or a metadata fields
    \item A match-action operation on a table expressed as the table’s \texttt{apply()} method
    \item An invocation of an action or a function that encapsulate a sequence of statements
    \item An invocation of an extern method that represents special, target- and architecture-specific processing, often involving additional state, preserved between packets
\end{itemize}

A sample implementation of basic L2 forwarding is provided in \fig{p4-sample-control}.

\begin{figure}[ht]
\begin{lstlisting}[mathescape=true]
control SampleControl(inout headers h, inout standard_metadata_t standard_metadata) {
    
    action l2_forward(egressSpec_t port) {
        standard_metadata.egress_spec = port;
    }
    
    table l2 {
        key = {
            h.ethernet.dstAddr: exact;
        }
        actions = {
            l2_forward; drop;
        }
        size = 1024;
        default_action = drop();
    }
    
    apply {
        if (h.ethernet.isValid()) {
            l2.apply();
        }
    }
}
\end{lstlisting}
\caption{Sample control block implementing basic L2 forwarding.}
\label{fig:p4-sample-control}
\end{figure}

\subsubsection{Actions}
Actions are code fragments that can read and write packet headers and metadata.
They work similarly to functions in other programming languages but have no return value.
Actions are typically invoked from \acp{MAT}.
They can receive parameters that are supplied by the control plane as action data in \ac{MAT} entries.

As in most general-purpose programming languages, the operations are written using expressions and the results are then assigned to the desired header or metadata fields. The operations available in P4 expressions include standard arithmetic and logical operations as well as more specialized ones such as bit slicing (\texttt{field[high:low]}), bit concatenation (\texttt{field1 ++ field2}), and saturated arithmetic (\texttt{|+|} and \texttt{|-|}).

Actions can also invoke methods of other objects, such as headers and architecture-specific externs, e.g., counters and meters.
Other actions can also be called, similar to nested function calls in traditional programming languages.

Action code is executed sequentially, although many hardware targets support parallel execution. In this case, the compiler can optimize the action code for parallel execution as long as its effects are the same as in case of the sequential execution.  

\subsubsection{Match-Action Tables (MATs)}
\label{sec:match-action-table}

\acp{MAT} are defined within control blocks and invoke actions depending on header and metadata fields of a packet.
The structure of a \ac{MAT} is declared in the P4 program and its table entries are populated by the control plane at runtime.
A packet is processed by selecting a matching table entry and invoking the corresponding action with appropriate parameters.

The declaration of a \ac{MAT} includes the match key, a list of possible actions, and additional attributes. 

The match key consists of one or more header or metadata fields (variables), each with the assigned \emph{match type}.
The P4 core library defines three standard match types: exact, ternary, and \ac{LPM}.
P4 architectures may define additional match types, e.g., the \emph{v1model} P4 architecture extends the set of standard match types with the range and selector match.

The list of possible actions includes the names of all actions that can be executed by the table.
These actions can have additional, directional parameters which are provided as action data in table entries.

Additional attributes may include the size of the \ac{MAT}, e.g., the maximum number of entries that can be stored in a table, a default action for a miss, or static table entries.

\begin{figure}[ht]
  \centering
  \includegraphics[width=.85\linewidth]{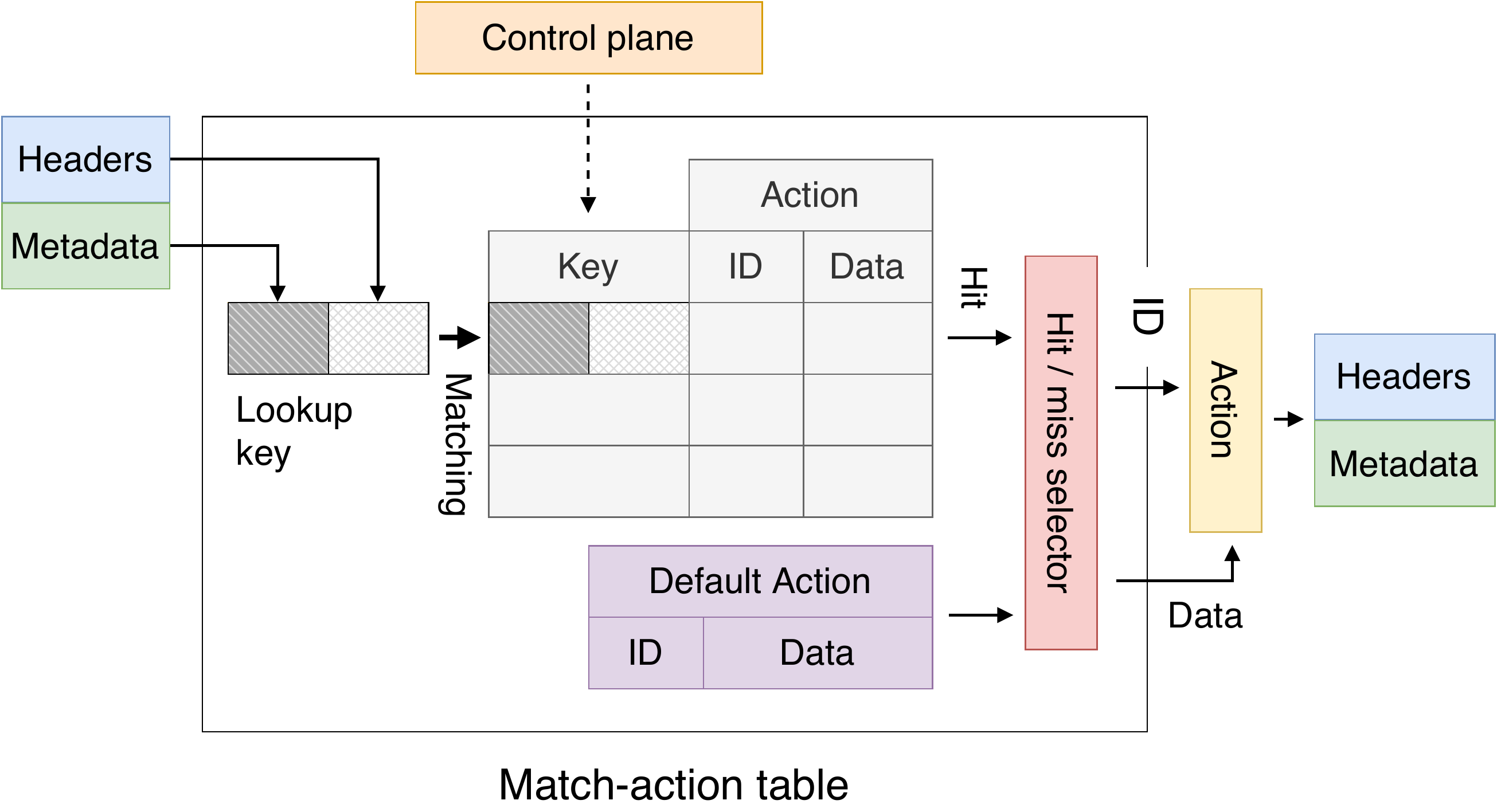}
  \caption{Structure of \acp{MAT} in P4.}
  \label{fig:p4_mat}
\end{figure}

\fig{p4_mat} illustrates the principle of \ac{MAT} operation. The \ac{MAT} contains entries with values for match keys, the ID of the corresponding action to be invoked, and action data that serve as parameters for action invocation.
For each packet, a lookup key is constructed from the set of header and metadata fields specified in the table definition.
It is matched against all entries of the \ac{MAT} using the rules associated with the individual field’s match type.
When the first match in the table is found, the corresponding action is called and the action data are passed to the action as directionless parameters.
If no match is found in the table, a default action is applied.

As a special case, tables without a specified key always invoke the default action. 

\subsection{Deparser}
The deparser is also defined as a control block. When packet processing by match-action control blocks is finished, the deparser serializes the packet.
It reassembles the packet header and payload back into a byte stream so that the packet can be sent out via an egress port or stored in a buffer.
Only valid headers are emitted, i.e., added to the packet. Thus, match-action control blocks can easily add and remove headers by manipulating their validity.
\fig{p4-sample-deparser} provides a sample implementation.

\begin{figure}[ht]
\begin{lstlisting}[mathescape=true]
control SampleDeparser(packet_out p, in headers h) {
    apply {
        p.emit(h.ethernet);
        p.emit(h.mpls);
        p.emit(h.ipv4);
        /* Normally, a packet can contain either 
         * a TCP or a UDP header (or none at all), 
         * but should never contain both
         */
        p.emit(h.tcp);
        p.emit(h.udp);
    }
}
\end{lstlisting}
\caption{Sample deparser implementation.}
\label{fig:p4-sample-deparser}
\end{figure}

\subsection{P4 Tutorials}
\label{sec:p4-programming-tutorials}
The P4 Language Consortium provides a GitHub repository with simple programming exercises and a development VM containing all required software \cite{p4-tutorial}.
A guide on GitHub lists useful information for P4 newcomers, e.g. demo programs, information about other GitHub repositories, and an overview of P4 \cite{p4-guide}.
The Networked Systems Group at ETH Zürich provides resources for people who want to learn programming in P4, including lecture slides, references to useful documentation, examples and exercises \cite{p4-learning}.

%% file: chapters/04_p4_deployment.tex
\section{P4 Architectures \& Compilers}
\label{sec:p4-architectures-compilers}

We present \psix architectures and introduce P4 compilers.

\subsection{\psix Architectures}
\label{sec:architectures}

We summarize the concept of \psix architectures, describe externs, and give an overview of the most common \psix architectures.

\subsubsection{Concept}
As described before, \psix introduces the concept of P4 architectures as an intermediate layer between the core P4 language and the targets.
A P4 architecture serves as programming models that represents the capabilities and the logical view of a target's P4 processing pipeline.
P4 programs are developed for a specific P4 architecture.
Such programs can be deployed on all targets that implement the same P4 architecture.
The manufacturers of P4 targets provide P4 compilers that compile architecture-specific P4 programs into target-specific configuration binaries.

\subsubsection{Externs}
P4 architectures may provide additional functionalities that are not part of the P4 language core.
Examples are checksum or hash computation units, random number generators, packet and byte counters, meters, registers, and many others.
To make such extern functionalities usable, \psix introduces so-called \emph{externs}.

Most of the externs have to be explicitly instantiated in P4 programs using their constructor method.
The other methods provided by these externs can then be invoked on the given extern instance.
Other externs (extern functions) do not require explicit instantiating. 

Along with tables and value sets, P4 externs are allowed to preserve additional state between packets. That state may be accessible by the control plane, the data plane, or both.
For example, the counter extern would preserve the number of packets or bytes that has been counted so that each new packet can properly increment it.  
The specifics of the state depend on the nature of the extern and cannot be specified in the language; this is done inside the vendor-specific API definitions.

While the P4 processing pipeline only allows packet header manipulation, extern functions may operate on packet payload as well.

\subsubsection{Overview of Common \psix Architectures}
We describe the four most common \psix architectures.

\begin{figure}[ht]
  \centering
  \includegraphics[width=.85\linewidth]{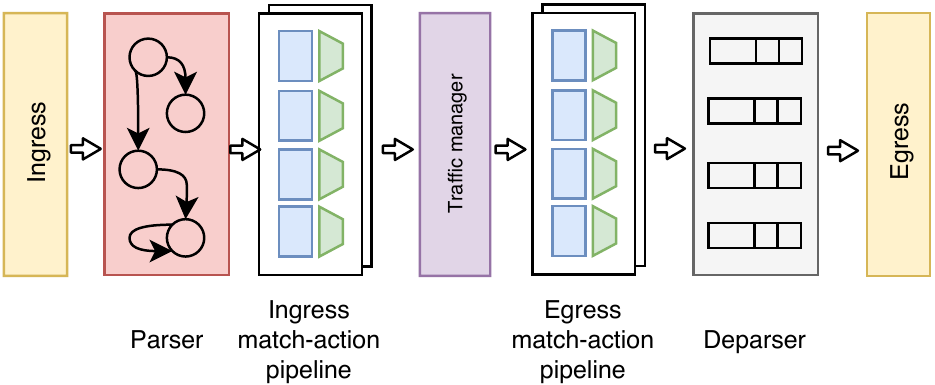}
  \caption{v1model architecture.}
  \label{fig:v1model}
\end{figure}

\paragraph{v1model}
The v1model mimics the processing pipeline of \pfour.
As depicted in \fig{v1model}, it consists of a programmable parser, an ingress match action pipeline, a traffic manager, an egress match-action pipeline, and a deparser.
It enables developers to convert \pfour programs into \psix programs.
Additional functionalities tracking the development of the reference P4 software switch \ac{bmv2} (see \sect{p4-targets}) are continuously added.
All P4 examples in this paper are written using v1model.

\begin{figure}[ht]
  \centering
  \includegraphics[width=\linewidth]{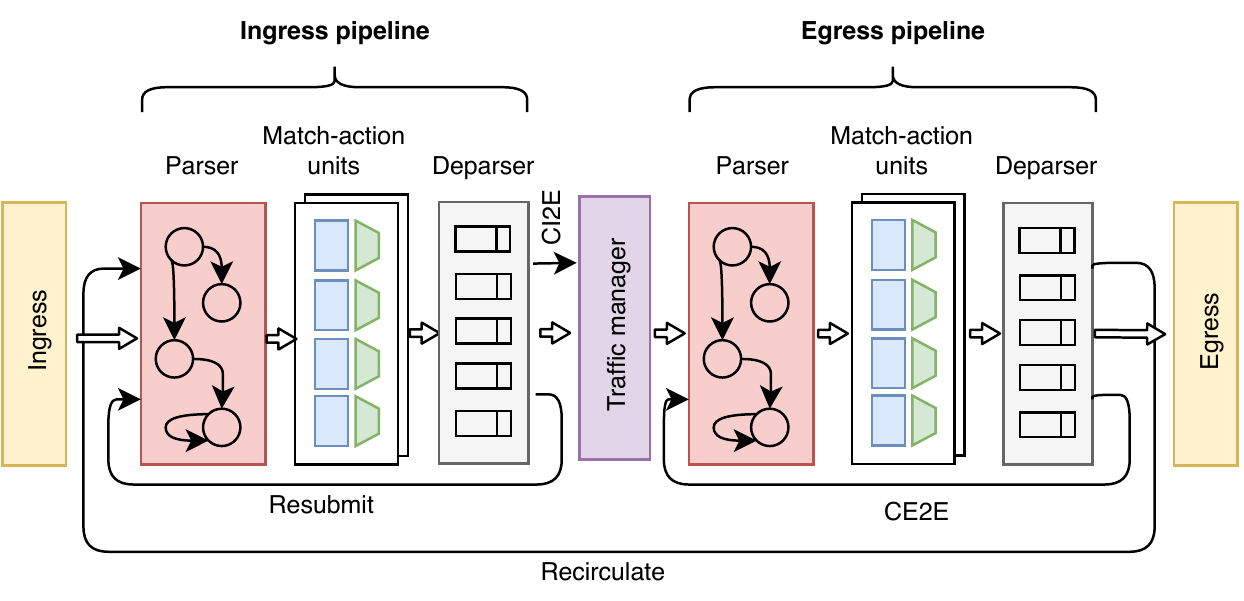}
  \caption{\acf{PSA} with programmable and fixed-function parts and special packet processing primitives.}
  \label{fig:pipeline-operations}
\end{figure}

\paragraph{\acf{PSA}}
The \ac{PSA} is a P4 architecture created and further developed by the Architecture \ac{WG} \cite{p4-architecture-wg} in the P4 Language Consortium.
Besides, the \ac{WG} also discusses standard functionalities, \acp{API}, and externs that every target mapping the \ac{PSA} should support.
Its last specification is Version 1.1 \cite{p4-16-psa} from November 2018.
\fig{pipeline-operations} illustrates the P4 processing pipeline of the \ac{PSA}. It is divided into an ingress and egress pipeline.
Each pipeline consists of the three programmable parts: parser, multiple control blocks, and deparser.
The architecture also defines configurable fixed-function components. 

\ac{PSA} specifies several packet processing primitives, such as:

\begin{itemize}
    \item Sending a packet to an unicast port
    \item Dropping a packet
    \item Sending the packet to a multicast group
    \item Resubmitting a packet, which moves the currently processed packet from the end of the ingress pipeline to the beginning of the ingress pipeline for the purpose of packet re-parsing
    \item Recirculating a packet, which moves the currently processed packet from the end of the egress pipeline to the beginning of the ingress pipeline for the purposes of recursive processing, e.g., tunneling
    \item Cloning a packet, which duplicates the currently processed packet. \emph{Clone ingress to egress (CI2E)} creates a duplicate of the ingress packet at the end of the ingress pipeline. \emph{Clone egress to egress (CE2E)} creates a duplicate of the deparsed packet at the end of the egress pipeline. In both cases, cloned instances start processing at the beginning of the egress pipeline. Cloning can be helpful to implement powerful applications such as mirroring and telemetry.
\end{itemize}

\begin{figure}[ht]
  \centering
  \includegraphics[width=.75\linewidth]{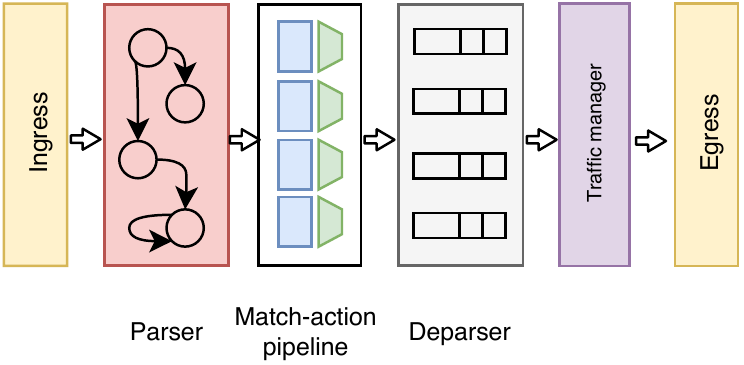}
  \caption{SimpleSumeArchitecture.}
  \label{fig:simplesume-architecture}
\end{figure}

\paragraph{SimpleSumeArchitecture}
The SimpleSumeArchitecture is a simplified P4 architecture that is implemented by \acs{FPGA}-based P4 targets.
As depicted in \fig{simplesume-architecture}, it features a parser, a programmable match-and-action pipeline, and a deparser.

\paragraph{\acf{TNA}}
\acs{TNA} is a proprietary \psix architecture designed for Intel Tofino switching \acp{ASIC} (see \sect{tofino}).
Intel has published the architecture definitions and allows developers to publish programs written by using it.

The architecture describes a very high-performance, ``industry-strength'' device that is relatively complex.
The basic programming unit is a so-called \texttt{Pipeline()} package that resembles an extended version of the \acf{PSA} pipeline and consists of 6 top-level programmable components: the ingress parser, ingress match-action control, ingress deparser, and their egress counterparts.
Since Tofino devices can have two or four processing pipelines, the final switch package can be formed anywhere from one to four distinct pipeline packages.
More complex versions of the \texttt{Pipeline()} package allow the programmer to specify different parsers for different ports.

\ac{TNA} also provides a richer set of externs compared to most other architectures.
Most notable is \ac{TNA} \texttt{RegisterAction()} which represents a small code fragment that can be executed on the register instead of simple read/write operations provided in other architectures.
\ac{TNA} provides a clear and consistent interface for mirroring and resubmit with additional metadata being passed via the packet byte stream.
The same technique is also used to pass intrinsic metadata which greatly simplifies the design.

Additional externs that are not present in other architectures include low-pass filters, weighted random early discard externs, powerful hash externs that can compute CRC based on user-defined polynomials, ParserCounter, and others.

The set of intrinsic metadata in Tofino is also larger than in most other P4 architectures as presented before.
Notable is support for two-level multicasting with additional source pruning, copy-to-cpu functionality, and support for IEEE 1588.

\subsection{P4 Compiler}
\label{sec:p4-compiler}

P4 compilers translate P4 programs into target-specific configuration binaries that can be executed on P4 targets. We first explain compilers based on the two-layer model which are most widely in use. Then we mention other compilers in less detail.

\subsubsection{Two-Layer Compiler Model}
Most P4 compilers use the two-layer model, consisting of a common frontend and a target-specific backend.

The frontend is common for all the targets and is responsible for parsing, syntactic and target-independent semantic analysis of the program. The program is finally transformed into an intermediate representation (IR) that is then consumed by the target-specific backend which performs target-specific transformations.

The first-generation P4 compiler for \pfour was written in Python and used the so-called \ac{HLIR} \cite{hlir} that represented \pfour program as a tree of Python objects. The compiler is referred to as p4-hlir.

\begin{figure}[ht]
    \begin{center}
    \includegraphics[width=0.95\linewidth]{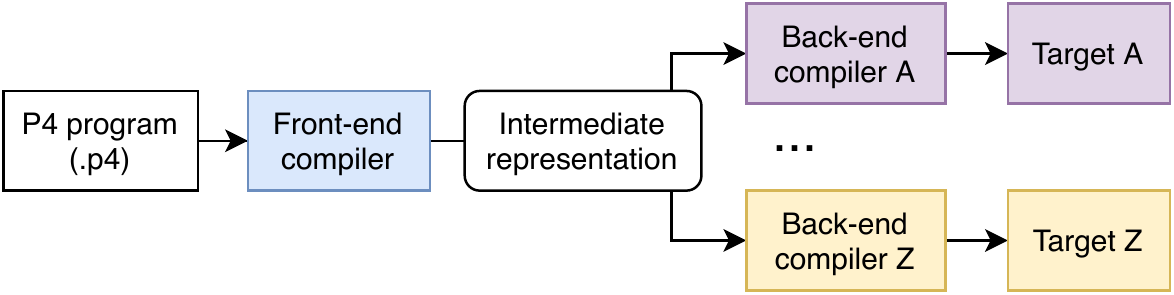}
    \end{center}
    \caption{Structure and operation principle of P4 compilers using the two-layer model.}
    \label{fig:compiler}
\end{figure}

The new P4 compiler (p4c) \cite{p4c} is written in C++ and uses C++-object-based IR. As an additional benefit, the IR can be output as a \psix program or a JSON file. The latter allows the developers and users to build powerful tools for program analysis without the need to augment the compiler.
\fig{compiler} visualizes its structure and operating principle.
The compiler consists of a generic frontend that accepts both \pfour and \psix code which may be written for any architecture. It furthermore has several reference backends for the \acs{bmv2}, eBPF, and uBPF P4 targets as well as a backend for testing purposes and a backend that can generate graphs of control flows of P4 programs.
In addition, p4c provides the so-called ``mid-end'' which is a library of generic transformation passes that are used by the reference backends and can also be used by vendor-specific backends.
The compiler is developed and maintained by P4.org. 

P4 target vendors design and maintain their own compilers that include the common frontend.  This ensures the uniformity of the language which is accepted by different compilers.

\subsubsection{Other Compilers}
MACSAD \cite{PaRo17} is a compiler that translates P4 programs into Open Data Plane (ODP) \cite{odp} programs.
Jose et al. \cite{JoYa15} introduce a compiler that maps P4 programs to FlexPipe and RMT, two common software switch architectures.
P4GPU \cite{LiLu16} is a multistage framework that translates a P4 program into intermediate representations and other languages to eventually generate GPU code.

%% file: chapters/05_p4_targets.tex
\section{P4 Targets}
\label{sec:p4-targets}

We describe P4 targets based on software, FPGA, ASIC, and NPU.
Table~\ref{tab:targets} compiles an overview of the targets, their supported architectures, and the current state of development.

\begin{table}[ht]
\caption{Overview of P4 targets.}
\begin{center}
\begin{tabularx}{\linewidth}{@{} X X X Y @{}}
\toprule
\textbf{Target}          & \textbf{P4 Version}                         & \centering\textbf{\psix\newline Architecture} & \textbf{Active Development}\\
\midrule
\textbf{Software} & & & \\
p4c-behavioral           & \pfour        & n.a.              & X \\
bmv2                     & \pfour, \psix & v1model, psa      & \checkmark \\
eBPF                     & \psix         & ebpf\_model.p4    & \checkmark \\
uBPF                     & \psix         & ubpf\_model.p4    & \checkmark \\
XDP                      & \psix         & xdp\_model.p4     & \checkmark \\
T4P4S                    & \pfour, \psix & v1model, psa      & \checkmark \\
Ripple                   & n.a           & n.a               & n.a \\
PISCES                   & \pfour        & n.a.              & X \\
PVPP                     & n.a.          & n.a.                 & X \\
ZodiacFX                 & \psix         & zodiacfx\_model.p4 & n.a. \\
\addlinespace
\textbf{FPGA} & & & \\
P4$\rightarrow$NetFPGA   & \psix         & SimpleSumeSwitch  & \checkmark \\
Netcope P4               & n.a.          & n.a.              & \checkmark \\
P4FPGA                   & \pfour, \psix & n.a.              & X \\
\addlinespace
\textbf{ASIC} & & & \\
Barefoot Tofino/Tofino 2 & \pfour, \psix & v1model, psa, TNA & \checkmark \\
Pensando Capri           & \psix         & n.a               & \checkmark \\
\addlinespace
\textbf{NPU} & & & \\
Netronome                & \pfour, \psix & v1model           & \checkmark \\ 
\bottomrule
\end{tabularx}
\end{center}

\label{tab:targets}
\end{table}

\subsection{Software-Based P4 Targets}
\label{sec:software-targets}

Software-based P4 targets are packet forwarding programs that run on a standard CPU.
We describe the 9 software-based P4 targets mentioned in Table~\ref{tab:targets}.

\subsubsection{p4c-behavioural}
p4c-behavioral \cite{p4c-behavioural-github} is a combined P4 compiler and P4 software target.
It was introduced with the first public release of P4.
p4c-behavioral translates the given \pfour program into an executable C program.

\subsubsection{\acf{bmv2}}
The second version of the P4 software switch Behavioral Model (\acs{bmv2}) \cite{bmv2} was introduced to address the limitations of p4c-behavioural (see also \cite{bmv2-replacement-reasons}).
In contrast to p4c-behavioral, the source code of \ac{bmv2} is static and independent of P4 programs.
P4 programs are compiled to a JSON representation that is loaded onto the \ac{bmv2} during runtime.
External functions and other extensions can be added by extending \ac{bmv2}'s C++ source code.
\ac{bmv2} is not a single target, but a collection of targets \cite{bmv2targets}:

\begin{itemize}
    \item \emph{simple\_switch} is the \ac{bmv2} target with the largest range of features. It contains all features from the \pfour specification and supports the v1model architecture of \psix. simple\_switch includes a program-independent Thrift \ac{API} for runtime control.
    \item \emph{simple\_switch\_grpc} extends simple\_switch by the P4Runtime \ac{API} that is based on gRPC (see \sect{p4-runtime-api}).
    \item \emph{psa\_switch} is similar to simple\_switch, but supports \ac{PSA} instead of v1model.
    \item \emph{simple\_router} and \emph{l2\_switch} support only parts of the standard metadata and do not support \psix. They are intended to show how different architectures can be implemented with \ac{bmv2}.
\end{itemize}

Although \ac{bmv2} is intended for testing purposes only, throughput rates up to \SI{1}{\giga\bit\per\s} for a P4 program with IPv4 \ac{LPM} routing have been reported \cite{bmv2-performance}.
\ac{bmv2} is under active development, i.e., new functionality is added frequently.

\subsubsection{BPF-based Targets}
\acp{BPF} add an interface on a UNIX system that allows sending and receiving raw packets via the data link layer.
User space programs may rely on \acp{BPF} to filter packets that are sent to it.
\ac{BPF}-based P4 targets are mostly intended for programming packet filters or basic forwarding in P4.

\paragraph{\ac{eBPF}}
\acp{eBPF} are an extension of \acp{BPF} for the Linux kernel.
\ac{eBPF} programs are dynamically loaded into the Linux kernel and executed in a \ac{VM}.
They can be linked to functions in the kernel, inserted into the network data path via iproute2, or bound to sockets or network interfaces.
\ac{eBPF} programs are always verified by the kernel before execution, e.g., programs with loops or backward pointers would not be executed.
Due to their execution in a \ac{VM}, \ac{eBPF} programs can only access certain regions in memory besides the local stack.
Accessing kernel resources is protected by a white list.
\ac{eBPF} programs may not block and sleep, and usage of locks is limited to prevent deadlocks.
The p4c compiler features the \emph{p4c-ebpf} back-end to compile \psix programs to \ac{eBPF} \cite{p4c-ebpf}.

\paragraph{\acs{uBPF}}
\acp{uBPF} relocate the \ac{eBPF} \ac{VM} from the kernel space to the user space.
\emph{p4c-ubpf} \cite{p4c-ubpf} is a backend for p4c that compiles P4 \ac{HLIR} for \ac{uBPF}.
In contrast to p4c-ebpf, it also supports packet modification, checksum calculation, and registers, but no counters.

\paragraph{\acs{XDP}}
\ac{XDP} is based on \ac{eBPF} and allows to load an \ac{eBPF} program into the RX queue of a device driver.
p4c-xdp \cite{p4c-xdp} is a backend for p4c that compiles P4 \ac{HLIR} for \ac{XDP}.
Similar to p4c-ubpf, it supports packet modification and checksum calculation.
In contrast to p4c-ebpf, it supports counters instead of registers.

\subsubsection{T\textsubscript{4}P\textsubscript{4}S}
T\textsubscript{4}P\textsubscript{4}S (pronounced "tapas") \cite{p4elte, LaHo16} is a software P4 target that relies on interfaces for accelerated packet processing such as \ac{DPDK} \cite{dpdk} or \ac{ODP} \cite{odp}.
T\textsubscript{4}P\textsubscript{4}S provides a compiler that translates P4 programs into target-independent C code that interfaces a network hardware abstraction library.
Hardware-dependent and hardware-independent functionalities are separated from each other.
Its source code is available on GitHub \cite{t4p4s-github}.
Bhardwaj et al. \cite{BhSh17} describe optimizations for improving T\textsubscript{4}P\textsubscript{4}S performance by up to 15\%.

\subsubsection{Ripple}
Ripple \cite{WuLi19} is a P4 target based on \ac{DPDK}.
It uses a static universal binary that is independent of the P4 program.
The data plane of the static binary is configured at runtime based on P4 \ac{HLIR}.
This results in a shorter downtime when updating a P4 program in contrast to targets like T\textsubscript{4}P\textsubscript{4}S.
Ripple uses vectorization to increase the performance of packet processing.

\subsubsection{PISCES}
PISCES \cite{ShCh16} transforms the \ac{OVS} \cite{ovs} into a software P4 target.
\ac{OVS} is a popular \ac{SDN} software switch that is designed for high throughput on virtualization platforms for flexible networking between \acp{VM}.
The PISCES compiler translates P4 programs into C code that replace parts of the source code of \ac{OVS}.
This makes \ac{OVS} dependent on the P4 program, i.e., \ac{OVS} must be recompiled with every modification of the P4 program.
PISCES does not support stateful components such as registers, counters, or meters.
The developers claim that PISCES does not add performance overhead to \ac{OVS}.
As the last commit in the public repository \cite{pisces-github} is from 2016, PISCES seems not to be under active development.

\subsubsection{PVPP}
PVPP \cite{ChLo17, ChLo17b} integrates P4 programs into plugins for \acf{VPP} (see \sect{data-flow-graph-abstractions}).
The P4-to-PVPP compiler comprises two stages.
First, a modified p4c compiler translates P4 programs into target-dependent JSON code.
Then, a Python compiler translates the JSON code into a \ac{VPP} plugin in C source code.
According to the authors, performance decreases by 5-17\% compared to \ac{VPP} but is still significantly better than \ac{OVS}.
Unfortunately, the source code and further information are not available for the public.

\subsubsection{ZodiacFX}
The ZodiacFX is a lightweight development and experimentation board originally designed as \ac{OF} switch featuring four Fast Ethernet ports.
It is based on an Atmel processor and an Ethernet switching chip \cite{zodiac}.
The authors provided an extension \cite{zodiac-p4, zodiac-p4c} to run P4 programs on the board.
P4 programs are compiled using an extended version of p4c and the p4c-zodiacfx backend compiler.
Then, the result of this compilation is used to generate a firmware image.
Zanna et al. \cite{ZaRa19} compare the performance of P4 and \ac{OF} on that target, and find out that differences among all test cases are small.

\subsection{FPGA-Based P4 Targets}
Several tool chains translate P4 programs into implementations for \acp{FPGA}.
The process includes logic synthesis, verification, validation, and placement/routing of the logic circuit for the \ac{FPGA}.
We describe the P4$\rightarrow$NetFPGA, Netcope P4, and P4FPGA tool chain. Finally, we mention research results for \ac{FPGA}-based P4 targets.

\subsubsection{P4$\rightarrow$NetFPGA}
The P4$\rightarrow$NetFPGA workflow \cite{p4netfga-wiki,IbBr19} provides a development environment for compiling and running P4 programs on the NetFPGA SUME board that provides four SFP+ ports \cite{ZiAu14}.
The development environment is built around the P4-SDnet compiler and the SDnet data plane builder from Xilinx, i.e., a full license for the Xilinx Vivado design suite is needed.
Custom external functions can be implemented in a \ac{HDL} such as Verilog and included in the final \ac{FPGA} program.
This also allows external IP cores to be integrated as P4 externs in P4 programs.
The P4$\rightarrow$NetFPGA tool chain supports \psix based on the P4 architecture SimpleSumeSwitch (see \sect{architectures}).

\subsubsection{Netcope P4}
Netcope P4 \cite{netcope} is a commercial cloud service that creates \ac{FPGA} firmware from P4 programs.
Knowledge of \ac{HDL} development is not needed and all necessary IP cores are provided by Netcope.
The cloud service can be used in conjunction with the Netcope \ac{SDK}.
This combination allows developers to combine the VHDL code of the cloud service with custom \ac{HDL} code, e.g., from an external function.
As target platform, Netcope P4 supports \ac{FPGA} boards from Netcope, Silicom, and Intel that are based on Xilinx or Intel \acp{FPGA}.

\subsubsection{P4FPGA}
P4FPGA \cite{WaSo17} is a \pfour and \psix compiler and runtime for the Bluespec programming language that can generate code for Xilinx and Altera \acp{FPGA}.
The last commit in the archived public repository \cite{p4fpga} is from 2017.

\subsubsection{Research Results}
Benácek and Kubátová \cite{BePu16, BePu17} present how P4 parse graph descriptions can be converted to optimized VHDL code for \acp{FPGA}.
The authors demonstrate how a complex parser for several header fields achieves a throughput of \SI{100}{\giga\bit\per\s} on a Xilinx Virtex-7 \ac{FPGA} while using 2.78\% slice \acp{LUT} and 0.76\% slice \acp{REG}.
In a follow-up work \cite{CaBe18}, the optimized parser architecture supports a throughput of \SI{1}{\tera\bit\per\s} on Xilinx UltraScale+ \acp{FPGA} and \SI{800}{\giga\bit\per\s} on Xilinx Virtex-7 \acp{FPGA}.
Da Silva et al. \cite{SiBo18} also investigate the high-level synthesis of packet parsers in \acp{FPGA}.
Kekely and Korenek \cite{KeKo17} describe how \acp{MAT} can be mapped to \acp{FPGA}.
Iša et al. \cite{IsBe18} describe a system for automated verification of \ac{RTL} generated from P4 source code.
Cao et al. \cite{CaSu20, CaSu20b} propose a template-based process to convert P4 programs to VHDL.
They use a standard P4 frontend compiler to compile the P4 program into an intermediate representation.
From this representation, a custom compiler maps the different elements of the P4 program to VHDL templates which are used to generate the \ac{FPGA} code.

\subsection{ASIC-Based P4 Targets}
\label{sec:tofino}

\subsubsection{Intel Tofino}
Intel Tofino is the world’s first user programmable Ethernet switch \ac{ASIC}. 
It is designed for very high throughput of \SI{6.5}{\tera\bit\per\s} (4.88 B pps) with 65 ports running at \SI{100}{\giga\bit\per\s}.
Its successor, the Tofino 2 \ac{ASIC}, supports throughput rates of up to \SI{12.8}{\tera\bit\per\s} with ports running at up to \SI{400}{\giga\bit\per\s}.
Tofino has been built by Barefoot Networks, a former startup company that was acquired by Intel in 2019.

The Tofino \ac{ASIC} implements the \ac{TNA}, a custom P4 architecture that significantly extends \ac{PSA} (see \sect{architectures}). It provides support for advanced device capabilities which are required to implement complex, industrial-strength data plane programs. The device comes with 2 or 4 independent packet processing pipelines (pipes), each capable of serving 16 \SI{100}{\giga\bit\per\s} ports. All pipes can run the same P4 program or each pipe can run its own program independently. Pipes can also be connected together, allowing the programmers to build programs requiring longer processing pipelines. 

The Tofino \ac{ASIC} processes packets at line rate irrespective of the complexity of the executed P4 program. This is achieved by a high degree of pipelining (each pipe is capable of processing hundreds of packets simultaneously) and parallelization. In addition to standard arithmetic and logical operations, Tofino provides specialized capabilities, often required by data plane programs, such as hash computation units and random number generators. For stateful processing Tofino offers counters, meters, and registers, as well as more specialized processing units. Some of them support specialized operations, such as approximate non-linear computations required to implement state-of-the-art data plane algorithms. Built-in packet generators allow the data plane designers to implement protocols, such as BFD, without using externally running control plane processes. These and other components are exposed through \ac{TNA} which is openly published by Intel \cite{open-tofino}. 

Tofino fixed-function components offer plenty of advanced functionality. The buffering engine has a unified  \SI{22}{\mega\byte} buffer, shared by all the pipes, that can be subdivided into several pools. Tofino Traffic Manager supports both store-and-forward as well as the cut-through mode, up to 32 queues per port, precise traffic shaping and multiple scheduling disciplines. Tofino provides nanosecond-precision timestamping that facilitates both the implementation of time synchronization protocols, such as IEEE 1588, as well as precise delay measurements. Additional intrinsic metadata support a variety of telemetry applications, such as INT.

The development is conducted using Intel P4 Studio which is a software development environment containing the P4 compiler, the driver, and other software necessary to program and manage the Tofino. A special interactive visualization tool (P4i) allows the developers to see the P4 program being mapped onto the specific hardware resources further assisting them in fitting and optimizing their programs. Intel P4 compiler for Tofino has special capabilities, allowing it to parallelize the code thereby taking advantage of the highly parallel nature of Tofino hardware. 

A number of \acp{ODM} produce open systems (white boxes) with the Tofino \ac{ASIC} that are used for research, development, and production of custom systems.
Examples include the EdgeCore Wedge 100BF-32X \cite{wedge-switch}, APS Networks BF2556-1T-A1F \cite{aps-bf2556} and BF6064-T-A2F \cite{aps-bf6064}, NetBerg Aurora 610 \cite{aurora-switch}, and others.

Most white box systems follow a modern, server-like design with a separate board management controller, responsible for handling power supplies, fans, LEDs, etc., and a main CPU, typically x86\_64, running a Linux operating system.
The main CPU is connected to the Tofino \ac{ASIC} via a PCIe interface.
Some boards also provide one or more high-speed on-board Ethernet connections for faster packet interface.
External Ethernet ports support speeds from \SI{10}{\giga\bit\per\s} to \SI{100}{\giga\bit\per\s} using standard QSFP28 cages although some systems offer lower-speed (\SI{1}{\giga\bit\per\s}) ports as well.
Most of these systems are also powerful enough to support running development tools natively, e.g., a P4 compiler, even though this is not necessarily required.

Tofino ASICs are also used in proprietary network switches, e.g., by Arista \cite{arista} and Cisco \cite{cisco}.
Some Tofino-based switches are supported by Microsoft SONiC \cite{sonic-devices}.

\subsubsection{Pensando Capri}
The Capri P4 Programmable Processor \cite{SeBa20,pensando} is an \ac{ASIC} that powers \acp{NIC} by Pensando Systems aimed for cloud providers.
It is coupled with fixed function components for cryptography operations like AES or compression algorithms and features multiple ARM cores.

\subsection{NPU-Based P4 Targets}
Network processing units (\acsp{NPU}) are software-programmable \acp{ASIC} that are optimized for networking applications.
They are part of standalone network devices or device boards, e.g., PCI cards.

Netronome \ac{NFP} silicons can be programmed with P4 \cite{netronome-p4} or C \cite{netronome-p4-c}.
A C-based programming model is available that supports program functions to access payloads and allows developing P4 externs.
The Agilio P4C \ac{SDK} consists of a tool chain including a backend compiler, host software, and a full-featured \ac{IDE}.
All current Agilio SmartNICs based on NFP-4000, NFP-5000, and NFP-6480 are supported.
Harkous et al. \cite{HaJa19} investigate the impact of basic P4 constructs on packet latency on Agilio SmartNICs.

%% file: chapters/06_p4_control_plane.tex
\section{P4 Data Plane \acsp{API}}
\label{sec:p4-data-plane-apis}

We introduce data plane \acp{API} for P4, present a characterization, describe the three most commonly used P4 data plane \acp{API}, and compare different control plane use cases.

\subsection{Definition \& Functionality}
\label{sec:control-plane-apis}
\begin{figure}[ht]
    \begin{center}
    \includegraphics[width=0.85\linewidth]{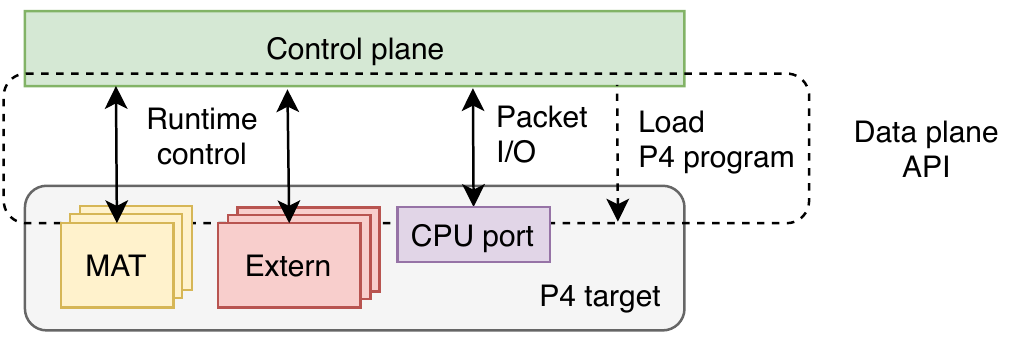}\\
    \end{center}
    \caption{Runtime management of a P4 target by the control plane through the data plane \ac{API}. The figure depicts the four most central operations: Runtime control of \acp{MAT} and extern objects, packet-in/out, and loading of P4 programs.}
    \label{fig:cp_api_concept}
\end{figure}

Control planes manage the runtime behavior of P4 targets via data plane \acp{API}.
Alternative terms are \emph{control plane \acp{API}} and \emph{runtime \acp{API}}.
The data plane \ac{API} is provided by a device driver or an equivalent software component.
It exposes data plane features to the control plane in a well-defined way.
Figure \ref{fig:cp_api_concept} shows the main control plane operations.
Most important, data plane \acp{API} facilitate runtime control of P4 entities (\acp{MAT} and externs).
They typically also comprise a packet I/O mechanism to stream packets to/from the control plane.
They also include reconfiguration mechanisms to load P4 programs onto the P4 target.
Control planes can control data planes only through data plane \acp{API}, i.e., if a data plane feature is not exposed via a corresponding \ac{API}, it cannot be used by the control plane.

It is important to note that P4 does not require a data plane \acp{API}.
P4 targets may also be used as a packet processor with a fixed behavior that is defined by the P4 program where static \ac{MAT} entries are part of the P4 program itself.

\subsection{Characterization of Data Plane APIs}
Data plane \acp{API} in P4 can be characterized by their level of abstraction, their dependency on the P4 program, and the location of the control plane.

\subsubsection{Level of Abstraction}
Data plane \acp{API} can be characterized by their level of abstraction.

\begin{itemize}
    \item \emph{Device access \acp{API}} provide direct access to hardware functionalities like device registers or memories. They typically use low-level mechanisms like DMA transactions. While this results in very low overhead, this type of \ac{API} can be neither vendor- nor device-independent.
    \item \emph{Data plane specific \acp{API}} are \acp{API} with a higher level of abstraction. They provide access to objects defined by the P4 program instead of hardware-specific parts. In contrast to device access \acp{API}, vendor- and device-independence is possible for this type of \ac{API}.
\end{itemize}

\subsubsection{Dependency on the P4 Program}
Data plane \acp{API} can be characterized by their dependency on the P4 program.

\begin{itemize}
    \item \emph{Program-dependent \acp{API}} have a set of functions, data structures, and other names that are derived from the P4 program itself. Therefore, they depend on the P4 program and are applicable to this P4 program only. If the corresponding P4 program is changed, function names, data structures, etc., might change, which requires a recompilation or modification of the control plane program.
    \item \emph{Program-independent \acp{API}} consist of a fixed set of functions that receives a list of P4 objects that are defined in the P4 program. Thus, the names of the \ac{API} functions, data structures, etc., do not depend on the program and are universally applicable. If the corresponding P4 program changes, neither the names, nor the definitions of the \ac{API} functions will change as long as the control plane “knows” the names of the right tables, fields and other object that need to be operated on. Program-independent \acp{API} model configurable objects either with the \emph{object-based} or the \emph{table-based} approach. As known from object-oriented programming, the object-based approach relies on methods that are defined for each class of data plane objects. In contrast, the table-based approach treats every class of data plane object as a variation of a table. This reduces the number of \ac{API} methods as only table manipulations need to be provided as methods.
\end{itemize}

\subsubsection{Control Plane Location}
Data plane \acp{API} can be characterized by the location of the control plane.

\begin{itemize}
    \item \emph{\acp{API} for local control} are implemented by the device driver and are executed on the local CPU of the device that hosts the programmable data plane. Usually, the \acp{API} are presented as set of C function calls just like for other devices that operating system are accessing.
    \item \emph{\acp{API} for remote control} add the ability to invoke \ac{API} calls from a separate system. This increases system stability and modularity, and is essential for SDN and other systems with centralized control. Remote control \acp{API} follow the base methodology of \acp{RPC} but rely on modern message-based frameworks that allow asynchronous communication and concurrent calls to the \ac{API}. Examples are Thrift \cite{apache-thrift} or gRPC \cite{grpc}. For example, gRPC uses HTTP/2 for transport and includes many functionalities ranging from access authentication, streaming, and flow control. The protocol's data structures, services, and serialization schemes are described with protocol buffers (protobuf) \cite{protobuf}.
\end{itemize}

\subsection{Data Plane API Implementations}
We introduce the three most common data plane \acp{API}: P4Runtime, Barefoot Runtime Interface (BRI), and BM Runtime.
All of them are data-plane specific and program-independent.
Table \ref{tab:control-plane-apis} lists their properties that have been introduced before.

\subsubsection{P4Runtime API}
\label{sec:p4-runtime-api}

P4Runtime is one of the most commonly used data plane \acp{API} that is standardized in the \ac{API} \ac{WG} \cite{p4-api-wg} of the P4 Language Consortium.
For implementing the \ac{RPC} mechanisms, it relies on the gRPC framework with protobuf.
Its most recent specification v1.3.0 \cite{p4-runtime-specification} was published in December 2020.

\begin{figure}[ht]
    \begin{center}
    \includegraphics[width=.9\linewidth]{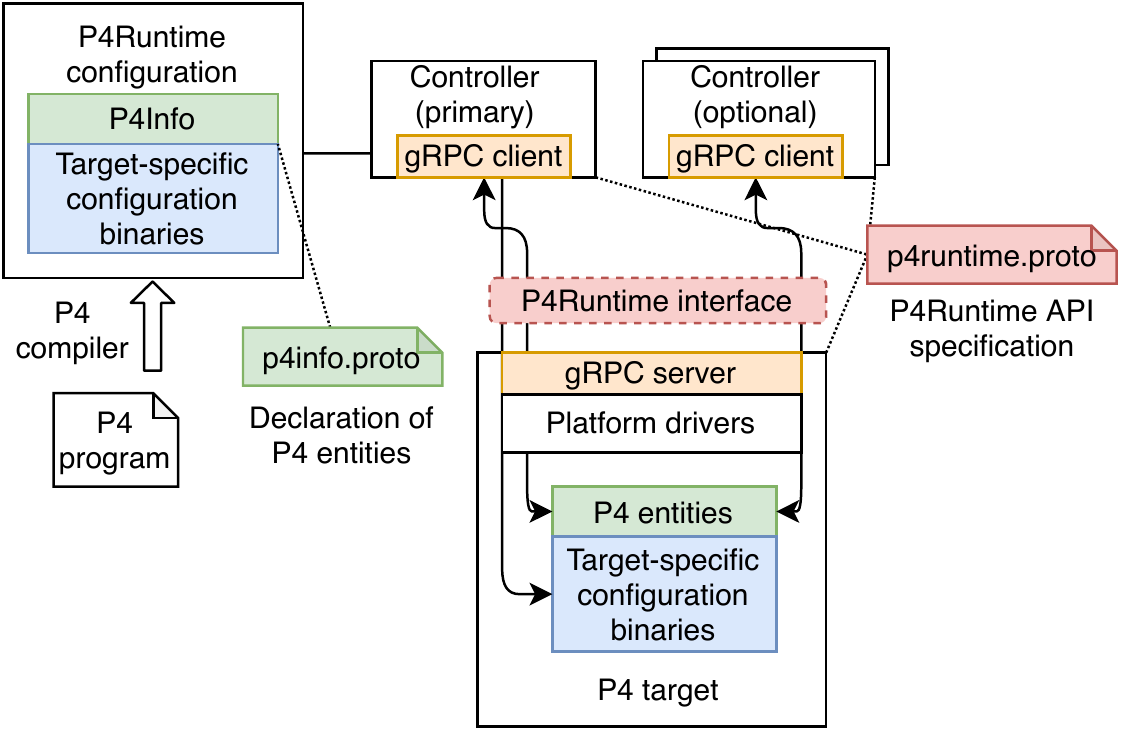}
    \end{center}
    \caption{P4Runtime architecture (similar to \cite{p4-runtime-specification}).}
    \label{fig:p4runtime}
\end{figure}

\paragraph{Operating Principle}
\fig{p4runtime} depicts the operating principle of P4Runtime.
P4 targets include a gRPC server, controllers implement a gRPC client.
To protect the gRPC connection, TLS with optional mutual certificate authentication can be enabled.
The \ac{API} structure of P4Runtime is described within the \texttt{p4runtime.proto} definition.
The gRPC server on P4 targets interacts with the P4-programmable components via platform drivers.
It has access to P4 entities (\acp{MAT} or externs) and can load target-specific configuration binaries.
The structure of the \ac{API} calls to access P4 entities are described in the \texttt{p4info.proto}.
It is part of the P4Runtime but developers can extend it to use custom data structures, e.g., to implement interaction with target-specific externs.
P4Runtime provides support for multiple controllers.
For every P4 entity, read access is provided to all controllers whereas write access is only provided to one controller.
To manage this access, P4 entities can be arranged in groups where each group is assigned to one primary controller with write access and arbitrary, secondary controllers with read access.
Interaction between controllers and P4 targets works as follows.
P4 compilers (see \sect{p4-compiler}) with support for P4Runtime generate a P4Runtime configuration.
It consists of the target-specific configuration binaries and P4Info metadata.
P4Info describes all P4 entities (\acp{MAT} and externs) that can be accessed by controllers via P4Runtime.
Then, the controllers establish a gRPC connection to the gRPC server on the P4 target.
The target-specific configuration is loaded onto the P4 target and P4 entities can be accessed.

\paragraph{Implementations}
gRPC and protobuf libraries are available for many high-level programming languages such as C++, Java, Go, or Python.
Thereby, P4Runtime can be implemented easily on both controllers and P4 targets.

\begin{itemize}
    \item \emph{Controllers}: P4Runtime is supported by most common \ac{SDN} controllers. P4 brigade \cite{onos} introduces support for P4Runtime on the Open Network Operating System (ONOS). OpenDaylight (ODL) introduces support for P4Runtime via a plugin \cite{opendaylight}. Stratum \cite{OcTs19} is an open-source network operating system that includes an implementation of the P4Runtime and OpenConfig interfaces. Custom controllers, e.g., for P4 prototypes, can be implemented in Python with the help of the p4runtime\_lib \cite{p4runtimelib}.
    \item \emph{Targets}: The \emph{PI Library} \cite{pi} is the open-source reference implementation of a P4Runtime gRPC server in C. It implements functionality for accessing \acp{MAT} and supports extensions for target-specific configuration objects, e.g., registers of a hardware P4 target. The PI Library is used by many P4 targets including \ac{bmv2} \cite{ssgrpc} and the Tofino.
\end{itemize}

\subsubsection{Barefoot Runtime Interface (BRI)}
The BRI consists of two independent \acp{API} that are available on Tofino-based P4 hardware targets.
The \emph{BfRt API} is an \ac{API} for local control.
It includes C, C++ and Python bindings that can be used to implement control plane programs.
The \emph{BF Runtime} is an \ac{API} for remote control.
As for P4Runtime, it is based on the gRPC \ac{RPC} framework and protobuf, i.e., bindings for different languages are available.
An additional Python library implements a simpler, BfRt-like interface for cases where simplicity is more essential than the performance of BF Runtime.

\subsubsection{BM Runtime API}
BM Runtime API is a program-independent data plane API for the \ac{bmv2} software target.
It relies on the Thrift \ac{RPC} framework.
\ac{bmv2} includes a \ac{CLI} program \cite{bmv2-cli} to manipulate \acp{MAT} and configure the multicast engine of the \ac{bmv2} P4 software target via this \ac{API}.

\begin{table}[ht]
\caption{Characterization of data plane specific \acp{API}.}
\begin{center}
\begin{tabularx}{\linewidth}{@{} l p{4em} X X @{}}
\toprule
\textbf{API} & \textbf{Program independence} & \textbf{Control plane location}\\
\midrule
P4Runtime & \checkmark & Remote (gRPC)\\
BF Runtime & \checkmark & Remote (gRPC)\\
BfRt API & \checkmark & Local (C, C++ and Python bindings)\\
BM Runtime & \checkmark & Remote (Thrift RPC)\\
\bottomrule
\end{tabularx}
\end{center}
\label{tab:control-plane-apis}
\end{table}

\subsection{Controller Use Case Patterns}
We present three use case patterns which are abstractions of the controller use cases introduced in the P4Runtime specification \cite{p4-runtime-specification}.
However, these are neither conclusive nor complete as derivations or extensions are possible.

\subsubsection{Embedded/Local Controller}
\begin{figure}[ht]
    \begin{center}
    \includegraphics[width=.6\linewidth]{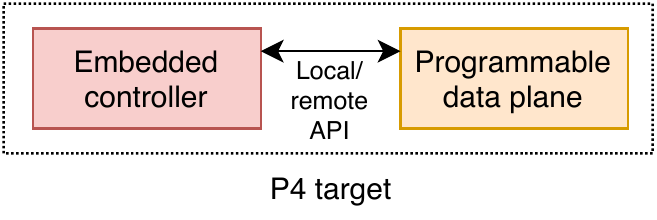}
    \end{center}
    \caption{Embedded/local controller use case pattern. The P4 target comprises an embedded controller that is running a control plane program.}
    \label{fig:embedded-controller}
\end{figure}

P4 hardware targets (see \sect{p4-targets}) comprise or are attached to a computing platform.
This facilitates running controllers directly on the P4 target.
\fig{embedded-controller} depicts this setup.
The controller application may either use a local \ac{API}, e.g., C calls, or just execute a controller application that interfaces the data plane via an \ac{RPC} channel.

\subsubsection{Remote Controllers}
\begin{figure}[ht]
    \begin{center}
    \includegraphics[width=.9\linewidth]{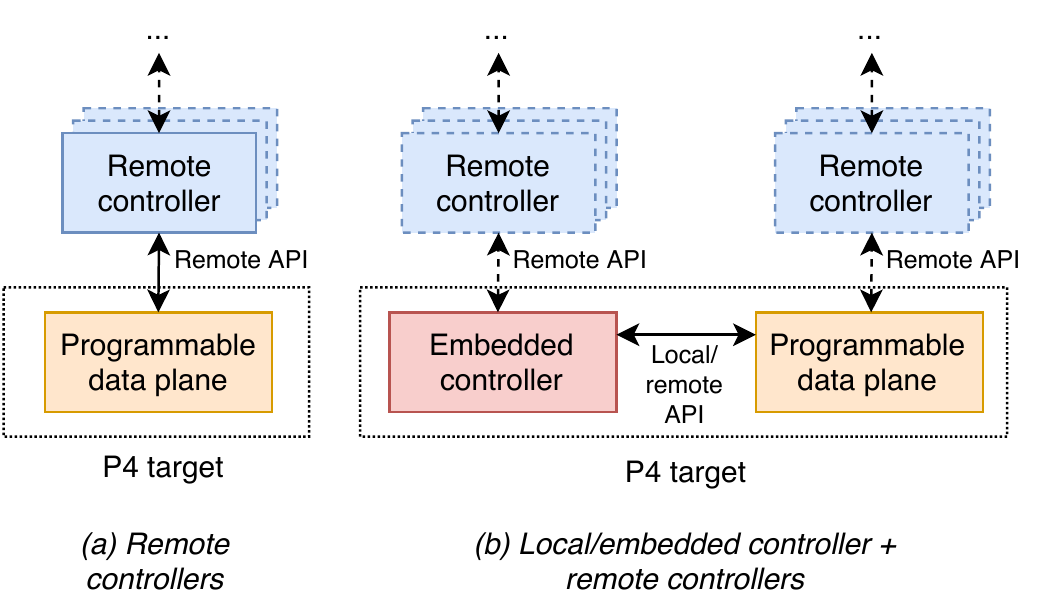}
    \end{center}
    \caption{Remote controller use case pattern.}
    \label{fig:remote-control}
\end{figure}

Remote controllers resemble the typical \ac{SDN} setup where data plane devices are managed by a centralized control plane with an overall view on the network.
Controllers need to be protected against outages and capacity overload, i.e., they need to be replicated for fail-safety and scalability.
\fig{remote-control} depicts two possible use cases.
In the first shown use case (a), the programmable data plane on the P4 target is managed by remote controllers.
In the second shown use case (b), the P4 target is managed by both, the embedded controller and remote controllers.
Remote controllers might be interfaced using the remote \ac{API} of the programmable data plane or an arbitrary \ac{API} that is provided by the embedded controller.
This option is often used for the implementation of so-called \emph{hierarchical control plane} structures where control plane functionality is distributed among different layers.
Control plane functions that do not require a global view of the network, e.g., link discovery, MAC learning for L2 forwarding, or port status monitoring, can be solely performed by the embedded/local controller. 
Other control plane functions that require an overall view of the network, e.g., routing applications, can be performed by the remote controller, possibly in cooperation with the embedded/local controller where the local controller acts as proxy, i.e., it relays control plane messages between the P4 target and the global controller.
Hierarchical control planes improve load distribution as many tasks can be performed locally, which reduces load on the remote controllers.
In particular, time-critical operations may benefit from local controllers as additional delays caused by the communication between a P4 target and a global controller are avoided.

%% file: chapters/07_p4_further_development.tex
\section{Advances in P4 Data Plane Programming}
\label{sec:development-debugging-testing}
\begin{figure}[htp]
    \begin{center}
    \includegraphics[width=.85\linewidth]{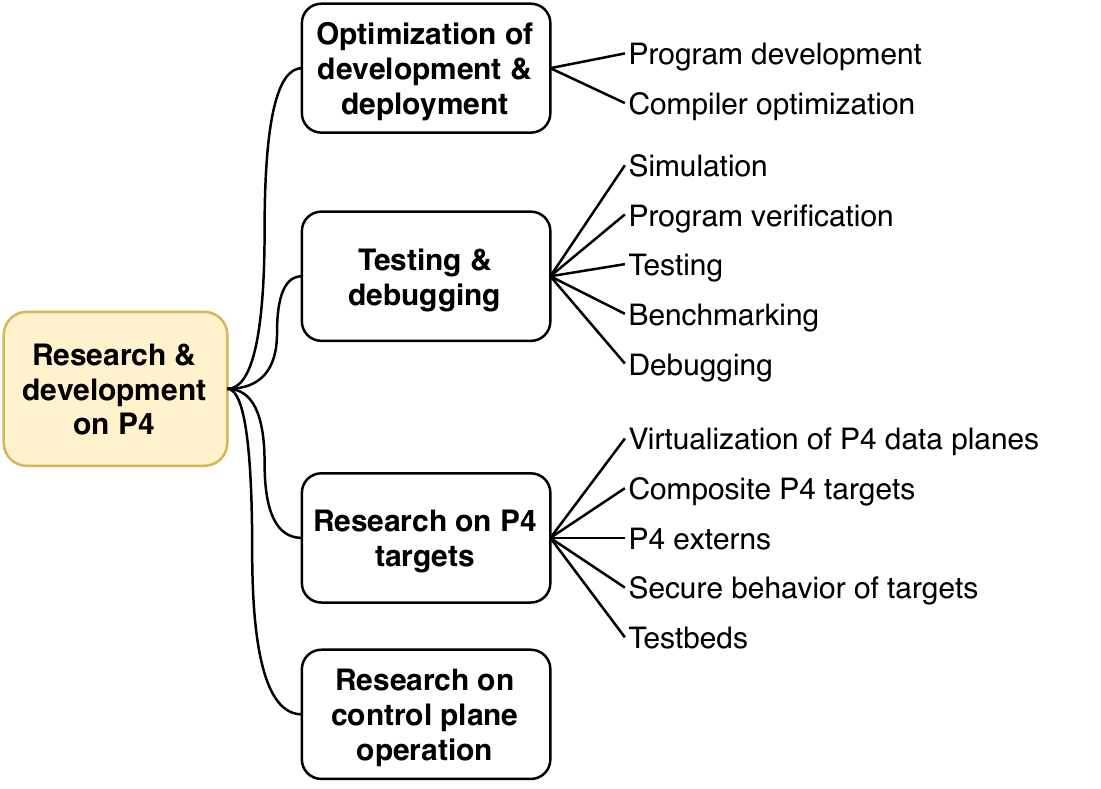}
    \end{center}
    \caption{Organization of Section \ref{sec:development-debugging-testing}.}
    \label{fig:organisation_reserach}
\end{figure}

We give an overview on research to improve P4 data plane programming.
\fig{organisation_reserach} depicts the structure of this section.
We describe related work on optimization of development and deployment, testing and debugging, research on P4 targets, and research on control plane operation.

\subsection{Optimization of Development and Deployment}
We describe research work on optimizing the development \& deployment process of P4.

\subsubsection{Program Development}
Graph-to-P4 \cite{ZaZh19} generates P4 program code for given parse graphs.
This introduces a higher abstraction layer that is particularly helpful for beginners.
Zhou et al. \cite{ZhBi17} introduce a module system for P4 to improve source code organization.
DaPIPE \cite{Ba19} enables incremental deployment of P4 program code on P4 targets.
SafeP4 \cite{EiCa19} adds type safety to P4.
P4I/O \cite{RiKu19} presents a framework for intent-based networking with P4.
Network operator describe their network functions with an \ac{IDL} and P4I/O generates a complete P4 program accordingly. 
To that end, P4I/O provides a P4 action repository with various network functions.
During reconfiguration, table and register state are preserved by applying backup mechanisms. 
P4I/O is implemented for a custom \ac{bmv2}.
Mantis \cite{YuSo20} is a framework to implement fast reactions to changing network conditions in the data plane without controller interaction.
To that end, annotations in the P4 code specify dynamic components and a quick control loop of those components ensure timely adjustments if necessary.
Lyra \cite{GaZh20} is a pipeline abstraction that allows developers to use simple statements to describe their desired data plane without low-level target-specific knowledge.
Lyra then compiles that description to target-specific code for execution.
GP4P4 \cite{RiOo19} is a programming framework for self-driven networks.
It generates P4 code from behavioral rules defined by the developer.
To that end, GP4P4 evaluates the quality of the automatically generated programs and improves them based on genetic algorithms.
FlowBlaze.p4 \cite{MoSa20, MoSa20b, MoSa20c} implements an executor for FlowBlaze, an abstraction based on an extended finite state machine for building stateful packet processing functions, in P4.
This library maps FlowBlaze elements to P4 components for execution on the \ac{bmv2}.
It also provides a GUI for defining the extended finite state machine.
Flightplan \cite{SuSo21} is a programming tool chain that disaggregates a P4 program into multiple P4 programs so that they can be executed on different targets.
The authors state that this improves performance, resource utilization, and cost.

\subsubsection{Compiler Optimization}
pcube \cite{ShSh18} is a preprocessor for P4 that translates primitive annotations in P4 programs into P4 code for common operations such as loops.
CacheP4 \cite{MaBi17} introduces a behavior-level cache in front of the P4 pipeline.
It identifies flows and performs a compound of actions to avoid unnecessary table matches.
The cache is filled during runtime by a controller that receives notifications from the switch.
P5 \cite{AbLe17} optimizes the P4 pipeline by removing inter-feature dependencies.
dRMT \cite{ChFi17} is a new architecture for programmable switches that introduces deterministic throughput and latency guarantees.
Therefore, it generates schedules for CPU and memory resources from a P4 program.
P2GO \cite{WiAp20} leverages monitored traffic information to optimize resource allocation during compilation.
It adjusts table and register size to reduce the pipeline length, and offloads rarely used parts of the program to the control plane. 
Yang et al. \cite{YaBa20} propose a compiler module that optimizes lookup speed by reorganizing flow tables and prioritization of popular forwarding rules.
Vass et al. \cite{VaBe20} analyze and discuss algorithmic aspects of P4 compilation.

\subsection{Testing and Debugging}
We describe research work on simulation, program verification, testing, benchmarking, and debugging.

\subsubsection{Simulation}
PFPSim \cite{AbAf16} is a simulator for validation of packet processing in P4.
NS4 \cite{BaBi18, FaBi17} is a network simulator for P4 programs that is based on the network simulator NS3.

\subsubsection{Program Verification}
McKeown et al. \cite{McTa16} introduce a tool to translate P4 to the Datalog declarative programming language.
Then, the Datalog representation of the P4 program can be analyzed for well-formedness. 
Kheradmand et al. \cite{KhRo18} introduce a tool for static analysis of P4 programs that is based on formal semantics.
P4v \cite{LiHa18} adapts common verification methods for P4 that are based on annotations in the P4 program code. 
Freire et al. \cite{FrNe18, NeFr18} introduce assertion-based verification with symbolic execution.
Stoenescu et al. \cite{StDu17} propose program verification based on symbolic execution in combination with a novel description language designed for the properties of P4. 
P4{\small{AIG}} \cite{NoHs19} proposes to use hardware verification techniques where developers have to annotate their code with First Order Logic (FOL) specifications.
P4{\small{AIG}} then encodes the P4 program as an Advanced-Inverter-Graph (AIG) which can be verified by hardware verification techniques such as circuit SAT solvers and bounded model checkers.
bf4 \cite{DuSt20} leverages static code verification and runtime checks of rules that are installed by the controller to confirm that the P4 program is running as intended. 
netdiff \cite{DuSt19} uses symbolic execution to check if two data planes are equivalent.
This can be useful to verify if a data plane behaves correctly by comparing it with a similar one, or to verify that optimizations of a data plane do not change its behavior.
Yousefi et al. \cite{YoAb20} present an abstraction for liveness verification of stateful \acp{NF}.
The abstraction is based on boolean formulae.
Further, they provide a compiler that translates these formulae into P4 programs.

\subsubsection{Testing}
P4pktgen \cite{NoKh18} generates test cases for P4 programs by creating test packets and table entries.
P4Tester \cite{ZhBi19b} implements a detection scheme for runtime faults in P4 programs based on probe packets. 
P4app \cite{p4app} is a partially automated open source tool for building, running, debugging, and testing P4 programs with the help of Docker images.
P4RL \cite{ShuHu19} is a reinforcement learning based system for testing P4 programs and P4 targets at runtime.
The correct behavior is described in a simple query language so that a reinforcement agent based on Double DQN can learn how to manipulate and generate packets that contradict the expected behavior.
P4TrafficTool \cite{JiJo19} analyzes P4 programs to produce plugin code for common traffic analyzers and generators such as Wireshark.

\subsubsection{Benchmarking}
Whippersnapper \cite{DaWa17} is a benchmark suite for P4 that differentiates between platform-independent and platform-specific tests.
BB-Gen \cite{RoPa18} is a system to evaluate P4 programs with existing benchmark tools by translating P4 code into other formats.
P8 \cite{HaJa20a} estimates the average packet latency at compilation time by analyzing the data path program.

\subsubsection{Debugging}
Kodeswaran et al. \cite{KoAr20} propose to use Ball-Larus encoding to track the packet execution path through a P4 program for more precise debugging capabilities.
p4-data-flow \cite{BiSi20} detects bugs by creating a control flow graph of a P4 program and then identifies incorrect behavior.
P4box \cite{NeHu19} extends the \psix reference compiler by so-called \emph{monitors} that insert code before and after programmable blocks, e.g., control blocks, for runtime verification.
P4DB \cite{ZhBi17c} \cite{ZhBi19c} introduces a runtime debugging system for P4 that leverages additional debugging snippets in the P4 program to generate reports during runtime.
Neves et al. \cite{NeLe17} propose a sandbox for P4 data plane programs for diagnosis and tracing.
P4Consist \cite{ShFa20} verifies the consistency between control and data plane.
Therefore, it generates active probe-based traffic for which the control and data plane generate independent reports that can be compared later. 
KeySight \cite{XiBi18} is a troubleshooting platform that analyzes network telemetry data for detecting runtime faults.
Gauntlet \cite{RuWa20} finds both crash bugs, i.e., abnormal termination of compilation operation, and semantic bugs, i.e., miscompilation, in compilers for programmable packet processors.

\subsection{Research on P4 Targets}
We describe research work on virtualization of P4 data planes, composite targets, P4 externs, secure behavior of targets, and testbeds.

\subsubsection{Virtualization of P4 Data Planes}
P4 targets are designed to execute one P4 program at any given time.
Virtualization aims at sharing the resources of P4 targets for multiple P4 programs.
Krude et al. \cite{KrHo19} provide theoretical discussions on how \ac{ASIC}- and \ac{FPGA}-based P4 targets can be shared between different tenants and how P4 programs can be made hot-pluggable.

HyPer4 \cite{HaMe16} introduces virtualization for P4 data planes.
It supports scenarios such as network slicing, network snapshotting, and virtual networking.
To that end, a compiler translates P4 programs into table entries that configure the HyPer4 \emph{persona}, a P4 program that contains implementations of basic primitives.
However, HyPer4 does not support stateful memory (registers, counters, meters), \ac{LPM}, range match types, and arbitrary checksums.
The authors describe an implementation for \ac{bmv2} and perform experiments that reveal 80 to 90\% lower performance in comparison to native execution.

HyperV \cite{ZhBi17a, ZhBi17b, p4-hypervdp-repo} is a hypervisor for P4 data planes with modular programmability.
It allows isolation and dynamic management of network functions.
The authors implemented a prototype for the \ac{bmv2} P4 target.
In comparison to Hyper4, HyperV achieves a 2.5x performance advantage in terms of bandwidth and latency while reducing required resources by a factor of 4.
HyperVDP \cite{ZhBi19} extends HyperV by an implementation of a dynamic controller that supports instantiating network functions in virtual data planes.

P4VBox \cite{SaBu20}, also published as VirtP4 \cite{SaBu19}, is a virtualization framework for the NetFPGA SUME P4 target.
It allows executing virtual switch instances in parallel and also to hot-swap them.
In contrast to HyPer4, HyperV and HyperVDP, P4VBox achieves virtualization by partially re-configuring the hardware.

P4Visor \cite{ZhBe18} merges multiple P4 programs. 
This is done by program overlap analysis and compiler optimization. 
Programming In-Network Modular Extensions (PRIME) \cite{PaCa20a} also allows combining several P4 programs to a single program and to steer packets through the specific control flows.

P4click \cite{ZaFr20a} does not only merge multiple P4 programs, but also combines the corresponding control plane blocks. 
The purpose of P4click is to increase the use of data plane programmability. 
P4click is currently in an early stage of development.

The Multi Tenant Portable Switch Architecture (MTPSA) \cite{StZi20} is a P4 architecture that offers performance isolation, resource isolation, and security isolation in a switch for multiple tenants.
MTPSA is based on the PSA.
It combines a \emph{Superuser} pipeline that acts as a hypervisor with multiple user pipelines.
User pipelines may only perform specific actions depending on their privileges.
MTPSA is implemented for \ac{bmv2} and NetFPGA-SUME \cite{mtpsa-repo}.

Han et al. \cite{HaJa20b} provide an overview of virtualization in programmable data planes with a focus on P4.
They classify virtualization schemes into hypervisor and compiler-based approaches, followed by a discussion of pros and cons of the different schemes.
The aforementioned works on virtualization of P4 data planes are described and compared in detail.

\subsubsection{Composite P4 Target}
Da Silva et al. \cite{SiSt18} introduce the idea of composite P4 targets.
This tries to solve the problem of target-dependent support of features.
The composed data plane appears as one P4 target; it is emulated by a P4 software target but relies on an \ac{FPGA} and \ac{ASIC} for packet processing.

eXtra Large Table (XLT) \cite{BeKr20} introduces gigabyte-scale \acp{MAT} by leveraging \ac{FPGA} and DRAM capabilities.
It comprises a P4-capable \ac{ASIC} and multiple \acp{FPGA} with DDR4 DRAM.
The P4-capable \ac{ASIC} pre-constructs the match key field and sends it with the full packet to the \ac{FPGA}.
The \ac{FPGA} sends back the original packet with the search results of the \ac{MAT} lookup.
The authors implement a \ac{DPDK} based prototype for the T\textsubscript{4}P\textsubscript{4}S P4 software target.

HyMoS \cite{AgXu17} is a hybrid software and hardware switch to support NFV applications.
The authors create a switch by using P4-enabled Smart \acp{NIC} as line cards and the PCIe interface of a computer as the switch fabric.
P4 is used for packet switching between the \acp{NIC}.
Additional processing may be done using \ac{DPDK} or applications running on a GPU.

\subsubsection{P4 Externs}
Laki et al. \cite{LaHo20, HoVo19} investigate asynchronous execution of externs.
In contrast to common synchronous execution, other packets may be processed by the pipeline while the extern function is running.
The authors implement and evaluate a prototype for T4P4S.
Scholz et al. \cite{ScOe19} propose that P4 targets should be extended by cryptographic hash functions that are required to build secure applications and protocols.
The authors propose an extension of the \ac{PSA} and discuss the PoC implementation for a CPU-, \ac{NPU}-, and \ac{FPGA}-based P4 target.
Da Silva et al. \cite{SiBo18b} investigate the implementation of complex operations as extensions to P4.
The authors perform a case study on integrating the Robust Header Compression (ROHC) scheme and conclude that an implementation as extern function is superior to an implementation as a new native primitive.

\subsubsection{Secure Behaviour of Targets}
Gray et al. \cite{GrGr19} demonstrate that hardware details of P4 targets influence their packet processing behavior.
The authors demonstrate this by sending a special traffic pattern to a P4 firewall.
It fills the cache of this target and results in a blocking behavior although the overall data rate is far below the capacity of the used P4 target.
Dumitru et al. \cite{DuDu20} investigate the exploitation of programming bugs in \ac{bmv2}, P4-NetFPGA, and Tofino.
The authors demonstrate attack scenarios by header field access on invalid headers, the creation of infinite loops and unintentionally processing of dropped packets in the P4 targets.

\subsubsection{Testbeds}
Large testbeds facilitate research and development on P4 programs.
The i-4PEN (International P4 Experimental Networks) \cite{MaCh19} is an international P4 testbed operated by a collaboration of network research institutions from the USA, Canada, and Taiwan.
Chung et al.\cite{ChTs19} describe how multi-tenancy is achieved in this testbed.
The 2STiC testbed \cite{2stic}, a national testbed in the Netherlands comprising six sites with at least one Tofino-based P4 target, is connected to i-4PEN.

\subsection{Research on Control Plane Operation}
When new forwarding entries are computed by the controller, the data plane has to be updated.
However, updating the targets has to be performed in a manner that prevents negative side effects.
For example, microloops may occur if packets are forwarded according to new rules at some targets while at other devices old rules are used because updates have to arrive yet.

Sukapuram et al. \cite{SuBa19, SuBa19b} introduce a timestamp in the packet header that contains the sending time of a packet.
When switches receive a packet during an update period, they compare the timestamp of both the packet and the update to determine whether a packet has been sent before the update, and thus, old rules should be used for forwarding.

Liu et al. \cite{LiBe19} introduce a mechanism where once a packet is matched against a specific forwarding rule, it cannot be matched downstream on a rule that is older. 
To that end, the packet header contains a timestamp field that records when the last applied forwarding rule has been updated.
If the packet is matched against an older rule, the packet is dropped, otherwise the timestamp is updated and the packet is forwarded.

Ez-Segway \cite{NgCh17} facilitates updating by including data plane devices in the update process.
When a data plane device receives an update, it determines which of its neighbors is affected by the update as well, and forwards the update to that neighbor. 
This prevents loops and black holes.

TableVisor \cite{GeHe19} is a transparent proxy-layer between the control plane and data plane. 
It provides an abstraction from heterogeneous data plane devices.
This facilitates the configuration of data plane switches with different properties, e.g., forwarding table size.

Molero et al. \cite{MoVi18} propose to offload tasks from the control plane to the data plane.
They show that programmable data planes are able to run typical control plane operations like failure detection and notification, and connectivity retrieval. 
They discuss trade-offs, limitations and future research opportunities. 

%% file: chapters/08-use-cases.tex
\section{Applied Research Domains: Classification \& Overview}
\label{sec:use-cases}
In the following sections, we give an overview of applied research conducted with P4.
In this section, we classify P4's core features that make it attractive for the implementation of data plane algorithms.
We define research domains, visualize them in a compact way, and explain our method to review corresponding research papers in the subsequent sections.
Finally, we delimit the scope of the surveyed literature.

\begin{figure*}[ht]
    \makebox[\textwidth][c]{
      \includegraphics[width=50em]{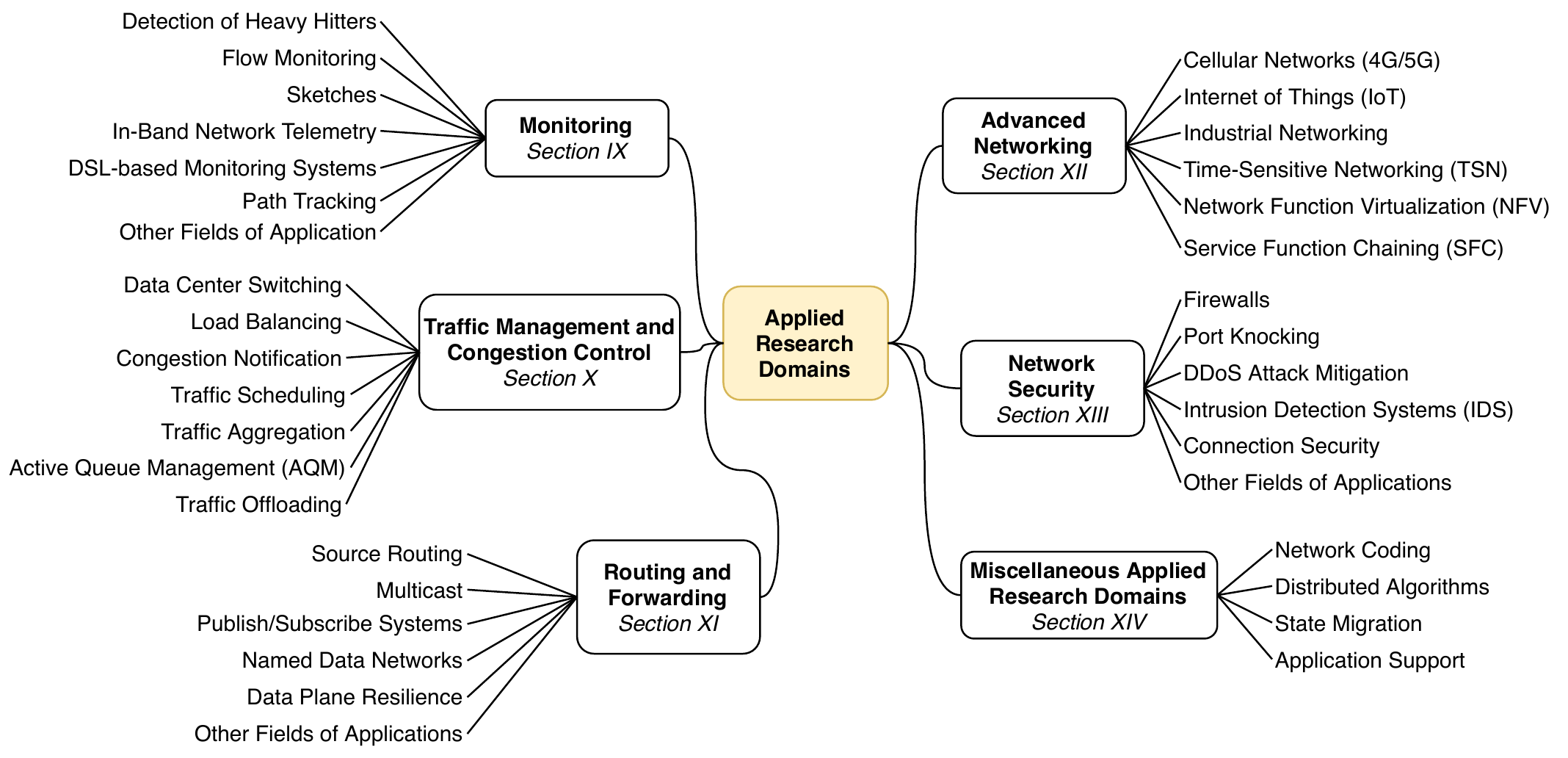}
    }
     \caption{Categorization of the surveyed works into applied research domains and subdomains -- they correspond to sections and subsections in the remainder of this paper.}
    \label{fig:use-cases-overview}
\end{figure*}

\subsection{Classification of P4's Core Features}
\label{sec:benefits}

We identify P4's core features for the implementation of prototypes.
We classify them in the following to effectively reason about P4's usefulness for the surveyed research works.

\subsubsection{Definition and Usage of Custom Packet Headers}
P4 requires the definition of packet headers (\sect{p4-parser}). These may be headers of standard protocols, e.g., TCP, use-case-specific protocols, e.g., 
\acs{GTP} in 5G, or new protocols.
As P4 supports the definition of custom headers, it is suitable for the implementation of data plane algorithms using new protocols or extensions of existing protocols, e.g., for in-band signalling.

\subsubsection{Flexible Packet Header Processing}
Control blocks with \acp{MAT} (\sect{p4-controls}) comprise the packet processing logic.
Packet processing includes default actions, e.g., forwarding and header field modifications, or custom, user-defined actions.
Both may be parameterized via \acp{MAT} or metadata.
Entries in the \acp{MAT} are maintained by a data plane \ac{API} (\sect{p4-data-plane-apis}). The flexible use of actions, the definition of new actions, and their parameterization offer high flexibility for header processing, which is often needed for research prototypes.

\subsubsection{Target-Specific Packet Header Processing Functions}
While the above-mentioned features are part of the P4 core language and supported by any P4-capable platform, devices may offer additional architecture- or target-specific  functionality which is made available as \emph{P4 extern} (\sect{p4-architectures-compilers}). 
Typical externs include components for stateful processing, e.g., registers or counters, operations to resubmit/recirculate the packet in the data plane, multicast operations, or more complex operations, e.g., hashing and encryption/decryption.
P4 software targets allow users to integrate custom externs and use them within P4 programs.
While this is also possible to some extent on some P4 hardware targets, e.g., the NetFPGA SUME board, high-throughput P4 targets based on the Tofino \ac{ASIC} have only a fixed set of externs (\sect{tofino}). Depending on the use case, the availability of externs may be essential for the implementation of prototypes. Thus, externs facilitate the implementation of more complex algorithms but make implementations platform-dependent.

\subsubsection{Packet Processing on the Control Plane}
Similar to control plane \ac{SDN} (e.g., \ac{OF}), more complex, and optionally centralized packet processing can be outsourced to an \ac{SDN} control plane; packet exchange and data plane control is performed via a data plane \ac{API} (\sect{p4-data-plane-apis}).
While \ac{OF} only allows the exchange of complete packets, P4 enables the end-users to define the packet formats.

\subsubsection{Flexible Development and Deployment}
Users are able to easily change the P4 programs on P4 targets that are installed in a network.
This facilitates agile development with frequent deployments and incremental functionality extensions by deploying new versions of a P4 programs.

\subsection{Categorization of Research Domains}
To organize the survey in the following sections, we define research domains and structure them in a two-level hierarchy as depicted in \fig{use-cases-overview}.
This categorization helps the reader to get a quick overview in certain applied areas and improves the readability of this survey.
The choice of the research domains is dominated by the fields of applications, but the summaries of the sections will show that the prototypes in these areas benefit from different core features of P4.

For each research domain, we provide a table that lists the publications with publication year, P4 target platforms, and source code availability.
This supports efficient browsing of the content and backs our conclusions in the section-specific summaries.

\subsection{Scope of the Surveyed Literature}
We consider the literature until the end of 2020 and selected papers from 2021, including journal papers, conference papers, workshop papers, and preprints.
Out of the \totalPapers scientific publications we surveyed in this work (see \sect{introduction}), \totalApplications fall in the area of applied research.
68 of those research papers were published in 2018 or before, 80 were published in 2019, 93 were published in 2020, and 4 were published in 2021.
60 out of all \totalApplications research publications released the source code of their prototype implementations.

\tbl{publication-statistics} depicts a statistic on major publication venues for the papers of applied research domains.
It helps the reader to identify potential venues for prospective own publications based on P4 technology.

\begin{table}[htp]
\caption{Statistics of scientific publications regarding applied research conducted with P4.}
\begin{center}
\begin{tabularx}{\linewidth}{X r}
\toprule
\textbf{Venue} & \textbf{\#Publications}\\
\midrule
\addlinespace
\textbf{Journals} & \textbf{41} \\
\addlinespace
\venueStat{IEEE ACCESS}{9}
\venueStat{IEEE/ACM ToN}{7}
\venueStat{IEEE TNSM}{6}
\venueStat{JNCA}{4}
\venueStat{Miscellaneous}{15}
\midrule
\addlinespace
\textbf{Conferences} & \textbf{168} \\
\addlinespace
\venueStat{ACM SOSR}{14}
\venueStat{IEEE NFV-SDN}{12}
\venueStat{IEEE ICNP}{12}
\venueStat{IEEE ICC}{10}
\venueStat{ACM SIGCOMM}{10}
\venueStat{IEEE/IFIP NOMS}{8}
\venueStat{ACM CoNEXT}{7}
\venueStat{IEEE NetSoft}{7}
\venueStat{USENIX NSDI}{6}
\venueStat{IEEE INFOCOM}{6}
\venueStat{ACM/IEEE ANCS}{5}
\venueStat{IFIP Networking}{5}
\venueStat{IEEE GLOBECOM}{4}
\venueStat{CNSM}{4}
\venueStat{IEEE CloudNet}{3}
\venueStat{APNOMS}{3}
\venueStat{IFIP/IEEE IM}{3}
\venueStat{Miscellaneous}{49}
\midrule
\addlinespace
\textbf{Workshops} & \textbf{36} \\
\addlinespace
\venueStat{EuroP4}{11}
\venueStat{Morning Workshop on In-Network Computing}{5}
\venueStat{SPIN}{3}
\venueStat{ACM HotNets}{3}
\venueStat{INFOCOM Workshops}{3}
\venueStat{Miscellaneous}{11}
\bottomrule
\end{tabularx}
\end{center}
\label{tbl:publication-statistics}
\end{table}

%% file: chapters/09-uc_monitoring.tex
\section{Applied Research Domains: Monitoring}
\label{sec:use-cases-monitoring}
We describe applied research on detection of heavy hitters, flow monitoring, sketches, in-band network telemetry, and other areas of application.
Table \ref{tab:implementations-monitoring} shows an overview of all the work described.
At the end of the section, we summarize the work and analyze it with regard to P4's core features described in \sect{benefits}.

\begin{table*}[htp]
\caption{Overview of applied research on monitoring (\sect{use-cases-monitoring}).}
\begin{tabularx}{\linewidth}{@{} l l X r @{}}
\toprule
\textbf{Research work} & \textbf{Year} & \textbf{Targets} & \textbf{Code}\\
\midrule
\addlinespace

\multicolumn{4}{@{\extracolsep{\fill}}l}{\noindent \textbf{Detection of Heavy Hitters} (\sect{use-cases-monitoring-heavy-hitters})} \\
\addlinespace
\paperTableSource{SiNa17}{2017}{HashPipe}{\ac{bmv2}}{p4-hashpipe}
\paperTable{LiHu19}{2019}{Lin et al.}{Tofino}
\paperTable{PoAn17}{2017}{Popescu et al.}{-}
\paperTable{HaCa18}{2018}{Harrison et al.}{Tofino}
\paperTable{KuPo20}{2020}{Kucera et al.}{\ac{bmv2}}
\paperTable{SiJa18}{2018}{IDEAFIX}{-}
\paperTable{TuOo19}{2019}{Turkovic et al.}{Netronome}
\paperTableSource{DiSa20b}{2020}{Ding et al.}{\ac{bmv2}}{p4-DiSa20b}
\midrule
\addlinespace

\multicolumn{4}{@{\extracolsep{\fill}}l}{\noindent \textbf{Flow Monitoring} (\sect{use-cases-monitoring-flows})}\\
\addlinespace
\paperTableSource{SoAv18}{2018}{TurboFlow}{Tofino, Netronome}{p4-turboflow}
\paperTableSource{SoMi18}{2018}{\textasteriskcentered\! Flow}{Tofino}{github-starflow}
\paperTable{HiAl18}{2018}{Hill et al.}{\ac{bmv2}}
\paperTable{CaPa19}{2019}{FlowStalker}{\ac{bmv2}}
\paperTable{PaCa20}{2020}{ShadowFS}{\ac{bmv2}}
\paperTableSource{BaSa21}{2021}{FlowLens}{\ac{bmv2}, Tofino}{BaSa21-repo}
\paperTable{WaTa20}{2020}{SpiderMon}{\ac{bmv2}}
\paperTable{ChLa19}{2019}{ConQuest}{Tofino}
\paperTable{ZhSh19}{2019}{Zhao et al.}{\ac{bmv2}, Tofino}
\midrule
\addlinespace

\multicolumn{4}{@{\extracolsep{\fill}}l}{\noindent \textbf{Sketches} (\sect{use-cases-monitoring-sketches})}\\
\addlinespace
\paperTableSource{HuLe18}{2018}{SketchLearn}{Tofino}{p4-sketchlearn}
\paperTableSource{TaHu20}{2020}{MV-Sketch}{\ac{bmv2}, Tofino}{p4-mv-sketch}
\paperTable{HaWe19}{2019}{Hang et al.}{Tofino}
\paperTable{LiMa16}{2016}{UnivMon}{p4c-behavioural}
\paperTableSource{YaJi18, YaJi19}{2018/19}{Yang et al.}{Tofino}{p4-ElasticSketch}
\paperTable{PeNe17}{2017}{Pereira et al.}{\ac{bmv2}}
\paperTable{MaVe18}{2018}{Martins et al.}{\ac{bmv2}}
\paperTable{LaSh19}{2019}{Lai et al.}{Tofino}
\paperTable{LiZh20}{2020}{Liu et al.}{Tofino}
\paperTableSource{TaHu20b}{2020}{SpreadSketch}{Tofino}{TaHu20b-repo}

\addlinespace

\bottomrule
\end{tabularx}
\label{tab:implementations-monitoring}
\end{table*}

\begin{table*}[htp]

\begin{tabularx}{\linewidth}{@{} l l X r @{}}
\toprule
\textbf{Research work} & \textbf{Year} & \textbf{Targets} & \textbf{Code}\\
\midrule
\addlinespace

\multicolumn{4}{@{\extracolsep{\fill}}l}{\noindent \textbf{In-Band Network Telemetry} (\sect{use-cases-monitoring-int})}\\
\addlinespace
\paperTable{VeKa19}{2019}{Vestin et al.}{Netronome}
\paperTable{WaCh19}{2019}{Wang et al.}{Tofino}
\paperTable{BhKa19}{2019}{IntOpt}{P4FPGA}
\paperTableSource{JiPa20}{2020}{Jia et al.}{\ac{bmv2}}{p4-JiPa20}
\paperTable{NiKo19}{2019}{Niu et al.}{Tofino, Netronome}
\paperTableSource{KaFi20}{2020}{CAPEST}{\ac{bmv2}}{p4-capest}
\paperTable{ChJa19}{2019}{Choi et al.}{\ac{bmv2}}
\paperTable{SgPa20}{2020}{Sgambelluri et al.}{\ac{bmv2}}
\paperTable{FePa20}{2020}{Feng et al.}{Netronome}
\paperTableSource{MaLe20}{2020}{IntSight}{bmv2, NetFPGA-SUME}{MaLe20-repo}
\paperTable{SuJa20}{2020}{Suh et al.}{-}
\midrule
\addlinespace

\multicolumn{4}{@{\extracolsep{\fill}}l}{\noindent \textbf{\acs{DSL}-Based Monitoring Systems} (\sect{use-cases-monitoring-dsl})}\\
\addlinespace
\paperTableSource{NaSi17, NaNa17}{2017}{Marple}{\ac{bmv2}}{p4-marple}
\paperTableSource{LaRo19}{2019}{MAFIA}{\ac{bmv2}}{p4-mafia}
\paperTableSource{GuHa18}{2018}{Sonata}{\ac{bmv2}, Tofino}{p4-sonata}
\paperTable{TeHa20}{2020}{Teixeira et al.}{\ac{bmv2}, Tofino}
\midrule
\addlinespace

\multicolumn{4}{@{\extracolsep{\fill}}l}{\noindent \textbf{Path Tracking} (\sect{use-cases-monitoring-path-tracking})}\\
\addlinespace
\paperTable{GaJi18}{2018}{UniRope}{\ac{bmv2}, PISCES}
\paperTable{KnHi19}{2019}{Knossen et al.}{Netronome}
\paperTable{BaRo20}{2020}{Basuki et al.}{\ac{bmv2}}
\midrule
\addlinespace

\multicolumn{4}{@{\extracolsep{\fill}}l}{\noindent \textbf{Other Areas of Application} (\sect{use-cases-monitoring-other})}\\
\addlinespace
\paperTableSource{JoQu18}{2018}{BurstRadar}{Tofino}{p4-burstradar}
\paperTable{GhBe17}{2017}{Dapper}{-}
\paperTable{HeCh18}{2018}{He et al.}{Tofino}
\paperTableSource{RiKi19}{2019}{Riesenberg et al.}{\ac{bmv2}}{p4-RiKi19}
\paperTable{WaHu20}{2020}{Wang et al.}{Tofino}
\paperTableSource{KuSi20}{2020}{P4STA}{\ac{bmv2}, Netronome}{p4-p4sta}
\paperTable{HaBh19}{2019}{Hark et al.}{-}
\paperTableSource{DiSa20}{2020}{P4Entropy}{\ac{bmv2}}{p4-p4entropy}
\paperTable{TaMe19}{2019}{Taffet et al.}{\ac{bmv2}}
\paperTable{LiZh20b}{2020}{NetView}{\ac{bmv2}, Tofino}
\paperTable{BaZh20}{2020}{FastFE}{Tofino}
\paperTable{KuBa20}{2020}{Unroller}{bmv2, Netcope P4-to-VHDL}
\paperTable{HaSh19}{2019}{Hang et al.}{Tofino}
\paperTable{GuSh19}{2019}{FlowSpy}{\ac{bmv2}}
\addlinespace

\bottomrule
\end{tabularx}

\end{table*}

\subsection{Detection of Heavy Hitters}
\label{sec:use-cases-monitoring-heavy-hitters}
Heavy hitters \cite{heavyhitters-slides} (or "elephant flows") are large traffic flows that are the major source of network congestion.
Detection mechanisms aim at identifying heavy hitters to perform extra processing, e.g., queuing, flow rate control, and traffic engineering.

HashPipe \cite{SiNa17} integrates a heavy hitter detection algorithm entirely on the P4 data plane.
A pipeline of hash tables acts as a counter for detected flows.
To fulfill memory constraints, the number of flows that can be stored is limited.
When a new flow is detected, it replaces the flow with the lowest count.
Thus, light flows are replaced, and heavy flows can be detected by a high count.
Lin et al. \cite{LiHu19} describe an enhanced version of the algorithm.

Popescu et al. \cite{PoAn17} introduce a heavy hitter detection mechanism.
The controller installs \ac{TCAM} entries for specific source IP prefixes on the switch.
If one of these entries matches more often than a threshold during a given time frame, the entry is split into two entries with a larger prefix size.
This procedure is repeated until the configured granularity is reached.

Harrison et al. \cite{HaCa18} presents a controller-based and distributed detection scheme for heavy hitters.
The authors make use of counters for the match key values, e.g., source and destination IP pair or 5-tuple, that are maintained by P4 switches.
If a counter exceeds a certain threshold, the P4 switch sends a notification to the controller.
The controller generates more accurate status reports by combining the notifications received from the switches.

Kucera et al. \cite{KuPo20} describe a system for detecting traffic aggregates.
The authors propose a novel algorithm that supports hierarchical heavy hitter detection, change detection, and super-spreader detection.
The complete mechanism is implemented on the P4 data plane and uses push notifications to a controller.

IDEAFIX \cite{SiJa18} is a system that detects elephant flows at edge switches of Internet exchange point networks.
The proposed system analyzes flow features, stores them with hash keys as indices in P4 registers, and compares them to thresholds for classification.

Turkovic et al. \cite{TuOo19} propose a streaming approach for detecting heavy hitters via sliding windows that are implemented in P4.
According to the authors, interval methods that are typically used to detect heavy hitters are not suitable for programmable data planes because of high hardware resources, bad accuracy, or a need for too much intervention by the control plane.

Ding et al. \cite{DiSa20b} propose an architecture for network-wide heavy hitter detection.
The authors' main focuses are hybrid \ac{SDN}/non-\ac{SDN} networks where programmable devices are deployed only partially.
To that end, they also present an algorithm for an incremental deployment of programmable devices with the goal of maximizing the number of network flows that can be monitored.

\subsection{Flow Monitoring}
\label{sec:use-cases-monitoring-flows}
In flow monitoring, traffic is analyzed on a per-flow level.
Network devices are configured to export per-flow information, e.g., packet counters, source and target IP addresses, ports, or protocol types, as flow records to a flow collector.
These flow records are often duplicates of network packets without payload data.
The flow collector then performs centralized analysis on this data.
The three most widely deployed protocols are Netflow \cite{rfc3954}, sFlow \cite{rfc3176}, and IPFIX \cite{rfc7011}.

TurboFlow \cite{SoAv18} is a flow record generator designed for P4 switches that does not have to make use of sampling or mirroring.
The data plane generates micro-flow records with information about the most recent packets of a flow.
On the CPU module of the switch, those micro-flow records are aggregated and processed into full flow records.

``\textasteriskcentered\! Flow'' \cite{SoMi18} partitions measurement queries between the data plane and a software component.
A switching ASIC computes grouped packet vectors that contain a flow identifier and a variable set of packet features, e.g. packet size and timestamps, while the software component performs aggregation.
``\textasteriskcentered\! Flow'' supports dynamic and concurrent measurement applications, i.e., measurement applications that operate on the same flows without impacting each other.

Hill et al. \cite{HiAl18} implement Bloom filters on P4 switches to prevent sending duplicate flow samples.
Bloom filters are a probabilistic data structure that can be used to check whether an entry is present in a set or not.
It is possible to add elements to that set, but it is not possible to remove entries from it.
For flow tracking, Bloom filters test if a flow has been seen before without control plane interaction.
Thereby, only flow data is forwarded to the collector from flows that were not seen before.

FlowStalker \cite{CaPa19} is a flow monitoring system running on the P4 data plane.
The monitoring operations on a packet are divided in two phases, a proactive phase that identifies a flow and keeps a per-flow packet counter and a reactive phase that runs for large flows only and gathers metrics of the flow, e.g., byte counts and packet sizes.
The controller gathers information from a cluster of switches by injecting a crawler packet that travels through the cluster at one switch.
ShadowFS \cite{PaCa20} extends FlowStalker with a mechanism to increase the throughput of the monitored flows.
It achieves this by dividing forwarding tables into two tables, a faster and a slower one.
The most utilized flows are moved to the faster table if necessary.

FlowLens \cite{BaSa21} is a system for traffic classification to support security network applications based on machine learning algorithms.
The authors propose a novel memory-efficient representation for features of flows called \emph{flow marker}.
A profiler running in the control plane automatically generates an application-specific flow marker that optimizes the trade-off between resource consumption and classification accuracy, according to a given criterion selected by the operator.

SpiderMon \cite{WaTa20} monitors network performance and debugs performance failures inside the network with little overhead.
To that end, SpiderMon monitors every flow in the data plane and recognizes if the accumulated latency exceeds a certain threshold. 
Furthermore, SpiderMon is able to trace back the path of interfering flows, allowing to analyze the cause of the performance degradation.

ConQuest \cite{ChLa19} is a data plane mechanism to identify flows that occupy large portions of buffers. 
Switches maintain snapshots of queues in registers to determine the contribution to queue occupancy of the flow of a received packet.

Zhao et al. \cite{ZhSh19} implement flow monitoring using hash tables.
Using a novel strategy for collision resolution and record promotion, accurate records for elephant flows and summarized records for other flows are stored.

\subsection{Sketches}
\label{sec:use-cases-monitoring-sketches}

Flow monitoring as described in \sect{use-cases-monitoring-flows} requires high sampling rates to produce sufficiently detailed data.
As an alternative, streaming algorithms process sequential data streams and are subject to different constraints like limited memory or processing time per item.
They approximate the current network status based on concluded summaries of the data stream.
The streaming algorithms output so-called sketches that contain summarized information about selected properties of the last $n$ packets of a flow.

SketchLearn \cite{HuLe18} is a sketch-based approach to track the frequency of flow records.
It features multilevel sketches that aim for small memory usage, fast per-packet processing, and real-time response.
Rather than finding the perfect resource configuration for measurement traffic and regular traffic, SketchLearn characterizes the statistical error of resource conflicts based on Gaussian distributions. 
The learned properties are then used to increase the accuracy of the approximated measurements.

Tang et al. \cite{TaHu20} present MV-Sketch, a fast and compact invertible sketch.
MV-Sketch leverages the idea of majority voting to decide whether a flow is a heavy hitter or heavy changer.
Evaluations show that MV-Sketch achieves a 3.38 times higher throughput than existing invertible sketches.

Hang et al. \cite{HaWe19} try to solve the problem of inconsistency when a controller needs to collect the data from sketches on one or more switches.
As accessing and clearing the sketches on the switches is always subject to latency, not all sketches are reset at the same time, and there might be some delay between accessing and clearing the sketches.
The authors propose to use two asymmetric sketches on the switches that are used in an interleaved way.
Furthermore, the authors propose to use a distributed control plane to keep latency low.

UnivMon \cite{LiMa16} is a flow monitoring system based on sketches.
After sampling the traffic, the data plane produces sketches and determines the top-$k$ heaviest flows by comparing the number of sketches for each flow.
Those flows are passed to the control plane which processes the data for the specific application. 

Yang et al. \cite{YaJi18, YaJi19} propose to adapt sketches according to certain traffic characteristics to increase data accuracy, e.g., during congestion or \ac{DDoS} attacks.
The mechanism is based on compressing and merging sketches when resources in the network are limited due to high traffic volume.
During periods with high packet rates, only the information of elephant flows is recorded to trade accuracy for higher processing speed. 

Pereira et al. \cite{PeNe17} propose a secured version of the Count-Min sketch.
They replace the function with a cryptographic hash function and provide a way for secret key renewal.

Martins et al. \cite{MaVe18} introduce sketches for multi-tenant environments.
The authors implement bitmap and counter-array sketches using a new probabilistic data structure called BitMatrix that consists of multiple bitmaps that are stored in a single P4 register.

Lai et al. \cite{LaSh19} use a sketch-based approach to estimate the entropy of network traffic.
The authors use CRC32 hashes of header fields as match keys for match-action tables and subsequently update k-dimensional data sketches in registers.
The content of the registers is then processed by the control plane CPU which calculates the entropy value.

Liu et al. \cite{LiZh20} use sketches for performance monitoring.
They introduce lean algorithms to measure metrics like loss or out-of-order packets.

SpreadSketch \cite{TaHu20b} is a sketch data structure to detect superspreaders.
The sketch data structure is invertible, i.e., it is possible to extract the identification of superspreaders from the sketch at the end of an epoch.

\subsection{In-Band Network Telemetry}
\label{sec:use-cases-monitoring-int}
Barefoot Networks, Arista, Dell, Intel and VMware specified \ac{INT} specifically for P4 \cite{KiBh16}.
It uses a pure data plane implementation to collect telemetry data from the network without any intervention by the control plane.
It was specified by 
INT is the main focus of the \emph{Applications \ac{WG}} \cite{p4-applications-wg} of the P4 Language Consortium.
Instructions for \ac{INT}-enabled devices that serve as traffic sources are embedded as header fields either into normal packets or into dedicated probe packets.
Traffic sinks retrieve the results of instructions to traffic sources.
In this way, traffic sinks have access to information about the data plane state of the \ac{INT}-enabled devices that forwarded the packets containing the instructions for traffic sources.
The authors of the \ac{INT} specification name network troubleshooting, advanced congestion control, advanced routing, and network data plane verification as examples for high-level use cases.

In two demos, \ac{INT} was used for diagnosing the cause of latency spikes during HTTP transfers \cite{KiSi15} and for enforcing QoS policies on a per-packet basis across a metro network \cite{CuGu19}.

Vestin et al. \cite{VeKa19} enhance \ac{INT} traffic sinks by event detection.
Instead of exporting telemetry items of all packets to a stream processor, exporting has to be triggered by an event.
Furthermore, they implement an \ac{INT} report collector for Linux that can stream telemetry data to a Kafka cluster.

Wang et al. \cite{WaCh19} design an \ac{INT} system that can track which rules in \acp{MAT} matched on a packet.
The resulting data is stored in a database to facilitate visualization in a web UI.

IntOpt \cite{BhKa19} uses \ac{INT} to monitor service function chains.
The system computes minimal monitoring flows that cover all desired telemetry demands, i.e., the number of INT-sources, sinks, and forwarding nodes that are covered by this flow is minimal.
IntOpt uses active probing, i.e., monitoring probes for the monitoring flows are periodically inserted into the network.

Jia et al. \cite{JiPa20} use \ac{INT} to detect gray failures in data center networks using probe packets.
Gray failures are failures that happen silently and without notification.

Niu et al. \cite{NiKo19} design a multilevel \ac{INT} system for IP-over-optical networks.
Their goal is to monitor both the IP network and the optical network at the same time.
To that end, they implement optical performance monitors for bandwidth-variable wavelength selective switches.
Their measurements can be queried by a P4 switch that is connected directly to it.

CAPEST \cite{KaFi20} leverages P4-enabled switches to estimate the network capacity and available bandwidth of network links.
The approach is passive, i.e., it does not disturb the network.
A controller sends \ac{INT} probe packets to trigger statistical analysis and export results.

Choi et al. \cite{ChJa19} leverage \ac{INT} for run-time performance monitoring, verification, and healing of end-to-end services.
P4-capable switches monitor the network based on \ac{INT} information and the distributed control plane verifies that SLAs and other metrics are fulfilled. 
They leverage metric dynamic logic (MDL) to specify formal assertions for SLAs.

Sgambelluri et at. \cite{SgPa20} propose a multi-layer monitoring system that uses an OpenConfig NETCONF agent for the optical layer an P4-based \ac{INT} for the packet layer.
In their prototype, they use \ac{INT} to measure the delay of packets by computing the processing time at each switch.

Feng et al. \cite{FePa20} implement an \ac{INT} sink for Netronome Smart \acp{NIC}.
After parsing the \ac{INT} headers using P4, they use algorithms written in C to perform \ac{INT} tasks like aggregation and notification.
Compared to a pure P4 implementation, this increases the performance.

IntSight \cite{MaLe20} is a system for detecting and analyzing violations of service-level objects (SLOs).
SLOs are performance guarantees towards a network, e.g., concerning bandwidth and latency.
IntSight uses INT to monitor the performance of the network during a specific period of time.
Egress devices gather this information and produce a report at the end of the period if an SLO has been violated.

Suh et al. \cite{SuJa20} explore how a sampling mechanism can be added to \ac{INT}.
Their solution supports rate-based and event-based sampling.
Based on these sampling strategies, \ac{INT} headers are only added to a fraction of the packets to reduce overhead.

\subsection{\acs{DSL}-Based Monitoring Systems}
\label{sec:use-cases-monitoring-dsl}
Monitoring tasks can often be broken down in a set of several basic operations, e.g., map, filter, or groupby.
A \ac{DSL} allows to combine these basic operations in more complex tasks.

Marple \cite{NaSi17, NaNa17} is a performance query language that supports existing constructs like map, filter, groupby, and zip.
A query compiler translates the queries either to P4 or to a simulator for programmable switch hardware.
Stateless constructs of the query language, e.g., filters, are executed on the data plane.
Stateful constructs, e.g., groupby, use a programmable key-value store that is split between a fast on-chip \ac{SRAM} cache and a large off-chip DRAM backing store.
The results are streamed from the switch to a collection server.

MAFIA \cite{LaRo19} is a \ac{DSL} to describe network measurement tasks.
They identify several fundamental primitive operations, examples are match, tag, timestamp, sketch, or counter.
MAFIA is a high-level language to describe more complex measurement tasks composed of those primitives.
The authors provide a Python-based compiler that translates MAFIA code into a P4 program in \pfour or \psix for a PISA-based P4 target.

Sonata \cite{GuHa18} is a query-driven telemetry system.
It provides a query interface that provides common operators like map and reduce that can be applied on arbitrary packet fields.
Sonata combines the capabilities of both programmable switches and stream processors.
The queries are partitioned between the programmable switches and the stream processors to reduce the load on the stream processors.
Teixeira et al. \cite{TeHa20} extend the Sonata prototype by functionalities to monitor the properties of packet processing inside switches, e.g., delay.

\subsection{Path Tracking}
\label{sec:use-cases-monitoring-path-tracking}
In path tracking, or packet trajectory tracing, information about the path a packet has taken in a network is gathered.

UniRope \cite{GaJi18} consists of two different algorithms for packet trajectory tracing that can be selected dynamically to be able to choose the trade-off between accuracy and efficiency.
These two algorithms are \emph{compact hash matching} and \emph{consecutive bits filling}.
With compact hash matching, the forwarding switch calculates a hash value and stores it in the packet.
With consecutive bits filling, the packet trajectory is recorded in the packet hop by hop and reconstructed at the controller.

Knossen et al. \cite{KnHi19} present two different approaches for path tracking in P4.
In \emph{hop recording}, all forwarding P4 nodes record their ID in the header of the target packet.
The last node can then reconstruct the path.
In \emph{forwarding state logging}, the first P4 node records the current version of the global forwarding state of the network and its node identifier in a header of the target packet.
If the version of the global forwarding state does not change while the packet flows through the network, the last P4 node in the network can reconstruct the path using the information in the header.

Basuki et al. \cite{BaRo20} propose a privacy-aware path-tracking mechanism.
Their goal is that the trajectory information in the packets cannot be used to draw conclusions about the network topology or routing information.
They achieve this by recording the information in an in-packet bloom filter.

\subsection{Other Fields of Application}
\label{sec:use-cases-monitoring-other}
BurstRadar \cite{JoQu18} is a system for microburst detection for data center networks that runs directly on P4 switches.
If queue-induced delay is above a certain threshold, BurstRadar reports a microburst and creates a snapshot of the telemetry information of involved packets. 
This telemetry information is then forwarded to a monitoring server.
As it is not possible to gather telemetry information of packets that are already part of the egress queue, the telemetry information of all packets and their corresponding egress port are temporarily stored in a ring buffer that is implemented using P4 registers. 

Dapper \cite{GhBe17} is a P4 tool to evaluate TCP.
It implements TCP in P4 and analyzes header fields, packets sizes, and timestamps of data and ACK packets to detect congestion.
Then, flow-dependent information are stored in registers.

He et al. \cite{HeCh18} propose an adaptive expiration timeout mechanism for flow entries in P4 switches.
The switches implement a mechanism to detect the last packet of a TCP flow.
In case of a match, it notifies the controller to delete the corresponding flow entries.

Riesenberg et al. \cite{RiKi19} implement alternate marking performance measurement (AM-PM) for P4.
AM-PM measures delay and packet loss in-band in a network using only one or two bit overhead per packet.
These bits are used for coordination and signalling between measurement points (MPs).

Wang et al. \cite{WaHu20} describe how TCP-friendly meters can be designed and implemented for P4-based switches.
According to their findings, meters in commercial switches interact with TCP streams in such a way that these streams can only reach about 10\% of the target rate.
The experimental evaluation of their TCP-friendly meters shows achieved rates of up to 85\% of the target rate.

P4STA \cite{KuSi20} is an open-source framework that combines software-based traffic load generation with accurate hardware packet timestamps.
Thereby, P4STA aggregates multiple traffic flows to generate high traffic load and leverage programmable platforms.

Hark et al. \cite{HaBh19} use P4 to filter data plane measurements.
To save resources, only relevant measurements are sent to the controller.
The authors implement a prototype and demonstrate the system by filtering measurements for a bandwidth forecast application.

P4Entropy \cite{DiSa20} presents an algorithm to estimate the entropy of network traffic within the P4 data plane.
To that end, they also developed two new algorithms, P4Log and P4Exp, to estimate logarithms and exponential functions within the data plane as well.

Taffet et al. \cite{TaMe19} describe a P4-based implementation of an in-band monitoring system that collects information about the path of a packet and whether it encountered congestion.
For this purpose, the authors repurpose previously unused fields of the IP header.

NetView \cite{LiZh20b} is a network telemetry framework that uses proactive probe packets to monitor devices.
Telemetry targets, frequency, and characteristics can be configured on demand by administrators.
The probe packets traverse arbitrary paths by using source routing.

FastFE \cite{BaZh20} is a system for offloading feature extraction, i.e., deriving certain information from network traffic, for \ac{ML}-based traffic analysis applications.
Policies for feature extraction are defined as sequential programs.
A policy enforcement engine translates these policies into primitives for either a programmable switch or a program running on a commodity server.

Unroller \cite{KuBa20} detects routing loops in the data plane in real-time.
It achieves this by encoding a subset of the path that a packet takes into the packet.

Hang et al. \cite{HaSh19} use a time-based sliding window approach to measure packet rates.
The goal is to record statistics entirely inside the data plane without having to use the CPU of a switch.
Their approach is able to measure traffic size without sampling.

FlowSpy \cite{GuSh19} is a network monitoring framework that uses load balancing.
Different monitoring tasks are distributed among all available switches by an ILP solver.
This reduces the workload on single switches in contrast to monitoring frameworks that perform all monitoring tasks on ingress or egress switches only.

\subsection{Summary and Analysis}
This research domain greatly benefits from all five core features described in \sect{benefits}.
\emph{Definition and usage of custom packet headers} enables new monitoring schemes where relevant information can be added to packets while it travels through a P4-enabled network.
One example is In-band Network Telemetry (INT) (Section \ref{sec:use-cases-monitoring-int}) that has been specified specifically for P4.
Another example are path tracking mechanisms (Section \ref{sec:use-cases-monitoring-path-tracking}) where the path of a packet is recorded in a dedicated header of the packet.
In the case of INT, this goes hand in hand with \emph{flexible packet header processing} as INT headers may contain instructions that other INT-enabled switches need to execute.
\emph{Target-specific packet header processing functions} in the form of stateful packet processing using, e.g., registers, is used by all areas of monitoring as it is necessary to gather data over a certain time frame instead of just looking at a single packet.
Because the register space is severely limited on most hardware targets, an efficient usage of the available resources is of great importance.
Sketches (Section \ref{sec:use-cases-monitoring-sketches}) is one approach to solve this.
After monitoring data is gathered on the control plane, the result is often \emph{processed on the control plane}.
This can range from simple notifications to splitting operations between data plane and control plane where the resources on the data plane are not sufficient.
Some DSL-based monitoring approaches (Section \ref{sec:use-cases-monitoring-dsl}) make use of \emph{flexible development and deployment}.
With these approaches, a P4 program is generated automatically on the basis of a monitoring workflow defined by an administrator.

%% file: chapters/10-uc_traffic_management.tex
\section{Applied Research Domains: Traffic Management and Congestion Control}
\label{sec:use-cases-traffic-management}

We describe applied research on data center switching, load balancing, congestion notification, traffic scheduling, traffic aggregation, active queue management (\acs{AQM}), and traffic offloading.
Table~\ref{tab:implementations-traffic-management} shows an overview of all the work described.
At the end of the section, we summarize the work and analyze it with regard to P4's core features described in \sect{benefits}.

\begin{table}[htp]
\caption{Overview of applied research on traffic management and congestion control (\sect{use-cases-traffic-management}).}
\begin{center}
\begin{tabularx}{\linewidth}{@{} l l X r @{}}
\toprule
\textbf{Research work} & \textbf{Year} & \textbf{Targets} & \textbf{Code}\\
\midrule
\addlinespace
\multicolumn{4}{@{\extracolsep{\fill}}l}{\noindent \textbf{Data Center Switching} (\sect{use-cases-data-center-switching})} \\
\addlinespace
\paperTableSource{trellis, trellis-p4-tutorial}{2019}{Trellis}{\ac{bmv2}}{p4-trellis}
\paperTableSource{SiKi15}{2015}{DC.p4}{\ac{bmv2}}{p4-SiKi15}
\paperTableSource{p4apps-onf}{2018}{Fabric.p4}{\ac{bmv2}}{fabricp4-github}
\paperTableSource{rare}{2019}{RARE}{\ac{bmv2}, Tofino}{rare-repo}
\midrule
\addlinespace
\multicolumn{4}{@{\extracolsep{\fill}}l}{\noindent \textbf{Load Balancing} (\sect{use-cases-data-center-load-balancing})}\\
\addlinespace
\paperTable{PiDe18}{2018}{SHELL}{NetFPGA-SUME}
\paperTable{MiZe17}{2017}{SilkRoad}{Tofino}
\paperTable{KaHi16}{2016}{HULA}{-}
\paperTable{BeKa18}{2018}{MP-HULA}{-}
\paperTable{ChWa19}{2019}{Chiang et al.}{\ac{bmv2}}
\paperTable{YeCh18}{2018}{W-ECMP}{\ac{bmv2}}
\paperTable{HsTa20}{2020}{DASH}{\ac{bmv2}}
\paperTable{PiSc18, PiSc20}{2018/20}{Pizzutti et al.}{\ac{bmv2}}
\paperTable{ZhWe20}{2020}{LBAS}{Tofino}
\paperTable{LiZh20c}{2020}{DPRO}{\ac{bmv2}}
\paperTable{KaKa19}{2019}{Kawaguchi et al.}{\ac{bmv2}}
\paperTable{CiCh17}{2017}{AppSwitch}{PISCES}
\paperTableSource{OlAg18}{2018}{Beamer}{\ac{bmv2}, NetFPGA-SUME}{github-beamer}
\midrule
\addlinespace
\multicolumn{4}{@{\extracolsep{\fill}}l}{\noindent \textbf{Congestion Notification} (\sect{use-cases-traffic-management-notification})}\\
\addlinespace
\paperTable{GeYa19}{2019}{P4QCN}{\ac{bmv2}}
\paperTable{JiZh19}{2019}{Jiang et al.}{-}
\paperTable{ShJu20}{2020}{EECN}{\ac{bmv2}}
\paperTable{ChFa20}{2020}{Chen et al.}{\ac{bmv2}}
\paperTable{LaFr20}{2020}{Laraba et al.}{\ac{bmv2}}
\midrule
\addlinespace
\multicolumn{4}{@{\extracolsep{\fill}}l}{\noindent \textbf{Traffic Scheduling} (\sect{use-cases-traffic-management-scheduling})}\\
\addlinespace
\paperTable{ShLi18}{2018}{Sharma et al.}{\ac{bmv2}}
\paperTable{CaBo17}{2017}{Cascone et al.}{-}
\paperTable{BhAn19}{2019}{Bhat et al.}{\ac{bmv2}}
\paperTable{KfCr19}{2019}{Kfoury et al.}{\ac{bmv2}}
\paperTable{ChYe19}{2019}{Chen et al.}{Tofino}
\paperTable{LeCh19}{2019}{Lee et al.}{\ac{bmv2}}
\midrule
\addlinespace
\multicolumn{4}{@{\extracolsep{\fill}}l}{\noindent \textbf{Traffic Aggregation} (\sect{use-cases-traffic-aggregation})}\\
\addlinespace
\paperTable{WaLi20}{2020}{Wang et al.}{Tofino}
\paperTable{ToSa19}{2019}{RL-SP-DRR}{\ac{bmv2}}
\bottomrule
\end{tabularx}
\end{center}
\label{tab:implementations-traffic-management}
\end{table}

\begin{table}[htp]
\begin{center}
\begin{tabularx}{\linewidth}{@{} l l X r @{}}
\toprule
\textbf{Research work} & \textbf{Year} & \textbf{Targets} & \textbf{Code}\\
\midrule
\multicolumn{4}{@{\extracolsep{\fill}}l}{\noindent \textbf{Active Queue Management (AQM)} (\sect{use-cases-traffic-management-aqm})}\\
\addlinespace
\paperTable{TuKu18}{2018}{Turkovic et al.}{\ac{bmv2}, Netronome}
\paperTableSource{KuBl18}{2018}{P4-Codel}{\ac{bmv2}}{codel-github}
\paperTable{MeMo19}{2019}{P4-ABC}{\ac{bmv2}}
\paperTable{TuKu20}{2020}{P4air}{\ac{bmv2}, Tofino}
\paperTable{FeCa20}{2020}{Fernandes et al.}{\ac{bmv2}}
\paperTable{WaCh18}{2018}{Wang et al.}{\ac{bmv2}, Tofino}
\paperTable{GrDi20}{2020}{SP-PIFO}{Tofino}
\paperTableSource{KuGu21}{2021}{Kunze et al.}{Tofino}{p4-KuGu21}
\paperTable{HaPa21}{2021}{Harkous et al.}{\ac{bmv2}, Netronome}
\midrule
\addlinespace
\multicolumn{4}{@{\extracolsep{\fill}}l}{\noindent \textbf{Traffic Offloading} (\sect{use-cases-traffic-management-offloading})}\\
\addlinespace
\paperTable{AnSa19}{2019}{Andrus et al.}{-}
\paperTable{IbAn19}{2019}{Ibanez et al.}{NetFPGA-SUME}
\paperTable{KfCr20}{2020}{Kfoury et al.}{Tofino}
\paperTable{KeUd20}{2020}{Falcon}{Tofino}
\paperTable{OsKo20}{2020}{Osiński et al.}{Tofino}
\bottomrule
\end{tabularx}
\end{center}
\end{table}

\subsection{Data Center Switching}
\label{sec:use-cases-data-center-switching}

Trellis \cite{trellis, trellis-p4-tutorial} is an open-source multipurpose L2/L3 spine-leaf switch fabric for data center networks.
It is designed to run on white box switches in conjunction with the ONOS controller where its main functionality is implemented.
It supports typical data center functionality such as bridging using VLANs, routing (IPv4/IPv6 unicast/multicast routing, MPLS segment routing), and vRouter functionality (BGBv4/v6, static routes, route black-holing).
Trellis is part of the CORD platform that leverages \ac{SDN}, \ac{NFV}, and Cloud technologies for building agile data centers for the network edge.

DC.p4 \cite{SiKi15} implements typical features of data center switches in P4.
The list of features includes support for VLAN, NVGRE, VXLAN, ECMP, IP forwarding, \acp{ACL}, packet mirroring, MAC learning, and packet-in/-out messages to the control plane.

Fabric.p4 is \cite{p4apps-onf, trellis-p4-tutorial} the underlying reference data plane pipeline implemented in P4.
By introducing support for P4 switches, the authors aim at increasing the platform heterogeneity for the CORD fabric.
Fabric.p4 is currently based on the V1Model switch architecture, but support for \ac{PSA} is planned.
It is inspired by the \acl{OF} data plane abstraction (OF-DPA) and currently supports L2 bridging, IPv4/IPv6 unicast/multicast routing, and MPLS segment routing.
Fabric.p4 comes with capability profiles such as \emph{fabric} (basic profile), \emph{spgw} (S/PGW), and \ac{INT}.
For control plane interaction, ONOS is extended by the P4Runtime.

RARE \cite{rare} (Router for Academia, Research \& Education) is developed in the G\'{E}ANT project GN4-3 and implements a P4 data plane for the FreeRouter open-source control plane.
Its feature list includes routing, bridging, \acp{ACL}, VLAN, VXLAN, MPLS, GRE, MLDP, and BIER among others.

\subsection{Load Balancing}
\label{sec:use-cases-data-center-load-balancing}

SHELL \cite{PiDe18} implements stateless application-aware load balancing in P4.
A load balancer forwards new connections to a set of randomly chosen application instances by adding a segment routing (SR) header.
Each application instance makes a local decision to either decline or accept the connection attempt.
After connection initiation, the client includes a previously negotiated identifier in all subsequent packets.
In the prototypical implementation, the authors use TCP time stamps for communicating the identifier, alternatives are identifiers of QUIC or TCP sequence numbers.

SilkRoad \cite{MiZe17} implements stateful load balancing on P4 switches.
SilkRoad implements two tables for stateful processing.
One table maps virtual IP addresses of services to server instances, another table records active connections identified by hashes of 5-tuples to forward subsequent flows.
It applies a Bloom filter to identify new connection attempts and to record those requests in registers to remember client requests that arrive while the pool of server instances changes.
In \cite{LeMi17}, the accompanying demo is described.

HULA \cite{KaHi16} implements a link load-based distance vector routing mechanism.
Switches in HULA do not maintain the state for every path but the next hops.
They send out probes to gather link utilization information.
Probe packets are distributed throughout the network on node-specific multicast trees. 
The probes have a header that contains a destination field and the currently best path utilization to that destination.
When a node receives a probe, it updates the best path utilization if necessary, sends one packet clone upstream back to the origin, and forwards copies along the multicast tree further downstream.
This way the origin will receive multiple probe packets with different path utilization to a specific destination.
Then, flowlets are forwarded onto the best currently available path to its destination.

MP-HULA \cite{BeKa18} extends HULA by using load information for $n$ best next hops and compatibility with multipath TCP (MP-TCP).
It tracks subflows of MP-TCP with individual flowlets per sub-flow. 
MP-HULA aims at distributing those subflows on different paths to aggregate bandwidth. 
To that end, it is necessary to keep track of the best $n$ next-hops which is done with additional registers and forwarding rules.

Chiang et al. \cite{ChWa19} propose a cost-effective congestion-aware load balancing scheme (CCLB).
In contrast to HULA, CCLB replaces only the leaf switches with programmable switches, and thus is more cost-effective.
They leverage \ac{ECN} information in probe packets to recognize congestion in the network and to adapt the load balancing.
CCLB further uses flowlet forwarding and is implemented for the \ac{bmv2}.

W-ECMP \cite{YeCh18} is an ECMP-based load balancing mechanism for data centers implemented for P4 switches.
Weighted probabilities based on path utilization, are used to randomly choose the best path to avoid congestion.
A local agent on each switch computes link utilization for the ports.
Regular traffic carries an additional custom packet header that keeps track of the current maximum link utilization on a path.
Based on the maximum link utilization, the switches update port weights if necessary.

DASH \cite{HsTa20} is an adaptive weighted traffic splitting mechanism that works entirely in the data plane. 
In contrast to popular weighted traffic splitting strategies such as WCMP, DASH does not require multiple hash table entries.
DASH splits traffic based on link weights by portioning the hash space into unique regions. 

Pizzutti et al. \cite{PiSc18, PiSc20} implement congestion-aware load balancing for flowlets on P4 switches.
Flowlets are bursts of packets that are separated by a time gap, e.g., as caused by factors such as TCP dynamics, buffer availability, or link congestion.
For distributing subflows on different paths, the congestion state of the last route is stored in a register.

LBAS \cite{ZhWe20} implements a load balancer to minimize the processing latency at both load balancers and application servers.
LBAS does not only reduce the processing latency at load balancers but also takes the application servers' state into account.
It is implemented for the Tofino and its average response time is evaluated.

DPRO \cite{LiZh20c} combines \ac{INT} with traffic engineering (TE) and reinforcement learning (RL). 
Network statistics, such as link utilization and switch load, are gathered using an adapted \ac{INT} approach.
An RL-agent inside the controller adapts the link weights based on the minimization of a max-link-utilization objective.

Kawaguchi et al. \cite{KaKa19} implement Unsplittable flow Edge Load factor Balancing (UELB).
A controller application monitors the link utilization and computes new optimal paths upon congestion.
The path computation is based on the UELB problem.
The forwarding is implemented in P4 for the \ac{bmv2}.

AppSwitch \cite{CiCh17} implements a load balancer for key-value storage systems.
However, the focus lies on a local agent and the control plane communication with the storage server.

Beamer \cite{OlAg18} operates in data centers and prevents interruption of connections when they are load-balanced to a different server.
To that end, the Beamer controller instructs the new target server to forward packets of the load-balanced connection to the old target server until the migration phase is over.

\subsection{Congestion Notification}
\label{sec:use-cases-traffic-management-notification}

P4QCN \cite{GeYa19} proposes a congestion feedback mechanism where network nodes check the egress ports for congestion before forwarding packets.
If a node detects congestion, it calculates a feedback value that is propagated upstream.
The mechanism clones the packet that caused the congestion, updates the feedback value in the header, changes the origin of the flow, and forwards it as a feedback packet to the sender.
The sender adjusts its sending rate to reduce congestion downstream.
The authors describe an implementation where \ac{bmv2} is extended by P4 externs for floating-point calculations.

Jiang et al. \cite{JiZh19} introduce a novel \ac{AWW} mechanism for TCP.
The authors argue that the current calculation of the advertised window in the TCP header is inaccurate because the source node does not know the actual capacity of the network.
\ac{AWW} dynamically updates the advertised window of ACK packets to feedback the network capacity indirectly to the source nodes.
Each P4 switch calculates the new \ac{AWW} value and writes it into the packet header. 

EECN \cite{ShJu20} presents an enhanced \ac{ECN} mechanism which piggybacks congestion information if the switch notices congestion. 
To that end, the \ac{ECN}-Echo bit is set for traversing ACKs as soon as congestion occurs for a given flow.
This enables fast congestion notification without the need for additional control traffic. 

Chen et al. \cite{ChFa20} present QoSTCP, a TCP version with adapted congestion window growth that enables rate limiting. 
QoSTCP is based on a marking approach similar to \ac{ECN}.
When a flow exceeds a certain rate, the packet gets marked with a so-called Rate-Limiting Notification (RLN) and the congestion window growth is adapted proportional to the RLN-marked packet rate.
Metering and marking is done using P4.

Laraba et al. \cite{LaFr20} detect \ac{ECN} misbehavior with the help of P4 switches.
They model \ac{ECN} as extended finite state machine (EFSM) and store states and variables in registers.
If end hosts do not conform to the specified \ac{ECN} state machine, packets are either dropped or, if possible, the misbehavior is corrected.

\subsection{Traffic Scheduling}
\label{sec:use-cases-traffic-management-scheduling}

Sharma et al. \cite{ShLi18} introduce a mechanism for per flow fairness scheduling in P4.
The concept is based on round-robin scheduling where each flow may send a certain number of bytes in each round.
The switch assigns a round number for each arriving packet that depends on the number of sent bytes of flow in the past.

Cascone et al. \cite{CaBo17} introduce bandwidth sharing based on sending rates between TCP senders.
P4 switches use statistical byte counters to store the sending rate of each user.
Depending on the recorded sending rate of the user, arriving packets are pushed into different priority queues.

Bhat et al. \cite{BhAn19} leverage P4 switches to translate application layer header information into link-layer headers for better QoS routing.
They use Q-in-Q tunneling at the edge to forward packets to the core network and present a \ac{bmv2} implementation for HTTP/2 applications, as HTTP/2 explicitly defines a Stream ID that can directly be translated in Q-in-Q tags.

Kfoury et al. \cite{KfCr19} present a method to support dynamic TCP pacing with the aid of network state information.
A P4 switch monitors the number of active TCP flows, i.e., they monitor the SYN, SYN-ACK, and ACK flags and notify senders about the current network state if a new flow starts or another terminates. 
To that end, they introduce a new header and show by simulations that the overall throughput increases.

Chen et al. \cite{ChYe19} present a design for bandwidth management for QoS with \ac{SDN} and P4-programmable switches. 
Their design classifies packets based on a two-rate three-color marker and assigns corresponding priorities to guarantee certain per flow bandwidth.
To that end, they leverage the priority queuing capabilities of P4-switches based on the assigned color.
Guaranteed traffic goes to a high-priority queue, best-effort traffic goes to a low-priority queue, and traffic that exceeds its bandwidth is simply dropped.

Lee et al. \cite{LeCh19} implement a multi-color marker for bandwidth guarantees in virtual networks. 
Their objective is to isolate bandwidth consumption of virtual networks and provide QoS for its serving flows.

\subsection{Traffic Aggregation}
\label{sec:use-cases-traffic-aggregation}

Wang et al. \cite{WaLi20} introduce aggregation and dis-aggregation capabilities for P4 switches.
To reduce the header overhead in the network, multiple small packets are thereby aggregated to a single packet.
They leverage multiple register arrays to store incoming small packets in 32 bit chunks. 
If enough small packets are stored, a larger packet gets assembled with the aid of multiple recirculations; each recirculation step appends a small packet to the aggregated large packet.

RL-SP-DRR \cite{ToSa19} is a combination of strict priority scheduling with rate limitation (RL-SP) and deficit round-robin (DRR).
RL-SP ensures prioritization of high-priority traffic while DRR enables fair scheduling among different priority classes.
They extend \ac{bmv2} to support RL-SP-DRR and evaluate it against strict priority queuing and no active queuing mechanism.

\subsection{Active Queue Management (AQM)}
\label{sec:use-cases-traffic-management-aqm}

Turkovic et al. \cite{TuKu18} develop an \ac{AQM} mechanism for programmable data planes.
The switches are programmed to collect metadata associated with packet processing, e.g., queue size and load, that are used to prevent, detect, and dissolve congestion by forwarding affected flows on an alternate path.
Two possible mechanisms for rerouting in P4 are described. 
In the first mechanism, primary and backup entries are installed in the forwarding tables and according to the gathered metadata, the suitable action is selected.
The second mechanism leverages a local controller on each switch that monitors flows and installs updated forwarding rules when congestion is noticed.

P4-CoDel \cite{KuBl18} implements the CoDel \ac{AQM} mechanism specified in RFC 8289 \cite{rfc8289}.
CoDel leverages a target and an interval parameter.
As long as the queuing delay is shorter than the target parameter, no packets are dropped.
If the queuing delay exceeds the target by a value that is at least as large as the interval, a packet is dropped, and the interval parameter is decreased.
This procedure is repeated until the queuing delay is under the target threshold again.
The interval is then reset to the initial value.
To avoid P4 externs, the authors use approximated calculations for floating-point operations.

P4-ABC \cite{MeMo19} implements activity-based congestion management (ABC) for P4. 
ABC is a domain concept where edge nodes measure the activity, i.e., the sending rate, of each user and annotate the value in the packet header.
Core nodes measure the average activity of all packets.
Depending on the current queue status, the average activity, and activity value in the packet header, a drop decision is made for each packet to prevent congestion. 
The P4$_{16}$ implementation for the \ac{bmv2} requires externs for floating-point calculations. 

P4air \cite{TuKu20} attempts to provide more fairness for TCP flows with different congestion control algorithms.
To that end, P4air groups flows into different categories based on their congestion control algorithm, e.g., loss-, delay- and loss-delay-based. 
Afterwards, the most aggressive flows are punished based on the previous categorization with packet drops, delay increase, or adjusted receive windows. 
P4air leverages switch metrics and flow reactions, such as queuing delay and sending rate, to determine the congestion control algorithm used by the flows.

Fernandes et al. \cite{FeCa20} propose a bandwidth throttling solution in P4. 
Incoming packets are dropped with a certain probability depending on the incoming rate of the flow and the defined maximum bandwidth. 
Rates are measured using time windows and byte counters.
Fernandes et al. extend the \ac{bmv2} for this purpose.

Wang et al. \cite{WaCh18} present an \ac{AQM} mechanism for video streaming. 
Data packets are classified as base packets (basic image information) or enhancement packets (additional information to improve the image quality). 
When the queue size exceeds a certain threshold, enhancement packets are preferably dropped.

SP-PIFO \cite{GrDi20} features an approximation of Push-In First-Out (PIFO) queues which enables programmable packet scheduling at line rate.
SP-PIFO dynamically adapts the mapping between packet ranks and available strict-priority queues.

Kunze et al. \cite{KuGu21} analyze the design of three popular \ac{AQM} algorithms (RED, CoDel, PIE). 
They implement PIE in three different variants for Tofino-based P4 hardware targets and show that implementation trade-offs have significant performance impacts.

Harkous et al. \cite{HaPa21} use virtual queues implemented in P4 for traffic management.
A traffic classifier in the form of \acp{MAT} assigns a data plane slice identifier to traffic flows.
P4 registers are used to implement virtual queues for each data plane slice for traffic management.

\subsection{Traffic Offloading}
\label{sec:use-cases-traffic-management-offloading}

Andrus et al. \cite{AnSa19} propose to offload video stream processing of surveillance cameras to P4 switches.
The authors propose to offload stream processing for storage to P4 switches.
In case the analytics software detected an event, it enables a multistage pipeline on the P4 switch.
In the first step, video stream data is replicated.
One stream is further sent to the analytics software, the other stream is dedicated to the video storage.
The P4 switch filters out control packets and rewrites the destination IP address of all video packets to the video storage.

Ibanez et al. \cite{IbAn19} try to tackle the problem of P4's packet-by-packet programming model.
Many tasks, such as periodic updates, require either hardware-specific capabilities or control-plane interaction.
Processing capabilities are limited to enqueue events, i.e., data plane actions are only triggered if packets arrive.
To eliminate this problem, the authors propose a new mechanism for event processing using the P4 language. 

Kfoury et al. \cite{KfCr20} propose to offload media traffic to P4 switches which act as relay servers.
A SIP server receives the connection request, replaces IP and port information with the relay server IP and port, and forwards the request to the receiver.
Afterwards, the media traffic is routed through the relay server.

Falcon \cite{KeUd20} offloads task scheduling to programmable switches. 
Job requests are sent to the switch and the switch assigns a task in first-come-first-serve order to the next executor in a pool of computation nodes.
Falcon reduces the scheduling overhead by a factor of 26 and increase scheduling throughput by a factor of 25 compared to state-of-the-art schedulers.

Osinski et al. \cite{OsKo20} present vBNG, a virtual Broadband Network Gateway (BNG).
Some components, such as PPPoE session handling, are offloaded to programmable switches.

\subsection{Summary and Analysis}
The research domain of traffic management and congestion control benefits from three core properties of P4: \emph{custom packet headers}, \emph{flexible header processing} and \emph{target-specific packet header processing functions}.
Data center switching mainly relies on packet header parsing of well-known protocols, such as IPv4/v6 or MPLS.
More advanced protocol solutions, such as VXLAN and BIER, can be implemented by leveraging the \emph{flexible packet header processing} property of P4.
The presented efforts on load balancing (Section \ref{sec:use-cases-data-center-load-balancing}) also use this property of P4 to implement novel approaches.
\emph{Target-specific packet header processing functions} such as externs are widely used in Section \ref{sec:use-cases-traffic-management-notification}.
Most works leverage externs such as metering and marking which may not be supported on all hardware targets.
A similar phenomenon appears in Section \ref{sec:use-cases-traffic-management-scheduling}. 
Here, many papers are based on priority queues.
The approaches on \ac{AQM} in Section \ref{sec:use-cases-traffic-management-aqm} encounter similar limitations.
Floating-point operations are not part of the P4 core.
Some targets may provide an extern for this functionality.
Multiple works avoid this problem by either using approximations or by relying on self-defined externs in software.

%% file: chapters/11-uc_forwarding_routing.tex
\section{Applied Research Domains: Routing and Forwarding}
\label{sec:use-cases-routing}
We describe applied research on source routing, multicast, publish-subscribe-systems, named data networking, data plane resilience, and other fields of application.
Table \ref{tab:implementations-routing} shows an overview of all the work described.
At the end of the section, we summarize the work and analyze it with regard to P4's core features described in \sect{benefits}.

\begin{table}[htbp]
\caption{Overview of applied research on routing and forwarding (\sect{use-cases-routing}).}
\begin{center}
\begin{tabularx}{\linewidth}{@{} l X X r @{}}
\toprule
\textbf{Research work} & \textbf{Year} & \textbf{Targets} & \textbf{Code}\\
\midrule
\addlinespace
\multicolumn{4}{@{\extracolsep{\fill}}l}{\noindent \textbf{Source Routing} (\ref{subsec:source_routing})} \\
\addlinespace
\paperTableSource{LeFa18}{2018}{Lewis et al.}{\ac{bmv2}}{p4-sourcerouting}
\paperTableSource{LuYu19}{2019}{Luo et al.}{\ac{bmv2}}{p4-paco}
\paperTable{KuSh20}{2020}{Kushwaha et al.}{Xilinx Virtex-7}
\paperTable{AbTu20}{2020}{Abdelsalam et al.}{\ac{bmv2}}
\midrule
\addlinespace
\multicolumn{4}{@{\extracolsep{\fill}}l}{\noindent \textbf{Multicast} (\ref{subsec:multicast})} \\
\addlinespace
\paperTableSource{BrHa17}{2017}{Braun et al.}{\ac{bmv2}}{p4-bier-demo}
\paperTableSource{MeLi20, MeLi21}{2020/21}{Merling et al.}{\ac{bmv2}, Tofino}{p4-bier-repo, p4-bier-tofino-repo}
\paperTableSource{ShSu19}{2019}{Elmo}{-}{p4-elmo}
\paperTable{LuYu20}{2020}{PAM}{\ac{bmv2}}
\midrule
\addlinespace
\multicolumn{4}{@{\extracolsep{\fill}}l}{\noindent \textbf{Publish/Subscribe Systems} (\ref{use-cases-routing-pub-sub})} \\
\addlinespace
\paperTable{WePa18, WePa19, WePa20a, WePa20b}{2018/19}{Wernecke et al.}{\ac{bmv2}}
\paperTable{JeMo18b}{2018}{Jepsen et al.}{Tofino}
\paperTableSource{KuGa20}{2020}{Kundel et al.}{\ac{bmv2}}{p4-pubsub}
\paperTable{VeKa20}{2020}{FastReact-PS}{-}
\midrule
\addlinespace
\multicolumn{4}{@{\extracolsep{\fill}}l}{\noindent \textbf{Named Data Networks} (\ref{use-cases-routing-ndn})} \\
\addlinespace
\paperTableSource{SiSt16, MiSi18}{2016/18}{NDN.p4}{\ac{bmv2}}{ndnp4-github, p4-netx-ndn}
\paperTable{KaSa20}{2020}{ENDN}{\ac{bmv2}}
\midrule
\addlinespace
\multicolumn{4}{@{\extracolsep{\fill}}l}{\noindent \textbf{Data Plane Resilience} (\ref{subsec:dataplane_resilience})} \\
\addlinespace
\paperTableSource{SeBo18}{2018}{Sedar et al.}{\ac{bmv2}}{p4-frr}
\paperTable{GiSh18}{2018}{Giesen et al.}{Tofino, Xilinx SDNet}
\paperTableSource{QuJo19}{2019}{SQR}{\ac{bmv2}, Tofino}{p4-sqr}
\paperTableSource{LiMe20}{2020}{P4-Protect}{\ac{bmv2}, Tofino}{p4-protect-bmv2, p4-protect-tofino}
\paperTable{HiTa19}{2019}{Hirata et al.}{-}
\paperTableSource{LiHa20}{2020}{Lindner et al.}{\ac{bmv2}, Tofino}{p4-source-protection-bmv2, p4-source-protection-tofino}
\paperTable{SuAb19}{2019}{D2R}{\ac{bmv2}}
\paperTable{ChSe19}{2019}{PURR}{\ac{bmv2}, Tofino}
\paperTableSource{HoCo19}{2019}{Blink}{\ac{bmv2}, Tofino}{github-blink}
\bottomrule
\end{tabularx}
\end{center}
\label{tab:implementations-routing}
\end{table}

\begin{table}[htbp]
\begin{center}
\begin{tabularx}{\linewidth}{@{} l X X r @{}}
\toprule
\textbf{Research work} & \textbf{Year} & \textbf{Targets} & \textbf{Code}\\
\midrule
\addlinespace
\multicolumn{4}{@{\extracolsep{\fill}}l}{\noindent \textbf{Other Fields of Applications} (\ref{subsec:other})} \\
\addlinespace
\paperTable{HsBe19}{2019}{Contra}{-}
\paperTable{MiKe16}{2016}{Michel et al.}{\ac{bmv2}}
\paperTable{BaOz18}{2018}{Baktir et al.}{\ac{bmv2}}
\paperTable{FrSa20}{2020}{Froes et al.}{\ac{bmv2}}
\paperTable{VaZh20}{2020}{QROUTE}{\ac{bmv2}}
\paperTable{GiGr20}{2020}{Gimenez et al.}{\ac{bmv2}}
\paperTable{FeTa19}{2019}{Feng et al.}{\ac{bmv2}}
\paperTable{GrLi20}{2020}{PFCA}{\ac{bmv2}}
\paperTable{McGo19}{2019}{McAuley et al.}{\ac{bmv2}}
\paperTableSource{KoPr19}{2019}{R2P2}{Tofino}{github-r2p2}

\bottomrule
\end{tabularx}
\end{center}
\end{table}

\subsection{Source Routing}
\label{subsec:source_routing}

With source routing, the source node defines the processing of the packet throughout the network.
To that end, a header stack is often added to the packet to specify the operations the other network devices should execute. 

Lewis et al. \cite{LeFa18} implement a simple source routing mechanism with P4 for the \ac{bmv2}. 
The authors introduce a header stack to specify the processing of the packet towards its destination. That header stack is constructed and pushed onto the packet by the source node.
Network devices match the header segments to determine how the packet should be processed.

Luo et al. \cite{LuYu19} implement segment routing with P4.
They introduce a header which contains segments that identify certain operations, e.g., forwarding the packet towards a specific destination or over a specific link, updating header fields, etc. 
Network nodes process packets according to the topmost segment in the segment routing header and remove it after successful execution.

Kushwaha et al. \cite{KuSh20} implement bitstream, a minimalistic programmable data plane for carrier-class networks, in P4 for FPGAs.
The focus of bitstream is to provide a programmable data plane while ensuring several carrier-grade properties, like deterministic latencies, short restoration time, and per-service measurements. 
To that end, the authors implement a source routing approach in P4 which leaves the configuration of the header stack to the control plane.

The authors of \cite{AbTu20} show a demo of segment routing over IPv6 data plane (SRv6) implementation in P4. 
It leverages the novel uSID instruction set for SRv6 to improve scalability and MTU efficiency.

\subsection{Multicast}
\label{subsec:multicast}

Multicast efficiently distributes one-to-many traffic from the source to all subscribers.
Instead of sending individual packets to each destination, multicast packets are distributed in tree-like structures throughout the network.

Bit Index Explicit Replication (BIER) \cite{MeMe18} is an efficient transport mechanism for IP multicast traffic. 
In contrast to traditional IP multicast, it prevents subscriber-dependent forwarding entries in the core network by leveraging a BIER header that contains all destinations of the BIER packet. 
To that end, the BIER header contains a bit string where each bit corresponds to a specific destination.
If a destination should receive a copy of the BIER packet, its corresponding bit is activated in the bit string in BIER header of the packet. 
Braun et al. \cite{BrHa17} present a demo implementation of BIER-based multicast in P4. 
Merling et al. \cite{MeLi20} implement BIER-based multicast with fast reroute capabilities in P4 for the \ac{bmv2} and for the Tofino \cite{MeLi21}.

Elmo \cite{ShSu19} is a system for scalable multicast in multi-tenant datacenters. 
Traditional IP multicast maintains subscriber dependent state in core devices to forward multicast traffic.
This limits scalability, since the state in the core network has to be updated every time subscribers change. 
Elmo increases scalability of IP multicast by moving a certain subscriber-dependent state from the core devices to the packet header.

Priority-based adaptive multicast (PAM) \cite{LuYu20} is a control protocol for data center multicast which is implemented by the authors in P4. 
Network administrators define different policies regarding priority, latency, completion time, etc., which are installed on the core switches.
The network devices than monitor link loads and adjust their forwarding to fulfill the policies.

\subsection{Publish/Subscribe Systems}
\label{use-cases-routing-pub-sub}

Publish/subscribe systems are used for data distribution.
Subscribers are able to subscribe to announced topics.
Based on the subscriptions, the data packets are distributed from the source to all subscribers.

Wernecke et al. \cite{WePa18, WePa19, WePa20a, WePa20b} implement a content-based publish/subscribe mechanism with P4.
The distribution tree to all subscribers is encoded directly in the header of the data packets.
To that end, the authors introduce a header stack which is pushed onto the packet by the source.
Each element in the stack consists of an ID and a value.
When a node receives a packet, it checks whether the header stack contains an element with its own ID.
If so, the value determines to which neighbors the packet has to be forwarded.

Jepsen et al. \cite{JeMo18b} introduce a description language to implement publish/subscriber systems.
The data plane description is translated into a static pipeline and dynamic filters. 
The static pipeline is a P4 program that describes a packet processing pipeline for P4 switches, the dynamic filters are the forwarding rules of the match-action tables that may change during operation, e.g., when subscriptions change.

Kundel et al. \cite{KuGa20} propose two approaches for attribute/value encoding in packet headers for P4-based publish/subscribe systems.
This reduces the header overhead and facilitates adding new attributes which can be used for subscription by hosts.

FastReact-PS \cite{VeKa20} is a P4-based framework for event-based publish/subscribe in industrial IoT networks.
It supports stateful and stateless processing of complex events entirely in the data plane. 
Thereby, the forwarding logic can be dynamically adjusted by the control plane without the need for recompilation. 

\subsection{Named Data Networking}
\label{use-cases-routing-ndn}
Named data networking (\acs{NDN}) is a content-centric paradigm where information is requested with resource identifiers instead of destinations, e.g., IP addresses.
Network devices cache recently requested resources.
If a requested resource is not available, network devices forward the request to other nodes.

\acs{NDN}.p4 \cite{SiSt16} implements \acs{NDN} without caching for P4.
However, the implementation cannot cache requests because of P4-related limitations with stateful storage.
Miguel et al. \cite{MiSi18} leverage the new functionalities of P4$_{16}$ to extend \acs{NDN}.p4 by a caching mechanism for requests and optimize its operation.
The caching mechanism is implemented with P4 externs.

Enhanced \acs{NDN} (ENDN) \cite{KaSa20} is an advanced \acs{NDN} architecture.
It offers a larger catalog of content delivery features like adaptive forwarding, customized monitoring, in-network caching control, and publish/subscribe forwarding.

\subsection{Data Plane Resilience}
\label{subsec:dataplane_resilience}

Sedar et al. \cite{SeBo18} implement a fast failover mechanism without control plane interaction for P4 switches.
The mechanism uses P4 registers or metadata fields for bit strings that indicate if a particular port is considered up or down.
In a match-action table, the port bit string provides an additional match field to determine whether a particular port is up or down.
Depending on the port status, default or backup actions are executed. 
The authors rely on a local P4 agent to populate the port bit strings.

Giesen et al. \cite{GiSh18} introduce a forward error correction (FEC) mechanism for P4.
Commonly, unreliable but not completely broken links are avoided.
As this happens at the cost of throughput, the proposed FEC mechanism facilitates the usage of unreliable links.
The concept features a link monitoring agent that polls ports to detect unreliable connections.
When a packet should be forwarded over such a port, the P4 switch calculates a resilient encoding for the packet which is then decoded by the receiving P4 switch.

Shared Queue Ring (SQR) \cite{QuJo19} introduces an in-network packet loss recovery mechanism for link failures. 
SQR caches recent traffic inside a queue with slow processing speed.
If a link failure is detected, the cached packets can be sent over an alternative path.
While P4 does not offer the possibility to store packets for a certain amount of time, the authors leverage the cloning operation of P4 to keep packets inside the buffer.
If a cached packet has not yet met its delay, it gets cloned to another egress port which takes some time. 
This procedure is repeated until the packet has been stored for a given time span.

P4-Protect \cite{LiMe20} implements 1+1 protection for IP networks. 
Incoming packets are equipped with a sequence number, duplicated, and sent over two disjoint paths.
At an egress point, the first version of each packet is accepted and forwarded.
As a result, a failure of a single path can be compensated without additional signaling or reconfiguration.
P4-Protect is implemented for the \ac{bmv2} and the Tofino.
Evaluations show that line-rate processing with 100 Gbit/s can be achieved with P4-Protect at the Tofino. 

Hirata et al. \cite{HiTa19} implement a data plane resilience scheme based on multiple routing configurations.
Multiple routing configurations with disjoint paths are deployed, and a header field identifies the routing configuration according to which packets are forwarded.
In the event of a failure, a routing configuration is chosen that avoids the failure. 

Lindner et al. \cite{LiHa20} present a novel prototype for in-network source protection in P4.
A P4-capable switch receives sensor data from a primary and secondary sensor, but forwards only the data from the primary sensor if available. It detects the failure of the primary sensor and then transparently forwards data from a secondary sensor to the application. 
Two different mechanisms are presented.
The \emph{counter-based} approach stores the number of packets received from the secondary sensor since the last packet from the primary sensor has been received. The \emph{timer-based} approach stores the time of the last arrival of a packet from the primary sensor and considers the time since then. If certain thresholds are exceeded, the P4-switch forwards the data from the secondary sensor.

D2R \cite{SuAb19} is a data-plane-only resilience mechanism. 
Upon a link failure, the data plane calculates a new path to the destination using algorithms like breadth-first search and iterative deepening depth-first search.
As one pipeline iteration has not enough processing stages to compute the path, recirculation is leveraged.
In addition, \emph{Failure Carrying Packets (FCP)} is used to propagate the link failure inside the network.
While the authors claim that their architecture works with hardware switches, e.g., the Tofino, they only present and evaluate a \ac{bmv2} implementation.

Chiesa et al. \cite{ChSe19} propose a primitive for reconfigurable fast ReRoute (PURR) which is a FRR primitive for programmable data planes, in particular for P4.
For each destination, suitable egress ports are stored in bit strings.
During packet processing, the first working suitable egress port is determined by a set of forwarding rules.
Encoding based on \emph{Shortest Common Supersequence} guarantees that only few additional forwarding rules are required.

Blink \cite{HoCo19} detects failures without controller interaction by analyzing TCP signals.
The core concept is that the behavior of a TCP flow is predictable when it is disrupted, i.e., the same packet is retransmitted multiple times. 
When this information is aggregated over multiple flows, it creates a characteristic failure signal that is leveraged by data plane switches to trigger packet rerouting to another neighbor.

\subsection{Other Fields of Applications}
\label{subsec:other}

Contra \cite{HsBe19} introduces performance-aware routing with P4.
Network paths are ranked according to policies that are defined by administrators.
Contra applies those policies and topology information to generate P4 programs that define the behavior of forwarding devices.
During runtime, probe packets are used to determine the current network state and update forwarding entries for best compliance with the defined policies.

Michel et al. \cite{MiKe16} introduce identifier-based routing with P4.
The authors argue that IP addresses are not fine-granular enough to enable adequate forwarding, e.g., in terms of security policies.
The authors introduce a new header that contains an identifier token. Before sending packets, applications transmit information on the process and user to a controller that returns an identifier that is inserted into the packet header.
P4 switches are programmed to forward packets based on that identifier.

Baktir et al. \cite{BaOz18} propose a service-centric forwarding mechanism for P4.
Instead of addressing locations, e.g., by IP addresses, the authors propose to use location-independent service identifiers.
Network hosts write the identifier of the desired service into the appropriate header field, the switches then make forwarding decisions based on the identifier in the packet header.
With this approach, the location of the service becomes less important since the controller simply updates the forwarding rules when a service is migrated or load balancing is desired.

Froes et al. \cite{FrSa20} classify different traffic classes which are identified by a label.
Packet forwarding is based on that controller-generated label instead of IP addresses.
The traffic classes have different QoS properties, i.e., prioritization of specific classes is possible.
To that end, switches leverage multiple queues to process traffic of different traffic classes.

QROUTE \cite{VaZh20} is a quality of service (QoS) oriented forwarding scheme in P4.
Network devices monitor their links and annotate values, e.g., jitter or delay, in the packet header so that downstream nodes can update their statistics.
Furthermore, packet headers contain constraints like maximum jitter or delay.
According to those values, forwarding decisions are made by the network devices.

Gimenez et al. \cite{GiGr20} implement the recursive internet-work architecture (RINA) in P4 for the \ac{bmv2}.
RINA is a networking architecture which sees computer networking as a type of inter-process communication where layering should be based on scope/scale instead of function. 
In general, efficient implementations require hardware support.
However, up to date only software-based implementations are available.
The authors hope that with the advance of programmable hardware in the form of P4, hardware-based RINA will soon be possible.

Feng et al. \cite{FeTa19} implement information-centric network (ICN) based forwarding for HTTP.
To that end, they propose mechanisms to convert packets from ICN to HTTP packets and vice-versa.

PFCA \cite{GrLi20} implements a forwarding information base (FIB) caching architecture in the data plane.
To that end, the P4 program contains multiple \acp{MAT} that are mapped to different memory, i.e., \ac{TCAM}, \ac{SRAM}, dynamic random access memory (DRAM), with different properties regarding lookup speed. 
Counters keep track of cache hits to move (un)popular rules to other tables.

McAuley et al. \cite{McGo19} present a hybrid error control booster (HEC) that can be deployed in wireless, mobile, or hostile networks that are prone to link or transport layer failures.
HECs increase the reliability by applying a modified Reed-Solomon code that adds parity packets or additional packet block acknowledgments.
P4 targets include an error control processor that implements this functionality.
It is integrated into the P4 program as P4 extern so that the data plane can exchange HEC packets with it.
A remote control plane includes the booster manager that controls HEC operations and parameters on the P4 targets via a data plane \ac{API}.

R2P2 \cite{KoPr19} is a transport protocol based on UDP for latency-critical RPCs optimized for datacenters or other distributed infrastructure.
A router module implemented in P4 or DPDK is used to relay requests to suitable servers and perform load balancing.
It may also perform queuing if no suitable server is available.
The goal of R2P2 is to overcome problems that typically come with TCP-based RPC systems, e.g., problems with load distribution and head-of-line-blocking.

\subsection{Summary and Analysis}
The research domain of routing and forwarding greatly benefits from P4's core features.
First, the \emph{definition and usage of custom packet headers} enables administrators to tailor the packet header to the specific use case.
Two examples are source routing (Section \ref{subsec:source_routing}) and multicast (Section \ref{subsec:multicast}). 
Both areas leverage custom headers to define lightweight mechanisms based on additional information in the packet header which are not part of any standard protocol. 
Although most of the projects were developed only for the \ac{bmv2}, they should be easily portable to hardware platforms as more complex, target specific operations are not required.
Second, users are able to define \emph{flexible packet header processing} depending on the information in the packet header, e.g., publish/subscribe systems (Section \ref{use-cases-routing-pub-sub}), named data networks (Section \ref{use-cases-routing-ndn}), and data plane resilience (Section \ref{subsec:dataplane_resilience}). 
Parametrized custom actions and (conditional) application of multiple MATs allow for adaptable packet processing for many specific use cases.
Similar to the previous P4 core feature, most projects were developed for the \ac{bmv2} but they should be easy to transfer if no target-specific actions are used. 
Third, we found that many papers in the area of data plane resilience (Section \ref{subsec:dataplane_resilience}) leverage \emph{target-specific packet header processing functions}. 
Often registers are used to store information whether egress ports are up or down to execute backup actions if necessary. 
Most projects were implemented for the hardware platform Tofino.
As a result, the implementations are highly target-specific and transferring them to other hardware platforms highly depends on the capabilities of the target platform and the used externs.

%% file: chapters/12-uc_advanced_networking.tex
\section{Applied Research Domains: Advanced Networking}
\label{sec:use-cases-advanced-networking}

We describe applied research on cellular networks (4G/5G), Internet of things (IoT), industrial networking, \ac{TSN}, \acf{NFV}, and \acfp{SFC}.
\tbl{implementations-advanced-networking} shows an overview of all the work described.
At the end of the section, we summarize the work and analyze it with regard to P4's core features described in \sect{benefits}.

\begin{table}[htbp]
    \caption{Overview of applied research on advanced networking (\sect{use-cases-advanced-networking}).}
    \begin{center}
    \begin{tabularx}{\linewidth}{@{} l l X r @{}}
    \toprule
    \textbf{Research work} & \textbf{Year} & \textbf{Targets} & \textbf{Code}\\

    \midrule
    \addlinespace
    \multicolumn{4}{@{\extracolsep{\fill}}l}{\noindent \textbf{Cellular Networks (4G/5G)} (\ref{use-cases-advanced-networking-5g})} \\
    \addlinespace
    \paperTable{HoLe20}{2020}{P4EC}{Tofino}
    \paperTableSource{p4apps-onf}{-}{Trellis}{-}{spgw-github}
    \paperTable{PaSi18}{2018}{SMARTHO}{\ac{bmv2}}
    \paperTable{AgHu18, AgXu19}{2018/19}{Aghdai et al.}{Netronome}
    \paperTable{XiQi19}{2019}{GRED}{\ac{bmv2}}
    \paperTable{XiGu20}{2020}{HDS}{-}
    \paperTable{ShLe19}{2019}{Shen et al.}{Xilinx SDNet}
    \paperTable{LeEb19}{2019}{Lee et al.}{Tofino}
    \paperTable{RiMa19}{2019}{Ricart-Sanchez et al.}{NetFPGA-SUME}
    \paperTable{SiRo19}{2019}{Singh et al.}{Tofino}
    \paperTable{ShKu20}{2020}{TurboEPC}{Netronome}
    \paperTable{VoPo20}{20200}{Vörös et al.}{Tofino}
    \paperTable{LiHu19a}{2019}{Lin et al.}{Tofino}

    \midrule
    \addlinespace
    \multicolumn{4}{@{\extracolsep{\fill}}l}{\noindent \textbf{Internet of Things} (\ref{use-cases-advanced-networking-iot})} \\
    \addlinespace

    \paperTable{UdMu17}{2017}{BLESS}{PISCES}
    \paperTable{UdMu18}{2018}{Muppet}{PISCES}
    \paperTable{WaWu19}{2019}{Wang et al.}{Tofino}
    \paperTable{MaAr20}{2020}{Madureira et al.}{\ac{bmv2}}
    \paperTable{EnZa19}{2019}{Engelhard et al.}{\ac{bmv2}}

    \midrule
    \addlinespace
    \multicolumn{4}{@{\extracolsep{\fill}}l}{\noindent \textbf{Industrial Networking} (\ref{use-cases-advanced-networking-industrial-networking})} \\
    \addlinespace

    \paperTable{VeKa18}{2018}{FastReact}{\ac{bmv2}}
    \paperTable{CeCs20}{2020}{Cesen et al.}{\ac{bmv2}}
    \paperTable{KuGl21}{2020}{Kunze et al.}{Tofino, Netronome}
    \midrule
    \addlinespace
    \multicolumn{4}{@{\extracolsep{\fill}}l}{\noindent \textbf{Time-Sensitive Networking (TSN)} (\ref{use-cases-advanced-networking-tsn})} \\
    \addlinespace
    
    \paperTable{RuGl18}{2018}{Rüth et al.}{Netronome}
    \paperTable{KaJo19}{2019}{Kannan et al.}{Tofino}
    \paperTable{KuSi19}{2019}{Kundel et al.}{Tofino}

    \bottomrule
    \end{tabularx}
    \end{center}
    \label{tbl:implementations-advanced-networking}
\end{table}

\begin{table}[htbp]
    \begin{center}
    \begin{tabularx}{\linewidth}{@{} l l X r @{}}
    \toprule
    \textbf{Research work} & \textbf{Year} & \textbf{Targets} & \textbf{Code}\\

    \midrule
    \addlinespace
    
    \multicolumn{4}{@{\extracolsep{\fill}}l}{\noindent \textbf{Network Function Virtualization (NFV)} (\ref{use-cases-advanced-networking-nfv})} \\
    \addlinespace
    
    \paperTable{BoIo18}{2018}{Kathará}{-}
    \paperTable{HeBa18}{2018}{P4NFV}{\ac{bmv2}}
    \paperTable{OsTa19}{2019}{Osiński et al.}{-}
    \paperTable{MoVe20}{2020}{Moro et al.}{-}
    \paperTable{OsTa19b}{2020}{DPPx}{\ac{bmv2}}
    \paperTable{MoPa19}{2019}{Mohammadkhan et al.}{Netronome}
    \paperTable{MoPe19a,MoPe19b}{2019}{FOP4}{\ac{bmv2}, eBPF}
    \paperTable{MaDo20}{2020}{PlaFFE}{Netronome}
    
    \midrule
    \addlinespace
    
    \multicolumn{4}{@{\extracolsep{\fill}}l}{\noindent \textbf{Service Function Chains (SFCs)} (\ref{use-cases-advanced-networking-sfc})} \\
    \addlinespace
    
    \paperTableSource{ChZh19,ZhCh19}{2019}{P4SC}{\ac{bmv2}, Tofino}{p4sc-github}
    \paperTable{LeLe19}{2019}{Re-SFC}{\ac{bmv2}}
    \paperTable{ZhBi20}{2020}{FlexMesh}{\ac{bmv2}}
    \paperTableSource{StHi20}{2019}{P4-SFC}{\ac{bmv2}, Tofino}{StHi20-repo}

    \bottomrule
    \end{tabularx}
    \end{center}
\end{table}

\subsection{Cellular Networks (4G/5G)}
\label{use-cases-advanced-networking-5g}

P4EC \cite{HoLe20} builds a local exit for LTE deployments with cloud-based EPC services.
A programmable switch distinguishes traffic and reroutes traffic for edge computing.
Non-critical traffic is forwarded to the cloud-based EPC.

The Trellis switch fabric (introduced in \sect{use-cases-data-center-switching}) features the spgw.p4 profile \cite{p4apps-onf, trellis-p4-tutorial}, an implementation of a Serving and PDN Gateway (SPGW) for 5G networking.
ONOS runs an SPGW-u application that implements the 3GPP control and user plane separation (CUPS) protocol to create, modify, and delete \ac{GTP} sessions.
It provides support for \ac{GTP} en- and decapsulation, filtering, and charging.

SMARTHO \cite{PaSi18} proposes a handover framework for 5G.
Distributed units (DUs) include real-time functions for multiple 5G radio stations.
Several DUs are controlled by a central unit (CU) that includes non-real-time control functions.
P4 switches are part of the CU and all DU nodes.
SMARTHO introduces a P4-based mechanism for preparing handover sequences for user devices that take a fixed path among 5G radio stations controlled by DUs.
This decreases the overall handover time, e.g., for users traveling in a train.

Aghdai et al. \cite{AgHu18} propose a P4-based transparent edge gateway (EGW) for mobile edge computing (MEC) in LTE or 5G networks.
Delay-sensitive and bandwidth-intense applications need to be moved from data centers in the core network to the edge of the radio access network (RAN).
5G networks rely on GTP-U for encapsulating IP packets from the mobile user to the core network.
IP routers in between forward packets based on the outer IP address of GTP-U frames.
The authors deploy EGWs as P4 switches at the edge of the IP transport network where service operators can deploy scalable network functions or services.
Each MEC service gets a virtual IP address, the P4-based EGWs parse the inner IP destination address of GTP-U.
If it sees traffic targeting a virtual IP address of a MEC service, it forwards it to the IP address of one of the serving instances of the MEC application.
In their follow-up work \cite{AgXu19}, the authors extend EGWs by a handover mechanism for migrating network state. 

GRED \cite{XiQi19} is an efficient data placement and retrieval service for edge computing.
It tries to improve routing path lengths and forwarding table sizes.
They follow a greedy forwarding approach based on DT graphs, where the forwarding table size is independent of the network size and the number of flows in the network.
GRED is implemented in P4, but the authors do not specify on which target.

HDS \cite{XiGu20} is a low-latency, hybrid, data sharing framework for hierarchical mobile edge computing.
The data location service is divided into two parts: intra-region and inter-region.
The authors present a data sharing protocol called Cuckoo Summary for fast data localization for the intra-region part. 
Further, they developed a geographic routing scheme to achieve efficient data location with only one overlay hop in the inter-region part.

Shen et al. \cite{ShLe19} present an FGPA-based \ac{GTP} engine for mobile edge computing in 5G networks.
Communication between the 5G back-haul and the conventional Ethernet requires de- and encapsulation of traffic with \ac{GTP}.
As most network entities do not have the capability to process \ac{GTP}, the authors leverage P4-programmable hardware for this purpose.

Lee et al. \cite{LeEb19} evaluate the performance of GTP-U and SRv6 stateless translation as GPT-U cannot be replaced by SRv6 without a transition period.
To that end, they implement \ac{GTP} and SRv6 on P4-programmable hardware.
They found that there are no performance drops if stateless translation is used and that SRv6 stateless translation is acceptable for the 5G user plane.

Ricart-Sanchez et al. \cite{RiMa19} propose an extension for the P4-NetFPGA framework for network slicing between different 5G users.
The authors extend the capabilities of the P4 pipeline and implement their mechanism on the NetFPGA-SUME. However, the authors do not provide any details about their implementation.

Singh et al. \cite{SiRo19} present an implementation for the Evolved Packet Gateway (EPG) in the Mobile Packet Core of 5G.
They show that they can offload the functionality to programmable switching ASICs and achieve line rate with low latency and jitter while scaling up to 1.7 million active users.

TurboEPC \cite{ShKu20} presents a redesign of the mobile packet core where parts of the control plane state is offloaded to programmable switches. 
State is stored in \acp{MAT}.
The switches then process a subset of signaling messages within the data plane itself, which leads to higher throughput and reduced latency.

Vörös et al. \cite{VoPo20} propose a hybrid approach for the next generation NodeB (gNB) where the majority of packet processing is done by a high-speed P4-programmable switch. 
Additional functions, such as ARQ or ciphering, are offloaded to external services such as DPDK implementations.

Lin et al. \cite{LiHu19a} enhance the Content Permutation Algorithm (eCPA) for secret permutation in 5G.
Packet payloads are split into code words and shuffled according to a secret cipher. 
They implement eCPA for switches of the Inventec D5264 series.

\subsection{Internet of Things (IoT)}
\label{use-cases-advanced-networking-iot}

BLESS \cite{UdMu17} implements a Bluetooth low energy (BLE) service switch based on P4 that acts as a proxy enabling flexible, policy-based switching and in-network operations of IoT devices.
BLE devices are strictly bound to a central device such as a smartphone or tablet.
IoT usage requires cloud-based solutions where central devices connect to an IoT infrastructure.
The authors propose a BLE service switch (BLESS) that is transparently inserted between peripheral and central devices and acts like a transparent proxy breaking up the peer-to-peer model.
It maintains BLE link layer connections to peripheral devices within its range.
A central controller implements functionalities such as service discovery, access policy enforcement, and subscription management so that features like service slicing, enrichment, and composition can be realized by BLESS.

Muppet \cite{UdMu18} extends BLESS by supporting the Zigbee protocol in parallel to BLE.
In addition to the features of BLESS, inter-protocol services between Zigbee and BLE and BLE/Zigbee and IP protocols are introduced.
An example for the latter are HTTP transactions that are automatically sent out by the switch if it sees a specified set of BLE/Zigbee transactions.
The data plane implementation of BLESS is extended by protocol-dependent packet parsers and processing and support for encrypted Zigbee packets via packet recirculation.

Wang et al. \cite{WaWu19} implement aggregation and disaggregation of small IoT packets on P4 switches.
For a small IoT packet, the header holds a large proportion of the packet's total size.
In large streams of IoT packets, this causes high overhead.
The current aggregation techniques for IoT packets are implemented by external servers or on the control plane of switches, both resulting in low throughput and added latency.
Therefore, the authors propose an implementation directly on P4 switches where IoT packets are buffered, aggregated, and encapsulated in UDP packets with a custom flag-header, type, and padding.
In disaggregation, the incoming packet is cloned to stripe out the single messages until all messages are separated.

Madureira et al. \cite{MaAr20} present the \emph{Internet of Things Protocol (IoTP)}, an L2 communication protocol for IoT data planes.
The main purpose of IoTP is data aggregation at the network level.
IoTP introduces a new, fixed header and is compatible with any forwarding mechanism.
The authors implemented IoTP for the \ac{bmv2} and store single packets of a flow in registers until the data can be aggregated.

Engelhard et al. \cite{EnZa19} present a system for massive wireless sensor networks.
They implement a physically distributed, and logically centralized wireless access systems to reduce the impairment by collisions. 
P4 is leveraged as connection between a physical access point and a virtual access point.
To that end, they extend the \ac{bmv2} to provide additional functionality. 
However, they give information about their P4 program only in form of a decision flow graph.

\subsection{Industrial Networking}
\label{use-cases-advanced-networking-industrial-networking}

FastReact \cite{VeKa18} outsources sensor data packet processing from centralized controllers to P4 switches.
The sensor data is recorded in variable-length time series data stores where an additional field holds the current moving average calculated on the time series.
Both data for all sensors can be polled by a central controller.
For controlling actuators directly on the data plane, FastReact supports the formulation of control logic in conjunctive normal form (CNF).
It is mapped to actions to either forward signal data to the controller, discard it, or directly send it to the actuator.
FastReact also features failure recovery directly on the switch.
For every sensor and actuator, timestamps for the last received packets along a timeout limit is recorded.
If failures are detected, sensor data are forwarded following failover rules with backup actuators for particular sensors.

Cesen et al. \cite{CeCs20} leverage P4-capable switches to move control logic to the network. 
Control applications reside in controllers that are responsible for emergency intervention, e.g., if a given threshold is exceeded.
The connection to the controller may be faulty and, therefore, controller intervention may not be fast enough.
In this work, the authors generate emergency packets, i.e., stop commands, directly in the data plane. 
The action is triggered if the switch receives a packet with a specific payload.

Kunze et al. \cite{KuGl21} investigate the applicability of in-network computing to industrial environments.
They offload a simple task, i.e., coordinate transformation, to different programmable P4 targets.
They come to the conclusion, that, while in general possible, even simple task have heavy demands on programmable network devices and that offloading may lead to inaccurate results.

\subsection{Time-Sensitive Networking (TSN)}
\label{use-cases-advanced-networking-tsn}

Rüth et al. \cite{RuGl18} introduce a scheme for implementing in-network control mechanisms for linear quadratic regulators (LQR).
LQRs can be described by a multiplication of a matrix and a vector.
The vector describes the control of the actuator, the matrix describes the current system state. 
The result of the multiplication is a control command.
The destination of a switch describes a specific actuator.
When a switch receives a control packet, it matches the destination of the packet onto a match-and-action table.
The lookup provides the control vector for the actuator.
The control vector from the lookup is then multiplied with the system state matrix that is stored in a register to calculate the control command for the actuator.
The resulting control command is written into the packet header and the packet is forwarded to the target actuator.

Kannan et al. \cite{KaJo19} introduce the Data Plane Time synchronization Protocol (DPTP) for distributed applications with computations directly on the P4 data plane.
DPTP follows a request-response model, i.e., all P4 switches request the global time from a designated master switch.
Therefore, each switch features a local control plane that generates time requests sent to the master switch.
Additionally, the control plane handles overflows in time calculation for administration.

Kundel et al. \cite{KuSi19} demonstrate timestamping with nanosecond accuracy.
They describe a simple setup with a Tofino-based switch and a breakout cable to connect two ports of the switch.
In the experiment, timestamps at the moment of sending and reception are recorded in the packet header.
The authors compare those two timestamps to show that very fine-grained measurements are possible.

\subsection{Network Function Virtualization (NFV)}
\label{use-cases-advanced-networking-nfv}

Kathará \cite{BoIo18} runs \acp{NF} as P4 programs either on software or hardware targets.
For software-based deployment, the framework leverages Docker containers that run \acp{NF} as container images or individual setups for Quagga, Open vSwitch, or \ac{bmv2} container images.
For hardware-based deployment on P4 switches, \acp{NF} are either replicated on every P4 switch or distributed on multiple P4 switches as needed.
In both cases, a load balancer or service classifier forwards flows to the appropriate P4 switch.
As a main advantage, P4 programs can be shifted between the \ac{bmv2}-based P4 software targets and hardware targets depending on the required performance.

P4NFV \cite{HeBa18} also deals with the idea of running \acp{NF} either on software- or hardware-based P4 targets.
The authors adopt the ETSI \ac{NFV} architecture with control and monitoring entities and add a layer that abstracts various types of software- and hardware-based P4 targets as P4 nodes.
For optimized deployment, the targets performance characteristics are part of the P4 node description.
For runtime reconfiguration, the authors propose two approaches.
In pipeline manipulation, the P4 program features multiple match-action pipelines that can be enabled or disabled by setting register flags.
In program reload, a new P4 program is compiled and loaded to the P4 target.
The authors propose to perform state management and migration either directly on the data plane or via a control plane.

Osiński et al. \cite{OsTa19} use P4 to offload the data plane of \acp{VNF} into a cloud infrastructure by allowing \acp{VNF} to inject small P4 programs into P4 devices like SmartNICs or top-of-rack switches.
This results in better performance and a microservice-based approach for the data plane.
A new P4 architecture model that integrates abstractions used to develop \ac{VNF} data planes was developed.

Moro et al. \cite{MoVe20} present a framework for \ac{NF} decomposition and deployment. 
They split \acp{NF} into components that can run on CPUs or that can be offloaded to specific programmable hardware, e.g., P4 programmable switches.
The presented orchestrator combines multiple functions into a single P4 program that can be deployed to programmable switches.

DPPx \cite{OsTa19b} implements a framework for P4-based data plane programmability and exposure which allows enhancing \ac{NFV} services.
They introduce data plane modules written in P4 which can be leveraged by the application plane.
As an example, a dynamic optimization of packet flow routing (DOPFR) is implemented using DPPx.

Mohammadkhan et al. \cite{MoPa19} provide a unified P4 switch abstraction framework where servers with software \acp{NF} and P4-capable SmartNICs are seen as one logical entity by the SDN controller.
They further leverage Mixed Integer Linear Programming (MILP) to determine partitioning of P4 tables for optimal placement of \acp{NF}.

FOP4 \cite{MoPe19a} \cite{MoPe19b} implements a rapid prototyping platform that supports container-based, P4-switch-based, and SmartNIC-based \acp{NF}.
They argue that a prototyping platform is needed to quickly develop and evaluate new \ac{NFV} use cases.

PlaFFE \cite{MaDo20} introduces \ac{NFV} offloading where some features of \acp{VNF} or embedded Network Functions (eNFs) are executed on SmartNICs using P4.
Additionally, P4 is used to steer traffic either through the eNFs or through \acp{VNF} using SR-IOV.

\subsection{Service Function Chains (SFCs)}
\label{use-cases-advanced-networking-sfc}

P4SC \cite{ChZh19} \cite{ZhCh19} implements a \ac{SFC} framework for P4 targets.
\acp{SFC} are described as directed acyclic graph of \acp{SF}.
In P4SC, \acp{SF} are represented by blocks.
Each block has a unique identifier, a P4 program for ingress processing, and a P4 program for egress processing. 
P4SC includes 15 \ac{SF} blocks, e.g., L2 forwarding, which are extracted from switch.p4.
After the user specified all \acp{SFC} for a particular P4 target, the P4SC converter merges the directed acyclic graphs of all \acp{SFC} with an LCS-based algorithm into an intermediate representation.
Then, the P4SC generator creates the final P4 program based on the intermediate representation to be deployed onto the P4 target.
P4 program generation includes runtime management, i.e., the generator creates one \ac{API} per \ac{SFC} while hiding \ac{SF}-specific details, e.g., names of particular match-and-action tables.

Re-SFC \cite{LeLe19} improves P4SC's resource usage by using resubmit operations.
If the specified order of \acp{SF} in an \ac{SFC} does not match the pre-embedded \ac{SF} of the P4 switch, incoming flows cannot be processed.
P4SC solves this problem by permitting redundant \ac{NF} embeds, i.e., if \acp{SF} of one \ac{SFC} are required by another \acp{SFC}, those \acp{SF} are just replicated.
To reduce the costly usage of match-and-action tables, Re-SFC introduces resubmit actions where packets are re-bounced to the ingress.

FlexMesh \cite{ZhBi20} tackles the problem of fixed \ac{SFC} flow control, i.e., when the specified order of \acp{SF} does not match the pre-embedded \ac{SF}, by leveraging \acp{MAT}.
\acp{SF} can be dynamically bypassed, and recirculation is used to build any desired \ac{SF} chain.

P4-SFC \cite{StHi20} is an \ac{SFC} framework based on MPLS segment routing and \ac{NFV}.
P4 is used to implement a traffic classifier.
A central orchestrator deploys service functions as \acp{VNF} and configures the traffic classifier based on definitions of \acp{SFC}.

\subsection{Summary and Analysis}
As the research domain of advanced networking covers different topics, almost all core properties of P4 are covered.
The area of cellular networks (Section \ref{use-cases-advanced-networking-5g}) greatly benefits from the \emph{definition and usage of custom packet headers} as many works are based on tunneling technologies, such as GTP.
Further, \emph{flexible packet header processing} allows implementing new 5G concepts such as gNB or EPG.
Some use cases still require offloading tasks to specialized hardware or software by leveraging the \emph{target-specific packet header processing function} property of P4, e.g., for ARQ or ciphering in the context of gNB.
\Acf{NFV} (Section \ref{use-cases-advanced-networking-nfv}) benefits from \emph{flexible development and deployment} as single \acfp{NF} can be replaced or relocated during operation.
New protocols and extensions to existing protocols presented in Section \ref{use-cases-advanced-networking-sfc} rely on \emph{definition and usage of custom packet headers} and \emph{flexible packet header processing}.

%% file: chapters/13-uc_network-security.tex
\section{Applied Research Domains: Network Security}
\label{sec:use-cases-network-security}

We describe applied research on firewalls, port knocking, \ac{DDoS} attack mitigation, intrusion detection systems, connection security, and other fields of application.
\tbl{implementations-network-security} shows an overview of all the work described.
At the end of the section, we summarize the work and analyze it with regard to P4's core features described in \sect{benefits}.

\begin{table}[htbp]
    \caption{Overview of applied research on network security (\sect{use-cases-network-security}).}
    \begin{center}
    \begin{tabularx}{\linewidth}{@{} l l X r @{}}
    \toprule
    \textbf{Research work} & \textbf{Year} & \textbf{Targets} & \textbf{Code}\\
    \midrule
    \addlinespace
    \multicolumn{4}{@{\extracolsep{\fill}}l}{\noindent \textbf{Firewalls} (\ref{use-cases-netsec-firewalls})} \\
    \addlinespace
    \paperTable{RiMa18,RiMa19a}{2018/19}{Ricart-Sanchez et al.}{NetFPGA-SUME}
    \paperTable{CaBi18}{2018}{CoFilter}{Tofino}
    \paperTable{DaCh18}{2018}{P4Guard}{\ac{bmv2}}
    \paperTable{VoKi16}{2016}{Vörös and Kiss}{p4c-behavioral}
    
    \midrule
    \addlinespace
    \multicolumn{4}{@{\extracolsep{\fill}}l}{\noindent \textbf{Port Knocking } (\ref{use-cases-netsec-portknocking})} \\
    \addlinespace
    \paperTable{ZaFr20}{2020}{P4Knocking}{\ac{bmv2}}
    \paperTable{AlAl19}{2019}{Almaini et al.}{\ac{bmv2}}
    
    \midrule
    \addlinespace
    \multicolumn{4}{@{\extracolsep{\fill}}l}{\noindent \textbf{DDoS Mitigation Mechanisms} (\ref{use-cases-netsec-ddos})} \\
    \addlinespace
    \paperTable{GrLi18}{2018}{LAMP}{\ac{bmv2}}
    \paperTable{FeXi18, FeXi19}{2018/19}{TDoSD@DP}{\ac{bmv2}}
    \paperTable{KuVo19}{2019}{Kuka et al.}{Xilinx UltraScale+, Intel Stratix 10}
    \paperTable{PaCu18, PaCi19}{2018/19}{Paolucci et al.}{\ac{bmv2}, NetFPGA-SUME}        
    \paperTable{MiWa19}{2019}{ML-Pushback}{-}        
    \paperTable{AfBr17}{2017}{Afek et al.}{p4c-behavioral}        
    \paperTableSource{LaMa19}{2019}{Cardoso Lapolli et al.}{\ac{bmv2}}{p4-ddosd}
    \paperTable{CaLa20}{2020}{Cai et al.}{-}
    \paperTable{LiWu20}{2020}{Lin et al.}{\ac{bmv2}}
    \paperTable{MuIo20}{2020}{Musumeci et al.}{\ac{bmv2}}
    \paperTable{KhCs20}{2020}{DIDA}{\ac{bmv2}}
    \paperTable{DiPa20}{2020}{Dimolianis et al.}{Netronome}
    \paperTableSource{ScGa20}{2020}{Scholz et al.}{\ac{bmv2}, \tapas, Netronome, NetFPGA SUME}{syn-proxy}
    \paperTable{FrKf20}{2020}{Friday et al.}{\ac{bmv2}}
    \paperTable{MeTs18}{2018}{NetHide}{-}
    \midrule
    \addlinespace
    \multicolumn{4}{@{\extracolsep{\fill}}l}{\noindent \textbf{Intrusion Detection Systems \& Deep Packet Inspection} (\ref{use-cases-netsec-ids})} \\
    \addlinespace
    \paperTable{LeBr19}{2019}{P4ID}{\ac{bmv2}}
    \paperTable{KaSa18}{2018}{Kabasele and Sadre}{\ac{bmv2}}
    \paperTableSource{HySo20}{2020}{DeepMatch}{Netronome}{deepmatch}
    \paperTableSource{QiPo20}{2020}{Qin et al.}{\ac{bmv2}, Netronome}{github-qin}
    \paperTable{AmSi20}{2020}{SPID}{\ac{bmv2}}
    \bottomrule
    \end{tabularx}
    \end{center}
    \label{tbl:implementations-network-security}
    \end{table}

\begin{table}[htbp]
    \begin{center}
    \begin{tabularx}{\linewidth}{@{} l l X r @{}}
    \toprule
    \textbf{Research work} & \textbf{Year} & \textbf{Targets} & \textbf{Code}\\
    \midrule
    \addlinespace
    \multicolumn{4}{@{\extracolsep{\fill}}l}{\noindent \textbf{Other Fields of Application} (\ref{use-cases-netsec-others})} \\
    \addlinespace
    \paperTable{ChSu19}{2019}{Chang et al.}{\ac{bmv2}}
    \paperTable{FeZh19}{2019}{Clé}{-}
    \paperTable{KuLi20}{2020}{P4DAD}{\ac{bmv2}}
    \paperTableSource{Chen20}{2020}{Chen}{Tofino}{aes-tofino-repo}
    \paperTable{GoSa20}{2020}{Gondaliya et al.}{NetFPGA SUME}
    \paperTableSource{KaXu20}{2020}{Poise}{Tofino}{github-poise}
    \midrule
    \addlinespace
    \multicolumn{4}{@{\extracolsep{\fill}}l}{\noindent \textbf{Connection Security} (\ref{use-cases-netsec-connectionsecurity})} \\
    \addlinespace
    \paperTableSource{HaSc20}{2020}{P4-MACsec}{\ac{bmv2}, NetFPGA-SUME}{p4-macsec}
    \paperTableSource{HaHa20}{2020}{P4-IPsec}{\ac{bmv2}, NetFPGA-SUME, Tofino}{p4-ipsec}
    \paperTableSource{DaFe19}{2019}{SPINE}{\ac{bmv2}}{spine}
    \paperTable{QiQu20}{2020}{Qin et al.}{\ac{bmv2}}
    \paperTableSource{LiQu20}{2020}{P4NIS}{\ac{bmv2}}{p4nis}
    \paperTable{LiGa19}{2020}{LANIM}{\ac{bmv2}}
    \bottomrule
    \end{tabularx}
    \end{center}
    \end{table}

\subsection{Firewalls}
\label{use-cases-netsec-firewalls}

Ricart-Sanchez et al. \cite{RiMa18} present a 5G firewall that analyzes \ac{GTP} data transmitted between edge and core networks.
P4 allows an implementation of parsing and matching \ac{GTP} header fields such as 5G user source IP, 5G user destination IP, and identification number of the \ac{GTP} tunnel.
The P4 pipeline implements an allow-by-default policy, DROP actions for specific sets of keys can be installed via a data plane \ac{API}.
In a follow-up work \cite{RiMa19a}, the authors extend the 5G firewall by support for multi-tenancy with VXLAN.

CoFilter \cite{CaBi18} implements an efficient flow identification scheme for stateful firewalls in P4.
To solve the problem of limited table sizes on SDN switches, flow identifiers are calculated by applying a hashing function to the 5-tuple of every packet directly on the switch.
The proposed concept includes a novel hash rewrite function that is implemented on the data plane.
It resolves hash commission and hash table optimization using an external server.

P4Guard \cite{DaCh18} replaces software-based firewalls by P4-based virtual firewalls in the VNGuard \cite{DeHu15} system.
VNGuard introduces controller-based deployment and management of virtual firewalls with the help of \ac{SDN} and NFV.
The P4-based firewall comprises a single \ac{MAT} that allows ALLOW/DROP decision for Layer 3/4 header fields as match keys.
The flow statistics are recorded with the help of counters.
Another \ac{MAT} allows enabling/disabling the firewall at runtime.

Vörös and Kiss \cite{VoKi16} present a firewall implemented in P4.
The parser supports Ethernet, IPv4/IPv6, UDP, and TCP headers.
A ban list comprises MAC address/IP address entries that represent network hosts. 
Packets matching this ban list are directly dropped.
To mitigate port scan or \ac{DDoS} attacks, counters track packet rate and byte transfer statistics. 
Another \ac{MAT} implements whitelist filtering.

\subsection{Port Knocking}
\label{use-cases-netsec-portknocking}

Port knocking is a simple authentication mechanism for opening network ports.
Network hosts send TCP SYN packets in predefined sequences to certain ports.
If the sequence is completed correctly, the server opens up a desired port.
Typically, port knocking is implemented in software on servers.

P4Knocking \cite{ZaFr20} implements port knocking on P4 switches.
The authors propose four different implementations for P4.
In the first implementation, P4 switches track the state of knock sequences in registers where the source IP address is used as an index.
The second implementation uses a CRC-hash of the source IP address as index for the knocking state registers.
To resolve the problem of hash collisions, the third implementation relies on identifiers that are calculated and managed by the controller.
The fourth implementation solely relies on the controller, i.e., P4 switches forward all knocking packets to the controller.

Almaini et al. \cite{AlAl19} implement port knocking with a ticket mechanism on P4 switches.
Traffic is only forwarded if the sender has a valid ticket.
Predefined trusted nodes have a ticket by default, untrustworthy nodes must obtain a ticket by successful authentication via port knocking.
The authors use the HIT/MISS construct of P4 as well as stateful P4 components to implement the concept.
Port knocking sequences and trusted/untrusted hosts can be maintained by the control plane.

\subsection{DDoS Attack Mitigation}
\label{use-cases-netsec-ddos}

LAMP \cite{GrLi18} presents a cooperative mitigation mechanism for \ac{DDoS} attacks that relies on information from the application layer.
Ingress P4 switches add a unique identifier to the IP options header field of any processed packet.
The last P4 switch ahead of the target host stores this mapping and empties the IP options header field.
If a network hosts, e.g., a database server, detects an ongoing \ac{DDoS} attack on the application layer, it adds an attack flag to the IP options header field and sends it back to the switch.
The switch forwards this packet to the ingress switch to enable dropping of all further packets of this flow.

TDoSD@DP \cite{FeXi18} is a P4-based mitigation mechanism for \ac{DDoS} attacks targeting SIP proxies.
Stateful P4 registers record the number of SIP INVITE and SIP BYE messages.
Then, a simple state machine monitors sequences of INVITE and BYE messages.
Many INVITES followed by zero BYE messages lead to dropping SIP INVITE packets where valid sequences of INVITE and BYE messages will keep the port open.
In a follow-up work \cite{FeXi19}, the authors present an alternative approach where P4 switches act as distributed sensors.
An \ac{SDN} controller periodically collects data from counters of P4 switches to perform centralized attack detection.
Then, attack mitigation is performed by installing DROP rules on the P4 switches.

Kuka et al. \cite{KuVo19} present a \ac{DDoS} mitigation system that targets volumetric \ac{DDoS} attacks called reflective amplification attacks.
The authors port an existing VHDL implementation into a P4 program that runs on \ac{FPGA} targets.
The implementation selects the affected subset of the incoming traffic, extracts packet data, and forwards it as a digest to an \ac{SDN} controller.
The \ac{SDN} controller continuously evaluates this information; a heuristic algorithm identifies aggressive IP addresses by looking at the volumetric contribution of source IP addresses to the attack.
In case of a detected attack, the SDN controller installs DROP rules.

Paolucci et al. \cite{PaCu18, PaCi19} present a stateful mitigation mechanism for TCP SYN flood attacks.
It is part of a P4-based edge packet-over-optical node that also comprises traffic engineering functionality.
P4 registers keep per-session statistics to detect TCP SYN flood attacks.
One register records the port number of the last TCP SYN packet, the another one records the number of attempts matching the TCP SYN flood behavior.
If the latter one exceeds a defined threshold, the packets are dropped.

ML-Pushback \cite{MiWa19} proposes an extension of the Pushback \ac{DDoS} attack mitigation mechanism by machine learning techniques.
P4 switches implement a data collector mechanism that collects dropped packets and forwards them as digest messages to the control plane.
On the control plane, a deep learning module extracts signatures and classifies the collected digest with a decision tree model.
Attack mitigation is performed by throttling attacker traffic via rate limits.

Afek et al. \cite{AfBr17} implement known mitigation mechanisms for SYN and DNS spoofing in \ac{DDoS} attacks for OpenFlow and P4 targets.
The OpenFlow implementation targets Open vSwitch and OpenFlow 1.5 where P4 implementations are compiled for p4c-behavioral without control plane involvement.
In addition, the authors implemented a set of algorithms and methods for dynamically distributing the rule space over multiple switches.

Cardoso Lapolli et al. \cite{LaMa19} describe an algorithmic approach to detect and stop \ac{DDoS} attacks on P4 data planes.
The algorithm was specifically created under the functional constraints of P4 and is based on the calculation of the Shannon entropy.

Cai et al. \cite{CaLa20} propose a novel method for collecting traffic information to detect TCP port scanning attacks.
The authors propose the "0-replacement" method as an efficient alternative to existing sampling and aggregation methods.
It introduces a pending request counter (PRcounter) and relies on registers to bind hashing identifiers of the attackers' IP addresses to PRcounter values.
The authors describe the concept as compliant to \ac{PSA}, but only simulation results are given.

Lin et al. \cite{LiWu20} present a comparison of \ac{OF}- and P4-based implementations of basic mitigation mechanisms against SYN flooding and ARP spoofing attacks.

Musumeci et al. \cite{MuIo20} present P4-assisted \ac{DDoS} attack mitigation using an \ac{ML} classifier.
An ML-based \ac{DDoS} attack detection module with a classifier is running on a controller. 
The P4 switch forwards traffic to the module; the \ac{DDoS} attack detection module responds with a decision.
The authors consider three use cases: packet mirroring + header mirroring + metadata extraction.
In metadata extraction, P4 switches implement counters that store occurrences of IP, UDP, TCP, and SYN packets.
In the case that one of the counters exceeds a defined threshold, the P4 switch inserts a custom header with the counter values and sends it to the \ac{DDoS} attack detection module.

DIDA \cite{KhCs20} presents a distributed mitigation mechanism against amplified reflection \ac{DDoS} attacks.
In this type of \ac{DDoS} attack, spoofed requests lead to responses that are by magnitude larger.
An example is a DNS ANY query.
The authors rely on count-min sketch data structures and monitoring intervals to put the number of requests and responses into relation.
In case of a detected \ac{DDoS} attack, \acp{ACL} are used to block the traffic near to the attacker.

Dimolianis et al. \cite{DiPa20} introduce a multi-feature \ac{DDoS} detection scheme for TCP/UDP traffic.
It considers the total number of incoming traffic for a particular network, the significance of the network, and the symmetry ratio of incoming and outgoing traffic for classifications.
The feature analysis is time-dependent and focuses on distinct time intervals.

Scholz et al. \cite{ScGa20} propose a SYN proxy that relies on SYN cookies or SYN authentication as protection against SYN flooding \ac{DDoS} attacks.
The authors present a software implementation based on \ac{DPDK} and compare it to a \ac{bmv2}-based P4 implementation that is ported to the \tapas P4 software target, Netronome P4 hardware target, and NetFPGA SUME P4 hardware target.
Evaluation results, benefits, and challenges for each platform are discussed.

Friday et al. \cite{FrKf20} present a two-part \ac{DDoS} detection and mitigation scheme.
In the first part, a P4 target applies a one-way traffic analysis using bloom filters and time-dependent statistics such as moving averages.
In the second part, the P4 target analyzes the bandwidth and transport protocols used by various applications to perform a volumetric analysis.
The processing pipeline then decides about malicious traffic to be dropped.
Administrators may supply custom network parameters used for dynamic threshold calculation that are then installed via an \ac{API} on the data plane.
The authors demonstrate the effectiveness of the proposed approach by three use cases: UDP amplification \ac{DDoS} attacks, SYN flooding \ac{DDoS} attacks, and slow \ac{DDoS} attacks.

NetHide \cite{MeTs18} prevents link-flooding attacks by obfuscating the topology of a network.
It achieves this by modifying path tracing probes in the data plane while preserving their usability.

\subsection{Intrusion Detection Systems (\acs{IDS}) \& Deep Packet Inspection (\acs{DPI})}
\label{use-cases-netsec-ids}

P4ID \cite{LeBr19} reduces \ac{IDS} processing load by apply pre-filtering on P4 switches (\ac{IDS} offloading/bypassing).
P4ID features a rule parser that translates Snort rules with a multistage mechanism into \ac{MAT} entries.
The P4 processing pipeline implements a stateless and a stateful stage.
In the stateless stage, TCP/ICMP/UDP packets are matched against a \ac{MAT} to decide if traffic should be dropped, forwarded to the next hop, or forwarded to the \ac{IDS}.
In the stateful stage, the first $n$ packets of new flows are forwarded to the \ac{IDS}.
This allows that traffic targeting well-known ports can be also analyzed.
Combining the feedback of the \ac{IDS} for packet samples with the stateless stage is future work.

Kabasele and Sadre \cite{KaSa18} present a two-level \ac{IDS} for industrial control system (ICS) networks.
The \ac{IDS} targets the Modbus protocol that runs on top of TCP in SCADA networks.
The first level comprises two whitelists: a flow whitelist for filtering on the TCP layer and a Modbus whitelist.
If no matching entry is found for a given packet, it is forwarded to the second layer.
This is in stark contrast to legacy whitelisting where packets are just dropped.
In the second level, a Zeek network security analyzer acts as deep packet inspector running on a dedicated host.
It analyzes the given packet, makes a decision, and instructs the controller to update filters on the switch.

DeepMatch \cite{HySo20} introduces \ac{DPI} for packet payloads.
The concept is implemented with the help of network processors; its prototype is built with the Netronome NFP-6000 SmartNIC P4 target.
The authors present regex matching capabilities that are executed in \SI{40}{\giga\bit\per\s} (line rate of the platform) for stateless intra-packet matching and about \SI{20}{\giga\bit\per\s} for stateful inter-packet matching.
The DeepMatch functionalities are natively implemented in Micro-C for the Netronome platform and integrated into the P4 processing pipeline with the help of P4 externs.

Qin et al. \cite{QiPo20} present an \ac{IDS} based on binarized neural networks (BNN) and federated learning.
BNNs compress neural networks into a simplified form that can be implemented on P4 data planes.
Weights are compressed into single bits and computations, e.g., activation functions, are converted into bit-wise operations.
P4 targets at the network edge then apply BNNs to classify incoming packets.
To continuously train the BNNs on the P4 targets, the authors propose a federated learning scheme.
Each P4 target is connected to a controller that trains an equally structured neural network with samples received from the P4 target. 
A cloud service aggregates local updates received from the controllers and responds with weight updates that are processed into the local model.

In the Switch-Powered Intrusion Detection (SPID) framework \cite{AmSi20}, switches compute and store flow statistics, and perform traffic change detection.
If a relevant change in traffic is detected, measurement data is pushed to the control plane.
In the control plane, the measurement data is fed to a \ac{ML}-based anomaly detection pipeline to detect potential attacks.

\subsection{Connection Security}
\label{use-cases-netsec-connectionsecurity}
P4-MACsec \cite{HaSc20} presents an implementation of IEEE 802.1AE (MACsec) for P4 switches.
A two-tier control plane with local switch controllers and a central controller monitor the network topology and automatically set up MACsec on detected links between P4 switches.
For link discovery and monitoring, the authors implement a secured variant of LLDP that relies on encrypted payloads and sequence numbers.
MACsec is directly implemented on the P4 data plane; encryption/decryption using AES-GCM is implemented on the P4 target and integrated in the P4 processing pipeline as P4 externs.

P4-IPsec \cite{HaHa20} presents an implementation of IPsec for P4 switches.
IPsec functionality is implemented in P4 and includes ESP in tunnel mode with support for different cipher suites.
As in P4-MACsec, the cipher suites are implemented on the P4 target and integrated as P4 externs.
In contrast to standard IPsec operation, IPsec tunnels are set up and renewed by an \ac{SDN} controller without IKE.
Site-to-site operation mode supports IPsec tunnels between P4 switches.
Host-to-site operation mode supports roadwarrior access to an internal network via a P4 switch.
To make the roadwarrior host manageable by the controller, the authors introduce a client agent tool for Linux hosts.

SPINE \cite{DaFe19} introduces surveillance protection in the network elements by IP address obfuscation against surveillance in intermediate networks.
In contrast to software-based approaches such as TOR, SPINE runs entirely on the data plane of two nodes with intermediate networks in between.
It applies a one-time-pad-based encryption scheme with key rotation to encrypt IP addresses and, if present, TCP sequence and acknowledgment numbers.
The SPINE nodes add a version number representing the encryption key index to each packet by which the receiving switch can select the appropriate key for decryption.
The key sets required for the key rotation are maintained by a central controller.

Qin et al. \cite{QiQu20} introduce encryption of TCP sequence numbers using substitution-boxes to protect traffic between two P4 switches.
An ONOS-based controller receives the first packet of each new flow and applies security policies to decide whether the protection should be enabled.
Then, it installs the necessary data in registers and updates \acp{MAT} to enable TCP sequence number substitution.

P4NIS \cite{LiQu20} proposes a scheme to protect against eavesdropping attacks.
It comprises three lines of defense.
In the first line of defense, packets that belong to one traffic flow are disorderly transmitted via various links.
In the second line of defense, source/destination ports and sequence/acknowledgment numbers are substituted via s-boxes similar to the approach of Qin et al. \cite{QiQu20}.
The third line of defense resembles existing encryption mechanisms that are not covered by P4NIS.

LANIM \cite{LiGa19} presents a learning-based adaptive network immune mechanism to prevent against eavesdropping attacks.
It targets the Smart Identifier Network (SINET) \cite{ZhQu16}, a novel, three-layer Internet architecture.
LANIM applies the minimum risk \ac{ML} algorithm to respond to irregular conditions and applies a policy-based encryption strategy focusing on the intent and application.

\subsection{Other Fields of Application}
\label{use-cases-netsec-others}

Chang et al. \cite{ChSu19} present IP source address encryption.
It accomplishes non-linkability of IP addresses as proactive defense mechanism.
Network hosts are connected to trusted P4 switches at the network edges.
In between, packets are exchanged via untrusted switches/routers.
The P4 switch next to the sender encrypts the sender IP address by applying an XOR operation with a hash calculated by a random number and a shared key.
The P4 switch next to the receiver decrypts the original sender IP address.
The mechanism includes a dynamic key update mechanism so that transformations are random.

Clé \cite{FeZh19} proposes to upgrade particular switches in a legacy network to P4 switches that implement security network functions (SNFs) such as rule-based firewalls or \ac{IDS} on P4 switches.
Clé comprises a smart device upgrade selection algorithm that selects switches to be upgraded and a controller that forwards traffic streams to the P4 switches that implement SNFs.

P4DAD \cite{KuLi20} presents a novel approach to secure duplicate address detection (DAD) against spoofing attacks.
Duplicate address detection is part of NDP in IPv6 where nodes check if an IPv6 address to be applied conflicts with another node.
As the messages exchanged in duplicate address detection are not authenticated or encrypted, it is vulnerable to message spoofing.
As simple alternative to authentication or encryption, P4DAD introduces a mechanism to filter spoofed NDP messages.
The P4 switch maintains registers to create bindings between IPv6 addresses, port numbers, and address states.
Thereby, it can detect and drop spoofed NDP messages.

Chen \cite{Chen20} shows how AES can be implemented on Tofino-based P4 targets in P4 using \acp{MAT} as lookup tables.
Expansion of the AES key is performed in the control plane.
MAT entries specific to the encryption keys are generated by a controller.

Gondaliya et al. \cite{GoSa20} implement six known mechanisms against IP address spoofing for the NetFPGA SUME P4 target.
Those are Network Ingress Filtering, Reverse Path Forwarding (Loose, Strict and Feasible), Spoofing Prevention Method (SPM), and Source Address Validation Improvement (SAVI).
The authors compare the different mechanisms with regard to resource usage on the FPGA and report that the implementations of all mechanisms achieve a throughput of about \SI{8.5}{\giga\bit\per\s} and a processing latency of about \SI{2}{\micro\s} per packet.

Poise \cite{KaXu20} introduces context-aware policies for securing P4-based networks in BYOD scenarios.
Instead of relying on a remote controller or software-based solution, Poise implements context-aware policy enforcement directly on P4 targets.
Network administrators define context-aware security policies in a declarative language based on Pyretic NetCore that are then compiled into P4 programs to be executed on P4 targets.
BYOD clients run a context collection module that adds context information headers to network packets.
The P4 program generated by Poise then parses and uses this information to enforce \acp{ACL} based on device runtime contexts.
P4 targets in Poise are managed by a Poise controller that compiles the P4 programs, installs them on the P4 targets, and provides configuration data to the collection modules.
The authors present a prototype including PoiseDroid, an implementation of the context collection module for Android devices.

\subsection{Summary and Analysis}
Several prototypes apply P4's \emph{custom packet headers}, e.g., for building a \ac{GTP} firewall for 5G networks, a DDoS attack mitigation mechanism for the SIP, or an \ac{IDS} for the Modbus protocol in industrial networks.
It is also used for in-band signaling, e.g., in cooperative DDoS attack detection.
All prototypes rely on \emph{flexible packet header processing}; outstanding for this section, many of them also rely on \emph{target-specific packet header processing functions} offered by the P4 target.
Some works require custom externs, e.g., for applying MACsec or IPsec on P4 data planes.
As for prototypes from the research area \emph{Monitoring} (\sect{use-cases-monitoring}), many prototypes rely on registers and counters for recording statistics, e.g., for detecting attacks in DDoS mitigation or in \acp{IDS}.
While custom packet headers and basic packet header processing are supported by all P4 hardware targets, the portability of prototypes using these specific functions is very limited.
Several prototypes also rely on \emph{packet processing on the control plane} where information (e.g., from blocking lists, \ac{IDS} rules) is translated into \ac{MAT} rules for data plane control or data received from the data plane (e.g., statistical data or packet digests) is used for runtime control.
\emph{Flexible deployment} allows to re-deploy network security programs on P4 switches in large networks.

%% file: chapters/14-uc_others.tex
\section{Miscellaneous Applied Research Domains}
\label{sec:uses-cases-others}

This section summarizes work that falls outside of the other application domains. 
We describe applied research on network coding, distributed algorithms, state migration, and application support.
Table~\ref{tab:implementations-others} shows an overview of all the work described.
At the end of the section, we summarize the work and analyze it with regard to P4's core features described in \sect{benefits}.

\begin{table}[htp]
\caption{Overview of applied research on miscellaneous research domains (\sect{uses-cases-others}).}
\begin{center}
\begin{tabularx}{\linewidth}{@{} l l X r @{}}
\toprule
\textbf{Research work} & \textbf{Year} & \textbf{Targets} & \textbf{Code}\\
\midrule
\addlinespace
\multicolumn{4}{@{\extracolsep{\fill}}l}{\noindent \textbf{Network Coding} (\sect{use-cases-nc})} \\
\addlinespace
\paperTableSource{KuBa18}{2018}{Kumar et al.}{\ac{bmv2}}{p4-KuBa18}
\paperTable{GoSi19}{2019}{Gonçalves et al.}{\ac{bmv2}}
\midrule
\addlinespace
\multicolumn{4}{@{\extracolsep{\fill}}l}{\noindent \textbf{Distributed Algorithm} (\sect{use-cases-distributed})}\\
\addlinespace
\paperTable{KoMa18}{2018}{P4CEP}{\ac{bmv2}, Netronome}
\paperTable{SaAb17}{2017}{DAIET}{-}
\paperTable{SaSi20}{2020}{Sankaran et al.}{-}
\paperTable{ZhHa17}{2017}{Zang et al.}{\ac{bmv2}}
\paperTableSource{DaCa16, DaBr20}{2016/20}{Dang et al.}{Tofino}{p4-DaBr20}
\paperTable{SaDe19, SaDe19b}{2019}{P4BFT}{\ac{bmv2}, Netronome}
\paperTable{ZePo20}{2020}{SwiShmem}{-}
\paperTableSource{HaJa20}{2020}{SC-BFT}{\ac{bmv2}}{p4-sc-bft-repo}
\paperTable{SvBo18}{2018}{LODGE}{\ac{bmv2}}
\paperTableSource{SvBo20}{2020}{LOADER}{}{loader-repo}
\paperTable{TaKe20}{2020}{FLAIR}{Tofino}
\midrule
\addlinespace
\multicolumn{4}{@{\extracolsep{\fill}}l}{\noindent \textbf{State Migration} (\sect{state_migration})}\\
\addlinespace
\paperTable{LuYu17}{2017}{Swing State}{\ac{bmv2}}
\paperTableSource{XiCh20}{2020}{P4Sync}{\ac{bmv2}}{p4sync-repo}
\paperTable{XuZh20}{2020}{Xue et al.}{\ac{bmv2}}
\paperTable{KuNe20}{2020}{Kurzniar et al.}{\ac{bmv2}}
\paperTable{SaSi20b}{2020}{Sankaran et al.}{NetFPGA-SUME}
\midrule
\addlinespace
\multicolumn{4}{@{\extracolsep{\fill}}l}{\noindent \textbf{Application Support} (\sect{application_support})}\\
\addlinespace
\paperTableSource{WoRa19}{2019}{P4DNS}{NetFPGA SUME}{p4-dns}
\paperTableSource{KuNo19}{2019}{P4-BNG}{\ac{bmv2}, Tofino, Netronome, NetFPGA-SUME}{p4-p4se-repo}
\paperTable{MaAl18}{2018}{ARP-P4}{\ac{bmv2}}
\paperTable{GlKr19}{2019}{Glebke et al.}{Netronome}
\paperTable{XiQi19b}{2019}{COIN}{-}
\paperTable{LuLi19}{2019}{Lu et al.}{Tofino}
\paperTable{YaPa19}{2019}{Yazdinejad et al.}{\ac{bmv2}}
\paperTableSource{OsTa20}{2020}{P4rt-OVS}{-}{p4-P4rt-OVS-repo}
\addlinespace
\bottomrule
\end{tabularx}
\end{center}
\label{tab:implementations-others}
\end{table}

\subsection{Network Coding}
\label{sec:use-cases-nc}

In Network Coding (NC) \cite{LiYe03}, linear encoding and decoding operations are applied on packets to increase throughput, efficiency, scalability, and resilience.
Network nodes apply primitive operations, e.g., splitting, encoding, or decoding packets, to implement NC mechanisms such as multicast, forward error correction, or rerouting (resilience).

Kumar et al. \cite{KuBa18} implement primitive NC operations such as splitting, encoding, and decoding for a PSA software switch.
This is the first introduction of NC for \ac{SDN}, as fixed-function data plane switches, e.g., as in \ac{OF}, did not support such operations.
The authors describe details of their implementation.
The open-source implementation \cite{p4-KuBa18} relies on clone and recirculate operations to generate additional packets for encoding and decoding operations and packet processing loops. 
Temporary packet buffers for gathering operations are implemented with P4 registers. 
However, P4 hardware targets are not considered.

Gonçalves et al. \cite{GoSi19} implement NC operations that may use information from multiple packets during processing. 
The authors implement their concept for \ac{PISA} in P4$_{16}$.
It features multiple complex NC operations that focus on multiplications in Galois fields used for encoding and decoding operations.
NC operations are implemented in P4 externs that extend the capabilities of the software switch to store a specific amount of received packets.
Again, hardware targets are not considered.

\subsection{Distributed Algorithms}
\label{sec:use-cases-distributed}
We describe related work on event processing and in-network consensus.

\subsubsection{Event Processing}
\label{use-cases-distributed-events}

Data with stream characteristics often require specific processing.
For example, sensor data may be analyzed to determine whether values are within certain thresholds, or chunks of data are aggregated and preprocessed.

P4CEP \cite{KoMa18} shifts complex event processing from servers to P4 switches so that event stream data, e.g., from sensors, is directly processed on the data plane.
The solution requires several workarounds to solve P4 limitations regarding stateful packet processing.

DAIET \cite{SaAb17} introduces in-network data aggregation where the aggregation task is offloaded to the entire network.
This reduces the amount of traffic and reliefs the destination of computational load. 
The authors provide a prototype implementation in \pfour but only a few details are disclosed.

Sankaran et al. \cite{SaSi20} increase the processing speed of packets by reducing the time that is required by forwarding nodes to parse the packet header.
To that end, ingress routers parse the header stack to compute a so-called unique parser code (UPC) which they add to the packet header. 
Downstream nodes need to parse only the UPC to make forwarding decisions. 

\subsubsection{In-Network Consensus}
\label{use-cases-distributed-consensus}

Distributed algorithms or mechanisms may require consensus to determine the right solution or processing.
This includes communication between participating entities and some ways to determine the right solution.

Zhang et al. \cite{ZhHa17} propose to offload parts of the Raft consensus algorithm to P4 switches.
However, the mechanisms require an additional client to run on the switch.
The authors implement their application for a P4 software switch, but details are not presented.

Dang et al. \cite{DaCa16, DaBr20} describe a P4 implementation of Paxos, a protocol that solves consensus for distributed algorithms in a network of unreliable processors based on information exchange between switches. 
This work contains a detailed description of a complex P4 implementation.
The authors explain all components, provide code snippets, and discuss their design choices. 

P4BFT \cite{SaDe19, SaDe19b} introduces a consensus mechanism against buggy or malicious control plane instances.
The controller responses are sent to trustworthy instances which compare the responses and establish consensus, e.g., by choosing the most common response. 
The authors propose to offload the comparison process to the data plane.

SwiShmem \cite{ZePo20} is a distributed shared state management layer for the P4 data plane to implement stateful distributed network functions.
In high-performance environments controllers are easily overloaded when consistency of write-intensive distributed network functions, like DDoS detection, or rate limiters, is required.
Therefore, SwiShmem offloads consistency mechanisms from the control plane to the data plane.
Then, consistency mechanisms operate at line rate because switches process traffic, and generate and forward state update messages without controller interaction.

Byzantine fault refers to a system where consensus between multiple entities has to be established where one or more entities are unreliable.
Byzantine fault tolerance (BFT) describes mechanisms that handle such faults.
However, BFTs often require significant time to reach consensus due to high computational overhead to reduce uncertainty. 
Switch-centric BFT (SC-BFT) \cite{HaJa20} proposes to offload BFT functionalities, i.e., time synchronization and state synchronization, into the data plane.
This significantly accelerates the consensus procedure since nodes process information at line rate. 

LODGE \cite{SvBo18} implements a mechanism for switches to make forwarding decisions based on global state without control of a central instance. 
Developers define global state variables which are stored by all stateful data plane devices.
When such a node processes a packet that changes a global state variable, the switch generates and forwards an update packet to all other stateful switches on a predefined distribution tree. 
LOADER \cite{SvBo20} introduces global state to the data plane.
Consensus is maintained by the data plane devices through distributed algorithms, i.e., the switches send notification messages when global state changes.
This increases scalability in comparison to mechanisms where consensus is managed by a central control entity. 

FLAIR \cite{TaKe20} accelerates read operations in leader-based consensus protocols by processing the read requests in the data plane. 
To that end, FLAIR devices in the core maintain persistent information about pending write operations on all objects in the system.
When a client submits a read request, the FLAIR switch checks whether the requested object is stable, i.e., if it has pending write operations. 
If the object is stable, the FLAIR switch instructs another client with a stable version of the object, to send it to the requesting client. 
If the object is not stable, the FLAIR switch forwards the write request to the leader.

\subsection{State Migration}
\label{sec:state_migration}

In Swing State \cite{LuYu17}, switches maintain state in registers that should be migrated to other nodes.
For migration, state information is carried by regular packets created by the P4 clone operation throughout the network.

P4Sync \cite{XiCh20} is a protocol to migrate data plane state between switches.
Thereby, it does not require controller interaction and provides guarantees on the authenticity of the transferred state.
To that end, it leverages the switch's packet generator to transfer the content of registers between devices.
Authenticity in a migration operation is guaranteed by a hash chain where each packet contains the hashed values of both the current payload and the payload of the previous packet. 

Xue et al. \cite{XuZh20} propose a hybrid approach for storing flow entries to address the issue of limited on-switch memory.
While some flow entries are still stored in the internal memory of the switch, some flow entries may be stored on servers.
Switches access them with only low latency via remote direct memory access (RDMA).

Kuzniar et al. \cite{KuNe20} propose to leverage programmable switches to act as in-network cache to speed up queries over encrypted data stores. 
Encrypted key-value pairs are thereby stored in registers.

Sankaran et al. \cite{SaSi20b} describe a system to relieve switches from parsing headers.
They propose to parse headers at an ingress switch only and add a \emph{unique parser code} to the packet that identifies the set of headers of the packet.
With this information, following switches can parse relevant information from the headers without having to parse the whole header stack.

\subsection{Application Support}
\label{sec:application_support}

This subsection describes work that focuses on support or implementation of existing applications and protocols.

P4DNS \cite{WoRa19} is an in-network DNS system.
The authors propose a hybrid architecture with performance-critical components in the data plane and components with flexibility requirements in the control plane.
The data plane responds to DNS requests and forwards regular traffic while cache management, recursive DNS requests, and uncached DNS responses are handled by the control plane. 

P4-BNG \cite{KuNo19} implements a carrier-grade broadband network gateway (BNG) in P4.
The authors aim to provide an implementation for many different targets.
To that end, they introduce a layer between data plane and control plane.
This hardware-specific BNG data plane controller runs directly on the targets to provide a uniform interface to the control plane. 
It then configures the data plane according to the control commands from the control plane. 

ARP-P4 \cite{MaAl18} implements MAC address learning based on ARP solely on the P4 data plane.
To substitute a control plane, the authors integrate MAC learning as an external function.

Glebke et al. \cite{GlKr19} propose to offload computer vision functionalities, in particular, time-critical computations, to the data plane.
To that end, the authors leverage convolution filters on a P4-programmable NIC. 
The necessary computations are distributed to various \acp{MAT}.

COordinate-based INdexing (COIN) \cite{XiQi19b} is a mechanism to ensure efficient access to data on multiple distributed edge servers. 
To that end, the authors introduce a centralized instance that indexes data and its associated location. 
When an edge server requires data that it has not cached itself, it requests the data index at the centralized instance which provides a data location.

Lu et al. \cite{LuLi19} propose intra-network inference (INI) and implement it in P4. It offloads neural network computations into the data plane.
To that end, each P4 switch communicates via USB with a dedicated neural compute stick which performs computations.

Yazdinejad et al. \cite{YaPa19} present a P4-based blockchain enabled packet parser.
The proposed architecture focuses on \acp{FPGA} and aims to bring the security characteristics of blockchains into the data plane to greatly increase processing speed. 

P4rt-OVS \cite{OsTa20} is an extension for the \ac{OVS} based on \acp{BPF} to combine the programmability of P4 and the well-known features of the \ac{OVS}.
P4rt-OVS enables runtime programming of the \ac{OVS}, in particular, the deployment of new network features without recompilation of the \ac{OVS}.
It contains a P4-to-\ac{BPF} compiler which allows developers to write data plane code for the \ac{OVS} in P4. 

\subsection{Summary and Analysis}

P4 facilitates the development of prototypes in the domain of network coding (see Subection \ref{sec:use-cases-nc}) by providing \emph{target-specific packet header processing functions}.
The prototypes heavily rely on externs to implement complex packet processing behavior, i.e., encoding and decoding operations, packet splitting and packet merging. 
Such prototypes were mainly developed for the \ac{bmv2} and portability to hardware platforms depends on the properties of the used externs and the capabilities of the hardware targets.
Distributed algorithms (see Section \ref{sec:use-cases-distributed}) leverage all sorts of P4's core features.
Some prototypes \emph{define and use custom packet headers} to transport information that are not available in standard protocols.
Others rely on \emph{flexible packet header processing} and \emph{target-specific packet header processing functions} to implement unconventional and complex packet processing behavior. 
Some prototypes require \emph{packet processing on the control plane} to resolve consistency issues or make network-wide configuration decisions.
In the context of state migration (see Section \ref{sec:state_migration}) the prototypes mainly leverage externs to enable stateful processing.
As a result, most projects were developed for the \ac{bmv2} with only limited portability to hardware platforms.
Finally, some prototypes reimplement traditional network protocols or network elements, e.g., DNS, BNG, or ARP. 
Those projects mainly \emph{define and use custom packet headers} for information transport, \emph{flexible packet header processing} to implement the functionality of the specific protocol or network element, \emph{target-specific packet header processing functions} for complex packet processing, and \emph{packet processing on the control plane} for corner cases.

%% file: chapters/15-discussion.tex
\section{Discussion \& Outlook}
\label{sec:discussion}

We discuss the findings of this survey and present an outlook.

\subsection{P4 as a Language for Programmable Data Planes}
From a variety of data plane programming approaches, P4 became the currently most widespread standard.
Learning resources (\sect{p4-programming-tutorials}) and the \ac{bmv2} P4 software target (\sect{software-targets}) consitute low entry barriers for P4 technology.
This is appealing for academia and hardware support on high speed platforms make P4 relevant for industry.
The large body of literature that we surveyed in this work demonstrates that P4 has the right abstractions to build prototypes for many use cases in different application domains.
Moreover, P4 allows simple and flexible definition of data plane \acp{API} (\sect{p4-data-plane-apis}) that can be used by simple control plane programs or complex, enterprise-grade \ac{SDN} controllers.
Thus, P4 allows practitioners and researchers to express their data plane and control plane algorithms in a simple way and thereby unleashes a great innovation potential.
As P4 is supported by multiple platforms, there is a potentially large user group.
In addition, P4 is an open programming language so that the source code can be published as open source.
Therefore, public P4 code can profit from a large user community, both in quantity and quality, which is a benefit for software maintenance and security.

\subsection{P4 Targets Revisited}
We have listed many available P4 targets in \sect{p4-targets}.
However, our literature overview showed that mostly the \ac{bmv2} development and testing platform and P4 hardware targets based on the Tofino \ac{ASIC} were applied in the reviewed papers.

The vast majority of prototypes runs on the software switch \ac{bmv2}.
One reason is that it is freely available for everyone.
In addition, the complexity of the code is not constrained by hardware restrictions.
And finally, any required extern can be customized.
Therefore, there is no limit on algorithmic complexity so that \ac{bmv2} can serve as a platform for any use case -- but only from a functional point of view.
As it is a pure software-based prototyping solution, it cannot provide high throughput and is, therefore, not suitable for deployment in productive environments.

The Tofino ASIC is the base for P4 hardware targets with high throughput on  many ports.
It is currently the only available programmable data plane platform with throughput rates over \SI{12.8}{\tera\bit\per\s} and ports running at up to \SI{400}{\giga\bit\per\s}.
Therefore, Tofino-based devices are appropriate programmable data planes for production environments like data centers or core networks. 
Tofino uses P4 as native programming language.
Therefore, comprehensive tools are offered to support the P4 development process on this platform.
Moreover, P4 gives access to all features of the Tofino chip so that there is no penalty of using P4 as a programming language.
Existing restrictions are due to the functional limitation of a high speed platform.
Thus, prototypes for Tofino are more challenging but prove the technical feasibility of a new concept at commercial scale.
Probably for these reasons the Tofino turned out to be the mostly used hardware platform in our survey.

P4 can be also used on \ac{FPGA}- or \ac{NPU}-based targets.
They come with only a few ports and lower throughput rates so that they may be used for special-purpose server applications but not for typical switching devices.
They excel through the possibility to extend the target functionality with user-defined externs.
These cards are typically programmed by vendor-specific languages.
P4 support is achieved by trans-compilers that translate P4 programs into the vendor-specific format.
P4 programmability might be limited to a restricted feature set while access to all features of a target is only possible through the vendor-specific programming language.
Whether the application of P4 for such targets is beneficial compared to vendor-specific programming languages or interfaces mainly depends on the use case, level of knowledge of the programmer, and if prospect target-independence is a goal.

\subsection{Target Independence and Portability}
Many of the surveyed works profited from P4's core features that we summarized in \sect{use-cases}.
Often P4 programs were developed only for the \ac{bmv2} target due to the complexity of their algorithm, required interaction with fixed-function blocks, or dependence on custom extern functions.
The portability of such programs is limited to platforms with similar externs and even then the code needs to be significantly adapted.

In some use cases, the authors even miss the original objectives of P4.
They suggest P4 for complex {\em packet processing} operations while P4 has been primarily conceived for {\em packet  header processing} with simple operations on high speed data planes. 

Although some of the presented prototypes may not be portable to current P4 hardware targets, they are close to modern switch architectures as their overall pipeline is described in P4.
Thereby, the conceptual feasibility of new data plane algorithms can be proven.
This is an advantage of \ac{bmv2}-based prototypes compared to general software implementations.

\subsection{A Business Perspective for P4-Programmable Data Planes}
Today, the most prevalent hardware network appliances are proprietary devices for which customized hardware and software are jointly developed.

Data plane programming breaks with this process.
Programmable packet processing \acp{ASIC} such as the Tofino may be sold by specialized manufacturers and integrated by other vendors with a motherboard, CPU, memory, and connectors in white box switches.
The accompanying software, i.e., data plane and control plane programs, might be provided by the same vendor, a third party, or implemented by the users themselves.

Because software is developed independently of hardware, the agility of the development process can be increased, which can reduce the time to market.
Hardware platforms become reusable; they can be leveraged for multiple purposes with the help of appropriate P4 programs.

Network solution providers may leverage the lowered entry barrier for customized hardware appliances to develop and sell P4 software for various P4-capable targets, at least with moderate adaptation effort.
A decade of implementation experience may no longer be a prerequisite for that business.

In addition, companies with large networks and particular use cases, e.g., special applications in data centers, may use customized algorithms to overcome inefficiencies of standardized protocols or mechanisms.

Large companies can avoid vendor lock-in by acquisition of programmable components instead of black boxes.
The components are assembled possibly with open-source software leveraging data plane
programming, \ac{SDN}, and \ac{NFV}.
The ACCESS 4.0 architecture \cite{access40} and the O-RAN Alliance \cite{oran} are examples.
This type of disaggregation also enables cost scaling effects where off-the-shelf components are bought at moderate cost instead of expensive specialized appliances.

\subsection{Outlook}
P4 is primarily a programming language for high-speed switches.
Currently, it is supported by Intel's Tofino \ac{ASIC}, but other manufacturers already announced support for P4 for the future.

The many prototypes surveyed in this paper showed that there is a need for more functionality on programmable switches, which may be provided by extern functions.
While they reduce portability, they enable more use cases.
Examples for such extern functions are features that have been used in some of the pure software-based P4 prototypes.
They encrypt and decrypt packet payload, support floating-point operations, provide flexible hash functions, or allow more complex calculations.
Those externs might be provided by the target manufacturers for common use cases or integrated by users.

Hardware with a vendor-specific programming language may benefit from offering interfaces and cross-compilers for P4 together with useful extern functions.
Although this may not give access to the full functionality of the platform, users with P4 programming knowledge can customize such devices for their needs without worrying about hardware details.

The biggest driver for P4 is possibly disaggregation.
While currently devices from different vendors can be orchestrated by a customized controller, P4 may have the potential to extend disaggregation towards specialized appliances based on off-the-shelf programmable hardware.
Hardware without an open programming interface cannot profit from that market.

%% file: chapters/16-conclusion.tex
\section{Conclusion}
\label{sec:conclusion}

In this paper, we first gave a tutorial on data plane programming with P4.
We delineated it from \ac{SDN} and introduced programming models with a special focus on  \ac{PISA} which is most relevant for P4.
We provided an overview of the current state of P4 with regard to programming language, architectures, compilers, targets, and data plane \acp{API}.
We reported research efforts to advance P4 that fall in the areas of optimization of development and deployment, research on P4 targets, and P4-specific approaches for control plane operation.

In the second part of the paper, we analyzed \totalApplications papers on applied research that leverage P4 for implementation purposes.
We categorized these publications into research domains, summarized their key points, and characterized them by prototype, target platform, and source code availability.
For each research domain, we presented an analysis on how works benefit from P4.
To that end, we identified a small set of core features that facilitate implementations.
The survey proved a tremendous uptake of P4 for prototyping in academic research from 2018 to 2021.
One reason is certainly the multitude of openly available resources on P4 and the \ac{bmv2} P4 software target.
They are an ideal starting point for creating P4-based prototypes, even for beginners.

The many P4-based activities which emerged only within short time show that P4 technology can speed up the evolution of computer networking.
While multiple hardware targets are available, most hardware-based prototypes leverage the Tofino \ac{ASIC} that is optimized for high throughput on many ports and particularly suited for data center and WAN applications.
However, the majority of P4-based prototypes was implemented with the \ac{bmv2} software switch.
Many of them were not ported to hardware, probably due to the complexity of their data plane algorithms and lack of required extern functions on current hardware.
This may change in the future if new P4 hardware targets are available.
We expect P4 to become a base technology for multiple hardware appliances, in particular in the context of disaggregation and for small-scale markets.

%% file: chapters/17-acknowledgement.tex
\section{Acknowledgement}
This work was partly supported by the Deutsche Forschungsgemeinschaft (DFG) under grant ME2727/1-2.
The authors alone are responsible for the content of this paper.

%% file: paper.bbl
\begin{thebibliography}{100}
\expandafter\ifx\csname url\endcsname\relax
  \def\url#1{\texttt{#1}}\fi
\expandafter\ifx\csname urlprefix\endcsname\relax\def\urlprefix{URL }\fi
\expandafter\ifx\csname href\endcsname\relax
  \def\href#1#2{#2} \def\path#1{#1}\fi

\bibitem{KoMo00}
E.~Kohler, R.~Morris, B.~Chen, J.~Jannotti, M.~F. Kaashoek, {The Click Modular
  Router}, {ACM Transactions on Computer Systems (TOCS)} 18 (2000) 217–231.

\bibitem{VPP}
{VPP/What is VPP?}, \url{https://bit.ly/2mrxVGE}, {accessed 01-20-2021} (2021).

\bibitem{Broadcom}
{GitHub: NPL-Spec}, \url{https://github.com/nplang/NPL-Spec}, {accessed
  01-20-2021} (2021).

\bibitem{Xilinx}
{Software Defined Specification Environment for Networking (SDNet)},
  \url{https://www.xilinx.com/support/documentation/backgrounders/sdnet-backgrounder.pdf},
  {accessed 01-20-2021} (2021).

\bibitem{BoDa14}
P.~Bosshart, D.~Daly, G.~Gibb, M.~Izzard, N.~McKeown, J.~Rexford,
  C.~Schlesinger, D.~Talayco, A.~Vahdat, G.~Varghese, D.~Walker, {P4:
  Programming Protocol-independent Packet Processors}, {ACM SIGCOMM Computer
  Communications Review (CCR)} 44 (2014) 87–95.

\bibitem{NuMe14}
B.~A.~A. Nunes, M.~Mendonca, X.-N. Nguyen, K.~Obraczka, T.~Turletti, {A Survey
  of Software-Defined Networking: Past, Present, and Future of Programmable
  Networks}, {IEEE Communications Surveys \& Tutorials (COMST)} 16 (2014)
  1617--1634.

\bibitem{JaMa14}
Y.~Jarraya, T.~Madi, M.~Debbabi, {A Survey and a Layered Taxonomy of
  Software-Defined Networking}, {IEEE Communications Surveys \& Tutorials
  (COMST)} 16 (2014) 1955--1980.

\bibitem{XiWe15}
W.~Xia, Y.~Wen, C.~H. Foh, D.~Niyato, H.~Xie, {A Survey on Software-Defined
  Networking}, {IEEE Communications Surveys \& Tutorials (COMST)} 17 (2015)
  27--51.

\bibitem{MaGu15}
D.~F. Macedo, D.~Guedes, L.~F.~M. Vieira, M.~A.~M. Vieira, M.~Nogueira,
  {Programmable Networks—From Software-Defined Radio to Software-Defined
  Networking}, {IEEE Communications Surveys \& Tutorials (COMST)} 17 (2015)
  1102--1125.

\bibitem{KrRa15}
D.~Kreutz, F.~M.~V. Ramos, P.~E. Veríssimo, C.~E. Rothenberg, S.~Azodolmolky,
  S.~Uhlig, {Software-Defined Networking: A Comprehensive Survey}, Proceedings
  of the IEEE 103 (2015) 14--76.

\bibitem{MaGh16}
R.~Masoudi, A.~Ghaffari, {Software defined networks: A survey}, {Journal of
  Network and Computer Applications (JNCA)} 67 (2016) 1--25.

\bibitem{TrFa16}
C.~Trois, M.~D. Del~Fabro, L.~C.~E. de~Bona, M.~Martinello, {A Survey on SDN
  Programming Languages: Toward a Taxonomy}, {IEEE Communications Surveys \&
  Tutorials (COMST)} 18 (2016) 2687--2712.

\bibitem{BrMe14}
W.~Braun, M.~Menth, {Software-Defined Networking Using OpenFlow: Protocols,
  Applications and Architectural Design Choices}, {MDPI Future Internet Journal
  (FI)} 6 (2014) 302–336.

\bibitem{HuHa14}
F.~Hu, Q.~Hao, K.~Bao, {A Survey on Software-Defined Network and OpenFlow: From
  Concept to Implementation}, {IEEE Communications Surveys \& Tutorials
  (COMST)} 16 (2014) 2181--2206.

\bibitem{LaKo14}
A.~Lara, A.~Kolasani, B.~Ramamurthy, {Network Innovation using OpenFlow: A
  Survey}, {IEEE Communications Surveys \& Tutorials (COMST)} 16 (2014)
  493--512.

\bibitem{BiRe18}
R.~Bifulco, G.~Rétvári, {A Survey on the Programmable Data Plane:
  Abstractions, Architectures, and Open Problems}, in: {IEEE International
  Conference on High Performance Switching and Routing (HPSR)}, 2018, pp. 1--7.

\bibitem{KaMa19}
E.~Kaljic, A.~Maric, P.~Njemcevic, M.~Hadzialic, {A Survey on Data Plane
  Flexibility and Programmability in Software-Defined Networking}, {IEEE}
  ACCESS 7 (2019) 47804--47840.

\bibitem{MiBi21}
O.~Michel, R.~Bifulco, G.~Rétvári, S.~Schmid, {The Programmable Data Plane:
  Abstractions, Architectures, Algorithms, and Applications}, {ACM Computing
  Surveys} 1 (2021).

\bibitem{KaKu21}
S.~Kaur, K.~Kumar, N.~Aggarwal, A review on p4-programmable data planes:
  Architecture, research efforts, and future directions, {Computer
  Communications} 170 (2021).

\bibitem{KfCr21}
E.~F. Kfoury, J.~Crichigno, E.~Bou-Harb, An exhaustive survey on p4
  programmable data plane switches: Taxonomy, applications, challenges, and
  future trends, {ArXiv} e-prints (2021).

\bibitem{McAn08}
N.~McKeown, T.~Anderson, H.~Balakrishnan, G.~Parulkar, L.~Peterson, J.~Rexford,
  S.~Shenker, J.~Turner, {OpenFlow: Enabling Innovation in Campus Networks},
  {ACM SIGCOMM Computer Communications Review (CCR)} 38 (2008) 69–74.

\bibitem{BESS}
{BESS: Berkeley Extensible Software Switch},
  \url{http://span.cs.berkeley.edu/bess.html}, {accessed 01-20-2021} (2021).

\bibitem{BoGi13}
P.~Bosshart, G.~Gibb, H.-S. Kim, G.~Varghese, N.~McKeown, M.~Izzard, F.~Mujica,
  M.~Horowitz, {Forwarding Metamorphosis: Fast Programmable Match-Action
  Processing in Hardware for SDN}, {ACM SIGCOMM Conference} 43 (2013) 99–110.

\bibitem{ChFi17}
S.~Chole, A.~Fingerhut, S.~Ma, A.~Sivaraman, S.~Vargaftik, A.~Berger,
  G.~Mendelson, M.~Alizadeh, S.-T. Chuang, I.~Keslassy, A.~Orda, T.~Edsall,
  {DRMT: Disaggregated Programmable Switching}, in: {ACM SIGCOMM Conference},
  2017, p. 1–14.

\bibitem{p4-tutorial-slides}
{Google Presentations: P4 Tutorial}, \url{http://bit.ly/p4d2-2018-spring},
  {accessed 01-20-2021} (2018).

\bibitem{p4-language-consortium}
{Website of the P4 Language Consortium}, \url{https://p4.org/}, {accessed
  01-20-2021} (2021).

\bibitem{p4_14}
{The P4 Language Specification},
  \url{https://p4.org/p4-spec/p4-14/v1.0.5/tex/p4.pdf}, {accessed 01-20-2021}
  (2018).

\bibitem{p4_16}
{P4 16 Language Specification (v.1.2.1},
  \url{https://p4.org/p4-spec/docs/P4-16-v1.2.1.html}, {accessed 01-20-2021}
  (2020).

\bibitem{MoBh14}
M.~Moshref, A.~Bhargava, A.~Gupta, M.~Yu, R.~Govindan, {Flow-level State
  Transition as a New Switch Primitive for SDN}, in: {ACM SIGCOMM Conference},
  2014, p. 61–66.

\bibitem{BiBo14}
G.~Bianchi, M.~Bonola, A.~Capone, C.~Cascone, {OpenState: Programming
  Platform-independent Stateful Openflow Applications Inside the Switch}, {ACM
  SIGCOMM Computer Communications Review (CCR)} 44 (2014) 44–51.

\bibitem{SiCh16}
A.~Sivaraman, A.~Cheung, M.~Budiu, C.~Kim, M.~Alizadeh, H.~Balakrishnan,
  G.~Varghese, N.~McKeown, S.~Licking, {Packet Transactions: High-Level
  Programming for Line-Rate Switches}, in: {ACM SIGCOMM Conference}, 2016, p.
  15–28.

\bibitem{PoBi19}
S.~Pontarelli, R.~Bifulco, M.~Bonola, C.~Cascone, M.~Spaziani, V.~Bruschi,
  D.~Sanvito, G.~Siracusano, A.~Capone, M.~Honda, F.~Huici, G.~Siracusano,
  {FlowBlaze: Stateful Packet Processing in Hardware}, in: {USENIX Symposium on
  Networked Systems Design \& Implementation (NSDI)}, 2019, p. 531–547.

\bibitem{So13}
H.~Song, {Protocol-Oblivious Forwarding: Unleash the Power of SDN Through a
  Future-proof Forwarding Plane}, in: {ACM Workshop on Hot Topics in Networks
  (HotNets)}, 2013, p. 127–132.

\bibitem{AnFo14}
C.~J. Anderson, N.~Foster, A.~Guha, J.-B. Jeannin, D.~Kozen, C.~Schlesinger,
  D.~Walker, {NetKAT: Semantic Foundations for Networks}, in: {ACM Symposium on
  Principles of Programming Languages (POPL)}, 2014, p. 113–126.

\bibitem{p4-tutorial}
{P4 Tutorial}, \url{https://github.com/p4lang/tutorials}, {accessed 05-05-2021}
  (2021).

\bibitem{p4-guide}
{P4 Guide}, \url{https://github.com/jafingerhut/p4-guide}, {accessed
  05-05-2021} (2021).

\bibitem{p4-learning}
{P4 Learning}, \url{https://github.com/nsg-ethz/p4-learning}, {accessed
  05-05-2021} (2021).

\bibitem{p4-architecture-wg}
{Charter of the P4 Architecture WG},
  \url{https://github.com/p4lang/p4-spec/blob/master/p4-16/psa/charter/P4_Arch_Charter.mdk},
  {accessed 01-20-2021} (2021).

\bibitem{p4-16-psa}
{P4\_16 PSA Specification (v1.1)},
  \url{https://p4lang.github.io/p4-spec/docs/PSA-v1.1.0.html}, {accessed
  01-20-2021} (2018).

\bibitem{hlir}
{P4-HLIR Specification v.0.9.30},
  \url{https://github.com/p4lang/p4-hlir/blob/master/HLIRSpec.pdf}, {accessed
  01-20-2021} (2016).

\bibitem{p4c}
{GitHub: p4c}, \url{https://github.com/p4lang/p4c}, {accessed 01-20-2021}
  (2021).

\bibitem{PaRo17}
P.~G. Patra, C.~E. Rothenberg, G.~Pongracz, {MACSAD: High Performance Dataplane
  Applications on the Move}, in: {IEEE International Conference on High
  Performance Switching and Routing (HPSR)}, 2017, pp. 1--6.

\bibitem{odp}
{Open Data Plane}, \url{https://opendataplane.org/}, {accessed 01-20-2021}
  (2021).

\bibitem{JoYa15}
L.~Jose, M.~R. N.~M. Lisa~Yan, Stanford University; George~Varghese, {Compiling
  Packet Programs to Reconfigurable Switches}, in: {USENIX Symposium on
  Networked Systems Design \& Implementation (NSDI)}, 2015, p. 103–115.

\bibitem{LiLu16}
P.~Li, Y.~Luo, {P4GPU: Accelerate Packet Processing of a P4 Program with a
  CPU-GPU Heterogeneous Architecture}, in: {ACM/IEEE Symposium on Architectures
  for Networking and Communications Systems (ANCS)}, 2016, pp. 125--126.

\bibitem{p4c-behavioural-github}
{GitHub: p4c-behavioural},
  \url{https://github.com/p4lang/p4c-behavioral/tree/master/p4c_bm}, {accessed
  01-20-2021} (2021).

\bibitem{bmv2}
{GitHub: Behavioural Model Version 2 (BMv2)},
  \url{https://github.com/p4lang/behavioral-model}, {accessed 01-20-2021}
  (2021).

\bibitem{bmv2-replacement-reasons}
{P4 Behaviour Model: Why did we need BMv2},
  \url{https://github.com/p4lang/behavioral-model\#why-did-we-replace-p4c-behavioral-with-bmv2},
  {accessed 01-20-2021} (2021).

\bibitem{bmv2targets}
{GitHub: Behavioral model targets},
  \url{https://github.com/p4lang/behavioral-model/blob/master/targets/README.md},
  {accessed 01-20-2021} (2021).

\bibitem{bmv2-performance}
{BMv2 Performance},
  \url{https://github.com/p4lang/behavioral-model/blob/master/docs/performance.md},
  {accessed 01-20-2021} (2021).

\bibitem{p4c-ebpf}
{GitHub: eBPF Backend for p4c},
  \url{https://github.com/p4lang/p4c/tree/master/backends/ebpf}, {accessed
  01-20-2021} (2021).

\bibitem{p4c-ubpf}
{p4c-ubpf: a New Back-end for the P4 Compiler},
  \url{https://p4.org/p4/p4c-ubpf.html}, {accessed 01-20-2021} (2021).

\bibitem{p4c-xdp}
{GitHub: p4c-xdp}, \url{https://github.com/vmware/p4c-xdp}, {accessed
  01-20-2021} (2021).

\bibitem{p4elte}
{P4@ELTE}, \url{http://p4.elte.hu/}, {accessed 01-20-2021} (2021).

\bibitem{LaHo16}
S.~Laki, D.~Horp\'{a}csi, P.~V\"{o}r\"{o}s, R.~Kitlei, D.~Lesk\'{o}, M.~Tejfel,
  High speed packet forwarding compiled from protocol independent data plane
  specifications, in: {ACM SIGCOMM Conference}, 2016, p. 629–630.

\bibitem{dpdk}
{Data Plane Development Kit (DPDK)}, \url{https://www.dpdk.org/}, {accessed
  01-20-2021} (2021).

\bibitem{t4p4s-github}
{GitHub: T4P4S}, \url{https://github.com/P4ELTE/t4p4s}, {accessed 01-20-2021}
  (2021).

\bibitem{BhSh17}
A.~Bhardwaj, A.~Shree, V.~B. Reddy, S.~Bansal, {A Preliminary Performance Model
  for Optimizing Software Packet Processing Pipelines}, in: {ACM SIGOPS
  Asia-Pacific Workshop on System (APSys)}, 2017, pp. 1--7.

\bibitem{WuLi19}
X.~{Wu}, P.~{Li}, T.~{Miskell}, L.~{Wang}, Y.~{Luo}, X.~{Jiang}, {Ripple: An
  Efficient Runtime Reconfigurable P4 Data Plane for Multicore Systems}, in:
  {International Conference on Networking and Network Applications (NaNA)},
  2019, pp. 142--148.

\bibitem{ShCh16}
M.~Shahbaz, S.~Choi, B.~Pfaff, C.~Kim, N.~Feamster, N.~McKeown, J.~Rexford,
  {PISCES: A Programmable, Protocol-Independent Software Switch}, in: {ACM
  SIGCOMM Conference}, 2016, p. 525–538.

\bibitem{ovs}
{Open vSwitch}, \url{https://www.openvswitch.org/}, {accessed 01-20-2021}
  (2021).

\bibitem{pisces-github}
{GitHub: PISCES}, \url{https://github.com/P4-vSwitch}, {accessed 01-20-2021}
  (2021).

\bibitem{ChLo17}
S.~Choi, X.~Long, M.~Shahbaz, S.~Booth, A.~Keep, J.~Marshall, C.~Kim, {The Case
  for a Flexible Low-Level Backend for Software Data Planes}, in: {Asia-Pacific
  Workshop on Networking (APnet)}, 2017, p. 71–77.

\bibitem{ChLo17b}
S.~Choi, X.~Long, M.~Shahbaz, S.~Booth, A.~Keep, J.~Marshall, C.~Kim, {PVPP: A
  Programmable Vector Packet Processor}, in: {ACM Symposium on SDN Research
  (SOSR)}, 2017, p. 197–198.

\bibitem{zodiac}
{Northbound Networks - Who are You?},
  \url{https://northboundnetworks.com/pages/about-us}, {accessed 01-20-2021}
  (2021).

\bibitem{zodiac-p4}
{GitHub: ZodiacFX-P4}, \url{https://github.com/NorthboundNetworks/ZodiacFX-P4},
  {accessed 01-20-2021} (2021).

\bibitem{zodiac-p4c}
{GitHub: p4c-zodiacfx},
  \url{https://github.com/NorthboundNetworks/p4c-zodiacfx}, {accessed
  01-20-2021} (2021).

\bibitem{ZaRa19}
P.~Zanna, P.~Radcliffe, K.~G. Chavez, {A Method for Comparing OpenFlow and P4},
  in: {International Telecommunication Networks and Applications Conference
  (ITNAC)}, 2019, pp. 1--3.

\bibitem{p4netfga-wiki}
{GitHub: P4-NetFPGA}, \url{https://github.com/NetFPGA/P4-NetFPGA-public/wiki},
  {accessed 01-20-2021} (2021).

\bibitem{IbBr19}
S.~Ibanez, G.~Brebner, N.~McKeown, N.~Zilberman, {The P4-NetFPGA Workflow for
  Line-Rate Packet Processing}, in: {ACM/SIGDA International Symposium on
  Field-Programmable Gate Arrays (FPGA)}, 2019, p. 1–9.

\bibitem{ZiAu14}
N.~Zilberman, Y.~Audzevich, G.~A. Covington, A.~W. Moore, {NetFPGA SUME: Toward
  100 Gbps as Research Commodity}, {IEEE Micro} 34 (2014) 32--41.

\bibitem{netcope}
{Netcope P4},
  \url{https://www.netcope.com/Netcope/media/content/NetcopeP4_2019_web.pdf},
  {accessed 01-20-2021} (2021).

\bibitem{WaSo17}
H.~Wang, R.~Soul{\'e}, H.~T. Dang, K.~S. Lee, V.~Shrivastav, N.~Foster,
  H.~Weatherspoon, {P4FPGA: A Rapid Prototyping Framework for P4}, in: {ACM
  Symposium on SDN Research (SOSR)}, 2017, p. 122–135.

\bibitem{p4fpga}
{GitHub: P4FPGA}, \url{https://github.com/p4fpga/p4fpga}, {accessed 01-20-2021}
  (2021).

\bibitem{BePu16}
P.~{Benácek}, V.~{Pu}, H.~{Kubátová}, {P4-to-VHDL: Automatic Generation of
  100 Gbps Packet Parsers}, in: {IEEE Annual International Symposium on
  Field-Programmable Custom Computing Machines (FCCM)}, 2016, pp. 148--155.

\bibitem{BePu17}
P.~{Benáček}, V.~{Puš}, J.~{Kořenek}, M.~{Kekely}, {Line Rate Programmable
  Packet Processing in 100Gb Networks}, in: {International Conference on Field
  Programmable Logic and Applications (FPL)}, 2017, pp. 1--1.

\bibitem{CaBe18}
J.~Cabal, P.~Ben\'{a}\v{c}ek, L.~Kekely, M.~Kekely, V.~Pu\v{s}, J.~Ko\v{r}enek,
  {Configurable FPGA Packet Parser for Terabit Networks with Guaranteed
  Wire-Speed Throughput}, in: {ACM/SIGDA International Symposium on
  Field-Programmable Gate Arrays (FPGA)}, 2018, p. 249–258.

\bibitem{SiBo18}
S.~da~Silva, Jeferson, Boyer, Fran\c{c}ois-Raymond, Langlois, J.~Pierre,
  {P4-Compatible High-Level Synthesis of Low Latency 100 Gb/s Streaming Packet
  Parsers in FPGAs}, in: {ACM/SIGDA International Symposium on
  Field-Programmable Gate Arrays (FPGA)}, 2018, p. 147–152.

\bibitem{KeKo17}
M.~{Kekely}, J.~{Korenek}, {Mapping of P4 Match Action Tables to FPGA}, in:
  {International Conference on Field Programmable Logic and Applications
  (FPL)}, 2017, pp. 1--2.

\bibitem{IsBe18}
R.~{Iša}, P.~{Benáček}, V.~{Puš}, {Verification of Generated RTL from P4
  Source Code}, in: {IEEE International Conference on Network Protocols
  (ICNP)}, 2018, pp. 444--445.

\bibitem{CaSu20}
Z.~{Cao}, H.~{Su}, Q.~{Yang}, J.~{Shen}, M.~{Wen}, C.~{Zhang}, {P4 to FPGA-A
  Fast Approach for Generating Efficient Network Processors}, {IEEE} ACCESS 8
  (2020) 23440--23456.

\bibitem{CaSu20b}
Z.~{Cao}, H.~{Su}, Q.~{Yang}, M.~{Wen}, C.~{Zhang}, {A Template-based Framework
  for Generating Network Processor in FPGA}, in: {IEEE Conference on Computer
  Communications Workshops (INFOCOM WKSHPS)}, 2019, pp. 1057--1058.

\bibitem{open-tofino}
{Open Tofino}, {https://github.com/barefootnetworks/open-tofino}, {accessed
  01-22-2021} (2021).

\bibitem{wedge-switch}
{EdgeCore Wedge 100BF-32X},
  \url{https://www.edge-core.com/productsInfo.php?cls=1&cls2=180&cls3=181&id=335},
  {accessed 01-20-2021} (2021).

\bibitem{aps-bf2556}
{APS Networks BF2556X-1T-A1F},
  \url{https://stordirect.com/shop/switches/25g-switches/aps-networks-bf2556x-1t-a1f/},
  {acessed 01-22-2021} (2021).

\bibitem{aps-bf6064}
{APS Networks BF6064X-T-A2F},
  \url{https://stordirect.com/shop/switches/100g-switches/aps-networks-bf6064x-t-a2f/},
  {acessed 01-22-2021} (2021).

\bibitem{aurora-switch}
{Netberg Aurora 610}, \url{https://netbergtw.com/products/aurora-610/},
  {accessed 01-20-2021} (2021).

\bibitem{arista}
{Arista Press Release: Arista Announces New Multi-function Platform for Cloud
  Networking},
  \url{https://www.arista.com/en/company/news/press-release/5148-pr-20180605},
  {accessed 01-20-2021} (2021).

\bibitem{cisco}
{Cisco Blog: Increase Flexibility with Cisco’s Programmable Cloud
  Infrastructure},
  \url{https://blogs.cisco.com/datacenter/increase-flexibility-with-ciscos-programmable-cloud-infrastructure},
  {accessed 01-20-2021} (2021).

\bibitem{sonic-devices}
{SONiC - Supported Platforms},
  \url{https://azure.github.io/SONiC/Supported-Devices-and-Platforms.html},
  {accessed 01-20-2021} (2021).

\bibitem{SeBa20}
A.~Seibulescu, M.~Baldi, {Leveraging P4 Flexibility to Expose Target-Specific
  Features}, in: {P4 Workshop in Europe (EuroP4)}, 2020, p. 36–42.

\bibitem{pensando}
{The Pensando Distributed Services Platform},
  \url{https://pensando.io/our-platform/}, {accessed 01-20-2021} (2021).

\bibitem{netronome-p4}
{Netronome: P4 Data Plane Programming},
  \url{https://netronome.com/media/documents/WP_P4_Data_Plane_Programming.pdf},
  {accessed 01-20-2021} (2018).

\bibitem{netronome-p4-c}
{Netronome: Programming with P4 and C},
  \url{https://www.netronome.com/media/documents/WP_Programming_with_P4_and_C.pdf},
  {accessed 09-20-2019} (2018).

\bibitem{HaJa19}
H.~Harkous, M.~Jarschel, M.~He, R.~Pries, W.~Kellerer, {Towards Understanding
  the Performance of P4 Programmable Hardware}, in: {ACM/IEEE Symposium on
  Architectures for Networking and Communications Systems (ANCS)}, 2019, pp.
  1--6.

\bibitem{apache-thrift}
{Apache Thrift}, \url{https://thrift.apache.org/}, {accessed 01-20-2021}
  (2021).

\bibitem{grpc}
{gRPC}, \url{https://grpc.io/}, {accessed 01-20-2021} (2021).

\bibitem{protobuf}
{Google Protocol Buffers},
  \url{https://developers.google.com/protocol-buffers/}, {accessed 01-20-2021}
  (2021).

\bibitem{p4-api-wg}
{Charter of the P4 API WG},
  \url{https://github.com/p4lang/p4-spec/blob/master/api/charter/P4_API_WG_charter.mdk},
  {accessed 01-20-2021} (2021).

\bibitem{p4-runtime-specification}
{P4 Runtime API Specification v.1.3.0 (2019-12-01)},
  \url{https://p4.org/p4runtime/spec/v1.3.0/P4Runtime-Spec.html}, {accessed
  01-20-2021} (2020).

\bibitem{onos}
{ONOS: P4 brigade}, \url{https://wiki.onosproject.org/display/ONOS/P4+brigade},
  {accessed 01-20-2021} (2021).

\bibitem{opendaylight}
{OpenDaylight: P4 brigade}, \url{P4 Plugin Developer Guide}, {accessed
  09-23-2019} (2019).

\bibitem{OcTs19}
B.~O’Connor, Y.~Tseng, M.~Pudelko, C.~Cascone, A.~Endurthi, Y.~Wang,
  A.~Ghaffarkhah, D.~Gopalpur, T.~Everman, T.~Madejski, J.~Wanderer, A.~Vahdat,
  {Using P4 on Fixed-Pipeline and Programmable Stratum Switches}, in: {P4
  Workshop in Europe (EuroP4)}, 2010, pp. 1--2.

\bibitem{p4runtimelib}
{GitHub: P4tutorial},
  \url{https://github.com/p4lang/tutorials/tree/master/utils/p4runtime_lib},
  {accessed 01-20-2021} (2021).

\bibitem{pi}
{GitHub: PI Library}, \url{https://github.com/p4lang/PI}, {accessed 01-20-2021}
  (2021).

\bibitem{ssgrpc}
{GitHub: Behavioural Model - simple\_switch\_grpc},
  \url{https://github.com/p4lang/behavioral-model/tree/master/targets/simple_switch_grpc},
  {accessed 01-20-2021} (2021).

\bibitem{bmv2-cli}
{GitHub: bmv2 Runtime CLI},
  \url{https://github.com/p4lang/behavioral-model/blob/master/tools/runtime_CLI.py},
  {accessed 01-20-2021} (2021).

\bibitem{ZaZh19}
E.~O. Zaballa, Z.~Zhou, {Graph-to-P4: A P4 Boilerplate Code Generator for Parse
  Graphs}, in: {P4 Workshop in Europe (EuroP4)}, 2019, pp. 1--2.

\bibitem{ZhBi17}
Y.~Zhou, J.~Bi, {ClickP4: Towards Modular Programming of P4}, in: {ACM SIGCOMM
  Conference Posters and Demos}, 2017, p. 100–102.

\bibitem{Ba19}
M.~Baldi, {daPIPE A Data Plane Incremental Programming Environment}, in: {P4
  Workshop in Europe (EuroP4)}, 2019, pp. 1--6.

\bibitem{EiCa19}
M.~Eichholz, E.~Campbell, N.~Foster, G.~Salvaneschi, M.~Mezini, {How to Avoid
  Making a Billion-Dollar Mistake: Type-Safe Data Plane Programming with
  SafeP4}, in: {European Conference on Object-Oriented Programming (ECOOP)},
  2019, pp. 1--28.

\bibitem{RiKu19}
M.~Riftadi, F.~Kuipers, {P4I/O: Intent-Based Networking with P4}, in: {IEEE
  Conference on Network Softwarization (NetSoft)}, 2019, pp. 438--443.

\bibitem{YuSo20}
L.~Yu, J.~Sonchack, V.~Liu, {Mantis: Reactive Programmable Switches}, in: {ACM
  SIGCOMM Conference}, 2020, p. 296–309.

\bibitem{GaZh20}
J.~Gao, E.~Zhai, H.~H. Liu, R.~Miao, Y.~Zhou, B.~Tian, C.~Sun, D.~Cai,
  M.~Zhang, M.~Yu, {Lyra: A Cross-Platform Language and Compiler for Data
  PlaneProgramming on Heterogeneous ASICs}, in: {ACM SIGCOMM Conference}, 2020,
  p. 435–450.

\bibitem{RiOo19}
M.~Riftadi, J.~Oostenbrink, F.~Kuipers, {GP4P4: Enabling Self-Programming
  Networks}, {ArXiv} e-prints (2019).

\bibitem{MoSa20}
D.~{Moro}, D.~{Sanvito}, A.~{Capone}, {FlowBlaze.p4: a library for quick
  prototyping of stateful SDN applications in P4}, in: {IEEE Conference on
  Network Function Virtualization and Software-Defined Networking (NFV-SDN)},
  2020, pp. 95--99.

\bibitem{MoSa20b}
D.~{Moro}, D.~{Sanvito}, A.~{Capone}, {Demonstrating FlowBlaze.p4: fast
  prototyping for EFSM-based data plane applications}, in: {IEEE Conference on
  Network Function Virtualization and Software-Defined Networking (NFV-SDN)},
  2020, pp. 116--117.

\bibitem{MoSa20c}
D.~Moro, D.~Sanvito, A.~Capone, {Developing EFSM-Based Stateful Applications
  with FlowBlaze.P4 and ONOS}, in: {P4 Workshop in Europe (EuroP4)}, 2020, p.
  52–53.

\bibitem{SuSo21}
N.~Sultana, J.~Sonchack, H.~Giesen, I.~Pedisich, Z.~Han, N.~Shyamkumar,
  S.~Burad, A.~DeHon, B.~T. Loo, Flightplan: Dataplane disaggregation and
  placement for p4 programs, in: {USENIX Symposium on Networked Systems Design
  \& Implementation (NSDI)}, 2021, pp. 571--592.

\bibitem{ShSh18}
R.~Shah, A.~Shirke, A.~Trehan, M.~Vutukuru, P.~Kulkarni, {pcube: Primitives for
  Network Data Plane Programming}, in: {IEEE International Conference on
  Network Protocols (ICNP)}, 2018, pp. 430--435.

\bibitem{MaBi17}
Z.~Ma, J.~Bi, C.~Zhang, Y.~Zhou, A.~B. Dogar, {CacheP4: A Behavior-level
  Caching Mechanism for P4}, in: {ACM SIGCOMM Conference Posters and Demos},
  2017, p. 108–110.

\bibitem{AbLe17}
A.~Abhashkumar, J.~Lee, J.~Tourrilhes, S.~Banerjee, W.~Wu, J.-M. Kang,
  A.~Akella, {P5: Policy-driven Optimization of P4 Pipeline}, in: {ACM
  Symposium on SDN Research (SOSR)}, 2017, p. 136–142.

\bibitem{WiAp20}
P.~Wintermeyer, M.~Apostolaki, A.~Dietmüller, L.~Vanbever, {P2GO: P4
  Profile-Guided Optimizations}, in: {ACM Workshop on Hot Topics in Networks
  (HotNets)}, 2020, p. 146–152.

\bibitem{YaBa20}
S.~Yang, L.~Baia, L.~Cui, Z.~Ming, Y.~Wu, S.~Yu, H.~Shen, Y.~Pan, {P4 Edge node
  enabling stateful traffic engineering and cyber security}, {Journal of
  Network and Computer Applications (JNCA)} 171 (2020) A84--A95.

\bibitem{VaBe20}
B.~Vass, E.~B\'{e}rczi-Kov\'{a}cs, C.~Raiciu, G.~R\'{e}tv\'{a}ri, {Compiling
  Packet Programs to Reconfigurable Switches: Theory and Algorithms}, in: {P4
  Workshop in Europe (EuroP4)}, 2020, p. 28–35.

\bibitem{AbAf16}
S.~Abdi, U.~Aftab, G.~Bailey, B.~Boughzala, F.~Dewal, S.~Parsazad, E.~Tremblay,
  {PFPSim: A Programmable Forwarding Plane Simulator}, in: {ACM/IEEE Symposium
  on Architectures for Networking and Communications Systems (ANCS)}, 2016, pp.
  55--60.

\bibitem{BaBi18}
J.~Bai, J.~Bi, P.~Kuang, C.~Fan, Y.~Zhou, C.~Zhang, {NS4: Enabling Programmable
  Data Plane Simulation}, in: {ACM Symposium on SDN Research (SOSR)}, 2018, pp.
  1--7.

\bibitem{FaBi17}
C.~Fan, J.~Bi, Y.~Zhou, C.~Zhang, H.~Yu, {NS4: A P4-Driven Network Simulator},
  in: {ACM SIGCOMM Conference Posters and Demos}, 2017, p. 105–107.

\bibitem{McTa16}
N.~McKeown, D.~Talayco, G.~Varghese, N.~P. Lopes, N.~Bj{\o}rner,
  A.~Rybalchenko, {Automatically Verifying Reachability and Well-Formedness in
  P4 Networks},
  \url{https://www.microsoft.com/en-us/research/wp-content/uploads/2016/09/p4nod.pdf},
  {accessed 01-20-2021} (2016).

\bibitem{KhRo18}
A.~Kheradmand, G.~Rosu, {P4K: A Formal Semantics of P4 and Applications},
  {ArXiv} e-prints (2018).

\bibitem{LiHa18}
J.~Liu, W.~Hallahan, C.~Schlesinger, M.~Sharif, J.~Lee, R.~Soul{\'e}, H.~Wang,
  C.~Ca\c{s}caval, N.~McKeown, N.~Foster, {P4V: Practical Verification for
  Programmable Data Planes}, in: {ACM SIGCOMM Conference}, 2018, p. 490–503.

\bibitem{FrNe18}
L.~Freire, M.~Neves, L.~Leal, K.~Levchenko, A.~Schaeffer-Filho, M.~Barcellos,
  {Uncovering Bugs in P4 Programs with Assertion-based Verification}, in: {ACM
  Symposium on SDN Research (SOSR)}, 2018, p. 1–7.

\bibitem{NeFr18}
M.~Neves, L.~Freire, A.~Schaeffer-Filho, M.~Barcellos, {Verification of P4
  Programs in Feasible Time using Assertions}, in: {ACM Conference on emerging
  Networking EXperiments and Technologies (CoNEXT)}, 2018, p. 73–85.

\bibitem{StDu17}
R.~Stoenescu, D.~Dumitrescu, M.~Popovici, L.~Negreanu, C.~Raiciu, {Debugging P4
  Programs with Vera}, in: {ACM SIGCOMM Conference}, 2018, p. 518–532.

\bibitem{NoHs19}
M.~A. Noureddine, A.~Hsu, M.~Caesar, F.~A. Zaraket, W.~H. Sanders, {P4AIG:
  Circuit-Level Verification of P4 Programs}, in: {IEEE/IFIP International
  Conference on Dependable Systems and Networks – Supplemental Volume
  (DSN-S)}, 2019, pp. 21--22.

\bibitem{DuSt20}
D.~Dumitrescu, R.~Stoenescu, L.~Negreanu, C.~Raiciu, {bf4: towards bug-free P4
  programs}, in: {ACM SIGCOMM Conference}, 2020, p. 571–585.

\bibitem{DuSt19}
D.~Dumitrescu, R.~Stoenescu, M.~Popovici, L.~Negreanu, C.~Raiciu, {Dataplane
  equivalence and its applications}, in: {USENIX Symposium on Networked Systems
  Design \& Implementation (NSDI)}, 2019, pp. 683--698.

\bibitem{YoAb20}
F.~Yousefi, A.~Abhashkumar, K.~Subramanian, K.~Hans, S.~Ghorbani, A.~Akella,
  {Liveness Verification of Stateful Network Functions}, in: {USENIX Symposium
  on Networked Systems Design \& Implementation (NSDI)}, 2020, pp. 257--272.

\bibitem{NoKh18}
A.~N\"{o}tzli, J.~Khan, A.~Fingerhut, C.~Barrett, P.~Athanas, {P4Pktgen:
  Automated Test Case Generation for P4 Programs}, in: {ACM Symposium on SDN
  Research (SOSR)}, 2018, pp. 1--7.

\bibitem{ZhBi19b}
Y.~Zhou, J.~Bi, Y.~Lin, Y.~Wang, D.~Zhang, Z.~Xi, J.~Cao, C.~Sun, {P4Tester:
  Efficient Runtime Rule Fault Detection for Programmable Data Planes}, in:
  {IEEE International Workshop on Quality of Service (IWQoS)}, 2019, pp. 1--10.

\bibitem{p4app}
{GitHub: P4app}, \url{https://github.com/p4lang/p4app}, {accessed 01-20-2021}
  (2021).

\bibitem{ShuHu19}
A.~Shukla, K.~N. Hudemann, A.~Hecker, S.~Schmid, {Runtime Verification of P4
  Switches with Reinforcement Learning}, in: {Workshop on Network Meets AI \&
  ML}, 2019, p. 1–7.

\bibitem{JiJo19}
D.~Jindal, R.~Joshi, B.~Leong, {P4TrafficTool: Automated Code Generation for P4
  Traffic Generators and Analyzers}, in: {ACM Symposium on SDN Research
  (SOSR)}, 2019, p. 152–153.

\bibitem{DaWa17}
H.~T. Dang, H.~Wang, T.~Jepsen, G.~Brebner, C.~Kim, J.~Rexford, R.~Soul{\'e},
  H.~Weatherspoon, {Whippersnapper: A P4 Language Benchmark Suite}, in: {ACM
  Symposium on SDN Research (SOSR)}, 2017, p. 95–101.

\bibitem{RoPa18}
F.~Rodriguez, P.~G.~K. Patra, L.~Csikor, C.~E. Rothenberg, P.~V{\"o}r{\"o}s,
  S.~Laki, G.~Pongr{\'a}cz, {BB-Gen: A Packet Crafter for P4 Target
  Evaluation}, in: {ACM SIGCOMM Conference Posters and Demos}, 2018, p.
  111–113.

\bibitem{HaJa20a}
H.~Harkous, M.~Jarschel, M.~He, R.~Pries, W.~Kellerer, {P8: P4 with Predictable
  Packet Processing Performance}, {{IEEE} Transactions on Network and Service
  Management (TNSM)} (2020) 1--1.

\bibitem{KoAr20}
S.~Kodeswaran, M.~T. Arashloo, P.~Tammana, J.~Rexford, {Tracking P4 Program
  Execution in the Data Plane}, in: {ACM Symposium on SDN Research (SOSR)},
  2020, p. 117–122.

\bibitem{BiSi20}
K.~Birnfeld, D.~C. da~Silva, W.~Cordeiro, B.~B.~N. de~França, {P4 Switch Code
  Data Flow Analysis: Towards Stronger Verification of Forwarding Plane
  Software}, in: {IEEE/IFIP Network Operations and Management Symposium
  (NOMS)}, 2020, pp. 1--8.

\bibitem{NeHu19}
M.~Neves, B.~Huffaker, K.~Levchenko, M.~Barcellos, {Dynamic Property
  Enforcement in Programmable Data Planes}, in: {IFIP-TC6 Networking Conference
  (Networking)}, 2019, pp. 1--9.

\bibitem{ZhBi17c}
C.~Zhang, J.~Bi, Y.~Zhou, J.~Wu, B.~Liu, Z.~Li, A.~B. Dogar, Y.~Wang, {P4DB:
  On-the-fly Debugging of the Programmable Data Plane}, in: {IEEE International
  Conference on Network Protocols (ICNP)}, 2017, pp. 1--10.

\bibitem{ZhBi19c}
Y.~Zhou, J.~Bi, C.~Zhang, B.~Liu, Z.~Li, Y.~Wang, M.~Yu, {P4DB: On-the-Fly
  Debugging for Programmable Data Planes}, IEEE/ACM Transactions on Networking
  (ToN) 27 (2019) 1714--1727.

\bibitem{NeLe17}
M.~Neves, K.~Levchenko, M.~Barcellos, {Sandboxing Data Plane Programs for Fun
  and Profit}, in: {ACM SIGCOMM Conference Posters and Demos}, 2017, p.
  103–104.

\bibitem{ShFa20}
A.~Shukla, S.~Fathalli, T.~Zinner, A.~Hecker, S.~Schmid, {P4Consist: Toward
  Consistent P4 SDNs}, {IEEE} Journal on Selected Areas in Communications
  (JSAC) 38 (2020) 1293--1307.

\bibitem{XiBi18}
Z.~Xia, J.~Bi, Y.~Zhou, C.~Zhang, {KeySight: A Scalable Troubleshooting
  Platform Based on Network Telemetry}, in: {ACM Symposium on SDN Research
  (SOSR)}, 2018, pp. 1--2.

\bibitem{RuWa20}
F.~Ruffy, T.~Wang, A.~Sivaraman, {Gauntlet: Finding Bugs in Compilers for
  Programmable Packet Processing}, in: {USENIX Symposium on Operating Systems
  Design and Implementation (OSDI)}, 2020, pp. 1--17.

\bibitem{KrHo19}
J.~Krude, J.~Hofmann, M.~Eichholz, K.~Wehrle, A.~Koch, M.~Mezini, {Online
  Reprogrammable Multi Tenant Switches}, in: {ACM CoNEXT Workshop on Emerging
  In-Network Computing Paradigms}, 2019, p. 1–8.

\bibitem{HaMe16}
D.~Hancock, J.~van~der Merwe, {HyPer4: Using P4 to Virtualize the Programmable
  Data Plane}, in: {ACM Conference on emerging Networking EXperiments and
  Technologies (CoNEXT)}, 2016, p. 35–49.

\bibitem{ZhBi17a}
C.~Zhang, J.~Bi, Y.~Zhou, A.~B. Dogar, J.~Wu, {HyperV: A High Performance
  Hypervisor for Virtualization of the Programmable Data Plane}, in: {IEEE
  International Conference on Computer Communications and Networks (ICCCN)},
  2017, pp. 1--9.

\bibitem{ZhBi17b}
C.~Zhang, J.~Bi, Y.~Zhou, A.~B. Dogar, J.~Wu, {MPVisor: A Modular Programmable
  Data Plane Hypervisor}, in: {ACM Symposium on SDN Research (SOSR)}, 2017, p.
  179–180.

\bibitem{p4-hypervdp-repo}
{GitHub: HyperVDP}, \url{https://github.com/HyperVDP}, {accessed 01-20-2021}
  (2021).

\bibitem{ZhBi19}
C.~Zhang, J.~Bi, Y.~Zhou, J.~Wu, {HyperVDP: High-Performance Virtualization of
  the Programmable Data Plane}, {IEEE} Journal on Selected Areas in
  Communications (JSAC) 37 (2019) 556--569.

\bibitem{SaBu20}
M.~{Saquetti}, G.~{Bueno}, W.~{Cordeiro}, J.~R. {Azambuja}, {P4VBox: Enabling
  P4-Based Switch Virtualization}, {IEEE} Communications Letters 24 (2020)
  146--149.

\bibitem{SaBu19}
M.~{Saquetti}, G.~{Bueno}, W.~{Cordeiro}, J.~R. {Azambuja}, {VirtP4: An
  Architecture for P4 Virtualization}, in: {IEEE International Parallel and
  Distributed Processing Symposium Workshops (IPDPSW)}, 2019, pp. 75--78.

\bibitem{ZhBe18}
P.~Zheng, T.~Benson, C.~Hu, {P4Visor: Lightweight Virtualization and
  Composition Primitives for Building and Testing Modular Programs}, in: {ACM
  Conference on emerging Networking EXperiments and Technologies (CoNEXT)},
  2018, p. 98–111.

\bibitem{PaCa20a}
R.~Parizotto, L.~Castanheira, F.~Bonetti, A.~Santos, A.~Schaeffer-Filho,
  {PRIME: Programming In-Network Modular Extensions}, in: {IEEE/IFIP Network
  Operations and Management Symposium (NOMS)}, 2020, pp. 1--9.

\bibitem{ZaFr20a}
E.~O. Zaballa, D.~Franco, M.~S. Berger, M.~Higuero, {A Perspective on P4-Based
  Data and Control Plane Modularity for Network Automation}, in: {P4 Workshop
  in Europe (EuroP4)}, 2020, p. 59–61.

\bibitem{StZi20}
R.~Stoyanov, N.~Zilberman, {MTPSA: Multi-Tenant Programmable Switches}, in: {P4
  Workshop in Europe (EuroP4)}, 2020, p. 43–48.

\bibitem{mtpsa-repo}
{GitHub: MTPSA}, \url{https://github.com/mtpsa}, {accessed 01-20-2021} (2021).

\bibitem{HaJa20b}
S.~{Han}, S.~{Jang}, H.~{Choi}, H.~{Lee}, S.~{Pack}, {Virtualization in
  Programmable Data Plane: A Survey and Open Challenges}, {IEEE Open Journal of
  the Communications Society} 1 (2020) 527--534.

\bibitem{SiSt18}
J.~{Santiago da Silva}, T.~{Stimpfling}, T.~{Luinaud}, B.~{Fradj},
  B.~{Boughzala}, {One for All, All for One: A Heterogeneous Data Plane for
  Flexible P4 Processing}, in: {IEEE International Conference on Network
  Protocols (ICNP)}, 2018, pp. 440--441.

\bibitem{BeKr20}
C.~Beckmann, R.~Krishnamoorthy, H.~Wang, A.~Lam, C.~Kim, {Hurdles for a
  DRAM-based Match-Action Table}, in: {Conference on Innovation in Clouds,
  Internet and Networks and Workshops (ICIN)}, 2020, pp. 13--16.

\bibitem{AgXu17}
A.~{Aghdai}, Y.~{Xu}, H.~J. {Chao}, {Design of a hybrid modular switch}, in:
  {IEEE Conference on Network Function Virtualization and Software-Defined
  Networking (NFV-SDN)}, 2017, pp. 1--6.

\bibitem{LaHo20}
S.~{Laki}, D.~{Horpacsi}, P.~{Voros}, M.~{Tejfel}, P.~{Hudoba}, G.~{Pongracz},
  L.~{Molnar}, {The Price for Asynchronous Execution of Extern Functions in
  Programmable Software Data Planes}, in: {Workshop on Flexible Network Data
  Plane Processing (NETPROC@ICIN)}, 2020, pp. 23--28.

\bibitem{HoVo19}
D.~Horpácsi, P.~Vörös, M.~Tejfel, S.~Laki, G.~Pongrácz, L.~Molnár,
  {Asynchronous Extern Functions in Programmable Software Data Planes}, in: {P4
  Workshop in Europe (EuroP4)}, 2019, pp. 1--2.

\bibitem{ScOe19}
D.~Scholz, A.~Oeldemann, F.~Geyer, S.~Gallenmüller, H.~Stubbe, T.~Wild,
  A.~Herkersdorf, G.~Carle, {Cryptographic Hashing in P4 Data Planes}, in: {P4
  Workshop in Europe (EuroP4)}, 2019, pp. 1--6.

\bibitem{SiBo18b}
J.~S. da~Silva, F.-R. Boyer, L.-O. Chiquette, J.~P. Langlois, {Extern Objects
  in P4: an ROHC Header Compression Scheme Case Study}, in: {IEEE Conference on
  Network Softwarization (NetSoft)}, 2018, pp. 517--522.

\bibitem{GrGr19}
N.~Gray, A.~Grigorjew, T.~Hosssfeld, A.~Shukla, T.~Zinner, {Highlighting the
  Gap Between Expected and Actual Behavior in P4-enabled Networks}, in:
  {IFIP/IEEE Symposium on Integrated Management (IM)}, 2019, pp. 731--732.

\bibitem{DuDu20}
M.~V. Dumitru, D.~Dumitrescu, C.~Raiciu, {Can We Exploit Buggy P4 Programs?},
  in: {ACM Symposium on SDN Research (SOSR)}, 2020, p. 62–68.

\bibitem{MaCh19}
J.~Mambretti, J.~Chen, F.~Yeh, S.~Y. Yu, {International P4 Networking Testbed},
  in: {ACM/IEEE Symposium on Architectures for Networking and Communications
  Systems (ANCS)}, 2019, pp. 1--2.

\bibitem{ChTs19}
B.~{Chung}, C.~{Tseng}, J.~H. {Chen}, J.~{Mambretti}, {P4MT: Multi-Tenant
  Support Prototype for International P4 Testbed}, in: {ACM/IEEE Symposium on
  Architectures for Networking and Communications Systems (ANCS)}, 2019, pp.
  1--2.

\bibitem{2stic}
{A national programmable infrastructure to experiment with next-generation
  networks},
  \url{https://www.2stic.nl/national-programmable-infrastructure.html},
  {accessed 01-20-2021} (2021).

\bibitem{SuBa19}
R.~Sukapuram, G.~Barua, {PPCU: Proportional Per-packet Consistent Updates for
  SDNs using Data Plane Time Stamps}, Computer Networks 155 (2019) 72--86.

\bibitem{SuBa19b}
R.~Sukapuram, G.~Barua, {ProFlow: Proportional Per-Bidirectional-Flow
  Consistent Updates}, {{IEEE} Transactions on Network and Service Management
  (TNSM)} 16 (2019) 675--689.

\bibitem{LiBe19}
S.~Liu, T.~A. Benson, M.~K. Reiter, {Efficient and Safe Network Updates with
  Suffix Causal Consistency}, in: {European Conference on Computer Systems
  (EUROSYS)}, 2019, p. 1–15.

\bibitem{NgCh17}
T.~D. Nguyen, M.~Chiesa, M.~Canini, {Decentralized Consistent Network Updates
  in SDN with ez-Segway}, {ArXiv} e-prints (2017).

\bibitem{GeHe19}
S.~Geissler, S.~Herrnleben, R.~Bauer, A.~Grigorjew, T.~Zinner, M.~Jarschel,
  {The Power of Composition: Abstracting aMulti-Device SDN Data Path Through a
  Single API}, {{IEEE} Transactions on Network and Service Management (TNSM)}
  (2019) 722–735.

\bibitem{MoVi18}
E.~C. Molero, S.~Vissicchio, L.~Vanbever, {Hardware-Accelerated Network Control
  Planes}, in: {ACM Workshop on Hot Topics in Networks (HotNets)}, 2018, p.
  120–126.

\bibitem{SiNa17}
V.~Sivaraman, S.~Narayana, O.~Rottenstreich, S.~Muthukrishnan, J.~Rexford,
  {Heavy-Hitter Detection Entirely in the Data Plane}, in: {ACM Symposium on
  SDN Research (SOSR)}, 2017, p. 164–176.

\bibitem{p4-hashpipe}
{GitHub: Hashpipe}, \url{https://github.com/vibhaa/hashpipe}, {accessed
  01-20-2021} (2021).

\bibitem{LiHu19}
Y.~{Lin}, C.~{Huang}, S.~{Tsai}, {SDN Soft Computing Application for Detecting
  Heavy Hitters}, IEEE Transactions on Industrial Informatics (ToII) 15 (2019)
  5690--5699.

\bibitem{PoAn17}
D.~A. Popescu, G.~Antichi, A.~W. Moore, {Enabling Fast Hierarchical Heavy
  Hitter Detection using Programmable Data Planes}, in: {ACM Symposium on SDN
  Research (SOSR)}, 2017, p. 191–192.

\bibitem{HaCa18}
R.~Harrison, Q.~Cai, A.~Gupta, J.~Rexford, {Network-Wide Heavy Hitter Detection
  with Commodity Switches}, in: {ACM Symposium on SDN Research (SOSR)}, 2018,
  pp. 1--7.

\bibitem{KuPo20}
J.~Ku\v{c}era, D.~A. Popescu, H.~Wang, A.~Moore, J.~Ko\v{r}enek, G.~Antichi,
  {Enabling Event-Triggered Data Plane Monitoring}, in: {ACM Symposium on SDN
  Research (SOSR)}, 2020, p. 14–26.

\bibitem{SiJa18}
M.~Silva, A.~Jacobs, R.~Pfitscher, L.~Granville, {IDEAFIX: Identifying Elephant
  Flows in P4-Based IXP Networks}, in: {IEEE Global Communications Conference
  (GLOBECOM)}, 2018, pp. 1--6.

\bibitem{TuOo19}
B.~Turkovic, J.~Oostenbrink, F.~Kuipers, {Detecting Heavy Hitters in the
  Data-plane}, {ArXiv} e-prints (2019).

\bibitem{DiSa20b}
D.~{Ding}, M.~{Savi}, G.~{Antichi}, D.~{Siracusa}, {An Incrementally-Deployable
  P4-Enabled Architecture for Network-Wide Heavy-Hitter Detection}, {{IEEE}
  Transactions on Network and Service Management (TNSM)} 17 (2020) 75--88.

\bibitem{p4-DiSa20b}
{GitHub: Network-Wide Heavy-Hitter Detection Implementation in P4 Language},
  \url{https://github.com/DINGDAMU/Network-wide-heavy-hitter-detection},
  {accessed 01-20-2021} (2021).

\bibitem{SoAv18}
J.~Sonchack, A.~J. Aviv, E.~Keller, J.~M. Smith, {Turboflow: Information Rich
  Flow Record Generation on Commodity Switches}, in: {European Conference on
  Computer Systems (EUROSYS)}, 2018, p. 1–16.

\bibitem{p4-turboflow}
{GitHub: TurboFlow}, \url{https://github.com/jsonch/TurboFlow}, {accessed
  01-20-2021} (2021).

\bibitem{SoMi18}
J.~Sonchack, O.~Michel, A.~J. Aviv, E.~Keller, J.~M. Smith, {Scaling Hardware
  Accelerated Network Monitoring to Concurrent and Dynamic Queries With *Flow},
  in: USENIX Annual Technical Conference (ATC), 2018, pp. 823--835.

\bibitem{github-starflow}
{GitHub: StarFlow}, \url{https://github.com/jsonch/starflow}, {accessed
  01-25-2021} (2021).

\bibitem{HiAl18}
J.~Hill, M.~Aloserij, P.~Grosso, {Tracking Network Flows with P4}, in:
  {IEEE/ACM Innovating the Network for Data-Intensive Science (INDIS)}, 2018,
  pp. 23--32.

\bibitem{CaPa19}
L.~{Castanheira}, R.~{Parizotto}, A.~E. {Schaeffer-Filho}, {FlowStalker:
  Comprehensive Traffic Flow Monitoring on the Data Plane using P4}, in: {IEEE
  International Conference on Communicaotions (ICC)}, 2019, pp. 1--6.

\bibitem{PaCa20}
R.~{Parizotto}, L.~{Castanheira}, R.~H. {Ribeiro}, L.~{Zembruzki}, A.~S.
  {Jacobs}, L.~Z. {Granville}, A.~{Schaeffer-Filho}, {ShadowFS: Speeding-up
  Data Plane Monitoring and Telemetry using P4}, in: {IEEE International
  Conference on Communicaotions (ICC)}, 2020, pp. 1--6.

\bibitem{BaSa21}
D.~Barradas, N.~Santos, L.~Rodrigues, S.~Signorello, F.~M.~V. Ramos,
  A.~Madeira, {FlowLens: Enabling Efficient Flow Classification for ML-based
  Network Security Applications}, in: {Network and Distributed Systems Security
  Symposium (NDSS)}, 2021, pp. 1--18.

\bibitem{BaSa21-repo}
{GitHub: FlowLens}, \url{https://github.com/dmbb/FlowLens}, {accessed
  04-14-2021} (2021).

\bibitem{WaTa20}
W.~Wang, P.~Tammana, A.~Chen, T.~S.~E. Ng, {Grasp the Root Causes in the Data
  Plane: Diagnosing Latency Problems with SpiderMon}, in: {ACM Symposium on SDN
  Research (SOSR)}, 2020, p. 55–61.

\bibitem{ChLa19}
X.~Chen, S.~Landau-Feibish, Y.~Koral, J.~Rexford, O.~Rottenstreich, S.~A.
  Monetti, T.-Y. Wang, {Fine-Grained Queue Measurement in the Data Plane}, in:
  {ACM Conference on emerging Networking EXperiments and Technologies
  (CoNEXT)}, 2019, p. 15–29.

\bibitem{ZhSh19}
Z.~Zhao, X.~Shi, X.~Yin, Z.~Wang, Q.~Li, {HashFlow for Better Flow Record
  Collection}, in: {IEEE International Conference on Distributed Computing
  Systems (ICDCS)}, 2019, pp. 1416--1425.

\bibitem{HuLe18}
Q.~Huang, P.~P.~C. Lee, Y.~Bao, {Sketchlearn: Relieving User Burdens in
  Approximate Measurement with Automated Statistical Inference}, in: {ACM
  SIGCOMM Conference}, 2018, p. 576–590.

\bibitem{p4-sketchlearn}
{GitHub: SketchLearn}, \url{https://github.com/huangqundl/SketchLearn},
  {accessed 01-20-2021} (2021).

\bibitem{TaHu20}
L.~Tang, Q.~Huang, P.~C. Lee, {A Fast and Compact Invertible Sketch for
  Network-Wide Heavy Flow Detection}, IEEE/ACM Transactions on Networking (ToN)
  28 (2020) 2350--2363.

\bibitem{p4-mv-sketch}
{GitHub: MV-Sketch}, \url{https://github.com/Grace-TL/MV-Sketch}, {accessed
  01-20-2021} (2021).

\bibitem{HaWe19}
Z.~{Hang}, M.~{Wen}, Y.~{Shi}, C.~{Zhang}, {Interleaved Sketch: Toward
  Consistent Network Telemetry for Commodity Programmable Switches}, {IEEE}
  ACCESS 7 (2019) 146745--146758.

\bibitem{LiMa16}
Z.~Liu, A.~Manousis, G.~Vorsanger, V.~Sekar, V.~Braverman, {One Sketch to Rule
  Them All: Rethinking Network Flow Monitoring with UnivMon}, in: {ACM SIGCOMM
  Conference}, 2016, p. 101–114.

\bibitem{YaJi18}
T.~Yang, J.~Jiang, P.~Liu, Q.~Huang, J.~Gong, Y.~Zhou, R.~Miao, X.~Li,
  S.~Uhlig, {Elastic Sketch: Adaptive and Fast Network-wide Measurements}, in:
  {ACM SIGCOMM Conference}, 2018, p. 561–575.

\bibitem{YaJi19}
T.~{Yang}, J.~{Jiang}, P.~{Liu}, Q.~{Huang}, J.~{Gong}, Y.~{Zhou}, R.~{Miao},
  X.~{Li}, S.~{Uhlig}, {Adaptive Measurements Using One Elastic Sketch},
  IEEE/ACM Transactions on Networking (ToN) 27 (2019) 2236--2251.

\bibitem{p4-ElasticSketch}
{GitHub: ElasticSketch}, \url{https://github.com/BlockLiu/ElasticSketchCode},
  {accessed 01-20-2021} (2021).

\bibitem{PeNe17}
F.~{Pereira}, N.~{Neves}, F.~M.~V. {Ramos}, {Secure network monitoring using
  programmable data planes}, in: {IEEE Conference on Network Function
  Virtualization and Software-Defined Networking (NFV-SDN)}, 2017, pp.
  286--291.

\bibitem{MaVe18}
R.~F.~T. Martins, F.~L. Verdi, R.~Villaça, L.~F.~U. Garcia, {Using
  Probabilistic Data Structures for Monitoring of Multi-tenant P4-based
  Networks}, in: {IEEE Symposium on Computers and Communications (ISCC)}, 2018,
  pp. 204--207.

\bibitem{LaSh19}
Y.-K. Lai, K.-Y. Shih, P.-Y. Huang, H.-P. Lee, Y.-J. Lin, T.-L. Liu, J.~H.
  Chen, {Sketch-based Entropy Estimation for Network Traffic Analysis using
  Programmable Data Plane ASICs}, in: {ACM/IEEE Symposium on Architectures for
  Networking and Communications Systems (ANCS)}, 2019, pp. 1--2.

\bibitem{LiZh20}
Z.~Liu, S.~Zhou, O.~Rottenstreich, V.~Braverman, J.~Rexford, {Memory-Efficient
  Performance Monitoring on Programmable Switches with Lean Algorithms}, in:
  {SIAM Symposium on Algorithmic Principles of Computer Systems (APOCS)}, 2020,
  pp. 31--44.

\bibitem{TaHu20b}
L.~{Tang}, Q.~{Huang}, P.~P.~C. {Lee}, {SpreadSketch: Toward Invertible and
  Network-Wide Detection of Superspreaders}, in: {IEEE International Conference
  on Computer Communications (INFOCOM)}, 2020, pp. 1608--1617.

\bibitem{TaHu20b-repo}
{GitHub: SpreadSketch},
  \url{http://adslab.cse.cuhk.edu.hk/software/spreadsketch/}, {accessed
  01-20-2021} (2021).

\bibitem{VeKa19}
J.~{Vestin}, A.~{Kassler}, D.~{Bhamare}, K.~{Grinnemo}, J.~{Andersson},
  G.~{Pongracz}, {Programmable Event Detection for In-Band Network Telemetry},
  in: {IEEE International Conference on Cloud Networking (IEEE CloudNet)},
  2019, pp. 1--6.

\bibitem{WaCh19}
S.~{Wang}, Y.~{Chen}, J.~{Li}, H.~{Hu}, J.~{Tsai}, Y.~{Lin}, {A
  Bandwidth-Efficient INT System for Tracking the Rules Matched by the Packets
  of a Flow}, in: {IEEE Global Communications Conference (GLOBECOM)}, 2019, pp.
  1--6.

\bibitem{BhKa19}
D.~{Bhamare}, A.~{Kassler}, J.~{Vestin}, M.~A. {Khoshkholghi}, J.~{Taheri},
  {IntOpt: In-Band Network Telemetry Optimization for NFV Service Chain
  Monitoring}, in: {IEEE International Conference on Communicaotions (ICC)},
  2019, pp. 1--7.

\bibitem{JiPa20}
C.~{Jia}, T.~{Pan}, Z.~{Bian}, X.~{Lin}, E.~{Song}, C.~{Xu}, T.~{Huang},
  Y.~{Liu}, {Rapid Detection and Localization of Gray Failures in Data Centers
  via In-band Network Telemetry}, in: {IEEE/IFIP Network Operations and
  Management Symposium (NOMS)}, 2020, pp. 1--9.

\bibitem{p4-JiPa20}
{GitHub: Gray Failures Detection and Localization},
  \url{https://github.com/graytower/INT_DETECT}, {accessed 01-20-2021} (2021).

\bibitem{NiKo19}
B.~{Niu}, J.~{Kong}, S.~{Tang}, Y.~{Li}, Z.~{Zhu}, {Visualize Your
  IP-Over-Optical Network in Realtime: A P4-Based Flexible Multilayer In-Band
  Network Telemetry (ML-INT) System}, {IEEE} ACCESS 7 (2019) 82413--82423.

\bibitem{KaFi20}
N.~S. {Kagami}, R.~I.~T. {da Costa Filho}, L.~P. {Gaspary}, {CAPEST: Offloading
  Network Capacity and Available Bandwidth Estimation to Programmable Data
  Planes}, {{IEEE} Transactions on Network and Service Management (TNSM)} 17
  (2020) 175--189.

\bibitem{p4-capest}
{GitHub: Capest}, \url{https://github.com/nicolaskagami/capest}, {accessed
  01-20-2021} (2021).

\bibitem{ChJa19}
N.~Choi, L.~Jagadeesan, Y.~Jin, N.~N. Mohanasamy, M.~R. Rahman, K.~Sabnani,
  M.~Thottan, {Run-time Performance Monitoring, Verification, and Healing of
  End-to-End Services}, in: {IEEE Conference on Network Softwarization
  (NetSoft)}, 2019, pp. 30--35.

\bibitem{SgPa20}
A.~{Sgambelluri}, F.~{Paolucci}, A.~{Giorgetti}, D.~{Scano}, F.~{Cugini},
  {Exploiting Telemetry in Multi-Layer Networks}, in: {International Conference
  on Transparent Optical Networks (ICTON)}, 2020, pp. 1--4.

\bibitem{FePa20}
Y.~{Feng}, S.~{Panda}, S.~G. {Kulkarni}, K.~K. {Ramakrishnan}, N.~{Duffield},
  {A SmartNIC-Accelerated Monitoring Platform for In-band Network Telemetry},
  in: {IEEE International Symposium on Local and Metropolitan Area Networks
  (LANMAN)}, 2020, pp. 1--6.

\bibitem{MaLe20}
J.~Marques, K.~Levchenko, L.~Gaspary, {IntSight: Diagnosing SLO Violations with
  in-Band Network Telemetry}, in: {ACM Conference on emerging Networking
  EXperiments and Technologies (CoNEXT)}, 2020, p. 421–434.

\bibitem{MaLe20-repo}
{GitHub: IntSight}, \url{https://github.com/jonadmark/intsight-conext},
  {accessed 01-20-2021} (2021).

\bibitem{SuJa20}
D.~Suh, S.~Jang, S.~Han, S.~Pack, X.~Wang, {Flexible sampling-based in-band
  network telemetry in programmable data plane}, {ICT Express} 6 (2020) 62--65.

\bibitem{NaSi17}
S.~Narayana, A.~Sivaraman, V.~Nathan, P.~Goyal, V.~Arun, M.~Alizadeh,
  V.~Jeyakumar, C.~Kim, {Language-Directed Hardware Design for Network
  Performance Monitoring}, in: {ACM SIGCOMM Conference}, 2017, p. 85–98.

\bibitem{NaNa17}
V.~Nathan, S.~Narayana, A.~Sivaraman, P.~Goyal, V.~Arun, M.~Alizadeh,
  V.~Jeyakumar, C.~Kim, {Demonstration of the Marple System for Network
  Performance Monitoring}, in: {ACM SIGCOMM Conference Posters and Demos},
  2017, p. 57–59.

\bibitem{p4-marple}
{GitHub: Marple}, \url{https://github.com/performance-queries/marple},
  {accessed 01-20-2021} (2021).

\bibitem{LaRo19}
P.~{Laffranchini}, L.~{Rodrigues}, M.~{Canini}, B.~{Krishnamurthy},
  {Measurements As First-class Artifacts}, in: {IEEE International Conference
  on Computer Communications (INFOCOM)}, 2019, pp. 415--423.

\bibitem{p4-mafia}
{GitHub: Mafia}, \url{https://github.com/paololaff/mafia-sdn}, {accessed
  01-20-2021} (2021).

\bibitem{GuHa18}
A.~Gupta, R.~Harrison, M.~Canini, N.~Feamster, J.~Rexford, W.~Willinger,
  {Sonata: Query-Driven Streaming Network Telemetry}, in: {ACM Symposium on SDN
  Research (SOSR)}, 2018, p. 357–371.

\bibitem{p4-sonata}
{GitHub: SONATA}, \url{https://github.com/Sonata-Princeton/SONATA-DEV},
  {accessed 01-20-2021} (2021).

\bibitem{TeHa20}
R.~Teixeira, R.~Harrison, A.~Gupta, J.~Rexford, {PacketScope: Monitoring the
  Packet Lifecycle Inside a Switch}, in: {ACM Symposium on SDN Research
  (SOSR)}, 2020, p. 76–82.

\bibitem{GaJi18}
Y.~Gao, Y.~Jing, W.~Dong, {UniROPE: Universal and Robust Packet Trajectory
  Tracing for Software-Defined Networks}, IEEE/ACM Transactions on Networking
  (ToN) 26 (2018) 2515–2527.

\bibitem{KnHi19}
S.~{Knossen}, J.~{Hill}, P.~{Grosso}, {Hop Recording and Forwarding State
  Logging: Two Implementations for Path Tracking in P4}, in: {IEEE/ACM
  Innovating the Network for Data-Intensive Science (INDIS)}, 2019, pp. 36--47.

\bibitem{BaRo20}
A.~{Indra Basuki}, D.~{Rosiyadi}, I.~{Setiawan}, {Preserving Network Privacy on
  Fine-grain Path-tracking Using P4-based SDN}, in: {International Conference
  on Radar, Antenna, Microwave, Electronics, and Telecommunications (ICRAMET)},
  2020, pp. 129--134.

\bibitem{JoQu18}
R.~Joshi, T.~Qu, M.~C. Chan, B.~Leong, B.~T. Loo, {BurstRadar: Practical
  Real-time Microburst Monitoring for Datacenter Networks}, in: {ACM SIGOPS
  Asia-Pacific Workshop on System (APSys)}, 2018, pp. 1--8.

\bibitem{p4-burstradar}
{GitHub: BurstRadar}, \url{https://github.com/harshgondaliya/burstradar},
  {accessed 01-20-2021} (2021).

\bibitem{GhBe17}
M.~Ghasemi, T.~Benson, J.~Rexford, {Dapper: Data Plane Performance Diagnosis of
  TCP}, in: {ACM Symposium on SDN Research (SOSR)}, 2017, p. 61–74.

\bibitem{HeCh18}
C.-H. He, B.~Y. Chang, S.~Chakraborty, C.~Chen, L.~C. Wang, {A Zero Flow Entry
  Expiration Timeout P4 Switch}, in: {ACM Symposium on SDN Research (SOSR)},
  2018, pp. 1--2.

\bibitem{RiKi19}
A.~Riesenberg, Y.~Kirzon, M.~Bunin, E.~Galili, G.~Navon, T.~Mizrahi,
  {Time-Multiplexed Parsing in Marking-Based Network Telemetry}, in: {ACM
  International Conference on Systems and Storage (SYSTOR)}, 2019, p. 80–85.

\bibitem{p4-RiKi19}
{GitHub: P4 Alternate Marking Algorithm},
  \url{https://github.com/AlternateMarkingP4/FlaseClase}, {accessed 01-20-2021}
  (2021).

\bibitem{WaHu20}
S.~Y. {Wang}, H.~W. {Hu}, Y.~B. {Lin}, {Design and Implementation of
  TCP-Friendly Meters in P4 Switches}, IEEE/ACM Transactions on Networking
  (ToN) 28 (2020) 1885--1898.

\bibitem{KuSi20}
R.~Kundel, F.~Siegmund, J.~Blendin, A.~Rizk, B.~Koldehofe, {P4STA: High
  Performance Packet Timestamping with Programmable Packet Processors}, in:
  {IEEE/IFIP Network Operations and Management Symposium (NOMS)}, 2020, p.
  1–9.

\bibitem{p4-p4sta}
{GitHub: P4STA}, \url{https://github.com/ralfkundel/P4STA}, {accessed
  01-20-2021} (2021).

\bibitem{HaBh19}
R.~{Hark}, D.~{Bhat}, M.~{Zink}, R.~{Steinmetz}, A.~{Rizk}, {Preprocessing
  Monitoring Information on the SDN Data-Plane using P4}, in: {IEEE Conference
  on Network Function Virtualization and Software-Defined Networking
  (NFV-SDN)}, 2019, pp. 1--6.

\bibitem{DiSa20}
D.~{Ding}, M.~{Savi}, D.~{Siracusa}, {Estimating Logarithmic and Exponential
  Functions to Track Network Traffic Entropy in P4}, in: {IEEE/IFIP Network
  Operations and Management Symposium (NOMS)}, 2020, pp. 1--9.

\bibitem{p4-p4entropy}
{GitHub: P4Entropy}, \url{https://github.com/DINGDAMU/P4Entropy}, {accessed
  01-20-2021} (2021).

\bibitem{TaMe19}
P.~{Taffet}, J.~{Mellor-Crummey}, {Lightweight, Packet-Centric Monitoring of
  Network Traffic and Congestion Implemented in P4}, in: {IEEE Symposium on
  High-Performance Interconnects (HOTI)}, 2019, pp. 54--58.

\bibitem{LiZh20b}
Y.~{Lin}, Y.~{Zhou}, Z.~{Liu}, K.~{Liu}, Y.~{Wang}, M.~{Xu}, J.~{Bi}, Y.~{Liu},
  J.~{Wu}, {NetView: Towards On-Demand Network-Wide Telemetry in the Data
  Center}, in: {IEEE International Conference on Communicaotions (ICC)}, 2020,
  pp. 1--6.

\bibitem{BaZh20}
J.~Bai, M.~Zhang, G.~Li, C.~Liu, M.~Xu, H.~Hu, {FastFE: Accelerating ML-Based
  Traffic Analysis with Programmable Switches}, in: {Workshop on Secure
  Programmable Network Infrastructure (SPIN)}, 2020, p. 1–7.

\bibitem{KuBa20}
J.~Ku\v{c}era, R.~B. Basat, M.~Kuka, G.~Antichi, M.~Yu, M.~Mitzenmacher,
  {Detecting Routing Loops in the Data Plane}, in: {ACM Conference on emerging
  Networking EXperiments and Technologies (CoNEXT)}, 2020, p. 466–473.

\bibitem{HaSh19}
Z.~{Hang}, Y.~{Shi}, M.~{Wen}, C.~{Zhang}, {TBSW: Time-Based Sliding Window
  Algorithm for Network Traffic Measurement}, in: IEEE International Conference
  on High Performance Computing and Communications; IEEE International
  Conference on Smart City; IEEE International Conference on Data Science and
  Systems (HPCC/SmartCity/DSS), 2019, pp. 1305--1310.

\bibitem{GuSh19}
B.~{Guan}, S.~{Shen}, {FlowSpy: An Efficient Network Monitoring Framework Using
  P4 in Software-Defined Networks}, in: {IEEE Semiannual Vehicular Technology
  Conference (VTC)}, 2019, pp. 1--5.

\bibitem{heavyhitters-slides}
{Heavy Hitter Detection: Guest lecture for CS344 at Stanford},
  \url{https://cs344-stanford.github.io/lectures/Lecture-4-HHD.pdf}, {accessed
  01-20-2021} (2018).

\bibitem{rfc3954}
B.~Claise, \href{http://www.rfc-editor.org/rfc/rfc3954.txt}{{Cisco Systems
  NetFlow Services Export Version 9}}, RFC 3954, RFC Editor (10 2004).
\newline\urlprefix\url{http://www.rfc-editor.org/rfc/rfc3954.txt}

\bibitem{rfc3176}
P.~Phaal, S.~Panchen, N.~McKee,
  \href{http://www.rfc-editor.org/rfc/rfc3176.txt}{{InMon Corporation's sFlow:
  A Method for Monitoring Traffic in Switched and Routed Networks}}, RFC 3176,
  RFC Editor (09 2001).
\newline\urlprefix\url{http://www.rfc-editor.org/rfc/rfc3176.txt}

\bibitem{rfc7011}
B.~Claise, B.~Trammell, P.~Aitken,
  \href{http://www.rfc-editor.org/rfc/rfc7011.txt}{{Specification of the IP
  Flow Information Export (IPFIX) Protocol for the Exchange of Flow
  Information}}, STD~77, RFC Editor (09 2013).
\newline\urlprefix\url{http://www.rfc-editor.org/rfc/rfc7011.txt}

\bibitem{KiBh16}
{In‐band Network Telemetry (INT)},
  \url{https://p4.org/assets/INT-current-spec.pdf}, {accessed 01-20-2021}
  (2021).

\bibitem{p4-applications-wg}
{Charter of the P4 Applications WG},
  \url{https://github.com/p4lang/p4-applications/blob/master/docs/charter.pdf},
  {accessed 01-20-2021} (2021).

\bibitem{KiSi15}
C.~Kim, A.~Sivaraman, N.~P. Katta, A.~Bas, A.~Dixit, L.~J. Wobker, {In-band
  Network Telemetry via Programmable Dataplanes},
  \url{https://nkatta.github.io/papers/int-demo.pdf} (2015).

\bibitem{CuGu19}
F.~Cugini, P.~Gunning, F.~Paolucci, P.~Castoldi, A.~Lord, {P4 In-Band Telemetry
  (INT) for Latency-Aware VNF in Metro Networks}, in: {Optical Fiber
  Communication Conference (OFC)}, 2019, pp. 1--3.

\bibitem{trellis}
{Open Networking Foundation: Trellis},
  \url{https://www.opennetworking.org/trellis/}, {accessed 01-20-2021} (2021).

\bibitem{trellis-p4-tutorial}
{Google Presentations: Trellis \& P4 Tutorial},
  \url{http://bit.ly/trellis-p4-slides}, {accessed 01-20-2021} (2018).

\bibitem{p4-trellis}
{GitHub: ONF Trellis},
  \url{https://github.com/opennetworkinglab/routing/tree/master/trellis},
  {accessed 01-20-2021} (2021).

\bibitem{SiKi15}
A.~Sivaraman, C.~Kim, R.~Krishnamoorthy, A.~Dixit, M.~Budiu, {DC.P4:
  Programming the Forwarding Plane of a Data-center Switch}, in: {ACM SIGCOMM
  Conference}, 2015, p. 1–8.

\bibitem{p4-SiKi15}
{GitHub: DC.p4}, \url{https://github.com/p4lang/papers/tree/master/sosr15},
  {accessed 01-20-2021} (2021).

\bibitem{p4apps-onf}
{Open Network Foundation: P4 apps at ONF},
  \url{https://github.com/p4lang/p4-applications/blob/master/meeting_slides/2018_04_19_ONF.pdf},
  {accessed 01-20-2021} (2018).

\bibitem{fabricp4-github}
{GitHub: fabric.p4},
  \url{https://github.com/opennetworkinglab/onos/blob/master/pipelines/fabric/impl/src/main/resources/fabric.p4},
  {accessed 01-20-2021} (2021).

\bibitem{rare}
{RARE (Router for Academia, Research \& Education)},
  {https://wiki.geant.org/display/RARE/Home}, {accessed 04-16-2021} (2021).

\bibitem{rare-repo}
{GitHub: RARE}, \url{https://github.com/frederic-loui/RARE}, {accessed
  04-16-2021} (2021).

\bibitem{PiDe18}
B.~Pit-Claudel, Y.~Desmouceaux, P.~Pfister, M.~Townsley, T.~Clausen, {Stateless
  Load-Aware Load Balancing in P4}, in: {IEEE International Conference on
  Network Protocols (ICNP)}, 2018, pp. 418--423.

\bibitem{MiZe17}
R.~Miao, H.~Zeng, C.~Kim, J.~Lee, M.~Yu, {SilkRoad: Making Stateful Layer-4
  Load Balancing Fast and Cheap using Switching ASICs}, in: {ACM SIGCOMM
  Conference}, 2017, p. 15–28.

\bibitem{KaHi16}
N.~Katta, M.~Hira, C.~Kim, A.~Sivaraman, J.~Rexford, {HULA: Scalable Load
  Balancing using Programmable Data Planes}, in: {ACM Symposium on SDN Research
  (SOSR)}, 2016, p. 1–12.

\bibitem{BeKa18}
C.~H. Benet, A.~J. Kassler, T.~Benson, G.~Pongracz, {MP-HULA: Multipath
  Transport Aware Load Balancing using Programmable Data Planes}, in: {Morning
  Workshop on In-Network Computing}, 2018, p. 7–13.

\bibitem{ChWa19}
B.~T. Chiang, K.~Wang, {Cost-effective Congestion-aware Load Balancing for
  Datacenters}, in: {International Conference on Electronics, Information, and
  Communication (ICEIC)}, 2019, pp. 1--6.

\bibitem{YeCh18}
J.-L. Ye, C.~Chen, Y.~H. Chu, {A Weighted ECMP Load Balancing Scheme for Data
  Centers using P4 Switches}, in: {IEEE International Conference on Cloud
  Networking (IEEE CloudNet)}, 2018, pp. 1--4.

\bibitem{HsTa20}
K.-F. Hsu, P.~Tammana, R.~Beckett, A.~Chen, J.~Rexford, D.~Walker, {Adaptive
  Weighted Traffic Splitting in Programmable Data Planes}, in: {ACM Symposium
  on SDN Research (SOSR)}, 2020, p. 103–109.

\bibitem{PiSc18}
M.~Pizzutti, A.~Schaeffer-Filho, {An Efficient Multipath Mechanism Based on the
  Flowlet Abstraction and P4}, in: {IEEE Global Communications Conference
  (GLOBECOM)}, 2018, pp. 1--6.

\bibitem{PiSc20}
M.~Pizzutti, A.~Schaeffer-Filho, {Adaptive Multipath Routing based on Hybrid
  Data and Control Plane Operation}, in: {IEEE International Conference on
  Computer Communications (INFOCOM)}, 2020, pp. 730--738.

\bibitem{ZhWe20}
J.~Zhang, S.~Wen, J.~Zhang, H.~Chai, T.~Pan, T.~Huang, L.~Zhang, Y.~Liu, F.~R.
  Yu, {Fast Switch-Based Load Balancer Considering Application Server States},
  IEEE/ACM Transactions on Networking (ToN) 28 (2020) 1391--1404.

\bibitem{LiZh20c}
Q.~Li, J.~Zhang, T.~Pan, T.~Huang, Y.~Liu, {Data-driven Routing Optimization
  based on Programmable Data Plane}, in: {IEEE International Conference on
  Computer Communications and Networks (ICCCN)}, 2020, pp. 1--9.

\bibitem{KaKa19}
E.~Kawaguchi, H.~Kasuga, N.~Shinomiya, {Unsplittable flow Edge Load factor
  Balancing in SDN using P4 Runtime}, in: {International Telecommunication
  Networks and Applications Conference (ITNAC)}, 2019, pp. 1--6.

\bibitem{CiCh17}
E.~Cidon, S.~Choi, S.~Katti, N.~McKeown, {AppSwitch: Application-layer Load
  Balancing withina Software Switch}, in: {Asia-Pacific Workshop on Networking
  (APnet)}, 2017, p. 64–70.

\bibitem{OlAg18}
V.~Olteanu, A.~Agache, A.~Voinescu, C.~Raiciu, {Stateless Datacenter
  Load-balancing with Beamer}, in: {USENIX Symposium on Networked Systems
  Design \& Implementation (NSDI)}, 2018, pp. 125--139.

\bibitem{github-beamer}
{GitHub: Beamer}, \url{https://github.com/Beamer-LB}, {accessed 01-25-2021}
  (2021).

\bibitem{GeYa19}
J.~Geng, J.~Yan, Y.~Zhang, {P4QCN: Congestion Control using P4-Capable Device
  in Data Center Networks}, {Electronics Journal} 8 (2019) 280.

\bibitem{JiZh19}
J.~Jiang, Y.~Zhang, {An Accurate Congestion Control Mechanism in Programmable
  Network}, in: {IEEE Annual Computing and Communication Workshop and
  Conference (CCWC)}, 2019, pp. 673--677.

\bibitem{ShJu20}
S.~Shahzad, E.~Jung, J.~Chung, R.~Kettimuthu, {Enhanced Explicit Congestion
  Notification (EECN) in TCP with P4 Programming}, in: {International
  Conference on Green and Human Information Technology (ICGHIT)}, 2020, pp.
  35--40.

\bibitem{ChFa20}
C.~Chen, H.~Fang, M.~S. Iqbal, {QoSTCP: Provide Consistent Rate Guarantees to
  TCP flows in Software Defined Networks}, in: {IEEE International Conference
  on Communicaotions (ICC)}, 2020, pp. 1--6.

\bibitem{LaFr20}
A.~Laraba, J.~François, I.~Chrisment, S.~R. Chowdhury, R.~Boutaba, {Defeating
  Protocol Abuse with P4: Application to Explicit Congestion Notification}, in:
  {IFIP-TC6 Networking Conference (Networking)}, 2020, pp. 431--439.

\bibitem{ShLi18}
N.~K. Sharma, M.~Liu, K.~Atreya, A.~Krishnamurthy, {Approximating Fair Queueing
  on Reconfigurable Switches}, in: {USENIX Symposium on Networked Systems
  Design \& Implementation (NSDI)}, 2018, p. 1–16.

\bibitem{CaBo17}
C.~Cascone, N.~Bonelli, L.~Bianchi, A.~Capone, B.~Sansò, {Towards Approximate
  Fair Bandwidth Sharing via Dynamic Priority Queuing}, in: {IEEE International
  Symposium on Local and Metropolitan Area Networks (LANMAN)}, 2017, pp. 1--6.

\bibitem{BhAn19}
D.~Bhat, J.~Anderson, P.~Ruth, M.~Zink, K.~Keahey, {Application-based QoE
  support with P4 and OpenFlow}, in: {IEEE Conference on Computer
  Communications Workshops (INFOCOM WKSHPS)}, 2019, pp. 817--823.

\bibitem{KfCr19}
E.~F. Kfoury, J.~Crichigno, E.~Bou-Harb, D.~Khoury, G.~Srivastava, {Enabling
  TCP Pacing using Programmable Data Plane Switches}, in: {International
  Conference on Telecommunications and Signal Processing (TSP)}, 2019, pp.
  273--277.

\bibitem{ChYe19}
Y.~Chen, L.~Yen, W.~Wang, C.~Chuang, Y.~Liu, C.~Tseng, {P4-Enabled Bandwidth
  Management}, in: {Asia-Pacific Network Operations and Management Symposium
  (APNOMS)}, 2019, pp. 1--5.

\bibitem{LeCh19}
S.~S.~W. Lee, K.~Chan, {A Traffic Meter Based on a Multicolor Marker for
  Bandwidth Guarantee and Priority Differentiation in SDN Virtual Networks},
  {{IEEE} Transactions on Network and Service Management (TNSM)} 16 (2019)
  1046--1058.

\bibitem{WaLi20}
S.-Y. Wang, J.-Y. Li, Y.-B. Lin, {Aggregating and disaggregating packets with
  various sizes of payload in P4 switches at 100 Gbps line rate}, {Journal of
  Network and Computer Applications (JNCA)} 165 (2020) 102676.

\bibitem{ToSa19}
K.~Tokmakov, M.~Sarker, J.~Domaschka, S.~Wesner, {A Case for Data Centre
  Traffic Management on Software Programmable Ethernet Switches}, in: {IEEE
  International Conference on Cloud Networking (IEEE CloudNet)}, 2019, pp.
  1--6.

\bibitem{TuKu18}
B.~Turkovic, F.~Kuipers, N.~van Adrichem, K.~Langendoen, {Fast Network
  Congestion Detection and Avoidance using P4}, in: {Workshop on Networking for
  Emerging Applications and Technologies (NEAT)}, 2018, p. 45–51.

\bibitem{KuBl18}
R.~Kundel, J.~Blendin, T.~Viernickel, B.~Koldehofe, R.~Steinmetz, {P4-CoDel:
  Active Queue Management in Programmable Data Planes}, in: {IEEE Conference on
  Network Function Virtualization and Software-Defined Networking (NFV-SDN)},
  2018, pp. 1--4.

\bibitem{codel-github}
{GitHub: P4-CoDel}, \url{https://github.com/ralfkundel/p4-codel}, {accessed
  01-20-2021} (2021).

\bibitem{MeMo19}
M.~Menth, H.~Mostafaei, D.~Merling, M.~Häberle, {Implementation and Evaluation
  of Activity-Based Congestion Management using P4 (P4-ABC)}, {MDPI Future
  Internet Journal (FI)} 11 (2019) 159.

\bibitem{TuKu20}
B.~Turkovic, F.~Kuipers, {P4air: Increasing Fairness among Competing Congestion
  Control Algorithms}, in: {IEEE International Conference on Network Protocols
  (ICNP)}, 2020, pp. 1--12.

\bibitem{FeCa20}
L.~B. Fernandes, L.~Camargos, {Bandwidth throttling in a P4 switch}, in: {IEEE
  Conference on Network Function Virtualization and Software-Defined Networking
  (NFV-SDN)}, 2020, pp. 91--94.

\bibitem{WaCh18}
G.~Wang, C.~Chen, C.~Chen, L.~Pan, Y.~Wang, C.~Fan, C.~Hsu, {Streaming Scalable
  Video Sequences with Media-Aware Network Elements Implemented in P4
  Programming Language}, in: {IEEE/IFIP Network Operations and Management
  Symposium (NOMS)}, 2018, pp. 1--2.

\bibitem{GrDi20}
A.~G. Alcoz, A.~Dietm{\"u}ller, L.~Vanbever, {SP-PIFO: Approximating Push-In
  First-Out Behaviors using Strict-Priority Queues}, in: {USENIX Symposium on
  Networked Systems Design \& Implementation (NSDI)}, 2020, pp. 59--76.

\bibitem{KuGu21}
I.~Kunze, M.~Gunz, D.~Saam, K.~Wehrle, J.~Rüth, {Tofino + P4: A Strong
  Compound for AQM on High-Speed Networks?}, in: {IFIP/IEEE International
  Symposium on Integrated Network Management}, 2021, pp. 72--80.

\bibitem{p4-KuGu21}
{GitHub: PIE for Tofino}, \url{https://github.com/COMSYS/pie-for-tofino},
  {accessed 04-15-2021} (2021).

\bibitem{HaPa21}
H.~Harkous, C.~Papagianni, K.~De~Schepper, M.~Jarschel, M.~Dimolianis,
  R.~Preis, Virtual queues for p4: A poor man’s programmable traffic manager,
  {{IEEE} Transactions on Network and Service Management (TNSM)} (2021) 1--1.

\bibitem{AnSa19}
B.~Andrus, S.~A. Sasu, T.~Szyrkowiec, A.~Autenrieth, M.~Chamania, J.~K.
  Fischer, S.~Rasp, {Zero-Touch Provisioning of Distributed Video Analytics in
  a Software-Defined Metro-Haul Network with P4 Processing}, in: {Optical Fiber
  Communication Conference (OFC)}, 2019, pp. 1--3.

\bibitem{IbAn19}
S.~Ibanez, G.~Antichi, G.~Brebner, N.~McKeown, {Event-Driven Packet
  Processing}, in: {ACM Workshop on Hot Topics in Networks (HotNets)}, 2019, p.
  133–140.

\bibitem{KfCr20}
E.~F. Kfoury, J.~Crichigno, E.~Bou-Harb, {Offloading Media Traffic to
  Programmable Data Plane Switches}, in: {IEEE International Conference on
  Communicaotions (ICC)}, 2020, pp. 1--7.

\bibitem{KeUd20}
I.~Kettaneh, S.~Udayashankar, A.~Abdel-hadi, R.~Grosman, S.~Al-Kiswany,
  {Falcon: Low Latency, Network-Accelerated Scheduling}, in: {P4 Workshop in
  Europe (EuroP4)}, 2020, p. 7–12.

\bibitem{OsKo20}
T.~Osi\'{n}ski, M.~Kossakowski, M.~Pawlik, J.~Palim\k{a}ka, M.~Sala,
  H.~Tarasiuk, {Unleashing the Performance of Virtual BNG by Offloading Data
  Plane to a Programmable ASIC}, in: {P4 Workshop in Europe (EuroP4)}, 2020, p.
  54–55.

\bibitem{LeMi17}
J.~Lee, R.~Miao, C.~Kim, M.~Yu, H.~Zeng, {Stateful Layer-4 Load Balancing in
  Switching ASICs}, in: {ACM SIGCOMM Conference Posters and Demos}, 2017, p.
  133–135.

\bibitem{rfc8289}
K.~Nichols, V.~Jacobson, A.~McGregor, J.~Iyengar,
  \href{https://tools.ietf.org/rfc/rfc8289.txt}{{Controlled Delay Active Queue
  Management}}, RFC 8289, RFC Editor (01 2018).
\newline\urlprefix\url{https://tools.ietf.org/rfc/rfc8289.txt}

\bibitem{LeFa18}
B.~Lewis, L.~Fawcett, M.~Broadbent, N.~Race, {Using P4 to Enable Scalable
  Intents in Software Defined Networks}, in: {IEEE International Conference on
  Network Protocols (ICNP)}, 2018, pp. 442--443.

\bibitem{p4-sourcerouting}
{GitHub: P4 Source Routing},
  \url{https://github.com/BenRLewis/P4-Source-Routing}, {accessed 01-20-2021}
  (2021).

\bibitem{LuYu19}
L.~Luo, H.~Yu, S.~Luo, Z.~Ye, X.~Du, M.~Guizani, {Scalable Explicit Path
  Control in Software-Defined Networks}, {Journal of Network and Computer
  Applications (JNCA)} 141 (2019) 86--103.

\bibitem{p4-paco}
{GitHub: P4 Paco}, \url{https://github.com/an15m/paco}, {accessed 01-20-2021}
  (2021).

\bibitem{KuSh20}
A.~Kushwaha, S.~Sharma, N.~Bazard, A.~Gumaste, B.~Mukherjee, {Design, Analysis,
  and a Terabit Implementation of a Source-Routing-Based SDN Data Plane}, {IEEE
  Systems Journal} (2020).

\bibitem{AbTu20}
A.~Abdelsalam, A.~Tulumello, M.~Bonola, S.~Salsano, C.~Filsfils, {Pushing
  Network Programmability to the limits with SRv6 uSIDs and P4}, in: {P4
  Workshop in Europe (EuroP4)}, 2020, p. 62–64.

\bibitem{BrHa17}
W.~Braun, J.~Hartmann, M.~Menth, {Demo: Scalable and Reliable Software-Defined
  Multicast with BIER and P4}, in: {IFIP/IEEE Symposium on Integrated
  Management (IM)}, 2017, pp. 905--906.

\bibitem{p4-bier-demo}
{Bitbucket: p4-bfr)}, \url{https://bitbucket.org/wb-ut/p4-bfr}, {accessed
  01-20-2021} (2021).

\bibitem{MeLi20}
D.~Merling, S.~Lindner, M.~Menth, {P4-Based Implementation of BIER and BIER-FRR
  for Scalable and Resilient Multicast}, {Journal of Network and Computer
  Applications (JNCA)} 169 (2020) 102764.

\bibitem{MeLi21}
D.~Merling, S.~Lindner, M.~Menth, Hardware-based evaluation of scalable and
  resilient multicast with bier in p4, {IEEE} ACCESS 9 (2021) 34500--34514.

\bibitem{p4-bier-repo}
{GitHub: P4-BIER}, \url{https://github.com/uni-tue-kn/p4-bier}, {accessed
  01-20-2021} (2021).

\bibitem{p4-bier-tofino-repo}
{GitHub: P4-BIER for Tofino},
  \url{https://github.com/uni-tue-kn/p4-bier-tofino}, {accessed 04-26-2021}
  (2021).

\bibitem{ShSu19}
M.~Shahbaz, L.~Suresh, J.~Rexford, N.~Feamster, O.~Rottenstreich, M.~Hira,
  {Elmo: Source Routed Multicast for Public Clouds}, in: {ACM Special Interest
  Group on Data Communication}, 2019, pp. 2587--2600.

\bibitem{p4-elmo}
{GitHub: Elmo MCast}, \url{https://github.com/Elmo-MCast/p4-programs},
  {accessed 01-20-2021} (2021).

\bibitem{LuYu20}
S.~Luo, H.~Yu, K.~Li, H.~Xing, {Efﬁcient File Dissemination in Data Center
  Networks with Priority-based Adaptive Multicast}, {IEEE} Journal on Selected
  Areas in Communications (JSAC) 38 (2020) 1161--1175.

\bibitem{WePa18}
C.~Wernecke, H.~Parzyjegla, G.~Mühl, P.~Danielis, D.~Timmermann, {Realizing
  Content-Based Publish/Subscribe with P4}, in: {IEEE Conference on Network
  Function Virtualization and Software-Defined Networking (NFV-SDN)}, 2018, pp.
  1--7.

\bibitem{WePa19}
C.~Wernecke, H.~Parzyjegla, G.~Mühl, E.~Schweissguth, D.~Timmermann, {Flexible
  Notiﬁcation Forwarding for Content-Based Publish/Subscribe Using P4}, in:
  {IEEE Conference on Network Function Virtualization and Software-Defined
  Networking (NFV-SDN)}, 2020, pp. 1--5.

\bibitem{WePa20a}
C.~Wernecke, H.~Parzyjegla, G.~Mühl, {Implementing Content-based
  Publish/Subscribe on the Network Layer with P4}, in: {IEEE Conference on
  Network Function Virtualization and Software-Defined Networking (NFV-SDN)},
  2020, pp. 144--149.

\bibitem{WePa20b}
C.~Wernecke, H.~Parzyjegla, G.~Mühl, P.~Danielis, E.~Schweissguth,
  D.~Timmermann, {Stitching Notification Distribution Trees for Content-based
  Publish/Subscribe with P4}, in: {IEEE Conference on Network Function
  Virtualization and Software-Defined Networking (NFV-SDN)}, 2020, pp.
  100--104.

\bibitem{JeMo18b}
T.~Jepsen, M.~Moshref, A.~Carzaniga, N.~Foster, R.~Soul{\'e}, {Packet
  Subscriptions for Programmable ASICs}, in: {ACM Workshop on Hot Topics in
  Networks (HotNets)}, 2018, p. 176–183.

\bibitem{KuGa20}
R.~Kundel, C.~Gaertner, M.~Luthra, S.~Bhowmik, B.~Koldehofe, {Flexible
  Content-based Publish/Subscribe over Programmable Data Planes}, in:
  {IEEE/IFIP Network Operations and Management Symposium (NOMS)}, 2020, pp.
  1--5.

\bibitem{p4-pubsub}
{GitHub: p4bsub}, \url{https://github.com/ralfkundel/p4bsub/}, {accessed
  01-20-2021} (2021).

\bibitem{VeKa20}
J.~Vestin, A.~Kassler, S.~Laki, G.~Pongrácz, {Towards In-Network Event
  Detection and Filtering for Publish/Subscribe Communication using
  Programmable Data Planes}, {{IEEE} Transactions on Network and Service
  Management (TNSM)} (2020) 415--428.

\bibitem{SiSt16}
S.~Signorello, R.~State, J.~François, O.~Festor, {NDN.p4: Programming
  Information-Centric Data-Planes}, in: {IEEE Conference on Network
  Softwarization (NetSoft)}, 2016, pp. 384--389.

\bibitem{MiSi18}
R.~Miguel, S.~Signorello, F.~M.~V. Ramos, {Named Data Networking with
  Programmable Switches}, in: {IEEE International Conference on Network
  Protocols (ICNP)}, 2018, pp. 400--405.

\bibitem{ndnp4-github}
{GitHub: NDN.p4}, \url{https://github.com/signorello/NDN.p4}, {accessed
  01-20-2021} (2021).

\bibitem{p4-netx-ndn}
{GitHub: NDN.p4-16}, \url{https://github.com/netx-ulx/NDN.p4-16}, {accessed
  01-20-2021} (2021).

\bibitem{KaSa20}
O.~Karrakchou, N.~Samaan, A.~Karmouch, {ENDN: An Enhanced NDN Architecture with
  a P4-programmable Data Plane}, in: {International Conference on Networking
  (ICN)}, 2020, p. 1–11.

\bibitem{SeBo18}
R.~Sedar, M.~Borokhovich, M.~Chiesa, G.~Antichi, S.~Schmid, {Supporting
  Emerging Applications With Low-Latency Failover in P4}, in: {Workshop on
  Networking for Emerging Applications and Technologies (NEAT)}, 2018, p.
  52–57.

\bibitem{p4-frr}
{GitHub: P4-FRR}, \url{https://bitbucket.org/roshanms/p4-frr/src/master/},
  {accessed 01-20-2021} (2021).

\bibitem{GiSh18}
H.~Giesen, L.~Shi, J.~Sonchack, A.~Chelluri, N.~Prabhu, N.~Sultana, L.~Kant,
  A.~J. McAuley, A.~Poylisher, A.~DeHon, B.~T. Loo, {In-Network Computing to
  the Rescue of Faulty Links}, in: {Morning Workshop on In-Network Computing},
  2018, pp. 1--6.

\bibitem{QuJo19}
T.~Qu, R.~Joshi, M.~Chan, B.~Leong, D.~Guo, Z.~Liu, {SQR: In-network Packet
  Loss Recovery from Link Failures for Highly Reliable Datacenter Networks},
  in: {IEEE International Conference on Network Protocols (ICNP)}, 2019, pp.
  1--12.

\bibitem{p4-sqr}
{GitHub: P4 SQR}, \url{https://git.io/fjbnV}, {accessed 01-20-2021} (2021).

\bibitem{LiMe20}
S.~Lindner, D.~Merling, M.~Häberle, M.~Menth, {P4-Protect: 1+1 Path Protection
  for P4}, in: {P4 Workshop in Europe (EuroP4)}, 2020, p. 21–27.

\bibitem{p4-protect-bmv2}
{GitHub: P4-Protect BMv2}, \url{https://github.com/uni-tue-kn/p4-protect},
  {accessed 01-20-2021} (2021).

\bibitem{p4-protect-tofino}
{GitHub: P4-Protect Tofino},
  \url{https://github.com/uni-tue-kn/p4-protect-tofino}, {accessed 01-20-2021}
  (2021).

\bibitem{HiTa19}
K.~Hirata, , T.~Tachibana, {Implementation of Multiple Routing Conﬁgurations
  on Software-Deﬁned Networks with P4}, in: {Asia-Pacific Signal and
  Information Processing Association Annual Summit and Conference (APSIPA
  ASC)}, 2019, pp. 13--16.

\bibitem{LiHa20}
S.~Lindner, M.~Häberle, F.~Heimgaertner, N.~Nayak, S.~Schildt, D.~Grewe,
  H.Loehr, M.~Ment, {P4 In-Network Source Protection for Sensor Failover}, in:
  {IFIP-TC6 Networking Conference (Networking)}, 2020, pp. 791--796.

\bibitem{p4-source-protection-bmv2}
{GitHub: P4 Source Protection BMv2},
  \url{https://github.com/uni-tue-kn/p4-source-protection}, {accessed
  01-20-2021} (2021).

\bibitem{p4-source-protection-tofino}
{GitHub: P4 Source Protection Tofino},
  \url{https://github.com/uni-tue-kn/p4-source-protection-tofino}, {accessed
  01-20-2021} (2021).

\bibitem{SuAb19}
K.~Subramanian, A.~Abhashkumar, L.~D'Antoni, A.~Akella, {D2R: Dataplane-Only
  Policy-Compliant Routing Under Failures} (2019).

\bibitem{ChSe19}
M.~Chiesa, R.~Sedar, G.~Antichi, M.~Borokhovich, A.~Kamisiński, G.~Nikolaidis,
  S.~Schmid, {PURR: A Primitive for Reconfigurable Fast Reroute}, in: {ACM
  Conference on emerging Networking EXperiments and Technologies (CoNEXT)},
  2019, p. 1–14.

\bibitem{HoCo19}
T.~Holterbach, E.~C. Molero, M.~Apostolaki, A.~Dainotti, S.~Vissicchio,
  L.~Vanbever, {Blink: Fast Connectivity Recovery Entirely in the Data Plane},
  in: {USENIX Symposium on Networked Systems Design \& Implementation (NSDI)},
  2019, pp. 161--176.

\bibitem{github-blink}
{GitHub: Blink}, \url{https://github.com/nsg-ethz/Blink}, {accessed 01-20-2021}
  (2021).

\bibitem{HsBe19}
K.-F. Hsu, R.~Beckett, A.~Chen, J.~Rexford, D.~Walker, {Contra: A Programmable
  System for Performance-aware Routing}, in: {USENIX Symposium on Networked
  Systems Design \& Implementation (NSDI)}, 2020, pp. 701--721.

\bibitem{MiKe16}
O.~Michel, E.~Keller, {Policy Routing using Process-Level Identifiers}, in:
  {IEEE International Conference on Cloud Engineering Workshop (IC2EW)}, 2016,
  pp. 7--12.

\bibitem{BaOz18}
A.~C. Baktir, A.~Ozgovde, C.~Ersoy, {Implementing Service-Centric Model with
  P4: A Fully-Programmable Approach}, in: {IEEE/IFIP Network Operations and
  Management Symposium (NOMS)}, 2018, pp. 1--6.

\bibitem{FrSa20}
W.~Froes, L.~Santos, L.~N. Sampaio, M.~Martinello, A.~Liberato, R.~S. Villaca,
  {ProgLab: Programmable Labels for QoS Provisioning on Software Defined
  Networks}, {Computer Communications} 161 (2020) 99--108.

\bibitem{VaZh20}
N.~VARYANI, Z.-L. ZHANG, D.~DAI, {QROUTE: An Efﬁcient Quality of Service
  (QoS) Routing Scheme for Software-Deﬁned Overlay Networks}, {IEEE} ACCESS 8
  (2020) 104109--104126.

\bibitem{GiGr20}
S.~Gimenez, E.~Grasa, S.~Bunch, {A Proof of Concept Implementation of a RINA
  Interior Router using P4-enabled Software Targets}, in: {Conference on
  Innovation in Clouds, Internet and Networks and Workshops (ICIN)}, 2020, pp.
  57--62.

\bibitem{FeTa19}
W.~Feng, X.~Tan, Y.~Jin, {Implementing ICN over P4 in HTTP Scenario}, in: {IEEE
  International Conference on Hot Information-Centric Networking (HotICN)},
  2019, pp. 37--43.

\bibitem{GrLi20}
G.~Grigoryan, Y.~Liu, M.~Kwon, {PFCA: A Programmable FIB Caching Architecture},
  IEEE/ACM Transactions on Networking (ToN) 28 (2020) 1872--1884.

\bibitem{McGo19}
A.~McAuley, Y.~M. Gottlieb, L.~Kant, J.~Lee, A.~Poylisher, {P4-Based Hybrid
  Error Control Booster Providing New Design Tradeoffs in Wireless Networks},
  in: {IEEE Military Communications Conference (MILCOM)}, 2019, pp. 731--736.

\bibitem{KoPr19}
M.~Kogias, G.~Prekas, A.~Ghosn, J.~Fietz, E.~Bugnion, {R2P2: Making RPCs
  first-class datacenter citizens}, in: USENIX Annual Technical Conference
  (ATC), 2019, pp. 863--880.

\bibitem{github-r2p2}
{GitHub: R2P2 - Request Response Pair Protocol},
  \url{https://github.com/epfl-dcsl/r2p2}, {accessed 01-25-2021} (2021).

\bibitem{MeMe18}
D.~Merling, M.~Menth, N.~Warnke, T.~Eckert, {An Overview of Bit Index Explicit
  Replication (BIER)}, {IETF Journal} (2018).

\bibitem{HoLe20}
M.~Hollingsworth, J.~Lee, Z.~Liu, J.~Lee, S.~Ha, D.~Grunwald, {P4EC: Enabling
  Terabit Edge Computing in Enterprise 4G LTE}, in: {USENIX Workshop on Hot
  Topics in Edge Computing (HotEdge)}, 2020, pp. 1--7.

\bibitem{spgw-github}
{GitHub: spgw.p4},
  \url{https://github.com/opennetworkinglab/onos/blob/master/pipelines/fabric/impl/src/main/resources/include/control/spgw.p4},
  {accessed 01-20-2021} (2021).

\bibitem{PaSi18}
P.~Palagummi, K.~M. Sivalingam, {SMARTHO: A Network Initiated Handover in
  NG-RAN using P4-based Switches}, in: {International Conference on Network and
  Services Management (CNSM)}, 2018, pp. 338--342.

\bibitem{AgHu18}
A.~Aghdai, M.~Huang, D.~Dai, Y.~Xu, J.~Chao, {Transparent Edge Gateway for
  Mobile Networks}, in: {IEEE International Conference on Network Protocols
  (ICNP)}, 2018, pp. 412--417.

\bibitem{AgXu19}
A.~Aghdai, Y.~Xu, M.~Huang, D.~H. Dai, H.~J. Chao, {Enabling Mobility in
  LTE-Compatible Mobile-edge Computing with Programmable Switches}, {ArXiv}
  e-prints (2019).

\bibitem{XiQi19}
J.~Xie, C.~Qian, D.~Guo, X.~Li, S.~Shi, H.~Chen, {Efficient Data Placement and
  Retrieval Services in Edge Computing}, in: {IEEE International Conference on
  Distributed Computing Systems (ICDCS)}, 2019, pp. 1029--1039.

\bibitem{XiGu20}
J.~Xie, D.~Guo, X.~Shi, H.~Cai, C.~Qian, H.~Chen, {A Fast Hybrid Data Sharing
  Framework for Hierarchical Mobile Edge Computing}, in: {IEEE International
  Conference on Computer Communications (INFOCOM)}, 2020, pp. 2609--2618.

\bibitem{ShLe19}
C.~Shen, D.~Lee, C.~Ku, M.~Lin, K.~Lu, S.~Tan, {A Programmable and
  FPGA-accelerated GTP Offloading Engine for Mobile Edge Computing in 5G
  Networks}, in: {IEEE Conference on Computer Communications Workshops (INFOCOM
  WKSHPS)}, 2019, pp. 1021--1022.

\bibitem{LeEb19}
C.~Lee, K.~Ebisawa, H.~Kuwata, M.~Kohno, S.~Matsushima, {Performance Evaluation
  of GTP-U and SRv6 Stateless Translation}, in: {International Conference on
  Network and Services Management (CNSM)}, 2019, pp. 1--6.

\bibitem{RiMa19}
R.~Ricart-Sanchez, P.~Malagon, J.~M. Alcaraz-Calero, Q.~Wang, {P4-NetFPGA-based
  network slicing solution for 5G MEC architectures}, in: {ACM/IEEE Symposium
  on Architectures for Networking and Communications Systems (ANCS)}, 2019, pp.
  1--2.

\bibitem{SiRo19}
S.~K. Singh, C.~E. Rothenberg, G.~Patra, G.~Pongracz, {Offloading Virtual
  Evolved Packet Gateway User Plane Functions to a Programmable ASIC}, in: {ACM
  CoNEXT Workshop on Emerging In-Network Computing Paradigms}, 2019, p. 9–14.

\bibitem{ShKu20}
R.~Shah, V.~Kumar, M.~Vutukuru, P.~Kulkarni, {TurboEPC: Leveraging Dataplane
  Programmability to Accelerate the Mobile Packet Core}, in: {ACM Symposium on
  SDN Research (SOSR)}, 2020, p. 83–95.

\bibitem{VoPo20}
P.~V\"{o}r\"{o}s, G.~Pongr\'{a}cz, S.~Laki, {Towards a Hybrid Next Generation
  NodeB}, in: {P4 Workshop in Europe (EuroP4)}, 2020, p. 56–58.

\bibitem{LiHu19a}
Y.~Lin, T.~Huang, S.~Tsai, {Enhancing 5G/IoT Transport Security Through Content
  Permutation}, {IEEE} ACCESS 7 (2019) 94293--94299.

\bibitem{UdMu17}
M.~Uddin, S.~Mukherjee, H.~Chang, T.~V. Lakshman, {SDN-Based Service Automation
  for IoT}, in: {IEEE International Conference on Network Protocols (ICNP)},
  2017, pp. 1--10.

\bibitem{UdMu18}
M.~Uddin, S.~Mukherjee, H.~Chang, T.~V. Lakshman, {SDN-Based Multi-Protocol
  Edge Switching for IoT Service Automation}, {IEEE} Journal on Selected Areas
  in Communications (JSAC) 36 (2018) 2775--2786.

\bibitem{WaWu19}
S.-Y. Wang, C.-M. Wu, Y.-B. Linm, C.-C. Huang, {High-Speed Data-Plane Packet
  Aggregation and Disaggregation by P4 Switches}, {Journal of Network and
  Computer Applications (JNCA)} 142 (2019) 98--110.

\bibitem{MaAr20}
A.~L.~R. Madureira, F.~R.~C. Araújo, L.~N. Sampaio, {On supporting IoT data
  aggregation through programmable data planes}, Computer Networks 177 (2020)
  107330.

\bibitem{EnZa19}
P.~Engelhard, A.~Zachlod, J.~Schulz-Zander, S.~Du, {Toward scalable and
  virtualized massive wireless sensor networks}, in: {International Conference
  on Networked Systems (NetSys)}, 2019, pp. 1--6.

\bibitem{VeKa18}
J.~Vestin, A.~Kassler, J.~{\AA}kerberg, {FastReact: In-Network Control and
  Caching for Industrial Control Networks using Programmable Data Planes}, in:
  {IEEE International Conference on Emerging Technologies and Factory
  Automation (ETFA)}, 2018, pp. 219--226.

\bibitem{CeCs20}
F.~E.~R. Cesen, L.~Csikor, C.~Recalde, C.~E. Rothenberg, G.~Pongrácz, {Towards
  Low Latency Industrial Robot Control in Programmable Data Planes}, in: {IEEE
  Conference on Network Softwarization (NetSoft)}, 2020, pp. 165--169.

\bibitem{KuGl21}
I.~Kunze, R.~Glebke, J.~Scheiper, M.~Bodenbenner, R.~H. Schmitt, K.~Wehrle,
  {Investigating the Applicability of In-Network Computing to Industrial
  Scenarios}, in: {International Conference on Industrial Cyber-Physical
  Systems (ICPS)}, 2021, pp. 334--340.

\bibitem{RuGl18}
J.~R\"{u}th, R.~Glebke, K.~Wehrle, V.~Causevic, S.~Hirche, {Towards In-Network
  Industrial Feedback Control}, in: {Morning Workshop on In-Network Computing},
  2018, p. 14–19.

\bibitem{KaJo19}
P.~G. Kannan, R.~Joshi, M.~C. Chan, {Precise Time-Synchronization in the
  Data-Plane using Programmable Switching ASICs}, in: {ACM Symposium on SDN
  Research (SOSR)}, 2019, p. 8–20.

\bibitem{KuSi19}
R.~Kundel, F.~Siegmund, B.~Koldehofe, {How to Measure the Speed of Light with
  Programmable Data Plane Hardware?}, in: {P4 Workshop in Europe (EuroP4)},
  2019, pp. 1--2.

\bibitem{BoIo18}
G.~Bonofiglio, V.~Iovinella, G.~Lospoto, G.~D. Battista, {Kathará: A
  Container-Based Framework for Implementing Network Function Virtualization
  and Software Defined Networks}, in: {IEEE/IFIP Network Operations and
  Management Symposium (NOMS)}, 2018, pp. 1--9.

\bibitem{HeBa18}
M.~He, A.~Basta, A.~Blenk, N.~Deric, W.~Kellerer, {P4NFV: An NFV Architecture
  with Flexible Data Plane Reconfiguration}, in: {International Conference on
  Network and Services Management (CNSM)}, 2018, pp. 90--98.

\bibitem{OsTa19}
T.~Osiński, H.~Tarasiuk, M.~Kossakowski, R.~Picard, {Offloading Data Plane
  Functions to the Multi-Tenant Cloud Infrastructure using P4}, in: {P4
  Workshop in Europe (EuroP4)}, 2019, pp. 1--6.

\bibitem{MoVe20}
D.~Moro, G.~Verticale, A.~Capone, {A Framework for Network Function
  Decomposition and Deployment}, in: {International Workshop on the Design of
  Reliable Communication Networks (DRCN)}, 2020, pp. 1--6.

\bibitem{OsTa19b}
T.~Osiński, H.~Tarasiuk, L.~Rajewski, E.~Kowalczyk, {DPPx: A P4-based Data
  Plane Programmability and Exposure framework to enhance NFV services}, in:
  {IEEE Conference on Network Softwarization (NetSoft)}, 2019, pp. 296--300.

\bibitem{MoPa19}
A.~Mohammadkhan, S.~Panda, S.~G. Kulkarni, K.~K. Ramakrishnan, L.~N. Bhuyan,
  {P4NFV: P4 Enabled NFV Systems with SmartNICs}, in: {IEEE Conference on
  Network Function Virtualization and Software-Defined Networking (NFV-SDN)},
  2019, pp. 1--7.

\bibitem{MoPe19a}
D.~Moro, M.~Peuster, H.~Karl, A.~Capone, {FOP4: Function Offloading Prototyping
  in Heterogeneous and Programmable Network Scenarios}, in: {IEEE Conference on
  Network Function Virtualization and Software-Defined Networking (NFV-SDN)},
  2019, pp. 1--6.

\bibitem{MoPe19b}
D.~Moro, M.~Peuster, H.~Karl, A.~Capone, {Demonstrating FOP4: A Flexible
  Platform to Prototype NFV Offloading Scenarios}, in: {IEEE Conference on
  Network Function Virtualization and Software-Defined Networking (NFV-SDN)},
  2019, pp. 1--2.

\bibitem{MaDo20}
D.~R. {Mafioletti}, C.~K. {Dominicini}, M.~{Martinello}, M.~R.~N. {Ribeiro},
  R.~d.~S.~{Villaça}, Piaffe: A place-as-you-go in-network framework for
  flexible embedding of vnfs, in: {IEEE International Conference on
  Communicaotions (ICC)}, 2020, pp. 1--6.

\bibitem{ChZh19}
X.~Chen, D.~Zhang, X.~Wang, K.~Zhu, H.~Zhou, {P4SC: Towards High-Performance
  Service Function Chain Implementation on the P4-Capable Device}, in:
  {IFIP/IEEE Symposium on Integrated Management (IM)}, 2019, pp. 1--9.

\bibitem{ZhCh19}
D.~Zhang, X.~Chen, Q.~Huang, X.~Hong, C.~Wu, H.~Zhou, Y.~Yang, H.~Liu, Y.~Chen,
  {P4SC: A High Performance and Flexible Framework for Service Function Chain},
  {IEEE} ACCESS 7 (2019) 160982--160997.

\bibitem{p4sc-github}
{GitHub: P4SC}, \url{https://github.com/P4SC/p4sc}, {accessed 01-20-2021}
  (2021).

\bibitem{LeLe19}
H.~Lee, J.~Lee, H.~Ko, S.~Pack, {Resource-Efficient Service Function Chaining
  in Programmable Data Plane}, in: {P4 Workshop in Europe (EuroP4)}, 2019.

\bibitem{ZhBi20}
Y.~Zhou, J.~Bi, C.~Zhang, M.~Xu, J.~Wu, {FlexMesh: Flexibly Chaining Network
  Functions on Programmable Data Planes at Runtime}, in: {IFIP-TC6 Networking
  Conference (Networking)}, 2020, pp. 73--81.

\bibitem{StHi20}
A.~Stockmayer, S.~Hinselmann, M.~H{\"a}berle, M.~Menth, {Service Function
  Chaining Based on Segment Routing Using P4 and SR-IOV (P4-SFC)}, in:
  {Workshop on Virtualization in High-Performance Cloud Computing (VHPC)},
  2020, pp. 297--309.

\bibitem{StHi20-repo}
{GitHub: P4-SFC}, \url{https://github.com/uni-tue-kn/p4-sfc-faas}, {accessed
  01-20-2021} (2021).

\bibitem{RiMa18}
R.~Ricart-Sanchez, P.~Malagon, J.~M. Alcaraz-Calero, Q.~Wang,
  {Hardware-Accelerated Firewall for 5G Mobile Networks}, in: {IEEE
  International Conference on Network Protocols (ICNP)}, 2018, pp. 446--447.

\bibitem{RiMa19a}
{Ruben Ricart-Sanchez and Pedro Malagon and Jose M. Alcaraz-Calero and Qi
  Wang}, {NetFPGA-Based Firewall Solution for 5G Multi-Tenant Architectures},
  in: {IEEE International Conference on Edge Computing (EDGE)}, 2019, pp.
  132--136.

\bibitem{CaBi18}
J.~Cao, J.~Bi, Y.~Zhou, C.~Zhang, {CoFilter: A High-Performance Switch-Assisted
  Stateful Packet Filter}, in: {ACM SIGCOMM Conference Posters and Demos},
  2018, p. 9–11.

\bibitem{DaCh18}
R.~Datta, S.~Choi, A.~Chowdhary, Y.~Park, {P4Guard: Designing P4 Based
  Firewall}, in: {IEEE Military Communications Conference (MILCOM)}, 2018, pp.
  1--6.

\bibitem{VoKi16}
P.~Vörös, A.~Kiss, {Security Middleware Programming Using P4}, in:
  {International Conference on Human Aspects of Information Security, Privacy,
  and Trust (HAS)}, 2016, pp. 277--287.

\bibitem{ZaFr20}
E.~O. Zaballa, D.~Franco, Z.~Zhou, M.~S. Berger, {P4Knocking: Offloading
  host-based firewall functionalities to the network}, in: {Conference on
  Innovation in Clouds, Internet and Networks and Workshops (ICIN)}, 2020, pp.
  7--12.

\bibitem{AlAl19}
A.~Almaini, A.~Al-Dubai, I.~Romdhani, M.~Schramm, {Delegation of Authentication
  to the Data Plane in Software-Defined Networks}, in: {IEEE International
  Conferences on Smart Computing, Networking and Services (SmartCNS)}, 2019,
  pp. 58--65.

\bibitem{GrLi18}
G.~Grigoryan, Y.~Liu, {LAMP: Prompt Layer 7 Attack Mitigation with Programmable
  Data Planes}, in: {ACM/IEEE Symposium on Architectures for Networking and
  Communications Systems (ANCS)}, 2018, pp. 1--4.

\bibitem{FeXi18}
A.~Febro, H.~Xiao, J.~Spring, {Telephony Denial of Service Defense at Data
  Plane (TDoSD@DP)}, in: {IEEE/IFIP Network Operations and Management Symposium
  (NOMS)}, 2018, pp. 1--6.

\bibitem{FeXi19}
A.~Febro, H.~Xiao, J.~Spring, {Distributed SIP DDoS Defense with P4}, in: {IEEE
  Wireless Communications and Networking Conference (WCNC)}, 2019, pp. 1--8.

\bibitem{KuVo19}
M.~Kuka, K.~Vojanec, J.~Kučera, P.~Benáček, {Accelerated DDoS Attacks
  Mitigation using Programmable Data Plane}, in: {ACM/IEEE Symposium on
  Architectures for Networking and Communications Systems (ANCS)}, 2019, pp.
  1--3.

\bibitem{PaCu18}
F.~Paolucci, F.~Cugini, P.~Castoldi, {P4-based Multi-Layer Traffic Engineering
  Encompassing Cyber Security}, in: {Optical Fiber Communication Conference
  (OFC)}, 2018, pp. 1--3.

\bibitem{PaCi19}
F.~Paolucci, F.~Civerchia, A.~Sgambelluri, A.~Giorgetti, F.~Cugini,
  P.~Castoldi, {An efficient pipeline processing scheme for programming
  Protocol-independent Packet Processors}, {IEEE/OSA Journal of Optical
  Communications and Networking} 11 (2019) 88--95.

\bibitem{MiWa19}
Y.~Mi, A.~Wang, {ML-Pushback: Machine Learning Based Pushback Defense Against
  DDoS}, in: {ACM Conference on emerging Networking EXperiments and
  Technologies (CoNEXT)}, 2019, p. 80–81.

\bibitem{AfBr17}
Y.~Afek, A.~Bremler-Barr, L.~Shafir, {Network Anti-Spoofing with SDN Data
  Plane}, in: {IEEE International Conference on Computer Communications
  (INFOCOM)}, 2017, pp. 1--9.

\bibitem{LaMa19}
A.~C. Lapolli, J.~A. Marques, L.~P. Gaspary, {Offloading Real-time DDoS Attack
  Detection to Programmable Data Planes}, in: {IFIP/IEEE Symposium on
  Integrated Management (IM)}, 2019, pp. 19--27.

\bibitem{p4-ddosd}
{GitHub: ddosd-p4}, \url{https://github.com/aclapolli/ddosd-p4}, {accessed
  01-20-2021} (2021).

\bibitem{CaLa20}
Y.-Z. Cai, C.-H. Lai, Y.-T. Wang, M.-H. Tsai, {Improving Scanner Data
  Collection in P4-based SDN}, in: {Asia-Pacific Network Operations and
  Management Symposium (APNOMS)}, 2020, pp. 126--131.

\bibitem{LiWu20}
T.-Y. Lin, J.-P. Wu, P.-H. Hung, C.-H. Shao, Y.-T. Wang, Y.-Z. Cai, M.-H. Tsai,
  {Mitigating SYN flooding Attack and ARP Spoofing in SDN Data Plane}, in:
  {Asia-Pacific Network Operations and Management Symposium (APNOMS)}, 2020,
  pp. 114--119.

\bibitem{MuIo20}
F.~Musumeci, V.~Ionata, F.~Paolucci, M.~Cugini, Filippo~Tornatore,
  {Machine-learning-assisted DDoS attack detection with P4 language}, in: {IEEE
  International Conference on Communicaotions (ICC)}, 2020, pp. 1--6.

\bibitem{KhCs20}
X.~Z. Khooi, L.~Csikor, D.~M. Divakaran, M.~S. Kang, {DIDA: Distributed
  In-Network Defense Architecture Against Amplified Reflection DDoS Attacks},
  in: {IEEE Conference on Network Softwarization (NetSoft)}, 2020, pp.
  277--281.

\bibitem{DiPa20}
M.~Dimolianis, A.~Pavlidis, V.~Maglaris, {A Multi-Feature DDoS Detection Schema
  on P4 Network Hardware}, in: {Workshop on Flexible Network Data Plane
  Processing (NETPROC@ICIN)}, 2020, pp. 1--6.

\bibitem{ScGa20}
D.~Scholz, S.~Gallenm\"{u}ller, H.~Stubbe, G.~Carle, {SYN Flood Defense in
  Programmable Data Planes}, in: {P4 Workshop in Europe (EuroP4)}, 2020, p.
  13–20.

\bibitem{syn-proxy}
{GitHub: syn-proxy}, \url{https://github.com/syn-proxy}, {accessed 01-20-2021}
  (2021).

\bibitem{FrKf20}
K.~Friday, E.~Kfoury, E.~Bou-Harb, J.~Crichigno, {Towards a Unified In-Network
  DDoS Detection and Mitigation Strategy}, in: {IEEE Conference on Network
  Softwarization (NetSoft)}, 2020, pp. 218--226.

\bibitem{MeTs18}
R.~Meier, P.~Tsankov, V.~Lenders, L.~Vanbever, M.~Vechev, {NetHide: Secure and
  Practical Network Topology Obfuscation}, in: USENIX Security Symposium, 2018,
  pp. 693--709.

\bibitem{LeBr19}
{Benjamin Lewis and Matthew Broadbent and Nicholas Race}, {P4ID: P4 Enhanced
  Intrusion Detection}, in: {IEEE Conference on Network Function Virtualization
  and Software-Defined Networking (NFV-SDN)}, 2019, pp. 1--4.

\bibitem{KaSa18}
{Gorby Kabasele Ndonda and Ramin Sadre}, {A Two-level Intrusion Detection
  System for Industrial Control System Networks using P4}, in: {International
  Symposium for ICS \& SCADA Cyber Security Research (ICS-CSR)}, 2018, pp.
  1--10.

\bibitem{HySo20}
J.~Hypolite, J.~Sonchack, S.~Hershkop, N.~Dautenhahn, A.~DeHon, J.~M. Smith,
  {DeepMatch: Practical Deep Packet Inspection in the Data Plane Using Network
  Processors}, in: {ACM Conference on emerging Networking EXperiments and
  Technologies (CoNEXT)}, 2020, p. 336–350.

\bibitem{deepmatch}
{GitHub: DeepMatch}, \url{https://github.com/jhypolite/DeepMatch}, {accessed
  01-20-2021} (2021).

\bibitem{QiPo20}
Q.~Qin, K.~Poularakis, K.~K. Leung, L.~Tassiulas, {Line-Speed and Scalable
  Intrusion Detection at the Network Edge via Federated Learning}, in:
  {IFIP-TC6 Networking Conference (Networking)}, 2020, pp. 352--360.

\bibitem{github-qin}
{GitHub: syn-proxy}, \url{https://github.com/vxxx03/IFIPNetworking20},
  {accessed 01-20-2021} (2021).

\bibitem{AmSi20}
J.~Amado, S.~Signorello, M.~Correia, F.~Ramos, Poster: Speeding up network
  intrusion detection, in: {IEEE International Conference on Network Protocols
  (ICNP)}, 2020, pp. 1--2.

\bibitem{ChSu19}
D.~Chang, W.~Sun, Y.~Yang, {A SDN Proactive Defense Mechanism Based on IP
  Transformation}, in: {International Conference on Safety Produce
  Informatization (IICSPI)}, 2019, pp. 248--251.

\bibitem{FeZh19}
W.~Feng, Z.-L. Zhang, C.~Liu, J.~Chen, {Clé: Enhancing Security with
  Programmable Dataplane Enabled Hybrid SDN}, in: {ACM Conference on emerging
  Networking EXperiments and Technologies (CoNEXT)}, 2019, p. 76–77.

\bibitem{KuLi20}
P.~Kuang, Y.~Liu, L.~He, {P4DAD: Securing Duplicate Address Detection Using
  P4}, in: {IEEE International Conference on Communicaotions (ICC)}, 2020, pp.
  1--7.

\bibitem{Chen20}
X.~Chen, Implementing aes encryption on programmable switches via scrambled
  lookup tables, in: {Workshop on Secure Programmable Network Infrastructure
  (SPIN)}, 2020, p. 8–14.

\bibitem{aes-tofino-repo}
{GitHub: Tofino AES encryption},
  \url{https://github.com/Princeton-Cabernet/p4-projects/tree/master/AES-tofino},
  {accessed 01-20-2021} (2021).

\bibitem{GoSa20}
H.~Gondaliya, G.~C. Sankaran, K.~M. Sivalingam, {Comparative Evaluation of IP
  Address Anti-Spoofing Mechanisms Using a P4/NetFPGA-Based Switch}, in: {P4
  Workshop in Europe (EuroP4)}, 2020, p. 1–6.

\bibitem{KaXu20}
Q.~Kang, L.~Xue, A.~Morrison, Y.~Tang, A.~Chen, X.~Luo, {Programmable
  In-Network Security for Context-aware {BYOD} Policies}, in: USENIX Security
  Symposium, 2020, pp. 595--612.

\bibitem{github-poise}
{GitHub: Poise}, \url{https://github.com/qiaokang92/poise}, {accessed
  01-20-2021} (2021).

\bibitem{HaSc20}
F.~Hauser, M.~Schmidt, M.~Häberle, M.~Menth, {P4-MACsec: Dynamic Topology
  Monitoring and Data Layer Protection With MACsec in P4-Based SDN}, {IEEE}
  ACCESS 8 (2020) 58845--58858.

\bibitem{p4-macsec}
{GitHub: P4-MACsec}, \url{https://github.com/uni-tue-kn/p4-macsec}, {accessed
  01-20-2021} (2021).

\bibitem{HaHa20}
F.~Hauser, M.~Häberle, M.~Schmidt, M.~Menth, {P4-IPsec: Site-to-Site and
  Host-to-Site VPN With IPsec in P4-Based SDN}, {IEEE} ACCESS 8 (2020)
  139567--139586.

\bibitem{p4-ipsec}
{GitHub: P4-IPsec}, \url{https://github.com/uni-tue-kn/p4-ipsec}, {accessed
  01-20-2021} (2021).

\bibitem{DaFe19}
T.~Datta, N.~Feamster, J.~Rexford, L.~Wang, {SPINE: Surveillance Protection in
  the Network Elements}, in: {USENIX Workshop on Free and Open Communications
  on the Internet (FOCI)}, 2019, pp. 1--7.

\bibitem{spine}
{GitHub: SPINE}, \url{https://github.com/SPINE-P4/spine-code}, {accessed
  01-20-2021} (2021).

\bibitem{QiQu20}
Y.~Qin, W.~Quan, F.~Song, L.~Zhang, G.~Liu, M.~Liu, C.~Yu, {Flexible Encryption
  for Reliable Transmission Based on the P4 Programmable Platform}, in:
  {Information Communication Technologies Conference (ICTC)}, 2020, pp.
  147--152.

\bibitem{LiQu20}
G.~Liu, W.~Quan, N.~Cheng, N.~Lu, H.~Zhang, X.~Shen, {P4NIS: Improving network
  immunity against eavesdropping with programmable data planes}, in: {IEEE
  Conference on Computer Communications Workshops (INFOCOM WKSHPS)}, 2020, pp.
  91--96.

\bibitem{p4nis}
{GitHub: P4NIS}, \url{https://github.com/KB00100100/P4NIS}, {accessed
  01-20-2021} (2021).

\bibitem{LiGa19}
M.~Liu, D.~Gao, G.~Liu, J.~He, L.~Jin, C.~Zhou, F.~Yang, Learning based
  adaptive network immune mechanism to defense eavesdropping attacks, {IEEE}
  ACCESS 7 (2019) 182814--182826.

\bibitem{DeHu15}
J.~Deng, H.~Hu, H.~Li, Z.~Pan, K.~Wang, G.~Ahn, J.~Bi, Y.~Park, {VNGuard: An
  NFV/SDN Combination Framework for Provisioning and Managing Virtual
  Firewalls}, in: {IEEE Conference on Network Function Virtualization and
  Software-Defined Networking (NFV-SDN)}, 2015, pp. 107--114.

\bibitem{ZhQu16}
H.~Zhang, W.~Quan, H.-c. Chao, C.~Qiao, {Smart identifier network: A
  collaborative architecture for the future internet}, {Networks Magazine}
  30~(3) (2016) 46--51.

\bibitem{KuBa18}
R.~Kumar, V.~Babu, D.~Nicol, {Network Coding for Critical Infrastructure
  Networks}, in: {IEEE International Conference on Network Protocols (ICNP)},
  2018, pp. 436--437.

\bibitem{p4-KuBa18}
{GitHub: AquaFlow}, \url{https://github.com/gopchandani/AquaFlow}, {accessed
  01-20-2021} (2021).

\bibitem{GoSi19}
D.~Goncalves, S.~Signorello, F.~M.~V. Ramos, M.~Medard, {Random Linear Network
  Coding on Programmable Switches}, in: {ACM/IEEE Symposium on Architectures
  for Networking and Communications Systems (ANCS)}, 2019, pp. 1--6.

\bibitem{KoMa18}
T.~Kohler, R.~Mayer, F.~D\"{u}rr, M.~Maa\ss, S.~Bhowmik, K.~Rothermel, {P4CEP:
  Towards In-Network Complex Event Processing}, in: {Morning Workshop on
  In-Network Computing}, 2018, p. 33–38.

\bibitem{SaAb17}
A.~Sapio, I.~Abdelaziz, M.~Canini, P.~Kalnis, {DAIET: A System for Data
  Aggregation Inside the Network}, in: {ACM Symposium on Cloud Computing
  (SoCC)}, 2017, p.~1.

\bibitem{SaSi20}
G.~C. Sankaran, K.~M. Sivalingam, {Design and Analysis of Fast IP
  Address-Lookup Schemes based on Cooperation among Routers}, in:
  {International Conference on COMmunication Systems and NETworks (COMSNETS)},
  2020, pp. 330--339.

\bibitem{ZhHa17}
Y.~Zhang, B.~Han, Z.-L. Zhang, V.~Gopalakrishnan, {Network-Assisted Raft
  Consensus Algorithm}, in: {ACM SIGCOMM Conference Posters and Demos}, 2017,
  p. 94–96.

\bibitem{DaCa16}
H.~T. Dang, M.~Canini, F.~Pedone, R.~Soul{\'e}, {Paxos Made Switch-y}, {ACM
  SIGCOMM Computer Communications Review (CCR)} 46 (2016) 18–24.

\bibitem{DaBr20}
H.~T. Dang, P.~Bressana, H.~Wang, K.~S. Lee, N.~Zilbermanand, H.~Weatherspoon,
  M.~Canini, F.~Pedone, R.~Soulé, {P4xos: Consensus as a Network Service},
  IEEE/ACM Transactions on Networking (ToN) 28 (2020) 1726--1738.

\bibitem{p4-DaBr20}
{GitHub: P4xos}, \url{https://github.com/P4xos/P4xos}, {accessed 01-20-2021}
  (2021).

\bibitem{SaDe19}
E.~Sakic, N.~Deric, E.~Goshi, W.~Kellerer, {P4BFT: Hardware-Accelerated
  Byzantine-Resilient Network Control Plane}, in: {IEEE Global Communications
  Conference (GLOBECOM)}, 2019, pp. 1--7.

\bibitem{SaDe19b}
E.~Sakic, N.~Deric, C.~B. Serna, E.~Goshi, W.~Kellerer, {P4BFT: A Demonstration
  of Hardware-Accelerated BFT in Fault-Tolerant Network Control Plane}, in:
  {ACM SIGCOMM Conference Posters and Demos}, 2019, p. 6–8.

\bibitem{ZePo20}
L.~Zeno, D.~R.~K. Ports, J.~Nelson, M.~Silberstein, {SwiShmem: Distributed
  Shared State Abstractions for Programmable Switches}, in: {ACM Workshop on
  Hot Topics in Networks (HotNets)}, 2020, p. 160–167.

\bibitem{HaJa20}
S.~Han, S.~Jang, H.~Lee, S.~Pack, {Switch-Centric Byzantine Fault Tolerance
  Mechanism in Distributed Software Defined Networks}, {IEEE} Communications
  Letters 24 (2020) 2236--2239.

\bibitem{p4-sc-bft-repo}
{GitHub: SC-BFT}, \url{https://github.com/MNC-KOR/SC-BFT}, {accessed
  01-20-2021} (2021).

\bibitem{SvBo18}
G.~Sviridov, M.~Bonola, A.~Tulumello, P.~Giaccone, A.~Bianco, G.~Bianchi,
  {LODGE: LOcal Decisions on Global statEs in Progrananaable Data Planes}, in:
  {IEEE Conference on Network Softwarization (NetSoft)}, 2018, pp. 257--261.

\bibitem{SvBo20}
G.~Sviridov, M.~Bonola, A.~Tulumello, P.~Giaccone, A.~Bianco, G.~Bianchi,
  {LOcAl DEcisions on Replicated States (LOADER) in programmable dataplanes:
  Programming abstraction and experimental evaluation}, Computer Networks 181
  (2020) 107637.

\bibitem{loader-repo}
{GitHub: LOADER}, \url{https://github.com/german-sv/loader}, {accessed
  01-20-2021} (2021).

\bibitem{TaKe20}
H.~Takruri, I.~Kettaneh, A.~Alquraan, S.~Al-Kiswany, {FLAIR: Accelerating Reads
  with Consistency-Aware Network Routing}, in: {USENIX Symposium on Networked
  Systems Design \& Implementation (NSDI)}, 2020, pp. 723--737.

\bibitem{LuYu17}
S.~Luo, H.~Yu, L.~Vanbever, {Swing State: Consistent Updates for Stateful and
  Programmable Data Planes}, in: {ACM Symposium on SDN Research (SOSR)}, 2017,
  p. 115–121.

\bibitem{XiCh20}
J.~Xing, A.~Chen, T.~E. Ng, {Secure State Migration in the Data Plane}, in:
  {Workshop on Secure Programmable Network Infrastructure (SPIN)}, 2020, p.
  28–34.

\bibitem{p4sync-repo}
{GitHub: P4Sync}, \url{https://github.com/jiarong0907/P4Sync}, {accessed
  01-20-2021} (2021).

\bibitem{XuZh20}
Y.~Xue, Z.~Zhu, {Hybrid Flow Table Installation: Optimizing Remote Placements
  of Flow Tables on Servers to Enhance PDP Switches for In-Network Computing},
  {{IEEE} Transactions on Network and Service Management (TNSM)} (2020)
  429--440.

\bibitem{KuNe20}
C.~Kuzniar, M.~Neves, I.~Haque, {POSTER: Accelerating Encrypted Data Stores
  Using Programmable Switches}, in: {IEEE International Conference on Network
  Protocols (ICNP)}, 2020, pp. 1--2.

\bibitem{SaSi20b}
G.~C. {Sankaran}, K.~M. {Sivalingam}, {Collaborative Packet Header Parsing in
  NetFPGA-Based High Speed Switches}, {IEEE Networking Letters} 2 (2020)
  124--127.

\bibitem{WoRa19}
J.~Woodruff, M.~Ramanujam, N.~Zilberman, {P4DNS: In-Network DNS}, in: {P4
  Workshop in Europe (EuroP4)}, 2019, pp. 1--6.

\bibitem{p4-dns}
{GitHub: P4DNS}, \url{https://github.com/cucl-srg/P4DNS}, {accessed 01-20-2021}
  (2021).

\bibitem{KuNo19}
R.~Kundel, L.~Nobach, J.~Blendin, H.-J. Kolbe, G.~Schyguda, V.~Gurevich,
  B.~Koldehofe, R.~Steinmetz, {P4-BNG: Central Office Network Functions on
  Programmable Packet Pipelines}, in: {International Conference on Network and
  Services Management (CNSM)}, 2019, pp. 1--9.

\bibitem{p4-p4se-repo}
{GitHub: p4se}, \url{https://github.com/opencord/p4se}, {accessed 01-20-2021}
  (2021).

\bibitem{MaAl18}
I.~Martinez-Yelmo, J.~Alvarez-Horcajo, M.~Briso-Montiano, D.~Lopez-Pajares,
  E.~Rojas, {ARP-P4: A Hybrid ARP-Path/P4Runtime Switch}, in: {IEEE
  International Conference on Network Protocols (ICNP)}, 2018, pp. 438--439.

\bibitem{GlKr19}
R.~Glebke, J.~Krude, I.~Kunze, J.~Rüth, F.~Senger, K.~Wehrle, {Towards
  Executing Computer Vision Functionality on Programmable Network Devices}, in:
  {ACM CoNEXT Workshop on Emerging In-Network Computing Paradigms}, 2019, p.
  15–20.

\bibitem{XiQi19b}
J.~Xie, C.~Qian, D.~Guo, M.~Wang, S.~Shi, H.~Chen, {Efficient Indexing
  Mechanism for Unstructured Data Sharing Systems in Edge Computing}, in: {IEEE
  International Conference on Computer Communications (INFOCOM)}, 2019, pp.
  820--828.

\bibitem{LuLi19}
Y.-S. Lu, K.~C.-J. Lin, {Enabling Inference Inside Software Switches}, in:
  {Asia-Pacific Network Operations and Management Symposium (APNOMS)}, 2020,
  pp. 1--4.

\bibitem{YaPa19}
A.~Yazdinejad, R.~M. Parizi, A.~Dehghantanha, K.-K.~R. Choo, {P4-to-blockchain:
  A secure blockchain-enabled packet parser for software deﬁned networking},
  {Computers \& Security Journal} 88 (2019) 101629.

\bibitem{OsTa20}
T.~Osiński, H.~Tarasiuk, P.~Chaignon, M.~Kossakowski, {P4rt-OVS: Programming
  Protocol-Independent,Runtime Extensions for Open vSwitch with P4}, in:
  {IFIP-TC6 Networking Conference (Networking)}, 2020, pp. 413--421.

\bibitem{p4-P4rt-OVS-repo}
{GitHub: P4rt-OVS}, \url{https://github.com/Orange-OpenSource/p4rt-ovs},
  {accessed 01-20-2021} (2021).

\bibitem{LiYe03}
S.~R. Li, R.~W. Yeung, N.~Cai, {Linear Network Coding}, IEEE Transactions on
  Information Theory 49 (2003) 371–381.

\bibitem{access40}
{Deutsche Telekom AG: Deutsche Telekom’s Access 4.0 platform goes live},
  \url{https://www.telekom.com/en/media/media-information/archive/deutsche-telekom-s-access-4-0-platform-goes-live-615974},
  {accessed 05-17-2021} (2021).

\bibitem{oran}
{O-RAN Alliance}, \url{https://www.o-ran.org/}, {accessed 05-17-2021} (2021).

\end{thebibliography}
